\documentclass[3p,11pt]{elsarticle}
\usepackage{amsfonts,amsmath,amssymb}
\usepackage{graphicx}
\usepackage{xcolor}
\usepackage{color}
\usepackage{xspace}
\usepackage{isotope}
\usepackage{float}
\usepackage{placeins} % used to keep floats in section
\usepackage{hyperref}
\usepackage{verbatim}{}{}
\usepackage{comment} % Wick used to comment out during editing
\usepackage{soul} % used for strikethrough \st def
\hypersetup{
   colorlinks=true, % false: boxed links; true: colored links
   linkcolor=blue,  % color of internal links
   citecolor=blue,  % color of links to bibliography
   filecolor=blue,  % color of file links
   urlcolor=blue    % color of external links
}
\usepackage{enumitem}
\usepackage{wrapfig}

\allowdisplaybreaks
\newcommand{\kf}{k_{\rm F}\xspace}

\newcommand{\MeV}{\,\mathrm{MeV}\xspace}
\newcommand{\keV}{\,\mathrm{keV}\xspace}

\newcommand{\fm}{\,\mathrm{fm}\xspace}

\newcommand{\fmiq}{\,\mathrm{fm}^{-3}\xspace}

\newcommand{\apriori}{\mbox{\emph{a~priori}}\xspace}
\newcommand{\abinitio}{\mbox{\emph{ab~initio}}\xspace}
\newcommand{\Abinitio}{\mbox{\emph{Ab~initio}}\xspace}

\newcommand{\etal}{\textit{et~al.}\xspace}

\newcommand{\Msol}{\,M_{\odot}\xspace}
\newcommand{\km}{\,\text{km}}

\newcommand\beq{\begin{eqnarray}}
\newcommand\eeq{\end{eqnarray}}

\newcommand*{\ket}[1]{\left|{#1}\right\rangle}
\newcommand*{\bra}[1]{\left\langle {#1} \right|}

\newcommand*{\braket}[2]{\left\langle  \left. {#1} \right| {#2} \right\rangle}
\newcommand*{\braopket}[3]{\left\langle {#1} \left| {#2} \right| {#3} \right\rangle}
\newcommand*{\ketbra}[2]{\ket{#1} \! \bra{#2}}

\newcommand{\eqnref}[1]{Eq.~\eqref{#1}}
\newcommand{\Eqnref}[1]{Equation~\eqref{#1}}
\newcommand{\figref}[1]{Fig.~\ref{#1}}
\newcommand{\Figref}[1]{Figure~\ref{#1}}
\newcommand{\secref}[1]{Sec.~\ref{#1}}
\newcommand{\tabref}[1]{Table~\ref{#1}}

\newcommand{\nxlo}[1]{N${}^{#1}$LO}
\newcommand*{\xSM}{\textrm{SM}}

\def\nd{\Delta\hskip-0.55em /}
\newcommand{\Dslash}[1]{D\hskip-#1em\slash}

\def\a{\alpha}
\def\b{\beta}
\def\d{\delta}
\def\e{\epsilon}
\def\g{\gamma}
\def\s{\sigma}

\def\D{\Delta}
\def\L{\Lambda}
\def\S{\Sigma}
\def\O{\Omega}

\def\eft{EFT/ET\xspace}
\def\efts{EFT/ETs\xspace}
\def\luscher{L{\"u}scher\xspace}
\def\znbb{0\nu\b\b\xspace}
\def\xpt{{$\chi$PT}\xspace}
\def\hbxpt{{HB$\chi$PT}\xspace}

\newcommand{\frib}{
	Facility for Rare Isotope Beams,
	Michigan State University,
	MI 48824, United States
}
\newcommand{\lanl}{
	Theoretical Division,
	Los Alamos National Laboratory,
	Los Alamos, NM 87545, United States
	}
\newcommand{\lblnsd}{
    Nuclear Science Division,
    Lawrence Berkeley National Laboratory,
	Berkeley, CA 94720, United States
	}
\newcommand{\llnl}{
	Physical and Life Sciences,
	Lawrence Livermore National Laboratory,
	Livermore, CA 94550, United States
	}
\newcommand{\ucb}{
	Department of Physics,
	University of California,
	Berkeley, CA 94720, United States
	}
\newcommand{\unc}{
	Department of Physics and Astronomy,
	University of North Carolina,
	Chapel Hill, NC 27516-3255, United States
	}

\setlist{leftmargin=*}

\def\new#1{{\color{blue}#1}}

\begin{document}
\title{Towards grounding nuclear physics in QCD}
%\thanks{Contribution to \textit{The tower of effective (field) theories and the emergence of nuclear phenomena}}

\author[1,2,3]{Christian~Drischler}
%\email{cdrischler@berkeley.edu}
\address[1]{\frib}
\address[2]{\ucb}
\address[3]{\lblnsd}

\author[2,3]{Wick~Haxton}
%\email{haxton@berkeley.edu}
%\affiliation{\ucb}
%\affiliation{\lblnsd}

\author[2,3]{Kenneth~McElvain}
%\email{kenmcelvain@berkeley.edu}
%\affiliation{\ucb}
%\affiliation{\lblnsd}

\author[4]{Emanuele~Mereghetti}
%\email{emereghetti@lanl.gov}
\address[4]{\lanl}

\author[5]{Amy~Nicholson}
%\email{annichol@email.unc.edu}
\address[5]{\unc}

\author[6,3]{Pavlos~Vranas}
%\email{vranas2@llnl.gov}
\address[6]{\llnl}
%\affiliation{\lblnsd}

\author[3,2,6]{Andr\'{e}~Walker-Loud}
%\email{walkloud@lbl.gov}
%\affiliation{\lblnsd}
%\affiliation{\ucb}
%\affiliation{\llnl}

%\date{Received: \today}% / Revised version: date}

%\preprint{LLNL-JRNL-786701}

\begin{abstract}
Exascale computing could soon enable a predictive theory of nuclear structure and reactions rooted in the Standard Model, with quantifiable and systematically improvable uncertainties. Such a predictive theory will help exploit experiments that use nucleons and nuclei as laboratories for testing the Standard Model and its limitations. Examples include direct dark matter detection, neutrinoless double beta decay, and searches for permanent electric dipole moments of the neutron and atoms. It will also help connect QCD to the properties of cold neutron stars and hot supernova cores. We discuss how a quantitative bridge between QCD and the properties of nuclei and nuclear matter will require a synthesis of lattice QCD (especially as applied to two- and three-nucleon interactions), effective field theory, and ab initio methods for solving the nuclear many-body problem. While there are significant challenges that must be addressed in developing this triad of theoretical tools, the rapid advance of computing is accelerating progress. In particular, we focus this review on the anticipated advances from lattice QCD and how these advances will impact few-body effective theories of nuclear physics by providing critical input, such as constraints on unknown low-energy constants of the effective (field) theories. We also review particular challenges that must be overcome for the successful application of lattice QCD for low-energy nuclear physics. We describe progress in developing few-body effective (field) theories of nuclear physics, with an emphasis on HOBET, a non-relativistic effective theory of nuclear physics, which is less common in the literature. We use the examples of neutrinoless double beta decay and the nuclear-matter equation of state to illustrate how the coupling of lattice QCD to effective theory might impact our understanding of symmetries and exotic astrophysical environments.
\end{abstract} %end of abstract
\maketitle

% TABLE OF CONTENTS
\setcounter{tocdepth}{4}
\setcounter{secnumdepth}{4}
\tableofcontents

%\newpage

%%% INTRO
\section{Motivation \label{sec:motivation}}
A number of high-profile, high-impact nuclear and particle physics experiments
are planned over the  next decade that will test the limits of the enormously
successful Standard Model (SM) of particle physics, probing for subtle
violations of symmetries at low energies or new particles and interactions at
the energy frontier. We are quite certain that the SM, with its many seemingly
arbitrary parameters, is incomplete, and hope the discovery of new physics will
provide the hints needed to deduce what lies beyond the SM. A more fundamental
theory could allow us to address some of the most important open questions in
physics:
\begin{itemize}
\item How did the Big Bang generate the matter-antimatter asymmetry, avoiding the fate of a universe filled only by radiation?
\item What is dark matter, and does it interact with visible matter other than gravitationally?
\item What is the origin of the neutrino mass, why are neutrinos so much lighter than other fermions, why are their mixing angles larger?
\item Are neutrinos their own  antiparticles, and is there a connection between neutrino properties and the universe's matter-antimatter asymmetry?
\item What is dark energy, and why do dark matter and dark energy play roughly comparable roles in the evolution of our universe?
\end{itemize}

Some of the experiments designed to explore these questions are prioritized in
the 2015 NSAC Long Range Plan for Nuclear Science~\cite{Geesaman:2015fha} and the High
Energy Physics P5 report~\cite{P5}.  They include large-scale, ultra-clean
detectors to search for the elastic scattering of heavy dark matter (DM)
particles off nuclei, employing targets such as Xe, Ar, Ge, and others
\cite{Mount:2017qzi,Akerib:2018lyp,Aprile:2018dbl,Cui:2017nnn,Liu:2017drf,Ajaj:2019imk,Agnes:2018fwg,Agnese:2018col,Baudis:2018bvr}, as well as
new technologies sensitive to lighter particles, should DM particle masses not
be associated with the weak scale \cite{Du:2018uak,Battaglieri:2017aum,Graham:2017ivz}.
They include neutrinoless double beta decay searches for lepton number violation and
Majorana masses, with ton-scale experiments use of elements such as Xe, Ge and Te planned for deep underground sites at the Gran Sasso National Laboratory,
SNOLAB, China Jinping Underground Laboratory and  Kamioka Observatory~\cite{Albert:2017hjq,Martin-Albo:2015rhw,KamLAND-Zen:2016pfg,Shirai:2017jyz,Aalseth:2017btx,Abgrall:2017syy,
Agostini:2013mzu,Albert:2014awa,Alfonso:2015wka,Andringa:2015tza,Alduino:2017ehq,Agostini:2017iyd,Aalseth:2017btx,Albert:2017owj,Azzolini:2018dyb,Agostini:2018tnm}.
The search for new sources of CP
violation among first-generation quarks are motivating new efforts to measure
electric dipole moments (EDM) of the neutron and of atomic nuclei, including
$^{199}$Hg~\cite{Graner:2016ses}, $^{225}$Ra~\cite{Bishof:2016uqx}, and potentially $^{229}$Pa:  the latter two are
long-lived unstable isotopes where the effects  of CP violation are expected to
be greatly enhanced due to nuclear level degeneracies and collectivity.   The
Facility for Rare Isotope Beams (FRIB) at Michigan State University (MSU) will
produce these isotopes at rates sufficient to allow high-statistics EDM
measurements in traps. Construction is beginning on the Deep Underground
Neutrino Experiment (DUNE),  a multi-kiloton liquid Ar detector that will record
neutrinos produced in Fermilab's Long-Baseline Neutrino Facility (LBNF), as they
arrive at the Sanford Underground Research Facility after traveling through 1300
kilometers of rock~\cite{Acciarri:2015uup,Acciarri:2016crz}.  Together with other long-baseline
experiments such as J-PARC's T2K, the results will determine the neutrino mass
hierarchy and the value of one of three neutrino CP phases.
Japan's megaton-scale water Cherenkov detector Hyper-Kamiokande will be used in
the J-PARC neutrino program, in searches for proton decay of unprecedented
sensitivity, and as an observatory for astrophysical neutrinos from a variety of
sources.   FermiLab and J-PARC are also planning extraordinarily sensitive tests
of flavor violation, probing $\mu \rightarrow e$ conversion in the nuclear field
\cite{Kuno:2013mha,Bartoszek:2014mya,Baldini:2018uhj}.

These are just a subset of the efforts to find new physics in the next decade.
Common to almost all of these experiments is the use of complex nuclei as
``laboratories" for probing beyond-the-Standard-Model (BSM) phenomena.  This
connects fundamental physics to one of the main subjects of this paper:
developing effective (field) theories (\efts{}) of nuclear physics, rooted in the SM,
that will make the nucleus into a more quantitative
laboratory for BSM searches.

Other exciting laboratories can be found in astrophysics; for instance, core-collapse supernovae or binary neutron-star mergers
produce matter at extremes of temperature, density, and isospin asymmetry.
As these conditions are difficult
to probe in terrestrial experiments~\cite{Danielewicz:2002pu} (see also Sec.~\ref{sec:intro_fs}), the emergence of multi-messenger astrophysics
is placing new demands on theory to provide a more predictive nuclear-matter equation of state (EOS).
Progress is
needed if we are to answer long-standing questions such as
\emph{What is the origin of elements heavier than iron?} Multi-messenger astronomy
encompasses electromagnetic, cosmic-ray, neutrino, and gravitational-wave (GW) detection.
KAGRA and other next-generation gravitational wave observatories will improve the sensitivity and extend
the frequency range of observations.  In addition, more will be learned about the maximum mass and
mass/radius relationship of neutron stars.  The need for a quantitative nuclear-matter EOS grounded
in first-principles QCD has never been greater.
At very high densities, $n \gtrsim 60 \,
n_0$, relative to
nuclear saturation density $n_0 \sim 0.16\fmiq$ ($\rho_0 \sim 2.7 \times 10^{14}$~g\,cm$^{-3}$),
perturbative QCD
(pQCD)~\cite{Freedman:1976xs,Freedman:1976ub,Baluni:1977ms,Kurkela:2009gj,Fraga:2013qra} has already been used to guide the construction of the EOS over the range of densities relevant for neutron stars~\cite{Fraga:2015xha,Annala:2017llu}.

Chiral effective field theory (EFT) has become the
standard approach for deriving nuclear (many-body) forces that respect the
symmetries of low-energy QCD.  A variety of many-body frameworks have been employed
in conjunction with chiral EFT
to generate low-density EOSs valid up to $\sim (1-2)\, n_0$.
Parameterizations of the EOS or the speed
of sound have been used to extrapolate the low-density EOS systematically to
intermediate densities relevant for neutron stars, $n \sim (2 - 10) \, n_0$,
without making assumptions on the composition of nuclear matter at these
densities (see, e.g., Ref.~\cite{Greif:2018njt}). The aforementioned stellar observations provide important constraints
on the extrapolated EOS; for instance, neutron-star radii are most sensitive to
the EOS at $\sim 2\, n_0$~\cite{Lattimer:2012xj}. To improve these extrapolations, constraints on the
EOS at nuclear densities must be tightened; for instance, by reducing quantified truncation errors in
the EFT expansion.

Furthermore, three-nucleon forces generally play an important role for nuclear
phenomena such as driplines and saturation (see Ref.~\cite{Hebeler:2015hla} for a review). Better constraints on the isospin
$\mathcal{T} = 3/2$ channel (e.g, three interacting neutrons) are particularly important
for neutron-rich matter.  Lattice QCD (LQCD)
calculations of three-neutron interactions can thus help to pin down the EOS at nuclear densities (see Sec.~\ref{sec:intro_fs}).
Alternatively, the functional renormalization group (FRG)~\cite{Wetterich:1992yh} is a
very promising non-perturbative method for computing the nuclear-matter EOS
at intermediate
densities directly from quark-gluon degrees of
freedom~\cite{Leonhardt:2019fua}. Clustering of quarks
into nucleons as well as the emergence of long-range correlations between
nucleons make the QCD-based FRG approach
increasingly difficult towards lower densities; especially,
since the average inter-nucleon distance is still larger than $1\fm$ at
$3\, n_0$.
We will discuss the details of such calculations in
Sec.~\ref{sec:neos}. For a review of nuclear-matter
calculations using FRG methods applied to chiral Lagrangians we refer the reader
to Ref.~\cite{Drews:2016wpi}.

The fundamental theory of nuclear physics is QCD, a well-accepted
cornerstone of the SM.  However, the nucleons, which make up atomic nuclei, are themselves
composite states of fundamental particles -- massless gluons and nearly massless quarks. The
deceptively simple equations of QCD describing the interactions between quarks and gluons give
rise to the strongly-coupled, non-perturbative strong interactions. At low temperatures and
densities, the quarks and gluons are confined into colorless, composite states of strongly
interacting matter, the hadrons. For example, the nucleons are composite states with the
quantum numbers of three quarks and a mass of $\sim 1$ GeV (see Ref.~\cite{Yang:2018nqn} for
the most comprehensive breakdown of the nucleon mass from QCD).  The emergence of nucleons,
whose masses are 95\% due to interactions, is a remarkable feature of QCD.

After confinement, there is a relatively small residual interaction binding protons and neutrons
into atomic nuclei with a typical binding energy per nucleon of $\sim8$~MeV.
While this energy
scale is two orders of magnitude smaller than the confinement scale, it is still very strong
compared to the other known forces. In conjunction with the other two interactions of the SM,
electromagnetism as well as the weak force, the entirety of the rich field of nuclear physics
\emph{emerges} from QCD: from the forces binding protons and neutrons into
the nuclear landscape, to
the fusion and fission reactions between nuclei, to the prospective interactions of nuclei with BSM
physics, and to the unknown state of matter at the cores of neutron stars.
\emph{How does this emergence take place exactly? How is the clustering of
quarks into nucleons and alpha particles realized? What are the mechanisms
behind collective phenomena in nuclei as strongly correlated many-body
systems? How does the extreme fine-tuning required to
reproduce nuclear binding energies proceed?} --~are big open
questions in nuclear physics. To give answers joint efforts of LQCD along with
\efts{} are required.

Exascale computing offers us the opportunity to develop a predictive theory of nuclear structure and reactions, rooted in the SM, with quantifiable theoretical uncertainties.
This is the focus of this review, with an emphasis placed on the connections between QCD and
\efts{} of nuclear physics.

The foundational layer of this theoretical framework is LQCD, a
formulation of QCD in which spacetime is discretized and truncated while
respecting gauge-invariance~\cite{Wilson:1974sk}, leading to a finite-dimensional path integral,
which may then be sampled numerically using Monte
Carlo methods, see Sec.~\ref{sec:lattice} for details. LQCD is the only non-perturbative regulator of QCD low-density regime in which all sources of systematic uncertainty can be
quantified, controlled, and improved. Therefore, LQCD offers the promise of
quantitatively understanding the emergence of nuclear physics from QCD and the
SM. However, because LQCD is formulated in terms of quark and gluon degrees of
freedom in Euclidean spacetime, our ability to determine properties of complex
systems directly from QCD is very limited. This sets the stage for \efts{}
constrained by LQCD.

In the exascale era, we can hope to compute the structure, spectrum, and
reactions of nucleons and light nuclei directly from the SM. In order to connect
to more complex systems and processes, we must couple the results from LQCD
calculations to EFTs and effective theories (ETs) of nuclear physics.
The concepts and rules behind EFTs and ET are very similar with the main difference that EFTs relate to quantum field theories
and ETs to quantum mechanics.  \efts{} form the basis of our modern understanding of nuclear physics and we will discuss them in some detail in Sec.~\ref{sec:net}.
The most popular realizations of \efts{}
are formulated in terms of nucleon (and pion and delta) degrees of freedom, whose
interactions are described by a series of operators of increasing irrelevancy based on the symmetries of
low-energy QCD and dictated by a power counting scheme. In principle,
\efts{} provide a complete, model-independent description of the underlying
physics over the range of energies for which they are valid. At each order in
the \eft{} expansion, the theory depends on a finite number of unknown couplings,
called low-energy constants (LECs), that encode the effects of unresolved short-distance physics.
Once these LECs are determined by experimental data of one
set of observables, the theory can be used to make predictions for other
observables. The advantage of the systematic expansion underlying \eft{} approaches is their
capability to provide theoretical uncertainties by estimating contributions from neglected higher orders
(truncation errors). In addition, uncertainties in the (experimental or lattice) data constraining the LECs
as well as approximations in the methods solving the nuclear many-body problem contribute to the
overall uncertainty budget of any predicted observable.

In nuclear physics, the LECs are commonly extracted from two- and few-body
observables by matching computed quantities to measured observables such as
binding energies or charge radii.  In order to root nuclear physics in QCD, it
is desirable to determine these LECs by requiring that
observables extracted from correlation functions
computed in the \efts{} and from QCD agree, at a given resolution scale and at a given order in the
\eft{}
expansion, much in the same way the coefficient of the four-fermion operators in
Fermi's theory of weak interactions can be determined by matching to the full
electro-weak SM, for example by matching the muon-decay amplitude in full SM and with the $W$ and $Z$ bosons integrated out, to determine Fermi's constant, $G_F$. The non-perturbative nature of QCD at low energies has so far
prevented this quantitative matching, but the growth in computational power and
algorithmic advances has brought us to the era in which it is beginning in
earnest. In practice, the best determination of the LECs will happen through a
combination of matching to LQCD calculations and also applying constraints from
physical processes, where LQCD in particular can help pin down LECs that are
very difficult to impossible to access from experimental measurements, such as
few-neutron or hyperonic systems as well as the LECs that accompany quark-mass dependent operators.

In addition to determining the LECs, by varying the quark masses in LQCD calculations, we can map out the convergence pattern of the \efts{} and determine their range of validity.
As an example, we will discuss constraints on chiral perturbation theory in the nucleon sector from
LQCD results in Sec.~\ref{sec:hbchipt}.

Another key aspect of nuclear \efts{} that necessitates LQCD input is the determination of matrix elements of BSM operators, for which there is no experimental information available. LQCD can be utilized to compute the hadronic and nuclear matrix elements of such operators, and thus make controlled predictions of the SM background and the possible BSM signal in fundamental symmetry tests of the SM.

In the next subsection, we describe in more detail the impact LQCD can have
for fundamental-symmetry tests of the SM and for observables which are extremely difficult or
inaccessible experimentally. We then provide a brief introduction to LQCD in
Sec.~\ref{sec:lattice}, emphasizing particular challenges in applying LQCD to nuclear physics,
reviewing the state of the field, and highlighting the most important challenges that must be
overcome. In Sec.~\ref{sec:net}, we introduce \efts of nuclear physics, chiral
EFT and Harmonic Oscillator Based Effective Theory (HOBET) and then discuss the particular
input from LQCD to these \efts{} that will enable the most progress. In Secs.~\ref{sec:0nubb} and
\ref{sec:neos}, we use neutrinoless double beta decay ($\znbb$) and the nuclear-matter EOS as specific
examples to elaborate upon the connection between LQCD and \efts{} of
nuclear physics. We then provide an outlook in Sec.~\ref{sec:outlook}

%%%%%%%%%%%%%%%%%%%%%%%%%%%%%%%%%%%%%%%%%%%
% Fundamental Symmetries and LQCD
\subsection{Fundamental Symmetry Tests and Experimentally difficult/inaccessible observables \label{sec:intro_fs}}

Low-energy tests of fundamental symmetries, such as searches for the violation
of CP, lepton number or baryon number, provide an important window on
physics beyond the SM, as they are competitive and complementary to
high-energy collider experiments, see for example Refs.~\cite{Bhattacharya:2011qm,Engel:2013lsa}.
In order to interpret bounds and, in the future, signals in low-energy precision experiments, to disentangle
different BSM models, and to quantitatively compare the reach of experiments at
the intensity and energy frontiers, it is necessary to develop a smooth
connection between physics at the microscopic level, often described in terms of
quark- and gluon-level effective operators, and theoretical predictions obtained
with hadronic and nuclear degrees of freedom. The last few years have witnessed
a growing awareness in the community of the need to replace estimates from
models of the strong interactions (e.g., large $N_c$ techniques, QCD sum rules,
naive dimensional analysis estimates) with controlled, first principle
calculations of mesonic, one- and few-nucleons observables, which can be input for nuclear \efts{}.

LQCD calculations in the mesonic sector are already mature and provide precise
input for SM and BSM contributions to, for example, meson weak decays \cite{Aoki:2019cca}, meson-
antimeson oscillations \cite{Aoki:2019cca}, and CPV in kaon decays \cite{Abbott:2020hxn}. LQCD calculations of hadronic
corrections to the muon anomalous magnetic moment are starting to compete
with phenomenological extractions, and promise a roadmap towards controlled
theoretical uncertainties, see the recent comprehensive theory review~\cite{Aoyama:2020ynm}.

The nuclear
calculations relevant for low-energy symmetry tests typically include the
evaluation of the matrix element of a weak operator between strongly interacting
states (nucleons or nuclei). The required LQCD input depends on the form of the
weak operator.
The simplest cases involve observables that are dominated by one-body operators.
For example, dark matter-nucleus scattering cross sections and non-standard corrections to
nuclear $\beta$
decays are mediated by the coupling of weakly interacting particles to quark bilinears that, below the hadronization
scale, translate in couplings to a single nucleon, possibly via pionic
contributions
\cite{Goodman:2010ku,Cirigliano:2013xha,Cirigliano:2013zta,Hoferichter:2015ipa,Baroni:2015uza,Krebs:2016rqz}.
In these cases, LQCD can provide nucleon charges and form factors. For the
vector and axial currents, which are induced by SM interactions, the comparison
between LQCD and experimental extractions of the form factors translates into
tests of possible BSM contributions, becoming more and more stringent as the accuracy of LQCD calculations increases \cite{Chang:2018uxx}.
LQCD calculations are even more important for currents that do not appear in the SM,
or are not easily extracted from data. Notable examples are the nucleon scalar
and tensor charges, the strange and charm content of the nucleon, and the
nucleon axial form factor.  We will discuss some of these calculations in Sec. \ref{sec:lattice}.
Two-body contributions to the vector, axial, scalar,
pseudoscalar and tensor currents are expected to be subleading. These operators
can be systematically constructed in chiral EFT
\cite{Cirigliano:2013zta,Hoferichter:2015ipa,Baroni:2015uza,Krebs:2016rqz} but
often involve LECs, in particular those corresponding to BSM currents, that cannot be experimentally
determined. The first LQCD evaluations of two-nucleon matrix elements have
appeared  \cite{Savage:2016kon,Chang:2017eiq}, allowing a cross-check of the size of
two-body effects, and to determine unknown LECs, albeit for heavy pion masses and the caveats discussed in Sec.~\ref{sec:nn_controversy}.

Baryon-number violating processes such as proton decay and neutron-antineutron oscillations
are also dominated by matrix elements between one incoming and one outgoing hadron. In these cases,
the processes are mediated by three- and six-quark operators, respectively, with the LQCD calculations under good control~\cite{Aoki:2017puj,Rinaldi:2018osy}.

Theoretical predictions of the EDM of the nucleon,
light nuclei, and diamagnetic atoms involve more complicated matrix elements.
While the nucleon EDM induced by the EDMs of the light quarks is related to the
well determined nucleon tensor charges \cite{Yamanaka:2018uud,Gupta:2018qil,Ottnad:2018fri,Aoki:2019cca}, in the
case of CP violation
from the QCD $\bar\theta$ term and from purely hadronic dimension-six operators
in the SM-EFT Lagrangian (e.g., the quark and gluon chromo-EDMs) one has to
evaluate a four-point function, with the insertion of the CP-violating source
and of the electromagnetic current on the nucleon. These calculations are very challenging, especially for the QCD $\bar\theta$ term, and
only recently some promising results have started to appear~\cite{Abramczyk:2017oxr,Bhattacharya:2018qat,Kim:2018rce,Dragos:2019oxn,Syritsyn:2019vvt}.
The intricate mixing pattern of BSM operators~\cite{Bhattacharya:2015rsa,Constantinou:2015ela,Rizik:2018lrz} further
complicates the extraction of the nucleon EDM, even as the application of
gradient flow techniques to this field has allowed for significant progress~\cite{Rizik:2018lrz,Kim:2018rce,Dragos:2019oxn}.
The calculation of EDMs of light nuclei and of the Schiff operator, relevant for atomic EDMs, require in addition the estimate of time-reversal-violating (TV) non-derivative
pion-nucleon couplings.
As shown in Refs.~\cite{Crewther:1979pi,Mereghetti:2010tp,deVries:2015una,Seng:2016pfd,deVries:2016jox,Cirigliano:2016yhc},
for all chiral-breaking TV operators these can be related to corrections to the meson and baryon spectrum, reducing the problem of TV pion-nucleon couplings to the evaluation of generalized pion and nucleon sigma terms, which should be tractable on the lattice~\cite{deVries:2016jox}.

Finally, predictions of $\znbb$ half-lives and hadronic
parity-violating (PV) observables involve the evaluation of two-body matrix
elements on hadronic/nuclear states. The form of the two-body operators can be
derived using nuclear EFTs and will in general include long-range, pion-range
and short-range contributions. In the case of light-Majorana neutrino exchange
(the so called ``standard mechanism''), the long-range piece of the
$0\nu\beta\beta$ transition operator is determined
in terms of the nucleon axial and vector form factors~\cite{Haxton:1985am,Simkovic:1999re}, while non-standard long-range mechanisms
involve the nucleon scalar, pseudoscalar and tensor form factors
\cite{Pas:1999fc,Cirigliano:2017djv,Cirigliano:2018yza}. The nucleon-nucleon PV
potential and corrections to the $0\nu\beta\beta$ transition operator from
dimension-nine quark-level operators have pion-range contributions~\cite{Desplanques:1979hn,Zhu:2004vw,Prezeau:2003xn}, demanding the
evaluation of new pion-pion and pion-nucleon matrix elements~\cite{Wasem:2011zz,Nicholson:2018mwc}.
Finally, the PV potential and the $0\nu\beta\beta$ transition operators from standard and non-standard mechanisms also have short-range pieces, which can be determined from LQCD calculations of nucleon-nucleon scattering amplitudes.
New insights using large $N_c$ indicate the most important matrix element to be determined for
understanding the observed strength of various PV amplitudes is the isospin-2 matrix
element~\cite{Gardner:2017xyl}, which is the easiest one to tackle with LQCD~\cite{Kurth:2015cvl}.

This brief and incomplete survey of fundamental symmetries tests highlights
how LQCD will play a more and more important role in connecting models of BSM
physics to nuclear observables, a connection that is necessary for a
quantitative understanding of the form of physics beyond the SM.
Another important role for LQCD is to help constrain components of \efts{} that are extremely
challenging
or impossible to isolate experimentally, such as those relevant to neutron stars.

%%%%%%%%%%%%%%%%%%%%%%%%%%%%%%%%%%%%%%%%%%%
% EOS and LQCD

Neutron stars are the densest objects in the observable universe (besides black holes)~\cite{Lattimer:2012nd,Lattimer:2017kni,Lattimer:2004pg,Ozel:2016oaf} and unique
laboratories for studying neutron-rich matter under extreme conditions. These
conditions are difficult (if not impossible) to reproduce in laboratory experiments~\cite{Danielewicz:2002pu}, even at rare
isotope facilities such as RIBF (RIKEN) or FRIB (MSU) and FAIR (GSI).
From the theory side, LQCD can pave the way for QCD-based calculations of the
nuclear-matter EOS as well as finite nuclei~\cite{Briceno:2014tqa,McIlroy:2017ssf}. Solving the full many-body problem with $A \geqslant 4$
directly from LQCD with physical quark masses, nonetheless, cannot be expected
in the foreseeable future despite exascale computing capabilities. Low-energy \efts{} of
QCD, on the other hand, are powerful approaches to derive nuclear potentials and
external currents as well as extrapolate lattice results to the physical regime. The
unknown LECs encoding unresolved short-distance physics can be
constrained by LQCD calculations of two- and few-nucleon systems.
Applying many-body frameworks to these microscopic
potentials combined with Renormalization Group (RG) methods~\cite{Bogner:2009bt,Furnstahl:2012fn} is a promising joint
approach towards truly \abinitio{} nuclear-structure and -reaction studies across
the nuclear chart.

Three-nucleon forces play an important role for nuclear
phenomena, such as driplines along isotopic chains or nuclear
saturation~\cite{Wienholtz:2013nya,Hebeler:2015hla}. The isospin $\mathcal{T} = 3/2$
components (e.g., three interacting neutrons), however, are only weakly constrained by
experimental
(few-body)
data
(see, e.g., Refs.~\cite{Lazauskas:2009gv,Viviani:2011ax,Ekstrom:2015rta,Witala:2016inj}) but are
crucial at the neutron-rich extremes. Particularly, constraints from LQCD
calculations of few-neutron systems, which are difficult to access
experimentally, may provide important insights that can lead to much
improved predictions for key quantities of nuclear physics and astrophysics,
such as the nuclear symmetry energy (especially beyond nuclear
saturation density)~\cite{Tsang:2012se,Hebeler:2015hla}.
A preliminary LQCD calculation of the $\mathcal{T} = 3/2$ system was presented two years ago~\cite{Wynen:2018zbt}, but not yet with enough statistics to determine the interaction energy.
We will discuss these points in more detail in Sec.~\ref{sec:nn_eft}.

The composition of neutron stars at the high densities near the inner core is an
open question (see Refs.~\cite{Lattimer:2012nd,Lattimer:2017kni,Lattimer:2004pg,Ozel:2016oaf,Watts:2016uzu,Alford:2007xm} for reviews).
Degrees of freedom heavier than nucleons may be favored due to the
occurring high chemical potentials, such as strange matter in the form of hyperons (Y).
However, scattering with hyperons is
experimentally very difficult to realize. The limited amount of data from
hyperon-nucleon (YN), yet alone hyperon-hyperon (YY), scattering makes the
development of hypernuclear potentials challenging. Dominant constraints, in
fact, come from binding and excitation energies of hypernuclei. Such potentials
have been derived from chiral
EFT~\cite{Polinder:2006zh,Polinder:2007mp,Polindera:2007uqg,Haidenbauer:2013oca, Haidenbauer:2015zqb,Petschauer:2015elq,Meissner:2016ood,Haidenbauer:2017sws,Haidenbauer:2019boi}) and applied to
hypernuclei as well as
matter~\cite{Nogga:2013pwa,Petschauer:2016pbn,Haidenbauer:2014uua,Petschauer:2015nea,
Haidenbauer:2016vfq,Wirth:2016iwn,Wirth:2017lso,Wirth:2014apa,Wirth:2017bpw,Wirth:2019cpp}
(for studies with phenomenological potentials see also
Refs.~\cite{Djapo:2008au,Lonardoni:2013gta,Lonardoni:2012rn,Lonardoni:2013rm,Lonardoni:2014hia,Lonardoni:2014bwa,Lonardoni:2017uuu,Gandolfi:2019dxa}).
However, these chiral EFT derivations rely on a perturbative expansion about the $SU(3)$ chiral
limit, and as we will discuss in Sec.~\ref{sec:hbchipt}, LQCD calculations have shown
that $SU(3)$ baryon chiral perturbation theory is not a converging expansion, even at
next-to-leading order. These chiral EFT potentials are therefore likely qualitative at best.

Obtaining more constraints on
hypernuclear potentials is essential for solving the \emph{hyperon puzzle} (see,
e.g., Refs.~\cite{Nishizaki:2001in,Nishizaki:2002ih,Hell:2014xva,Bombaci:2016xzl,Watts:2016uzu}): the presence of
hyperons in neutron stars, although probable given the high central densities,
causes a softening of the EOS that seems to be incompatible with observations of
$\sim 2\Msol$ neutron stars~\cite{Demorest:2010bx,Antoniadis:2013pzd,Fonseca:2016tux,Cromartie:2019kug}. Lattice results for baryon systems
with at least one strange quark, such as YN phase shifts, may help to
overcome these experimental limitations directly from
QCD~\cite{Beane:2012ey,Sasaki:2013zwa}.
Such observables are computationally less expensive compared to pure nucleon systems as the
reduced number of light quarks in the system leads to a milder signal-to-noise ratio in the LQCD
calculations, see Sec.~\ref{sec:lattice_challenges}.
This will shed light on many-body hypernuclear forces (i.e., YNN, YYN, etc.) and consequently the role of hyperons in neutron stars.

Very neutron rich nuclei are short-lived in the laboratory, but form the ground state of nuclear matter under the
conditions that prevail in core collapse and in neutron star mergers, where neutron fluxes are so high that neutron
capture is fast compared to beta decay.  It is thought that many of the heavy nuclei we find on Earth are
the children of the species synthesized in such neutron-rich astrophysical explosions, in the process of rapid
neutron capture (or the $r$-process)~\cite{Kasen:2017sxr,Thielemann:2017acv,Cowan:2019pkx}. One of the main purposes of RIBs such as RIKEN RIBF~\cite{Sakurai:2018hxi} and FRIB~\cite{Bala14FRIB} is to create such nuclei in the laboratory,
determining their masses and decays rates, and thereby allowing us to more quantitatively describe the $r$-process~\cite{Horowitz:2018ndv}.  This in turn
will make nucleosynthetic yields a more powerful test of the astrophysical environments that sustain the
nucleosynthesis.  The properties of such nuclei may help us test our understanding of nuclear interactions within
neutron-rich material: very weakly bound nuclei found near the drip-line~\cite{Forssen:2012yn, Neufcourt:2018syo, Holt:2019gmc, Neufcourt:2019qvd} tend to have extended neutron halos~\cite{Hammer:2017tjm,Platter:2017ikf}.
Properties of those halos can be related to the pressure of a neutron gas at subnuclear densities, providing an
additional constraint on the EOS.

%%% Lattice QCD
\section{Lattice QCD \label{sec:lattice}}
LQCD is a non-perturbative regularization of QCD that is amenable to a numerical implementation.
The QCD path integral is implemented on a discretized spacetime lattice with the quark fields defined on the sites and the gauge fields defined on the links between sites.
The lattice spacing $a$ serves as an ultraviolet (UV) regulator while the finite spatial and temporal extents serve as an infrared (IR) regulator.
Even with such a truncated space, it is not possible to perform the path integral exactly with classical computers.
By Wick rotating to Euclidean spacetime,
\begin{equation}
Z = \int [dU] e^{-S_G(U)}
	\int [d\bar\psi d\psi]
	e^{-S_\psi(U,\bar\psi,\psi)}
\end{equation}
the path integral can be evaluated stochastically, provided the fermion action
\begin{equation*}
S_{\psi} = a^4 \sum_x \bar\psi_x \left[ \Dslash{0.65}(U) + m \right] \psi_x\, ,
\end{equation*}
is real valued.  In this case, $0 <  e^{-S_{\psi}}$ and so this term can be treated as a Boltzmann weight.
$S_G(U)$ is the gauge action which is also real valued.
A Hybrid Monte Carlo (HMC) algorithm~\cite{Duane:1987de} is used to generate a stochastic sampling of gauge fields $U$ with a probability according to these weights.
If the fermion action were complex, such as with a finite chemical potential, the known HMC methods will not work, except for perturbatively small (phenomenologically uninteresting) values of the chemical potential.
There is a substantial effort to explore methods of simulating at finite density, but that is beyond the scope of this article.  For a nice reviwe of finite density, see Ref.~\cite{Banuls:2019rao} and another promissing new idea, the review in Ref.~\cite{Bedaque:2017epw}.
We also do not address processes involving real-time dynamics, focusing instead on observables that we expect to be accessible to lattice QCD in the exascale era.

The discretized Dirac operator couples fermions at neighboring sites.  Therefore, a gauge link field
\beq
	U_\mu(x) = e^{i a A_\mu(x)},
\eeq
is used which transforms under gauge rotations $\O(x)$ as $U_\mu(x) \rightarrow \O(x) U_\mu(x) \O^\dagger(x+\hat\mu)$ with $\hat\mu = a n_\mu$ a vector of length $a$ in the $\mu$ direction.
Then, $\bar\psi(x) \g_\mu U_\mu \psi(x+\hat\mu)$ and similar operators that appear in $S_\psi$, are gauge invariant~\cite{Wilson:1974sk}.

The simplest gauge action is given by a sum over the smallest ``plaquettes'' that tile the lattice,
\beq
	S_G(U) = \b \sum_{n} \sum_{\mu<\nu} \textrm{Re}\left[
		1 - \frac{1}{N_c} \textrm{Tr} U_{\mu\nu}(n)
		\right]\, ,
\eeq
where $\sum_n=$~sum over all sites and the plaquette is
\begin{align}
U_{\mu\nu}(n) &=
	U_\mu(n) U_\nu(n+\hat\mu) U_{-\mu}(n+\hat\mu +\hat\nu) U_{-\nu}(n+\hat\nu)
\nonumber\\&=
	U_\mu(n) U_\nu(n+\hat\mu) U_\mu^\dagger(n+\hat\nu) U_\nu^\dagger(n)\, .
\end{align}
Expanding the gauge links for small lattice spacing, the gauge action can be shown to be
\beq
S_G(U) = \frac{\b}{2N_c} a^4 \sum_{n,\mu,\nu}
	\frac{1}{2} \textrm{Tr} \left[G_{\mu\nu}(n) G_{\mu\nu}(n) \right]
	+\mathrm{O}(a^6),
\eeq
where $G_{\mu\nu}$ is the standard gluonic field strength tensor, which in the continuum is given by $G_{\mu\nu} = \partial_\mu A_\nu - \partial_\nu A_\mu -i [A_\mu, A_\nu]$.
This defines the parameter
\beq
	\beta = \frac{2N_c}{g_0^2}\, ,
\eeq
to match the lattice onto QCD in the continuum limit with the gauge coupling, $g_0$.

The fermion fields are Grassmann variables, which are not simple to handle on computers.  However, the QCD action is quadratic in the fermion fields, allowing for an analytic manipulation of the fermionic integral to a form suitable for computation.  This involves the introduction of pseudo-fermions, which are bosonic degrees of freedom
\begin{align}
	Z_\psi &= \int [d\bar\psi d\psi]
		e^{-a^4 \bar\psi_x [ \Dslash{0.55}(U) + m]_{xy} \psi_y}
\nonumber\\&=
	\textrm{Det}[ \Dslash{0.65}(U) + m ]
\nonumber\\&=
	\int [d \phi^\dagger d\phi]
		e^{-a^4 \phi^\dagger_x [\Dslash{0.55}(U) + m]^{-1}_{xy} \phi_y}\, .
\end{align}
In this expression, we have explicitly written the Dirac operator as a spacetime matrix, which happens to be very sparse.
One of the most demanding aspects of LQCD is to solve the inverse Dirac operator for small quark masses, as the Dirac matrix becomes increasingly ill-conditioned with decreasing quark mass: the smallest eigenvalue is set by the light quark mass which is much smaller than a typical QCD scale.
The full dimension of the Dirac matrix is $12V \times 12V$ where $12=3\textrm{ colors} \times 4\textrm{ spins}$ and $V=N_L^3 \times N_T$ with $N_L$ spatial sites and $N_T$ temporal sites.
A typical state-of-the-art computation has $N_L = L/a =64$ and $N_T = T/a =128$ resulting in a matrix of
size $\mathcal{O}(400 \,\rm M\times400 \,\rm M)$.
Significant effort has gone into optimizing the solution of the inverse Dirac operator, as historically this step has represented 90\% or more of the effort required for LQCD calculations (see for example~\cite{Boyle:2017wul}).
The advent of GPU accelerated compute nodes has been a disruptive innovation for LQCD, as GPUs are particularly fast at performing matrix-vector and matrix-matrix multiplications, two key kernels for solving $[ \Dslash{0.65}(U) + m ]^{-1}$.
This, for example, has led to the optimized \texttt{QUDA} library for performing key LQCD routines on NVIDIA GPUs~\cite{Clark:2009wm,Babich:2011np}.

Once a set of gauge field configurations are generated using a given discretization of the QCD action, any number of observables may be computed on this set.
A stochastic estimate of the observable is then given by
\beq\label{eq:stoch_O}
\langle \mathcal{O} \rangle =
	\frac{1}{N_{\rm cfg}} \sum_{i=1}^{N_{\rm cfg}} \mathcal{O}(U_i)
	+ \mathcal{O}\left( N_{\rm cfg}^{-1/2} \right)\, ,
\eeq
where the stochastic error vanishes with the square-root of the number of
configurations, $N_{\mathrm{cfg}}$.  This assumes each configuration is statistically
independent, which is generally not the case in LQCD calculations as the configurations
are generated from an ergodic Markov chain Monte Carlo; however, the auto-correlation
effects are typically accounted for when results are presented.
These autocorrelation effects are thought to only become problematic for very fine lattices, with
$a\lesssim 0.05$~fm, when the autocorrelation length becomes so large that $\mathcal{O}(1\,\rm M)$
configurations with standard algorithms are required to produce enough stochastically independent
configurations to properly sample the path integral (see for example
Ref.~\cite{Luscher:2010we}).
Typical calculations involve $\mathcal{O}(1000)$ nearly statistically independent configurations.

In order to recover QCD, there are three extrapolations/interpolations that must be carried out: the continuum limit, $a\rightarrow0$, the infinite volume limit, $V\rightarrow\infty$, and the physical quark mass limit, $m_q\rightarrow m_q^{\rm phys}$.
Uncertainties from each of these are systematically improvable, enabling predictions of nuclear and hadronic processes with quantifiable uncertainties.

The Flavour Lattice Averaging Group (FLAG), which is the lattice community equivalent to the PDG, provides a set of criteria (Sec.~II in Ref.~\cite{Aoki:2016frl}) which may be used as a rule of thumb for the minimum number and value ranges of lattice spacings, volumes, and quark masses necessary for controlling or quantifying the systematics associated with each parameter. For example, at least three values of each of the extrapolation parameters, $a$, $L$, and $m_{\pi}$, are generally preferred (although this can vary in certain scenarios). For the lattice spacings and pion (quark) masses, the lowest values should be below 0.1~fm and 200~MeV, respectively, All volumes are expected to be greater than 2~fm, and if fewer than 3 volumes are studied, a more stringent requirement, depending on the pion mass, is placed on the size of the volumes. A rule of thumb is to hold $m_\pi L \gtrsim 4$~\cite{Luscher:1985dn,Colangelo:2005gd}, which corresponds to $L\sim5.8$~fm at physical pion mass. It should be noted that the numbers and ranges of parameters necessary for a given calculation depends strongly on the observable,  target precision, method, analysis, etc., so these criteria are presented here merely as guidelines based on the collective experience of the community.

The use of many different discretized versions of QCD in LQCD computations is
beneficial, and a particular choice is often referred to as a \textit{lattice action}.  For
example, there are substantial challenges to include fermions which respect chiral
symmetry in LQCD calculations~\cite{Nielsen:1981hk}.  This has led to the development
of a number of different discretizations of the fermion action with varying degrees of
complexity.  The most commonly used ones are Wilson, Staggered, Domain Wall,
Overlap and their variants.  We refer the reader to the nice review in
Ref.~\cite{Faessler:2004yf} for more details.
Each lattice action defines a different UV theory which can be expanded as an effective local action close to the continuum limit~\cite{Symanzik:1983dc,Symanzik:1983gh}.
For all lattice actions, at mass-dimension four the only operators are those of QCD, and so there is a universality of the continuum limit which should be observed if all calculations are performed sufficiently close to the continuum limit and a continuum extrapolation is performed.

Some of the simplest observables are derived from two-point correlation functions, in which creation and
annihilation operators of fields (source and sink operators) are chosen with the quantum numbers of the
system of interest and separated in Euclidean time $t$.
The Euclidean time behavior of such correlation functions, $C(t)$, may be shown to behave as,
\beq
\label{eq:corr}
C(t) = \sum_n Z^{\mathrm{snk}}_{n} Z^{\dagger\mathrm{src}}_{n}
	e^{-E_n t} \ ,
\eeq
where the sum runs over the infinite tower of states having the quantum numbers of the operators, and $Z^{\mathrm{src}}_{n}(Z^{\dagger\mathrm{snk}}_{n})$ represents the overlap between the chosen source (sink) operator and the wavefunction of eigenstate $n$.
In the large Euclidean time limit, the lowest energy state for quantum numbers specified by the interpolating fields dominates the correlation function, allowing for a precise determination of the ground state hadron spectrum, for example.

Matrix elements are traditionally computed using three point correlation functions in which the quark level operator of interest is inserted at a third Euclidean time between the source and sink, requiring a second time to be made large, which increases the numerical cost and the complexity of properly isolating the ground state matrix elements.
Additional methods exist in which the effects of the operators are instead folded into modified quark propagators, which are then inserted into correlation function calculations~\cite{Maiani:1987by,Bulava:2011yz,deDivitiis:2012vs,Bernardoni:2014kla,Chambers:2014qaa,Chambers:2015bka,Savage:2016kon,Bouchard:2016heu}.

In practice, particularly for calculations involving nucleons as discussed in the following subsections, the limit of large Euclidean time can be difficult to reach due to computational limitations, particularly as the physical pion mass is lowered towards its physical value.
Thus, while calculations at the physical pion mass are necessary to lend confidence in the ultimate results from LQCD predictions, calculations at heavier than physical pion masses can be used in conjunction to improve the precision of results interpolated to the physical pion mass point, provided the chiral extrapolation is well behaved.

%%%%%%%%%%%% LQCD Challenges
\subsection{Challenges for nuclear physics applications \label{sec:lattice_challenges}}

\begin{figure}
\includegraphics[width=0.49\textwidth]{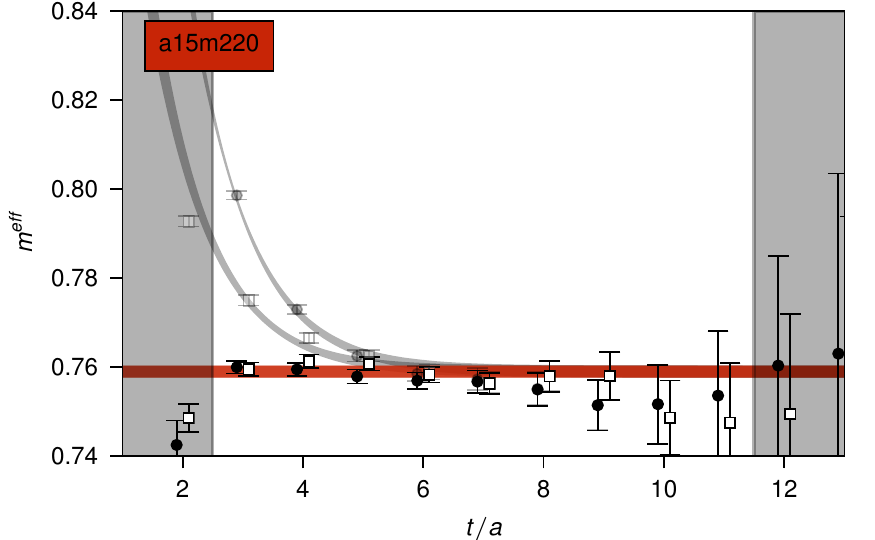}
\includegraphics[width=0.49\textwidth]{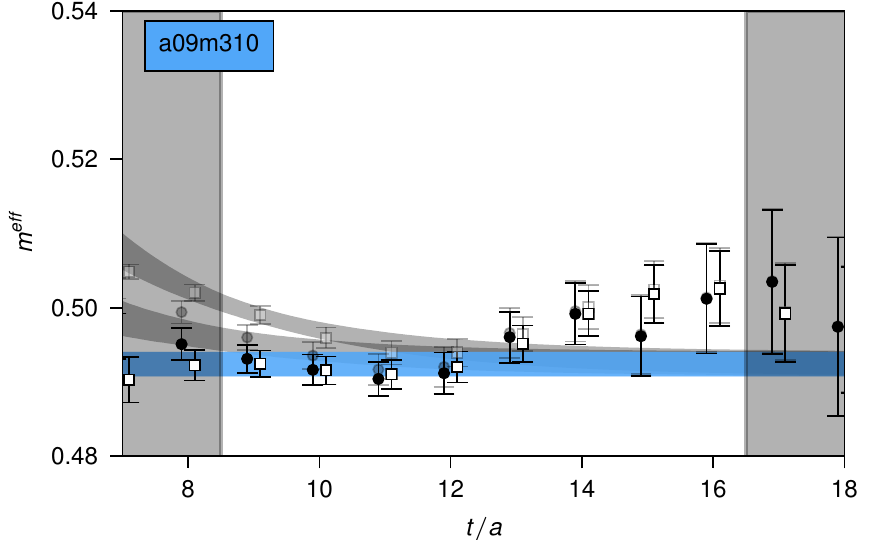}
\caption{\label{fig:meff_nucleon}
Effective mass plots of the nucleon from Ref.~\cite{Chang:2018uxx} which suffer from correlated, late time fluctuations, making it more challenging to identify the ground state.
The top plot is from a calculation with $a\sim0.15$~fm and $m_\pi\sim220$~MeV while the bottom is for $a\sim0.09$~fm and $m_\pi\sim310$~MeV.
}
\end{figure}

While simple in concept, the application of LQCD to problems of interest in nuclear physics is hindered by a few well-known challenges beyond those discussed above.
The first serious challenge one encounters is an exponentially bad signal-to-noise (S/N) problem that manifests from the stochastic evaluation of the path integral, Eq.~\eqref{eq:stoch_O}.
The variance of the correlation function of a nucleon is dominated at late times by a three-pion state.  Therefore, in the long time limit, the S/N ratio decays exponentially as~\cite{Lepage:1989hd}
\beq
\label{eq:s_to_n}
\mathcal{R}(t) \underset{t\to\infty}{\longrightarrow}
	\sqrt{N_{\textrm{stoch}}} e^{-\left(M_N - \frac{3}{2} m_{\pi}\right) t} \ ,
\eeq
where $M_N$ and $m_\pi$ are the nucleon and pion masses and $N_{\textrm{stoch}}$
is the number of stochastic samples.
Overcoming this S/N problem at late times requires exponentially more statistics, while at early times, the signal is polluted by excited states.
Furthermore, the correlation function at neighboring times is highly correlated, and so the late time behavior, as the noise grows, is susceptible to correlated fluctuations, making it non-trivial to robustly identify the ground state of the correlation function.
For example, Fig.~\ref{fig:meff_nucleon} shows two sample effective mass plots from Ref.~\cite{Chang:2018uxx} of
the nucleon mass which suffer from these correlated late time fluctuations.
%%%%%%%%%%%%%%%%%%%%%%%%%%%%%%%%%%%%%%%%
% MN vs Mpi figure
%%%%%%%%%%%%%%%%%%%%%%%%%%%%%%%%%%%%%%%%
%%%%%%%%%%%%%%%%%%%%%%%%%%%%%%%%%%%%%%%%

These problems become exacerbated when computing the structure of the nucleon or multi-nucleon systems.
For nucleon structure calculations, there are two challenges.  First, there are two times which must be taken large as the current insertion must be far away from both the source and the sink time.
Second, achieving high-momentum transfers is difficult since the ground state energy grows with momentum but the variance couples to zero-momentum states, increasing the energy gap in Eq.~\eqref{eq:s_to_n}.
To alleviate this issue, a modified quark smearing algorithm with a momentum profile centered about non-zero momentum has been shown to work well up to momentum transfers of the order $Q\sim3$~GeV~\cite{Bali:2016lva}.

For multi-nucleon systems, the argument of the exponential of the S/N ratio scales as $A(M_N-\frac{3}{2}m_\pi)t$, where $A$ is the number of nucleons.
Fortunately, it has been observed that for early Euclidean times up to a little over $t=1$~fm, the S/N is not degrading exponentially, but is roughly constant for up to a few baryons~\cite{Beane:2009kya,Beane:2009gs}: a region in time which has been coined the ``Golden Window''~\cite{Beane:2009py}.  It was also observed that shortly after the Golden Window, the characteristic energy scale associated with the noise became worse than $(M_N - \frac{3}{2}m_\pi)$.  Whether this is a real effect or just a signal of the grossly degraded signal (challenging the ability to reliably estimate the mean and variance of the correlation functions), this further emphasizes the need to determine the few-nucleon correlation functions early in Euclidean time before the S/N becomes overwhelming.

Beyond the S/N challenge, additional complications for multi-nucleon systems include:
\begin{itemize}[label=$\bullet$]
\item The nuclear physics of interest resides in the small energy difference $\D E = E_{AN} - A\,m_N
\ll A \,m_N$. For the deuteron ($A=2$), the interaction energy is a
mere 0.1\% of the total mass at the physical pion mass.

\item
Multi-nucleon systems exhibit complex excitations, creating a large density of states with splittings of order $\D E$.
Additionally, each relative momentum mode of the system corresponds to an additional excited state, and these energy gaps are also $\mathrm{O}(\D E)$.
Resolving these energy gaps without sophisticated operators requires that the correlation function be determined at a time given roughly by $t_{\D E}\sim 1/\D E$, which is well into the region where the noise has swamped the signal.  For a volume of $L=6.4$~fm at the physical pion mass, the first non-zero back-to-back momentum mode (each nucleon moving with one unit of equal and opposite momentum) has an energy gap to the first elastic scattering state of $\D E\sim 40$~MeV leading to a time-scale of $t_{\D E} \sim 5$~fm.  However, in existing calculations at heavy pion masses, the two-nucleon correlation function becomes swamped by the noise before 2~fm.  The ability to resolve these energy gaps will become exponentially worse as the pion mass is reduced.

\item There is a factorial growth in the number of quark-level Wick contractions required to construct multi-nucleon systems which scales as $N_u! \times N_d!$ where $N_u$ and $N_d$ are the number of $up$ and $down$ quarks required for the system of interest.  While important algorithmic methods have been developed to dramatically reduce the number of required contractions by exploiting symmetries of the system of interest~\cite{Yamazaki:2009ua,Doi:2012xd,Detmold:2012eu,Gunther:2013xj}, these methods only work for unrealistically simple ``wave-functions'' in which all the quarks originate from the same space-time point or are otherwise identical.  For example, a proton $p$ or neutron $n$ requires two contractions. With the simplest contraction routines, $pp$, the triton $pnn$, and ${}^4$He $ppnn$ require 48, 2880 and 518,400 contractions, respectively.  These costs exceeds the cumulative sum of all other parts of the LQCD calculations.

\end{itemize}

\smallskip\noindent
The challenges facing LQCD computations of two or more baryons are greater than imagined even a few years ago.
The first calculation of two-nucleons with dynamical sea-quarks was carried out in 2006 with a single lattice spacing and pion masses $m_\pi \sim\{350, 490, 590\}$~MeV~\cite{Beane:2006mx}.
Shortly after that, the first implementation of the HAL QCD potential was introduced with quenched LQCD calculations~\cite{Ishii:2006ec}.
This was followed by several calculations including hyperon-nucleon interactions over a period of a few years, culminating in the first identification of a bound di-baryon system with LQCD~\cite{Beane:2010hg,Inoue:2010es}, the H-dibaryon~\cite{Jaffe:1976yi}.

Based upon the progress at the time, the first exascale Scientific Grand Challenge report for nuclear physics~\cite{np_exa} (2009) estimated that sustained 10-Petaflop-years would yield an understanding of two-nucleon interactions at the physical pion mass.
However, over the lifetime of the Titan Supercomputer (a 27 Petaflop peak computer operating from 2012--2017 at the Oak Ridge Leadership Computing Facility),
there has only been a single calculation of NN interactions with a lighter pion mass $m_\pi\sim300$~MeV~\cite{Yamazaki:2015asa}, which is at the upper edge of the optimistic range where the two-nucleon EFT can be extrapolated to~\cite{Epelbaum:2002gb,Beane:2002xf}.
In the follow up 2016 DOE Exascale Requirement Review for Nuclear Physics~\cite{np_exa:2017}, it was estimated that in 2020, we would have complete calculations at the physical quark masses (and continuum and infinite volume limits) of two-nucleon and hyperon-nucleon interactions.

The present situation (in early 2021) is that it is still unclear whether or not two nucleons form a bound state with pion masses as heavy as $m_\pi\sim800$~MeV, there has been no continuum scaling study of two-nucleon interactions and there are still no results with pion masses lighter than $m_\pi\sim300$~MeV, though there are preliminary results (conference proceedings) with low statistics from HAL QCD using their alternative potential method.
While the actual availability of near-exascale computing is behind the projected availability when the first 2009 exascale report was written, and perhaps a year behind when the 2016 report was prepared~\cite{Geesaman:2015fha,np_exa:2017} this alone does not explain the lag in LQCD results.
This review describes some of the progress the field has made that has led us to recognize that LQCD treatments of systems of two or more nucleons are more challenging that originally anticipated.
The importance of creating more accurate projections of results, and understanding how to maximize the scientific output per node hour, has become more acute as the cost of the supercomputers has become very non-trivial with the Sierra and Summit supercomputer costing a few hundred million US Dollars and Frontier, the Exascale computer to be delivered to OLCF, costing a projected \$600M US dollars.%
% FOOTNOTE
\footnote{See for example \url{https://en.wikipedia.org/wiki/Summit_(supercomputer)} and the Frontier press release linked at the url \url{https://www.olcf.ornl.gov/frontier/}.}
%---------
An excellent place to start is the two-nucleon bound state controversy at $m_\pi\sim800$~MeV.
While we can not offer a resolution, which will require extensive computations that do not currently exist, we can use this controversy to highlight the challenges of these LQCD computations.
First, we begin with a brief review of single nucleon properties.

%%%%%%%%%%%%%%%%%%%%%%%%%%%%%%%%%%%%%%%%%%%%%%%%%%
%%%%         SINGLE NUCLEONS
\subsection{Status and challenges for single nucleons \label{sec:lattice_n}}

As discussed above, there are in principle an infinite number of valid lattice actions, which are expected to give different results for a given lattice spacing; only in the continuum limit do we expect results from the different actions to reproduce QCD.
Therefore, a lattice QCD calculation performed on a single lattice spacing should be interpreted as a model, at best a controlled approximation to QCD, whose systematic errors cannot be quantitatively measured.
In this review, we bias our highlights to those computations which have a controlled extrapolation to the physical point in all three systematics (the physical pion mass, continuum and infinite volume limits).

In the last few years, we have witnessed the emergence of the first LQCD results of properties of single nucleons with controlled extrapolations to the continuum, infinite volume and physical pion mass limits, so we focus on such calculations in this section. A reproduction of the ground state spectrum of octet and decuplet baryons was first achieved in 2008~\cite{Durr:2008zz}.
Seven years later, a reproduction of the neutron-proton mass splitting was achieved including dynamical
QED effects with an accuracy of $\sim 300\keV$~\cite{Borsanyi:2014jba}.
More recently, there have been determinations of the static charges of the nucleon~\cite{Bhattacharya:2015wna,Bhattacharya:2016zcn,Chang:2018uxx,Gupta:2018qil,Gupta:2018lvp,Harris:2019bih} and the elastic form factors~\cite{Jang:2019jkn} which have included full continuum, infinite volume and physical pion mass extrapolations.
In contrast, determinations of light and heavy meson properties have progressed to the point that they are routinely reviewed and averaged by FLAG~\cite{Aoki:2016frl}.  The precision of these results rivals what can be achieved from experiment and phenomenology and the LQCD results play a central role in unitarity tests of the CKM Matrix.
Signaling a paradigm shift, for the first time, properties of nucleons were included in the most recent FLAG review~\cite{Aoki:2019cca}.

The most ubiquitous nucleon quantity appearing in weak interactions is the nucleon axial coupling (often referred to as the axial charge, or $g_A$).
It was initially thought this would be a ``gold-plated'' quantity for benchmarking the success of LQCD applications to nucleon matrix elements, but it has proven to be a challenging quantity to determine, leading the field to estimate that a determination with a 2\% uncertainty would be possible by 2020 with near-exascale computers (such as Summit at ORNL)~\cite{usqcd_doe_2016,Bhattacharya:2016zcn}.
However, the use of an unconventional strategy~\cite{Bouchard:2016heu} that included the use of early Euclidean time results, where the stochastic noise is exponentially smaller, allowed for a determination of $g_A$ with 1\% precision in 2018 utilizing Titan era supercomputers~\cite{Chang:2018uxx}. Whether one can fully control and quantify excited state contributions when including early Euclidean times has been cause for concern within the community and merits stringent checks on the results. Such checks are discussed extensively in Ref.~\cite{Chang:2018uxx}; a comprehensive discussion of excited state contamination in calculations of $g_A$ will be a focus for a future review~\cite{Chang:2020xxx}.
Here, we make two comments: first, a point that continues to be misunderstood in much of the literature is that the method utilized~\cite{Maiani:1987by,Bulava:2011yz,deDivitiis:2012vs,Bernardoni:2014kla,Chambers:2014qaa,Chambers:2015bka,Savage:2016kon,Bouchard:2016heu} suppresses excited state contamination above and beyond the typical exponential suppression governed by the mass gap which is what enables the use of early time results.
The point has not been entirely lost however, as we have seen the use of early time results (as well as late) in other nucleon structure calculations~\cite{Orginos:2017kos,Hasan:2019noy,Alexandrou:2019brg} as originally advocated in Ref.~\cite{Detmold:2011bp} to the best of our knowledge.

Second, the tension between the results from CalLat and PNDME, which use many of the same HISQ configurations but different valence actions, arises from the continuum extrapolation with the results on the finest ensemble used by PNDME pulling their result to a lower value~\cite{Gupta:2018qil,Bhattacharya:2020zjf}.
It has been pointed out that the span of source-sink separation times used by PNDME is particularly small on the finest ensemble, just under $0.25$~fm, which is also the most susceptible to statistical correlations in neighboring time slices~\cite{Walker-Loud:latt2019}.
Further, PNDME has uncovered a fitting systematic in the analysis of their correlation functions which was not previously incorporated, which is the inclusion of the nucleon-pion excited state.  This state has not been systematically included in any calculations utilizing the standard fixed source-sink separation method, though it has been included in the analysis of the Feynman-Hellmann calculations~\cite{Bouchard:2016heu}.  However, it's inclusion both by PNDME and the CalLat groups is sub-optimal as it is only included in the anlysis of operators which are still local hadronically, meaning, they have not incorporated an explicity nucleon-pion source-sink creation/annihilation operator which can be used to diagonalize the correlation functions in terms of the eigenstates of the Hamiltonian.

What is most important for $g_A$, or any matrix element, is to demonstrate that the ground state matrix element is not sensitive to the various models of excited states used in the correlation function analysis~\cite{Chang:2018uxx}.
Nevertheless, when the nucleon-pion state is explicitly included in the analysis of the PNDME results, they observe a significant increase in their value of $g_A$ on ensembles with close to the physical pion mass~\cite{Jang:2019vkm}.  When these larger values of $g_A$ results are incorporated into the chiral extrapolation, the tension between the PNDME extrapolation and that of CalLat~\cite{Chang:2018uxx} is alleviated~\cite{Jang:2020ygs}, though the specific formulation they use is in some contradiction with the $\chi$PT inspired theory estimate~\cite{Bar:2018wco}.

Theoretical studies of quantities which may be precisely measured experimentally, such
as $g_A$, can signal new physics when discrepancies are found. Calculations of
quantities that have not yet been observed in nature, such as tensor and scalar
isovector currents, may also be used to constrain potential BSM interactions. Lattice
QCD results for these quantities have been performed by several groups, with global
averages reported in Ref.~\cite{Aoki:2019cca}.  See Ref.~\cite{Bhattacharya:2011qm} for
a nice review of how precision neutron beta-decay measurements constrains heavy BSM
physics.

Isoscalar charges represent processes mediated via electromagnetic, weak neutral, or dark matter interactions. For example, the tensor isoscalar and isovector charges quantify contributions to the nucleon electric dipole moment from quark electric dipole moments, and are therefore relevant for probes of CP violation. Lattice calculations of these quantities present additional technical difficulties due to the presence of so-called ``disconnected" diagrams, involving a quark loop at the operator insertion. These contributions require calculating propagators between all possible lattice points, and therefore scale with the lattice volume. In practice, they are often estimated stochastically or sometimes disregarded altogether.

There is currently significant tension between LQCD results and phenomenological determinations of one interesting quantity,
 the pion-nucleon sigma term $\s_{\pi N}$, the source of the primary uncertainty in estimating the spin-independent coupling of dark matter to nuclei.
The LQCD results tend to favor smaller values of $\s_{\pi N}\sim40$~MeV~\cite{Bali:2012qs,Durr:2015dna,Yang:2015uis,Bali:2016lvx,Alexandrou:2019brg} (see also the latest determination with the most complete continuum, infinite volume and physical quark mass extrapolations/interpolations~\cite{Borsanyi:2020bpd}) while the best phenomenological determination is $\s_{\pi N}=59.1\pm3.5$~MeV~\cite{Hoferichter:2015dsa}.
The phenomenological extraction makes very few assumptions and relies upon a dispersive understanding of pion-nucleon scattering, while the LQCD calculations provide a direct determination of the matrix elements.  It has been proposed~\cite{Hoferichter:2016ocj} that LQCD be used to directly determine the pion-nucleon scattering lengths such that they can be compared to those that are input to Ref.~\cite{Hoferichter:2015dsa}.

%%%%%%%%%%%%%%%%%%%%%%%%%%%%%%%%%%%%%%%%%%%%%%%%%%
%%%%         two-nucleonS
\subsection{Status and challenges for two-nucleons \label{sec:lattice_nn}}
As discussed earlier in Sec.~\ref{sec:lattice}, LQCD calculations are necessarily performed in a finite Euclidean volume typically with periodic spatial boundary conditions.
The finite volume means there are no asymptotic states and so direct access to scattering amplitudes, as one would perform experimentally,  is not possible~\cite{Maiani:1990ca}.
However, there is substantial literature on the ``\luscher{}
method''~\cite{Luscher:1986pf,Luscher:1990ux} to relate the finite volume spectrum to the infinite volume scattering
amplitudes~\cite{Beane:2003yx,Beane:2003da,Ishizuka:2009bx,Luu:2011ep,Leskovec:2012gb,Li:2012bi,Briceno:2013lba,Briceno:2013bda},
including boosted systems~\cite{Rummukainen:1995vs,Li:2003jn,Feng:2004ua,Kim:2005gf,Bour:2011ef,Davoudi:2011md,Briceno:2012yi,Lee:2017igf}, coupled channels~\cite{Li:2012bi,Briceno:2012yi} and states of arbitrary spin~\cite{Briceno:2014oea}.  There is also significant effort in extending the method to handle three-particle systems which is nearly complete~\cite{Polejaeva:2012ut,Briceno:2012rv,Hansen:2014eka,Hansen:2015zga,Briceno:2017tce,Hammer:2017uqm,Hammer:2017kms,Mai:2017bge,Doring:2018xxx,Hansen:2020zhy}
(see Sec.~\ref{sec:hobetbox} for the construction of an effective theory in finite volume, which has also been performed in Refs.~\cite{Bernard:2007cm,Beane:2012ey,Hall:2013qba,Wu:2014vma}).

The \luscher{} method provides a mapping between the finite-volume energy spectrum, in particular the
shift in the energy levels from their non-interacting values, to the infinite volume scattering phase shifts.
The underlying assumptions, which do not rely on perturbation theory or non-relativistic expansions, are:
\begin{enumerate}
	\item The theory is unitary;
	\item The range of the interaction is smaller than the spatial dimensions of the volume and the exponentially suppressed corrections, scaling as $e^{-m_{\pi} L}$, are negligibly small;
	\item The total energy of the two-particle system is below the first inelastic threshold not taken into account;
\end{enumerate}

In periodic volumes, the boundary conditions induce an unphysical mixing of various partial waves due to the reduced symmetry of the cubic volume.  In practice, one must truncate the basis of partial waves considered, which is a controlled approximation for low-energy scattering.  In the limiting case of a single channel in the most symmetric state (the $A_1$ cubic representation for a spin-0 system) where one truncates all but the $S$-wave interactions, one has the relation~\cite{Beane:2003yx,Beane:2003da}
\begin{align}
&p \cot \delta(p) =
	\frac{1}{\pi L} S\left(\left(\frac{pL}{2\pi}\right)^2\right)\, ,&
%\nonumber\\
&S(\eta) \equiv
	\lim_{\L\rightarrow\infty}\sum_\mathbf{j}^{\L} \frac{1}{|\mathbf{j}|^2 -\eta}-4\pi\L\, ,&
\end{align}
where $p$ is the relative momentum of the two-particle system satisfying $E = 2 \sqrt{m^2 + p^2}$ (for
non-interacting systems, $\mathbf{p}_n = 2\pi\mathbf{n}/L$, but the interactions distort this relation),
$L$ is the size of the periodic box, and $\delta$ is the corresponding phase shift at that momentum.
The sum is over all integer three-vectors such that $|\mathbf{j}|< \L$.
The function $S(\eta)$ has poles precisely at the non-interacting energies, showing consistency with the poles on the left hand side corresponding to $\delta \to 0$.

A significant challenge for these calculations is that the interaction energies are small compared to the
total energy of the system and so the shift in the value of $p$ from the non-interacting values
$|\mathbf{p}_n|$ are small.  Thus, one requires statistical precision at the multiple sigma level on the
difference in the total energy from the non-interacting levels to ensure that the uncertainty in the
scattering phase shifts does not become grossly magnified by the proximity to these poles.
Phase shifts for two-meson systems have been studied extensively due to their exponentially milder statistical noise compared to nucleons.
For a nice review, see Ref.~\cite{Briceno:2017max}.
Systems containing a single baryon are considerably noisier than purely mesonic systems which has limited calculations to meson-baryon scattering to just a few~\cite{Torok:2009dg,Lang:2012db,Mohler:2012nh,Detmold:2015qwf,Lang:2016hnn,Andersen:2017una,Paul:2018yev,Khan:2019yck}.

Baryon-baryon calculations include NN~\cite{Beane:2006mx,Beane:2011iw,Beane:2012vq,Yamazaki:2012hi,Beane:2013br,Orginos:2015aya,Yamazaki:2015asa,Berkowitz:2015eaa,Wagman:2017tmp,Berkowitz:2019yrf}, YN and YY~\cite{Beane:2006gf,Nemura:2008sp,Beane:2009py,Beane:2010hg,Inoue:2010es,Beane:2012ey,Beane:2012vq,Wagman:2017tmp,Francis:2018qch} and light (hyper-)nuclei up to $A=4$~\cite{Beane:2009gs,Yamazaki:2012hi,Beane:2012vq,Yamazaki:2015asa}, in all cases with $m_\pi \gtrsim 300$~MeV.
The two-nucleon calculations are of particular interest in the $s$-wave channels, where fine-tuning and the appearance of real or virtual bound states may be studied.
The hyperon-nucleon calculations may be important for understanding the EOS in neutron stars and can be directly compared with hyper-nuclei being produced at J-PARC, the upcoming JLab, and FAIR experiments.

% --  HAL QCD METHOD -------------
The challenges with these calculations have also inspired new approaches, beginning with the HAL QCD potential~\cite{Ishii:2006ec,Aoki:2009ji,Murano:2011nz,Aoki:2011gt,HALQCD:2012aa,Aoki:2012tk,Aoki:2012bb,Aoki:2013tba,Iritani:2018zbt} and, more recently, an idea to extract the infinite volume scattering amplitudes from finite volume spectral functions~\cite{Bulava:2019kbi}.
The HAL QCD potential method was inspired by a successful application of $I=2\ \pi\pi$ wavefunctions to determine the scattering phase shifts~\cite{Aoki:2005uf}.
It has been used to study NN, YN, and YY interactions~\cite{Nemura:2008sp,Inoue:2010hs,Inoue:2010es,Etminan:2014tya,Sasaki:2015ifa}.
They have also studied the $\Lambda_c N$ interactions, parity odd interactions~\cite{Murano:2013xxa} and a first exploration of the three-nucleon potential~\cite{Doi:2011gq}.
Most recently, they have some first results at the physical pion mass in YN and YY channels~\cite{Gongyo:2017fjb,Sasaki:2019qnh,Iritani:2018sra} as well as a first look at the two-nucleon interactions at the physical pion mass~\cite{Doi:2017zov}.

Before discussing the controversy, we first provide a brief high-level summary of the HAL QCD potential method~\cite{Ishii:2006ec,Aoki:2009ji,Murano:2011nz,Aoki:2011gt,HALQCD:2012aa,Aoki:2012tk,Aoki:2012bb,Aoki:2013tba,Iritani:2018zbt}.
Instead of directly relating the energy levels to the scattering amplitude via the \luscher{} quantization condition, a
coordinate space ratio correlation function is constructed
\beq
\label{eq:nn_n_ratio}
R(\mathbf{r},t) = \frac{C_{NN}(\mathbf{r},t)}{(C_N(t))^2}\, ,
\eeq
where the two particle correlation function is given by
\beq
\label{eq:c_nn}
C_{NN}(\mathbf{r},t) = \sum_\mathbf{x} \langle
	N(\mathbf{x+r},t) N(\mathbf{x},t)
	N^\dagger(0)N^\dagger(0)
	\rangle\, .
\eeq
$N(\mathbf{x},t)$ is an interpolating field with quantum numbers of the nucleon.
The standard correlation functions used in the \luscher{} method can be obtained from
Eq.~\eqref{eq:c_nn} by a Fourier transform over the relative coordinate.
Alternatively, Eq.~\eqref{eq:nn_n_ratio} can be used as a Nambu-Bethe-Salpeter wavefunction to extract scattering phase shifts~\cite{Lin:2001ek,Aoki:2005uf,Ishii:2006ec}.
A time-dependent version has also been developed~\cite{HALQCD:2012aa}
\beq
\label{eq:halqcd_improved}
\left[\frac{\partial_t^2}{4M} - \partial_t -H_0\right] R(\mathbf{r},t)
=
\int d^3 s \, U(\mathbf{r},\mathbf{s}) R(\mathbf{s},t)\, ,
\eeq
where $H_0$ is the kinetic-energy operator.
The strategy is to determine $U(\mathbf{r},\mathbf{s})$, which then provides a complete description of the elastic NN scattering states below the particle production threshold.
There are several required assumptions for this strategy to be valid, and if satisfied, it provides an alternative means of determining the infinite volume elastic scattering amplitudes.

First, one must assume that in the range of $t$ used to determine $U(\mathbf{r},\mathbf{s})$, the inelastic single-nucleon states appearing in the numerator and denominator of Eq.~\eqref{eq:nn_n_ratio} exactly cancel and that the two-nucleon inelastic states are numerically negligible  within the numerical precision to which $R(\mathbf{r},t)$ can be determined.
Otherwise, $U(\mathbf{r},\mathbf{s})$ is not energy independent, which invalidates the method~\cite{Aoki:2013tba}.
Such an energy dependence should, in principle, manifest as a $t$ dependence of the extracted potential $U$.
In Ref.~\cite{Iritani:2018zbt}, HAL QCD constructs the potential for the $\Xi \Xi$ system at several values of $t$, and within their statistical precision, do not observe any $t$ dependence.
However, it should be noted that the same expectation, that systematic errors should manifest as time dependence in the extracted result, is true when extracting energy levels via the \luscher{} method, thus, similar scrutiny may need to be applied.

Assuming these systematics are under control, the time-dependent method, Eq.~\eqref{eq:halqcd_improved} allows for the relaxation of the requirement of ground-state saturation of the correlation function as the same potential describes all of the elastic two-nucleon scattering states. In practice, the engineering of good single-nucleon operators has a long history, however, it is generally observed that contamination from inelastic states can persist into the typical time ranges used for two-nucleon calculations~\cite{Berkowitz:2017smo}, unless multiple single-nucleon operators are combined~\cite{Beane:2009kya,Beane:2009py,Beane:2013br,Berkowitz:2017smo,Berkowitz:2019yrf}. Such engineered operators have not been utilized within the potential method.

A practical approximation that must be made to determine $U(\mathbf{r},\mathbf{s})$ is that one must assume it can be well approximated by a
local gradient expansion
\begin{align}\label{eq:halqcd_potential}
U(\mathbf{r},\mathbf{s}) = \Big[&
	V_C(r) + V_\sigma(r)\mathbf{\s}_1 \cdot \mathbf{\s}_2
		+V_{LS}(r) \mathbf{L}\cdot\mathbf{S}
		+\cdots
%\nonumber\\&
		+V_{{\rm N}^2{\rm LO}}(\mathbf{r}) \frac{\nabla^2}{\Lambda^2} + \cdots
	\Big]\delta^3(\mathbf{r-s})\, ,
\end{align}
where the potential has been decomposed into a central term, $V_C$, a spin-dependent term $V_\s$, spin-orbit coupling terms, $V_{LS}$, etc.
The gradient expansion then continues with higher order terms appearing at NLO in $\nabla^2/\L^2$ (denoted as \nxlo{2}).
The suppression scale, $\L$ in this gradient expansion is not clearly defined, making the task of
rigorously showing whether all neglected orders are small compared to the leading terms and whether
the series converges at low orders, a difficult one.

If all of these systematic uncertainties are shown to be under control, the HAL QCD potential offers a nice alternative to the \luscher{} method for determining two-nucleon (two-baryon) interactions at low-energies.

%%%%%%%%%%%%%%%%%%%%%%%%%%%%%%%%%%%%%%%%%%%%
%
%  NN Controversy
\subsubsection{Two-nucleon controversy \label{sec:nn_controversy}}

An unresolved controversy in the literature is that results generated using the HAL QCD potential method do not agree, even qualitatively, with results generated from a more common method:
even with heavy pion masses in the $SU(3)$ flavor limit ($m_\pi\sim800$~MeV)~\cite{Beane:2012vq,Berkowitz:2015eaa} as well as intermediate pion masses of $m_\pi\sim450-510$~MeV~\cite{Yamazaki:2012hi,Orginos:2015aya}, where the S/N problem is exponentially less severe than at light pion masses,
LQCD calculations using \luscher's method have observed that the deuteron becomes more deeply
bound and the di-neutron also becomes bound, while applications with the HAL QCD potential have
concluded there are no bound states in either channel for similar pion
masses~\cite{HALQCD:2012aa,Inoue:2011ai}.
Complicating the issue, calculations coming from different groups utilize different sets of gauge configurations, so it is often not possible to isolate the issue to be due only to a given method, and not also from underlying discretization effects or other lattice systematics. The HALQCD group has performed comparisons between the two methods on a single set of configurations, focusing largely on the $\Xi\Xi$ system, and have pointed to operator dependence as being the leading source of discrepancy between the methods~\cite{Iritani:2016jie,Iritani:2018vfn}.

%-- FIGURE HAL QCD MIRAGE ----------------------
\begin{figure*}
\includegraphics[width=\textwidth]{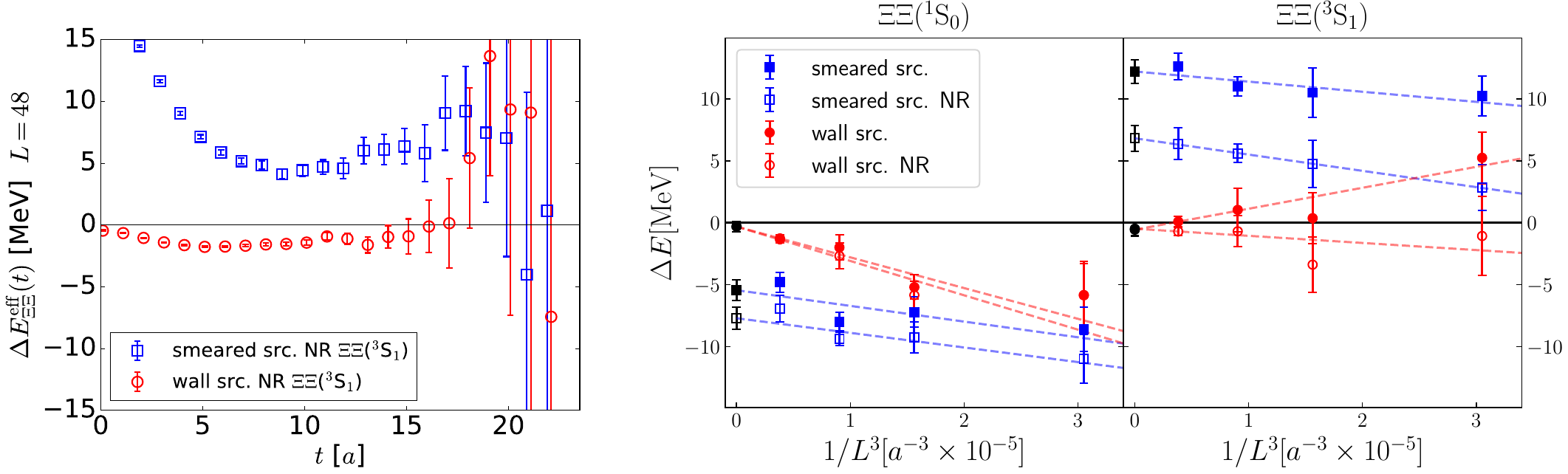}
\caption{\label{fig:halqcd_mirage_xixi}
LEFT: Effective mass of the $\Xi\Xi (^3{\rm S}_1)$ ratio correlation function with a non-relativistic (NR) Gaussian smeared source (top) and wall source (bottom) on the $L=48$ volume~\cite{Iritani:2016jie}.
RIGHT: Extrapolation of ground state energies in both channels on all four volumes comparing the Gaussian smeared and wall sources.
The interaction energies were extracted from a late time-region where the single baryon effective mass is dominated by the ground state ($t\gtrsim 1$~fm).
Note, for scattering states (vs bound states) the interaction energy must extrapolate to 0 in the infinite volume limit ($1/L^3\rightarrow0$).
In this work, HALQCD determined that the smeared source exhibited evidence of a fake plateau at early times, and therefore the energies extracted using this source were unphysical. Yamazaki et al.~\cite{Yamazaki:2017euu,Yamazaki:2017jfh} responded to these findings with an order of magnitude increase in statistics, and concluded that the wall source agreed at late times with the early time plateau of the smeared source, contrary to the HALQCD conclusions.  However, the study was performed on a different set of quenched configurations and so a direct comparison with the work of HAL QCD can not be performed.
Further, the results from the smeared source in the $\Xi\Xi (^3{\rm S}_1)$ channel simply can not be correct on the physical ground noted for the $L\rightarrow\infty$ limit above.
See the text for more detail.
}
\end{figure*}
%-----------------------------------------------

In more detail, calculations which use compact, hexa-quark creation operators at the source in which all 6 quarks originate from the same spacetime point, observe deeply bound states (and sometimes two) in the various two-baryon channels including the $nn$ (${}^1{\rm S}_0$) and deuteron (${}^3{\rm S}_1$)~\cite{Beane:2012vq,Yamazaki:2012hi,Orginos:2015aya}.
In contrast, calculations which use spatially diffuse (displaced nucleons~\cite{Berkowitz:2015eaa,Berkowitz:2017smo,Berkowitz:2019yrf}) or momentum based~\cite{Francis:2018qch} creation operators observe more shallow bound states or no bound states.  The HAL QCD potential is generally constructed from a very diffuse wall-source creation operator for the quark fields and finds no bound states.

If the broader nuclear physics community is to have confidence in LQCD results of two-nucleon interactions, two-nucleon electroweak and BSM matrix elements and more, it is critical to resolve this discrepancy and robustly identify which techniques faithfully reproduce the strong interactions that emerge from QCD.
We will review the state of the LQCD results and the current discrepancy.

It is tempting to brush off this controversy as arising from some unresolved systematic uncertainty with the HAL QCD method.
However, HAL QCD has explored criticisms of their method, finding no obvious issues~\cite{Iritani:2018zbt}.
Further, they have highlighted deficiencies in the application of the standard method which require further scrutiny.
The extreme cost of the numerical calculations is a significant contributor to the slow progress in resolving the discrepancy and identifying which of the results are correct, if any.

The controversy, and its resolution, essentially boil down to two questions:
\begin{enumerate}
	\item In the application of \luscher{}'s method, have the true eigenstates of the system been resolved or are there unrealized systematic uncertainties that have led to the false identification of the spectrum?
	\item Are all of the extra assumptions and systematic uncertainties in the HAL QCD method fully controlled?
\end{enumerate}
At this time, the literature contains suggestive evidence, from both sides, but no conclusive proof.  We begin with the first point.

%-- FIGURE k cot d consistency ----------------------
\begin{figure*}
\includegraphics[width=\textwidth]{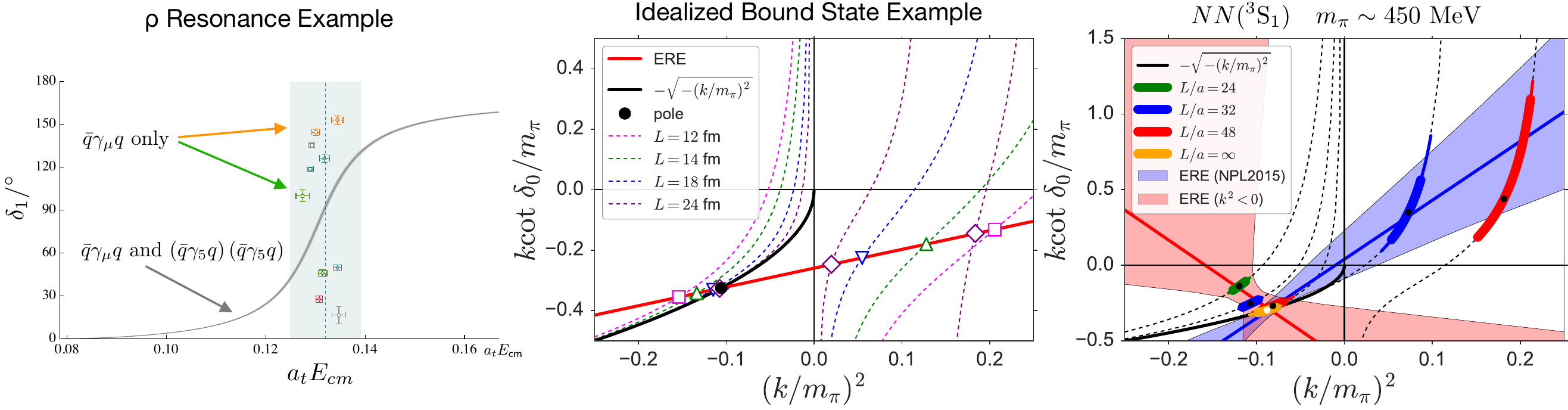}
\caption{\label{fig:kcot_consistency}
LEFT (figure adapted from Ref.~\cite{Wilson:2015dqa}): The gray Breit-Wigner-like curve represents a determination of the phase shift near the $\rho$-resonance when both local $\rho$-like ($\bar{q}\g_\mu q$) and $\pi\pi$- and $KK$-like ($(\bar{q}\g_5q)(\bar{q}\g_5q)$) operators are used to determine the spectrum.  The scattered energy levels in the vertical light-blue band are determined from a set of operators that only include the local ones, which do not lie upon the phase shift curve as is required by the \luscher{} quantization condition.
MIDDLE (figure adapted from Ref.~\cite{Iritani:2017rlk}): An idealized bound state example.
The dashed lines are determined from the \luscher{} quantization condition upon which all allowed values of $k\cot \delta$ and $k^2$ must lie for a particular volume.  In this idealized example, one can see at low-momentum, the effective range expansion (ERE) extrapolates smoothly through the extracted values.
RIGHT (figure adapted from Ref.~\cite{Iritani:2017rlk}): HAL QCD analyzed the $k^2, k\cot \d$ values from the calculations showing bound states.  Plotted is an example from the NPLQCD calculation with $m_\pi\sim450$~MeV~\cite{Orginos:2015aya}.  In this case, they observed the ERE determined from the negative shifted energy levels is not consistent with that determined from the positively shifted levels.}
\end{figure*}
%-----------------------------------------------

HAL QCD speculates that in the application of \luscher{}'s method, the calculations suffer from the false identification of the ground state plateau due to contamination from excited two-nucleon elastic scattering states that are not properly resolved~\cite{Iritani:2016jie}.
Central to their speculation are a few key points:
\begin{enumerate}[label=\Alph*.]
\item With one exception~\cite{Francis:2018qch} (which was not criticized), all LQCD calculations of two-baryon systems utilize only a single source in any given analysis, such that an operator basis diagonalization to more faithfully project onto the eigenstates of the system is not possible, and therefore, they rely upon the Euclidean time dependence to filter the excited states before the noise swamps the signal.  Complicating the analysis, the correlation functions are not positive definite, so the excited states can contribute with opposite sign as the ground state;
\item\label{item:source_dependence} The lowest energy of the two-nucleon system depends upon the creation operator used (wall source versus Gaussian smeared source for the quarks, or spatially displaced two-nucleon operators versus local), leading to extracted energies which are multiple sigma apart.  At face value, this observation may be troubling as the true ground state of the theory can not depend upon the the basis of creation/annihilation operators;
\item Using the criteria advocated by NPLQCD (Refs.~\cite{Beane:2017edf,Wagman:2017tmp} for example), which state that the ratio correlation function can be used provided the fitting window does not begin until the single nucleon correlator has been saturated by the ground state, leads to a scenario in the $\Xi\Xi (^3{\rm S}_1)$ system with $m_\pi\sim510$~MeV where the ground state energy, determined from a fit in the observed plateau region generated from Gaussian smeared sources, is a positively shifted scattering state which, when extrapolated to infinite volume, results in a finite interaction energy.
This is not physically possible for a scattering state, as asymptotically the interaction energy scales as $\D E \propto L^{-3}$.  When wall sources are used, the energies extrapolate to zero in both channels, which is consistent with physical scattering states.  See \figref{fig:halqcd_mirage_xixi} adapted from Ref.~\cite{Iritani:2016jie};

\item HAL QCD proposed ``consistency checks'' to perform on the extracted energy levels which can diagnose if they are consistent with expectations from the \luscher{} quantization condition~\cite{Iritani:2017rlk}.
Such consistency checks are commonly used (but not by any name) in two-meson scattering calculations, see for example Ref.~\cite{Wilson:2015dqa}.  If a particular eigenenergy was poorly fit, or there were missing operators that led to a false plateau, when these energy levels are then passed through the \luscher{} quantization with the appropriate effective range or K-matrix or similar parameterization, these energy levels typically``stick out like sore thumbs''~\cite{raul_luscher_check}, see \figref{fig:kcot_consistency}.
The analysis of HAL QCD found issues in nearly all calculations with deep bound states~\cite{Yamazaki:2011nd,Yamazaki:2012hi,Yamazaki:2015asa,Beane:2011iw,Beane:2012vq,Beane:2013br,Orginos:2015aya,Berkowitz:2015eaa}, though NPLQCD has refuted the observation~\cite{Beane:2017edf,Wagman:2017tmp} for the $m_\pi\sim800$~MeV data~\cite{Beane:2013br}.
In a similar vein, low-energy theorems~\cite{Baru:2015ira} were applied to the $m_\pi\sim450$~MeV data~\cite{Orginos:2015aya} and it was found that the LQCD results are self-inconsistent~\cite{Baru:2016evv}: the binding energies in the deuteron and di-neutron channels, when processed with the low-energy theorems, lead to scattering lengths and effective ranges that are incompatible with the effective range expansion phase shift analysis results from the same LQCD computation.
\end{enumerate}

%-- FIGURE False energy levels  ----------------------
\begin{figure*}
\includegraphics[width=\textwidth]{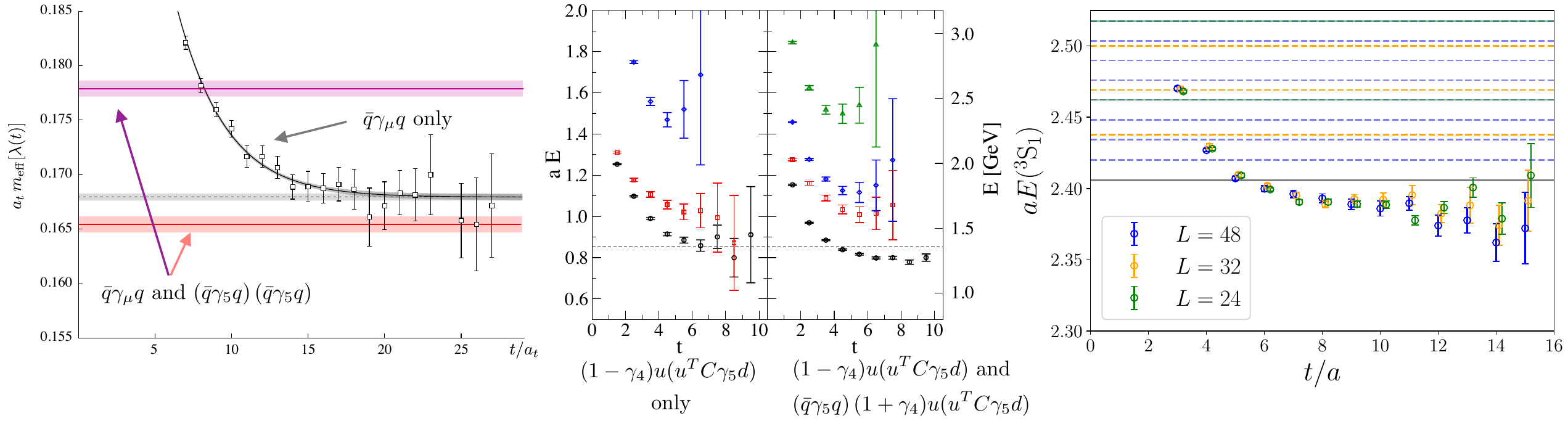}
\caption{\label{fig:op_spec}
LEFT ($I=1\ \pi\pi$ scattering~\cite{Dudek:2012xn}):
If only local $\rho = \bar{q}\gamma_\mu q$ operators are used, one obtains a very reasonable looking spectrum with no obvious by-eye or by-analysis indication the ground state is wrong.  If one expands the basis of operators to include both the local $\rho$-like operator and also two-pion operators, $(\bar{q}\gamma_5 q)(\bar{q}\gamma_5 q)$, then one obtains a different spectrum with eigenvalues that are different by multiple sigma.
MIDDLE (figure adapted from Ref.~\cite{Lang:2012db}):
The same observation has been made in the negative parity nucleon channel  when using only the local negative parity nucleon operator or when also using non-local $\pi N$ operators.
RIGHT (Effective mass of the NN$(^3{\rm S}_1)$ system at $m_\pi\sim800$~MeV generated from a hexa-quark source operator~\cite{Beane:2017edf}): Superimposed on the effective mass plot are the threshold (solid gray line) and non-interacting two-nucleon energy levels (dashed lines) for the three volumes used.}
\end{figure*}
%-----------------------------------------------

A key component of the HAL QCD criticism is that energy levels extracted, including the negatively shifted energy levels, are not the true spectrum, but are false plateaus arising from a pollution of the two-nucleon elastic scattering excited states.
How plausible is this speculation?

One observation from the two-meson LQCD calculations that has been acknowledged for more than a decade is the need to use a large basis of interpolating operators that overlap with the complete set of states in the system of study in the energy range of interest, otherwise the resulting spectrum can be systematically biased without a clear indication of the problem from the correlation function analysis.
For example, it has been clearly demonstrated that for $I=1$ $\pi\pi$ scattering,
a basis of operators that is purely local (``single-particle"),
such as $\bar{q}(x)\gamma_\mu q(x)$, or purely non-local, $\bar{q}(x)\gamma_5 q(x)
\bar{q}(y)\gamma_5 q(y)$ (``two-particle"), both fail to individually produce the correct spectrum.
Either set of operators on its own produces effective mass plateaus that appear like well defined energy levels, but in fact, they are linear combinations of the true eigenstates, which lie multiple sigma away~\cite{Dudek:2012xn}.
If both sets of operators are used, a more complete spectrum is recovered, which we believe to be the correct spectrum, see \figref{fig:op_spec}.
For example, the $\bar{q}\g_\mu q$ energy level determined in the left panel of \figref{fig:op_spec} is like one of the energy levels appearing in the light-blue vertical band in the left panel of \figref{fig:kcot_consistency}.
The same kind of false-plateaus also appear in the negative parity $\pi N$ scattering system when only local three-quark operators are used and not also the pion-nucleon-like operators~\cite{Lang:2012db}.
This issue is discussed in more detail in the review article Ref.~\cite{Briceno:2017max}.

Without a large basis of operators that can be used to diagonalize the correlation function and project onto eigenstates of the system, the rough estimate for the length in time one needs to examine the correlation function is set by the inverse gap to the lowest lying excited state, $t\sim1/\D_{10}$.
At the physical pion mass, with $m_\pi L=4$, the excited state of the single nucleon is expected to look roughly like a nucleon-pion system in a $P$-wave
\begin{align}
E_1^{N\pi} &= \sqrt{ m_N^2 + p^2} + \sqrt{m_\pi^2 + p^2}
	+\d E_{\pi N}(p)
\nonumber\\
	&\simeq m_N + \frac{p^2}{2m_N} + \sqrt{m_\pi^2 + p^2} + \d E^{\pi N}(p)
\nonumber\\
	&\simeq 1.03 m_N + 1.9 m_\pi + \d E_{\pi N}(p)\, , \textrm{ for $p\sim\frac{2\pi}{L}$}
\end{align}
where the momentum $p$ will be distorted from a quantized level due to the interactions and in the last line, we have approximated $p$ by the lowest quantized mode.
One observes the excited state gap can be relatively large, $\D_{10}\simeq 2m_\pi$, if the interaction energy is small compared to the pion mass.
Contrast this with the excited state gap to the first two-nucleon scattering state
\begin{align}
E_1^{NN} -2m_N&\simeq \frac{p^2}{m_N} + \d E^{NN}(p)
\nonumber\\
	&\simeq \frac{(2\pi)^2}{m_NL^2} + \d E^{NN}(p)\, ,
	\textrm{ for $p\sim\frac{2\pi}{L}$}\, ,
\end{align}
where again, we have approximated the momentum by the first quantized mode, in which case numerically, $\D_{10}\sim m_\pi / 3$.
Some in the literature advocate the need for having $m_\pi L\sim8$ rather than 4~\cite{Beane:2010hg,Beane:2011iw,Briceno:2013bda}, which would lead to another factor of 4 suppression in the gaps between excited states.
As the volume is increased, not only are the excited state gaps reduced but the number of low-lying excited states below the inelastic threshold increases: above, we only considered the first elastic scattering state.
At the physical pion mass, for $m_\pi L=4$, there are three states (including the ground state) below the inelastic pion-production threshold ($E=2m_N + m_\pi$) while for $m_\pi L=8$, there are 12 states below the inelastic threshold.

Responding to Ref.~\cite{Iritani:2016jie}, Yamazaki et al.~\cite{Yamazaki:2017euu,Yamazaki:2017jfh} carried out a high statistics calculation with two orders of magnitude larger statistics, since, as they comment, the wall source results take the longest Euclidean time for the ground state to saturate the correlation function.  They show a clear consistency between both the wall and smeared sources, and notably it is the wall source which becomes consistent at late times with the smeared source value that is reached at earlier times, provided enough statistics are used to resolve the correlation function at late time.
Unfortunately, the results were not determined on the same set of correlation functions as in Ref.~\cite{Iritani:2016jie} and so a direct comparison can not be made.

NPLQCD responded~\cite{Beane:2017edf,Wagman:2017tmp} by pointing out that the scattering states in their calculation have a large volume dependence, while the ground state energies (which are negatively shifted from threshold) are essentially volume independent, see \figref{fig:op_spec}. This is the expected finite volume behavior for physical bound states versus scattering states.
While very compelling, these arguments are not proof of correctness.
If the number of elastic states creating the fake plateau were fixed as the volume increased, then it would require a practically unbelievable conspiracy for the fake plateau to be volume independent.  But, as we discussed above, as the volume is increased, both the gap to the excited elastic scattering states reduces and the number of states that appear in a fixed energy window, and therefore the number of states potentially contributing to a ``false plateau", grows significantly.
In the $m_\pi\sim800$~MeV NPLQCD results~\cite{Beane:2012vq,Beane:2017edf,Wagman:2017tmp}, with quoted masses in MeV of~\cite{Beane:2012vq} $M_N = 1633.43(55)$ and $m_\pi=806.93(22)$, the approximate gap to the first excited state, $\D_{10}$ and the number of excited states below the inelastic pion production threshold ($2M_N + m_\pi$) are given by
\begin{equation*}
\addtolength{\tabcolsep}{4pt}
\begin{tabular}{cccc}
$L$& $n_{\rm states}$& $\D_{10}[{\rm MeV}]$ & $1 / \D_{10}[{\rm fm}]$\\
\hline
24& 12& 76.5& 2.58\\
32& 21& 43.4& 4.56\\
48& 47& 19.3& 10.2
\end{tabular}
\addtolength{\tabcolsep}{-4pt}
\end{equation*}
The lowest lying non-interacting excited state levels are superimposed on the NPLQCD effective mass data in the right panel of \figref{fig:op_spec}, indicating a very dense set of low-lying elastic scattering excited states.

This observation is also not proof that the NPLQCD results are wrong.  Rather, we are providing a plausibility explanation for how the effective masses in these systems can be misleading.
Take a similar situation as an example, that of excited state contamination to single nucleon matrix elements.
Using $\chi$PT to provide a model of the excited states, B\"{a}r demonstrated that the excited state contamination results in an over-estimation of the axial, scalar and tensor charges for typical source-sink separation times utlized in the computations~\cite{Bar:2016uoj,Bar:2017kxh}.  This result utilized the first pion-nucleon excited state summed over the first few momentum modes allowed in the finite volume.
In contrast, Hansen and Meyer demonstrated that if one uses a larger tower of excited states then the extracted matrix elements will be underestimated by a comparable amount~\cite{Hansen:2016qoz}.  Two different theoretical models of excited states, both rooted in the underlying physics, predict the opposite sign volume correction and the ``truth'' likely lies somewhere in between.  These results are a demonstration that the change in the number of relevant excited states can have a dramatic impact on the inferred long-time value of the matrix element.  As the volume is increased and more excited states become relevant, it is plausible that the systematic bias would be statisticaly volume independent.

Clearly, in order to determine the two-nucleon interactions, one must suppress the single nucleon inelastic states, such that a signal can be obtained early in Euclidean time before the noise overwhelms the correlation functions, but perhaps more important, it is imperative to have interpolating operators that highly suppress all but one of the allowed elastic scattering momentum modes either through a diagonalization of an operator basis or from very good operators.
Otherwise, one would have to determine with sub-percent level precision the two-nucleon energy at late Euclidean time, but this is practically impossible given the exponentially degrading S/N for the two-nucleon system.
What is required are more dedicated studies of these two-nucleon calculations with enough control over the overlap of the operators onto the states so that more definitive conclusions can be drawn.

There are two additional points of note.
First, HAL QCD has demonstrated that, if instead of individually projecting the final state nucleons to definite momentum at the sink, one instead uses the spatial wavefunction determined from solving their potential, then the spectrum from the \luscher analysis agrees with the spectrum obtained by their potential~\cite{Iritani:2018vfn}.
While this is interesting, it could also be a symptom of the unresolved systematic uncertainty associated with the creation/annihilation operators used that HAL QCD has belabored and in this case, you get out what you put in.

Second, related to this apparent source operator dependence of the spectrum (key point \ref{item:source_dependence}, above), it has been known for some time that the local hexa-quark interpolating operator used as a source, in which all 6 quarks originate from the same spacetime location, is sub-optimal for coupling to the two-nucleon scattering states.  One example of this is that it is observed that the overlap of such operators onto the excited states is comparable or larger than the overlap of the operator onto the ground state.
Such sources also do not allow one to compute most of the higher partial waves, which in the NN systems all have large phase shifts.
In order to address these issues, CalLat performed a calculation at $m_\pi\sim800$~MeV (the same configurations with $L=24,32$ as NPLQCD~\cite{Beane:2012vq}) in which the two nucleons were displaced from each other and arranged in such a way to try and maximize the overlap with the maximal number of low-lying representations of the cubic group (low-partial waves)~\cite{Berkowitz:2015eaa}.
They observed:
\begin{enumerate}
\item When the two-nucleons at the source were displaced, the overlap factor for the lowest energy state became dominant;
\begin{enumerate}
\item The lowest energy state in the ${}^3{\rm S}_1$ channel had a significantly smaller negative energy shift and was consistent with either a shallow bound state or a continuum scattering state;
\item The lowest energy state in the ${}^1{\rm S}_0$ channel was consistent with a scattering state and not a bound state;
\end{enumerate}
\item When the source nucleons were created at the same space-time location as in the NPLQCD calculation~\cite{Beane:2012vq}, the lowest energy state is consistent with a deeply bound state in both channels.
\end{enumerate}
CalLat speculated the deep bound state might have a large overlap with a compact hexa-quark operator but not with the diffuse non-local operator and conversely, the shallow-bound state has a large overlap with the non-local operator but not the hexa-quark operator.  This speculation was motivated by the $\D t\simeq1$~fm length of time in which there was a multi-sigma splitting between the energy levels associated with the different operators.
HAL QCD pointed out that the phase shift analysis performed by CalLat~\cite{Berkowitz:2015eaa} was flawed as it imposed an unphysical sign of the residue of the more deeply bound state (which may also lay outside the range of applicability of the effective range expansion).  While the application of the effective range was flawed, this does not alter the conclusion about the possible existence of a second negatively shifted energy level.  Without new results, more definitive conclusions can not be drawn.

\bigskip\noindent
Regarding potential issues with the HAL QCD method, the collaboration has been addressing
some of the criticisms about the lack of quantitative control over the various systematics in the potential method. These are highlighted in a paper studying the systematics of a calculation of the potential and scattering parameters for the $\Xi\Xi$ system~\cite{Iritani:2018zbt}.
Away from the $SU(3)$ flavor-symmetric point, assuming the light quark masses are smaller than the strange quark mass, the $\Xi\Xi$ channels are stochastically less noisy than the two-nucleon channels, which should make it easier to study the systematic effects.
The main systematics to be addressed stem from excited state contamination, finite volume, and the cutoff in the gradient expansion, \eqnref{eq:halqcd_potential}. In their work, these points are each studied and no clear problems are noted, however, there is still room for additional debate as to whether these systematics have been fully controlled.

\begin{figure*}
\includegraphics[width=\textwidth]{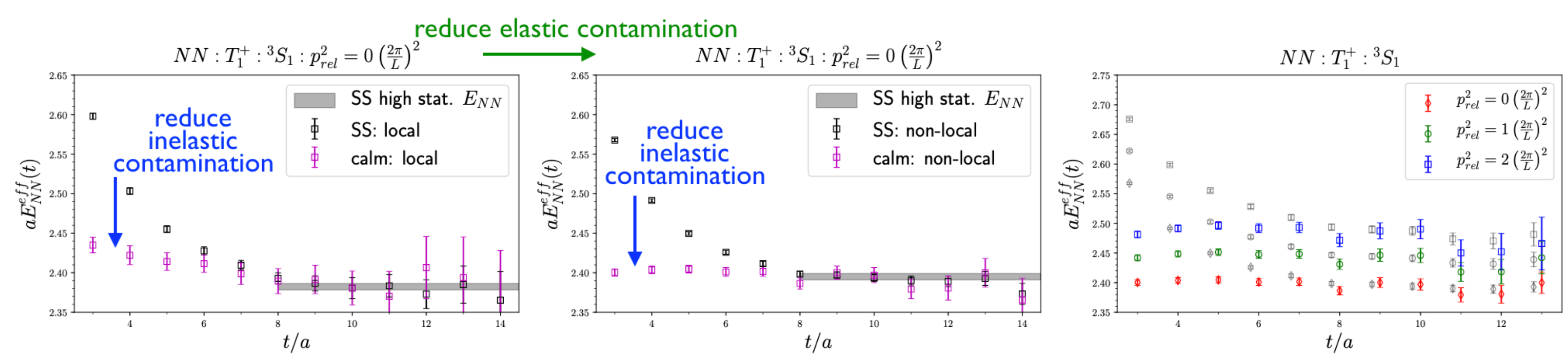}
\caption{\label{fig:nn_mp}
Studies of the effects of two-nucleon elastic excited state and inelastic single-nucleon excited state contamination and source operator dependence for two-nucleon correlators from Ref.~\cite{Berkowitz:2017smo}. In the two plots on the left, the gray data shows the time dependence of the two-nucleon correlators using only one optimized single-nucleon operator, while the purple data shows the time dependence of the two-nucleon correlators using instead a linear combination of single-nucleon operators produced from the Matrix Prony technique for diagonalizing operators~\cite{Beane:2009kya,Beane:2009gs,Beane:2009py}. Clear contamination from single-nucleon inelastic excited states is seen even at late times in the absence of diagonalization techniques for using multiple single-nucleon operators. Reduction of two-nucleon excited states can be seen when comparing the left (local two-nucleon operator) to the right (spatially displaced operator). The plot on the far right shows the clear separation of multiple two-nucleon excited states using the local (gray data) and spatially displaced (colored) operators projected onto different relative momenta. While the displaced operators give a far cleaner signal early in time, the local operator agrees with these energy levels at late times, showing that the energy levels also display operator independence.}
\end{figure*}

For the issue of excited states, as discussed above, the time-dependent potential method, \eqnref{eq:halqcd_improved} in principle offers the relaxed requirement of only needing to sufficiently suppress inelastic excited states with Euclidean time, as opposed to the \luscher method requirement of single-state saturation. While this method certainly reduces worry about one's ability to correctly resolve the closely spaced two-nucleon elastic scattering states, in practice, while the community has been exploring the engineering of good single-nucleon operators for well over a decade, it has been difficult to sufficiently suppress single-nucleon inelastic excited states, even for Euclidean times exceeding ~$\sim1$~fm, except through the linear combination of multiple operators~\cite{Beane:2009kya,Beane:2009gs,Beane:2009py}. It has been shown by the CalLat collaboration~\cite{Berkowitz:2015eaa} that clean projections onto elastic scattering states might be made simply through the use of different geometric configurations of the two-nucleons at the source (to the extent at which we trust that these are the correct energy levels, see above), while linear algebra methods are necessary to reduce the much larger inelastic state contamination, even at late times~\cite{Berkowitz:2017smo,Berkowitz:2019yrf} (see Figs.~\ref{fig:nn_mp}).

%-- FIGURE NN MP  ----------------------
%-----------------------------------------------

HAL QCD is also investigating the application of their method to more challenging channels, such as $I=1\ \pi\pi$ with the $\rho$ resonance.
Their first application uncovered issues with at least the application of smeared sinks when trying to map out the phase shift in this resonant channel~\cite{Kawai:2018hem}.
However, a follow up publication found that with heavy pion masses, for which the $\rho$ meson is deeply bound, the potential method and traditional methods agree on the spectrum~\cite{Akahoshi:2020ojo}, albeit it with a large quoted systematic uncertainty.
More recently, they also studied the $S$-wave kaon-nucleon interactions~\cite{Murakami:2020yzt}.
For the existing applications of the potential to two-baryon systems, the time at which the potential is determined is earlier than the time at which the effective mass of the single nucleon correlation function plateaus, and therefore, there is an implicit assumption that in the ratio correlation function, \eqnref{eq:nn_n_ratio}, the single nucleon excited states in the numerator and denominator cancel to the point of being numerically negligible when determining the potential.  So far, it has not been explicitly demonstrated that this cancellation is exact enough as when the potential is constructed with larger Euclidean times, the noise becomes sufficiently large that it is not possible to detect a difference.  This does not rule out the possibility of undetected inelastic pollution of the more precisely determined  potential from earlier times.

With the potential method, failure to fully suppress inelastic excited states could show up as source operator dependence, or time dependence of the potential. HAL QCD explores both of these in Ref.~\cite{Iritani:2018zbt}. They find that the overall potential appears independent of time using their wall sources for the single-nucleons by plotting the full potential for several different times and noting no significant deviations. However, no attempt is made to extrapolate to infinite times, nor are data points which are clearly trending in one direction with time taken into account. Points nearby in Euclidean times are highly correlated, and error bars tend to grow with Euclidean time, such that an observable with slow Euclidean time dependence can be easily misinterpreted, much the same as when studying plateaus in the \luscher method. Furthermore, both the sizes of the time deviations and error bars vary with the spatial distance measured in the potential. A full study of excited state contamination might look much like it does for the \luscher method: plateaus are demonstrated for each data point, $V(r)$ over several time ranges with no clear trend in the data pointing to a slow Euclidean time dependence, and/or extrapolations to infinite Euclidean time are performed for each point.

A second demonstration of the lack of excited state contamination in Ref.~\cite{Iritani:2018zbt} comes from comparing the use of two different source operators. It is noted that at earlier times the potential produced using the two sources do not agree, but that they begin to at later times. However, the error bars on the potential from the smeared source have grown significantly at this point, so it is difficult to draw strong conclusions.

A similar argument may be made about studies of finite volume effects. Here again, the potentials are overlaid and no significant difference is seen. While worries about correlated statistical fluctuations are absent when comparing different volumes, the same principle that was presented in the previous paragraph for a satisfactory study of excited state dependence may be applied to finite volume. In particular, a study of the volume dependence for different spatial separations in $V(r)$ is crucial, and this must be done for each physical system of interest. Unfortunately, the tail of the potential at large $r$, which may be crucial for correctly identifying finely-tuned bound states in nucleon-nucleon scattering and is most likely to be affected by finite volume effects, often has the largest statistical error bars. Therefore, establishing volume independence by simply comparing multiple volumes is difficult in this region.
See Sec.~\ref{sec:hobetbox} for a nice treatment of incorporating the periodic boundary conditions into a two-nucleon interaction framework.  Such a construction could be adopted in the HAL QCD potential analysis.

Finally, HAL QCD has investigated the effects of the cutoff in the gradient expansion. In Ref.~\cite{Iritani:2018zbt} they use differences in the potentials calculated using different source operators to calculate the potential to the next order beyond LO (N$^2$LO). From this they conclude that, at least for the wall source, the LO potential is unchanged. An important assumption made here is that deviations from using different sources are due only to differences stemming from the cut off in the gradient expansion, and not to residual contamination from inelastic excited states (i.e. the potentials calculated using two different sources should agree as long as the gradient expansion has converged).
While suggestive, this does not satisfactorily demonstrate convergence.
It would be ideal to independently show control over the inelastic excited state contamination and cut off in the gradient expansion: even if one could demonstrate the correlation function was free of all inelastic excited state contamination at early time, there would still be systematic uncertainties arising from the truncation of the gradient expansion.
Furthermore, particularly because the gradient expansion does not have a well-defined small expansion parameter, showing convergence of a series necessitates at least the explicit calculation of multiple orders, compared independently to each other.

\bigskip
A resolution of the controversy remains unfortunately murky.
Both methods have been defended against the criticism presented about them, though not in a completely satisfactory way.
Irrespective of any potential issues within the HAL QCD potential method, HAL QCD has uncovered issues with the standard \luscher calculations that must be resolved, and which are very suggestive that there are larger systematic uncertainties in the calculations than are currently reported.
It is clear that new ideas and methods are needed to resolve the discrepancy, and additionally, cooperation to coordinate a set of gauge configurations on which all methods are performed, such that the discrepancy can be isolated to the method and its application.
Some examples of further reading on the issues can be found in Refs.~\cite{Yamazaki:2017gjl,Davoudi:2017ddj,Aoki:2017yru,Namekawa:2017sxs,Yamazaki:2018qut,Nicholson:2018laj,Berkowitz:2019yrf}.

Many members of the community agree that a resolution of the issue will require more sophisticated interpolating operators that allow for the construction of a hermitian, positive definite set of correlation functions including operators for which the individual momentum projection is performed at both the source and sink~\cite{Peardon:2009gh,Morningstar:2011ka}.%
%FOOTNOTE
\footnote{A step in this direction was proposed by CalLat, inspired by the Matrix Prony analysis methods used to analyze high statistics NPLQCD results~\cite{Beane:2009kya,Beane:2009py}.  Rather than perform the Matrix Prony rotation after producing the correlation functions, one could first find an optimal single-nucleon through Matrix Prony and insert that nucleon into the two-nucleon contraction routine~\cite{Berkowitz:2017smo,Berkowitz:2019yrf}.
This showed great promise, particularly when the displaced nucleon source operator was used~\cite{Berkowitz:2015eaa}.
However, this only suppresses contamination from the inelastic single-nucleon excited state and does not address the issue of the closely packed elastic scattering states.}
%----------

Such methods have not been used for two-baryon systems, with one recent exception~\cite{Francis:2018qch}, due to the significant increase in numerical cost associated with the method.
The benefit of the method is that it generates a positive definite matrix of correlation functions that can be diagonalized by solving a generalized eigenvalue problem~\cite{Blossier:2009kd}, with each eigenstate given by a linear combination of interpolating operators.
This is the common practice in the LQCD calculations of two-meson systems~\cite{Prelovsek:2013ela,Prelovsek:2013cra,Wilson:2015dqa,Bali:2015gji,Briceno:2016mjc,Briceno:2017qmb,Bali:2017pdv,Alexandrou:2017mpi,Cheung:2017tnt,Culver:2019qtx} (and now a first look at the three meson system~\cite{Woss:2019hse,Horz:2019rrn}).
The momentum projection at the source further provides an additional volume averaging as compared to the methods with spatially fixed nucleon sources, whether they are local or non-local, such that the increased cost of the computations may not be as high as otherwise one might presume.
Such a basis will allow for a much more confident determination of the true spectrum from the LQCD calculations.
Of course, with finite statistics, and a finite operator basis, one can never be 100\% certain of the results, but it will be a significant step forward from the current methods. Furthermore, one will be able to quantitatively address how much the momentum project at the sink works to isolate a single elastic scattering state.

Such a computation of the H-dibaryon was performed with a basis of momentum-space to momentum-space operators, and also compared the results to those arising from a hexa-quark interpolating operator, at a pion mass of $m_\pi\sim960$~MeV~\cite{Francis:2018qch}.
A bound state was observed in the H-dibaryon channel using the momentum space operators just as NPLQCD~\cite{Beane:2012vq} and HAL QCD~\cite{Inoue:2010es,Inoue:2011ai} observe at this heavy pion mass, and the value of the binding energy is consistent with that reported by HAL QCD and not with the deep bound state observed by NPLQCD with $B\sim80$~MeV.
It was observed that the hexa-quark creation operator on its own also gave rise to a bound state, was consistent with the shallow bound state result, but had so much more stochastic noise that conclusions are difficult to be drawn.
While suggestive that the deep bound state may be fictitious, one can not draw a strong conclusion, as the hexa-quark must be included in the same basis as the momentum space operators in order to rule out a deep bound state coupling strongly to the local hexa-quark operator. Furthermore, the lattice ensembles were different, and in this calculation, the strange quark was quenched, so that a direct comparison cannot be made to existing calculations.%
% FOOTNOTE
\footnote{In LQCD, if a quark is quenched, this is equivalent to sending its mass to infinity such that the fermion determinant associated with this flavor becomes a trivial multiplicative constant.  Such techniques were common more than a decade ago as the majority of the HMC cost is associated with computing the fermion determinants.  Quenching introduces an uncontrolled systematic error, but with a possible rigorous connection in the large $N_c$ limit in which the fermion loops decouple~\cite{Chen:2002mj}.}
%----------
In Ref.~\cite{Beane:2002np}, it was shown in quenched and partially quenched LQCD calculations, the long-range part of the potential has an extra term (which vanishes in the QCD limit, $m_q^{\rm val}\rightarrow m_q^{\rm sea}$) that scales as $(m_q^{\rm sea} - m_q^{\rm val.})e^{-m_\pi r}$ rather than the well known Yukawa potential, $\frac{1}{r}e^{-m_\pi r}.$

In order for the broader nuclear physics community to have confidence in LQCD results of two-nucleon interactions, two-nucleon electroweak and BSM matrix elements and more, it is critical to resolve this discrepancy.

% Nuclear Response functions
\subsubsection{Nuclear Structure and Reactions}

One of the main goals of this big endeavor is to be able to predict the structure and reactions of nuclei with a quantitative connection to the Standard Model.  Such matrix elements are related to the search for dark matter, permanent electric dipole moments in light and heavy nuclei, the search for $0\nu\b\b$, hadronic parity violation and more.
Such calculations are numerically more challenging than the two-body interactions and they require additional formalism first introduced by Lellouch and \luscher~\cite{Lellouch:2000pv} to relate the finite volume matrix elements to the infinite volume transition amplitudes, a topic which has gained considerable attention as of late~\cite{Detmold:2004qn,Bernard:2012bi,Briceno:2012yi,Briceno:2014uqa,Briceno:2015csa,Briceno:2015tza,Baroni:2018iau,Briceno:2019nns,Briceno:2019opb,Briceno:2020xxs}.

The main challenge as of now for computing even two-nucleon matrix elements is that first, one must have sufficient control over the two-nucleon interaction calculations, which as we have just discussed, have a number of issues that must be resolved.  In order to relate the finite volume matrix elements to their infinite volume counterparts, one need not only know the phase shift at a given energy, but one must also know the derivative of the phase shift.  For transitions near a resonance or generically, when the interaction is strong, these correction factors can be $\mathrm{O}(1)$~\cite{Briceno:2014uqa,Briceno:2015csa,Briceno:2015tza,Baroni:2018iau,Briceno:2019nns,Briceno:2019opb,Briceno:2020xxs} or larger and so even if all that is required is a 20\% uncertainty on some matrix element, these finite volume effects must be understood and controlled.

Despite these challenges, it is important to push the boundaries, and in particular, the NPLQCD Collaboration has performed a number of proof-of-principle calculations extracting two-nucleon matrix elements from LQCD:
magnetic moments~\cite{Beane:2014ora};
$np\rightarrow d\g$~\cite{Beane:2015yha};
$pp$ fusion~\cite{Savage:2016kon};
the isotensor axial polarizability~\cite{Shanahan:2017bgi};
$2\nu\b\b$ matrix elements~\cite{Tiburzi:2017iux};
the isoscalar scalar, axial and tensor matrix elements~\cite{Chang:2017eiq};
and a first calculation of gluonic matrix elements~\cite{Winter:2017bfs}.
CalLat has performed a preliminary computation of the $I=2$ hadronic parity violating amplitude for the NN system, also with heavy pion masses~\cite{Kurth:2015cvl}.
In each of these cases, the calculations were performed with sufficiently heavy pion masses that NPLQCD has found the initial and final two-nucleon states to be deeply bound.  In this limit, the finite volume formalism mentioned above is not needed, and so it is still not understood exactly how challenging the computations will be.
However, it is encouraging to see such calculations being performed.
Hopefully soon, they will exist with sufficiently light pion masses that a direct connection with nuclear Effective (Field) Theories can be made, as discussed in the next section.

A more complicated application one can envision is the nuclear response to an axial current, as is needed to understand the neutrino-nuclear cross sections for the DUNE experiment~\cite{Acciarri:2015uup,Acciarri:2016crz}.
This program is quite challenging as what is needed is an understanding of the cross section with a momentum transfer of $\mathrm{O}(1 {\rm GeV})$.  This is precisely the region where both perturbative QCD and Effective (Field) Theory are not applicable.
Nevertheless, there are still computations from LQCD that can be instrumental in reducing the systematic uncertainty in the modeling of the nuclear response functions.

First, LQCD can be used to determine the nucleon axial form factor for which a handful of calculations exist at the physical pion mass~\cite{Alexandrou:2017hac,Capitani:2017qpc,Alexandrou:2017hac,Ishikawa:2018rew,Shintani:2018ozy} and extrapolated to the continuum limit~\cite{Rajan:2017lxk,Capitani:2017qpc}.
There remain non-trivial subtleties in understanding the excited state contamination in these calculations, in particular, the nucleon-pion state~\cite{Bar:2015zwa,Bar:2016uoj,Bar:2017kxh,Bar:2019igf}.
The lattice community is beginning to try and incorporate such corrections~\cite{Bali:2018qus,Jang:2019vkm} though only the most recent is in close agreement with the strategy motivated by chiral symmetry constraints~\cite{Bali:2019yiy}.
The second step is to perform calculations of the $N\rightarrow N\pi$ transition form factors (which has been demonstrated with mesons~\cite{Briceno:2016kkp,Alexandrou:2018jbt}) as well as
control over the $\pi N$ scattering states~\cite{Andersen:2017una}.
Next, LQCD calculations of two-nucleon form factors must be performed.
While a full LQCD calculation of the Argon response function will likely never happen, the one-, two- and few-nucleon matrix element calculations can be used as benchmarks that the Monte Carlo event generators can be calibrated against for light nuclear systems, such that some of the model dependence can be removed in the estimate of the full nuclear response function.
They can also be used to constrain few-body corrections used in the nuclear EFT calculations, such as Ref.~\cite{Payne:2019wvy}.

This is a very ambitious program, but we should see exciting progress towards these goals in the exascale computing era.

%%% Nuclear EFT
\section{Nuclear Effective (Field) Theories \label{sec:net}}

The evaluation of properties of light nuclei directly from LQCD will be extremely challenging due to the fermion sign problem, the exponential growth in noise with increasing nucleon number (see Secs.~\ref{sec:lattice_challenges}--\ref{sec:lattice_nn}).   But there is an attractive alternative for creating a controlled theory of finite nuclei based on QCD -- an \eft{} formulated in terms of nucleons (and possibly pions and deltas), for which the onset of the S/N problem can be postponed, if not mitigated all together, based upon the particular many-body method used.
This approach can be successful because nuclei are relatively dilute fermi systems -- the probability of finding clusters of nucleons acting at one time through strong, short-range interactions declines rapidly with the number of nucleons in the cluster.

\efts{} are naturally organized in terms of short-range terms, capturing the unresolved UV physics, from the long-range pion exchange effects and medium range effects arising from multi-pion exchanges etc. The short-range physics can be added back in through an effective-operator expansion, with coefficients that can be fitted either to experimental data, or alternatively taken from LQCD, once LQCD calculations reach the necessary precision in the exascale era. Low-momentum observables can then be predicted with quantifiable errors, determined from the rate of convergence of the operator expansion. This coupling of LQCD and \efts{} is important for rooting our understanding of nuclear physics in the SM but becomes acutely important for observables related to BSM searches for which there is no alternative means of assessing the unknown short-distance physics.

\efts{} are powerful theoretical tools to describe physical processes over a specified range of energy
scales in terms of effective instead of the fundamental degrees of freedom.
The procedure is to construct the most general Lagrangian consistent with the
symmetries of the underlying theory and then compute all Feynman graphs which
contribute to a given process.  ``The result will simply be the most general possible $S$-matrix
consistent with analyticity, perturbative unitarity, cluster decomposition, and
the assumed symmetry principles''~\cite{Weinberg:1978kz}. The EFT
will only be effective if certain conditions are met. First, one
should choose the effective degrees of freedom that are relevant at the energy
scale of interest since this may impact the rate of convergence of the EFT considerably:
\begin{quote}
	``You may use any degrees of freedom you like to describe a physical system, but
	if you use the wrong ones, you'll be sorry.''~\cite{Weinberg:1981qq}
\end{quote}
For example, the small excitation energy of the delta resonance leads to resonance saturation of
pion-nucleon couplings in the delta-less ($\nd$) EFT for nucleons, which renders these LECs
(somewhat) unnaturally large and thus affects the convergence. For nuclear-physics
applications, one might expect a faster convergence of delta-full EFT,
where the delta resonance is  treated as an explicit degree of
freedom in addition to nucleons and pions.

Second, there must be a well-defined perturbative expansion to organize the infinite set of operators and their resulting Feynman diagrams.
The parameter or parameters controlling this perturbative expansion most often arise from ratios
of IR and UV scales, which set the bounds of validity of the EFT.
For example, in chiral perturbation theory (\xpt)~\cite{Gasser:1983yg}, a low-energy EFT of QCD,
the pion mass naturally emerges as an IR scale.

\xpt{} is formulated to respect the approximate chiral symmetry of the QCD
Lagrangian for the light quark flavors. Considering the up and down quarks, and neglecting their masses, the QCD action has a
global $SU(2)_L \otimes SU(2)_R$ chiral symmetry. While the small light quark
masses explicitly break this symmetry, the dominant chiral symmetry breaking
arises from the spontaneous breaking from the QCD vacuum down to the vector
subgroup, which is also known as isospin symmetry for the two light flavors.
This spontaneous chiral symmetry breaking leads to a large splitting of the
degeneracy of the parity partner states, such as the nucleon and the negative
parity nucleon. It also gives rise to the relatively light pions that emerge as
pseudo-Nambu-Goldstone bosons. Their masses, $m_\pi^2 \propto m_l$, vanish as
the average light quark mass, $m_l$, goes to zero (named the chiral limit).

The masses of all other hadrons composed of light quarks, such as the rho-meson
or the nucleons, do not vanish in the chiral limit and are of $\mathcal{O}(1)$~GeV. This
emergent scale of QCD is also
observed to be approximately the UV scale associated with chiral symmetry
breaking, which has been phenomenologically determined to be $\L_{\chi}\sim4\pi
F_\pi$ ($F_\pi\sim92$~MeV). For sufficiently small
external pion momentum of order of the pion mass, \xpt\ gives rise to a
perturbative EFT, where the small parameter controlling the expansion is given by
	\beq\label{eq:eps_pi}
	\e_\pi^2 = \frac{m_\pi^2}{\L_\chi^2}\, .
	\eeq
This results in a rapidly converging perturbative expansion. One can then define a power counting scheme and work to a fixed order in $\e_\pi^2$,
retaining all Feynman diagrams that contribute to a given process up to the
specified order, $n \geqslant 0$. The order one works to, i.e., leading order (LO), next-to-leading
order (NLO) etc., is set by the desired
precision of the result as the theoretical uncertainty arising from the
neglected contributions is expected to scale roughly as $(\e_\pi^2)^{n+1}$,
assuming the asymptotic nature of the expansion has not set in and that there
are no unnaturally large coefficients in the expansion.

The strange quark mass is sufficiently light that it can be advantageous to
formulate a three-flavor variant of \xpt{} based on an approximate $SU(3)$ chiral
symmetry incorporating kaons and the eta pseudo-scalar
meson~\cite{Gasser:1984gg}. $SU(3)$ \xpt{} is less convergent than the $SU(2)$
variant as the kaon mass is significantly heavier than the pion mass and the number of virtual states propagating in loop diagrams grows
with the number of flavors $N_F$.

While the form of \eft{} operators are restricted by the underlying symmetries,
their scale- as well as scheme-dependent coefficients, the LECs, must be determined either
through comparisons with the fundamental theory or by constraining the predicted amplitudes to reproduce experimental data.
A critical aspect of EFTs is that the LECs are process
independent. Therefore, assuming a converging expansion, if the LECs are
constrained by one set of observables, they can be used to predict other
observables. This makes \efts{} powerful and predictive.

Many of the LECs accompany operators which explicitly depend upon the quark
masses and so they cannot be reliably determined through a comparison with
experimental results alone. This is precisely where the interplay between LQCD
and EFT allows us to build a quantitative bridge between QCD and low-energy EFTs
of nuclear physics as we are free to vary the quark masses in LQCD
calculations. Practically, LQCD calculations get more expensive as the light
quark mass is reduced towards the physical value, for the reasons discussed in
Sec.~\ref{sec:lattice}. The common practice has therefore been to perform
calculations with the strange quark fixed near its physical value and with the
light quark masses approaching their physical value from above.%
% FOOTNOTE
\footnote{Another approach has been to hold the sum of the up, down, and strange
quark masses fixed at their physical value and to approach the physical masses
by reducing the light quark masses while simultaneously increasing the strange
quark mass~\cite{Bietenholz:2010jr,Bietenholz:2011qq}.  This approach is also
utilized by the CLS Collaboration~\cite{Bruno:2014jqa}.  While this method has
advantages related to renormalization of Wilson fermions and smooth
extrapolations to the physical mass point, from the \xpt{} point of view, it
forces the use of an extrapolation that depends upon the kaon mass as well as
the pion mass. }
%%%%%%%%%%
In the process, the LECs of $SU(2)$ and $SU(3)$ \xpt{} have been determined, the
results of which are a central part of the FLAG
reviews~\cite{Aoki:2016frl,Aoki:2019cca}.

For simple quantities, LQCD has been reliably used to explore in detail the
range of convergence of the various formulations of \xpt.  This has been
performed for $SU(2)$ \xpt, see for
example Refs.~\cite{Beane:2011zm,Borsanyi:2012zv,Boyle:2015exm} and the
review in Ref.~\cite{Durr:2014oba}. It has been determined that an NLO extrapolation of
pion quantities works well up to $m_\pi^{\textrm{max}}~\sim350$~MeV, provided the
minimum pion mass used is near the physical value.  The \nxlo{2} contributions
can raise the range of convergence a little, but a complete breakdown occurs
beyond $m_\pi\sim500$~MeV~\cite{Durr:2014oba}. For $SU(3)$ \xpt, it has been
observed that the expansion to \nxlo{2} is sufficient to interpolate quantities
about the physical strange quark mass, but that for a reliable determination of
the $SU(3)$ LECs, such as the decay constant, $F_0$ and condensates $\langle
\bar{q} q \rangle$ in the $SU(3)$ chiral limit, one is restricted to values of
the strange quark mass $m_s \lesssim 0.7 \, m_s^{\rm phys}$~\cite{Bernard:2015wda}.
To achieve the desired theoretical precision, $SU(2)$ \xpt{} with external
kaons has been developed~\cite{Roessl:1999iu,Allton:2008pn,Bae:2013tca}.

The inclusion of matter fields in \xpt, such as nucleons,
is the
subject of the next section.
After discussing what we have learned from LQCD about the convergence pattern of baryon \xpt{} (Sec.~\ref{sec:nn_eft}) we turn to the EFT of the two-nucleon sector and its extension to
many-body nuclear physics.
NN EFT and nuclear chiral EFT are often reviewed in the literature.
We therefore in this article focus not on an extensive review, but rather try and highlight specific places where we anticipate NN/chiral EFT making contact with LQCD and specifically, how LQCD may help improve our understanding of the EFTs.  We instead refer the reader to these selected reviews~\cite{Bedaque:2002mn,Epelbaum:2008ga,Machleidt:2011zz,Hammer:2019poc} as well as a very recent and more in depth discussion of the interplay between NN/chiral EFT and many body methods as well as connections to LQCD~\cite{Tews:2020hgp}.
We then provide a more in depth introduction to Harmonic Oscillator Based Effective Theory
(HOBET), a non-relativistic theory of nuclear interactions (Sec.~\ref{sec:hobet}).  We discuss the power counting of HOBET and compare and contrast it with other well-known strategies for many-body nuclear physics calculations
such as the no-core-shell-model (NCSM), Quantum Monte Carlo (QMC), self-consistent Greens functions and lattice EFT, which are extensively discussed in the literature.  We refer the reader to Refs.~\cite{Barrett:2013nh}, \cite{Lynn:2019rdt,Carlson:2014vla}, \cite{Dickhoff:2004xx,Barbieri:2016uib} and \cite{Lee:2016fhn,Lahde:2019npb}, respectively, as well as references therein for more in depth reviews of these methods.

%% Single Nucleon EFT

\subsection{Single Nucleon Effective Field Theory \label{sec:hbchipt}}
In building an EFT of nuclear structure and reactions, it is critical to verify
that the single-nucleon EFT upon which it is built is converging. While a
determination of the NLO LECs of \xpt{} with LQCD is now a standard component of
the FLAG reviews~\cite{Aoki:2016frl}, the situation is different for baryons:
the most recent FLAG review~\cite{Aoki:2019cca} is the first to include baryon
quantities.

The first challenge in incorporating nucleons in an EFT is that their masses do not vanish in the
chiral limit, which complicates the
power counting and the renormalization of the LECs~\cite{Gasser:1987rb}.
Multi-loop corrections can generate finite contributions to the $S$-matrix
elements that scale with arbitrarily large powers of $M_0 / \L_\chi$, where
$M_0$ is $\lim_{m_\pi \rightarrow 0} M_N$.

To overcome this challenge, Jenkins and Manohar developed heavy-baryon \xpt{}
(\hbxpt)~\cite{Jenkins:1990jv}, inspired by heavy-quark effective
theory~\cite{Georgi:1990um}, where a non-perturbative expansion is formulated
about the rest mass of the nucleon in the chiral limit.  In this way, the
power counting is restored, as the nucleon mass only appears in inverse powers
in the chiral Lagrangian, as well as in radiative loop corrections. The original
formulation utilized the three-flavor expansion about the $SU(3)$ chiral limit,
incorporating the interactions of the pions, kaons, and eta with the hyperons.
It also incorporated the decuplet resonances as explicit degrees of freedom,
given that the mass gap between the octet and decuplet baryons is comparable to
the pseudoscalar meson masses.

A substantive difference in the chiral expansion of matter fields from the
pseudoscalar mesons is that the loop corrections for matter fields generate odd
powers of the meson masses.  For example, the first non-trivial correction to
the nucleon mass from virtual pions is given by
\begin{equation} \label{eq:mn_mpi3}
	\d M_N^{(3)} = -\frac{3\pi g_A^2}{2} \frac{m_\pi^3}{(4\pi F_\pi)^2}
                 = -\frac{3\pi g_A^2}{2} \e_\pi^3\ \L_\chi  \, ,
\end{equation}
which is non-analytic in the quark mass as $m_\pi^2 \propto m_l$.
We also observe that the coefficient of this NLO correction to the nucleon mass
is large, $3\pi g_A^2 / 2 \simeq 7.6$.  This is a common feature of baryon \xpt{} such that the
convergence is challenged, not only by a larger expansion
parameter, $\e_\pi$, as compared to $\e_\pi^2$ for pions, but also the
coefficients appearing in the expansion tend to be larger than $\mathcal{O}(1)$.

When considering non-static quantities, such as form factors with some
characteristic soft momentum scale $q$, it is common to denote $Q$ as a generic
small expansion parameter representing both, the pion mass and the soft momentum scale in this
double expansion. The
chiral expansion is then expressed in powers of
\begin{equation} \label{INTRO:eq:Q_n}
	Q = \frac{q}{\L_\chi} \sim \frac{m_\pi}{\L_\chi}\, ,
\end{equation}
where typically, the soft momentum scale is counted as the same order as the
pion mass. Working to $\mathcal{O}(n)$ implies determining all corrections which
scale as $Q^n$ with respect to the LO contribution ($n=0$). The corrections at different orders
do not
necessarily map to the same order for different observables. For example, the LO
correction to the nucleon mass scales as $m_\pi^2 / \L_\chi = \L_\chi Q^2$, the
NLO, Eq.~\eqref{eq:mn_mpi3}, as $\L_\chi Q^3$, the \nxlo{2} as $\L_\chi Q^4$
etc. In the literature, it is common to encounter results described as order
$Q^n$ and/or N$^n$LO.

The non-relativistic treatment of the nucleon in HB$\chi$PT leads to a power
series expansion of all operators in inverse powers of the nucleon mass in the
chiral limit, $M_0$, as well as the chiral symmetry breaking scale, $\L_\chi$. Each new operator in this expansion is accompanied by an
unknown LEC that must be determined.  Some of the LECs can be exactly
determined by enforcing Lorentz invariance perturbatively, which goes under the
name reparameterization invariance~\cite{Luke:1992cs}: the
coefficient of the NLO kinetic operator is exactly fixed by the LO kinetic
operator
\begin{equation}
	\mathcal{L}_{\partial} = \overline{N} i v \cdot D N
        -\overline{N} \frac{D_\perp^2}{2M_0} N\, ,
\end{equation}
where $v_\mu$ is the four-velocity of the nucleon, $D_\mu$ is the chiral
covariant derivative, and $D_\perp^2 = D^2 - (v\cdot
D)^2$~\cite{Jenkins:1990jv,Luke:1992cs}. In the rest frame of the nucleon
$v^T_\mu = (1,0,0,0)$. This reparameterization invariance also ensures that the
propagator for the nucleon has the expected non-relativistic kinetic correction,
when such terms need to be included:
\begin{equation}\label{eq:nlo_hb_prop}
	S_N(p) = \frac{i}{v\cdot p - \frac{\mathbf{p}^2}{2M_0} +i\varepsilon}\, .
\end{equation}

The original formulation of HB$\chi$PT was constructed for
both the octet and decuplet baryons, as an expansion about the $SU(3)$ chiral
limit. LQCD calculations of the baryon
spectrum~\cite{WalkerLoud:2008bp,Ishikawa:2009vc} and meson-baryon scattering
lengths~\cite{Torok:2009dg} have established that this $SU(3)$ chiral expansion
does not converge for these observables at the physical value of the strange-quark mass for a
range of light-quark masses spanning the physical value. Given
the larger value of the kaon and eta masses, and an expansion parameter that
scales linearly rather than quadratically in these masses, this may not be
surprising. However, this theory has enjoyed much phenomenological success and
so it is instructive to use LQCD results to understand when the patterns of
$SU(3)$ flavor symmetry breaking are expected to yield more than a qualitative
guide to physical phenomena.

One approach has been to formulate observables based upon the combined
expansions of large $N_c$~\cite{tHooft:1973alw,Witten:1979kh} and $SU(3)$ flavor~\cite{Dashen:1993as,Dashen:1993ac,Jenkins:1993zu,Dashen:1993jt}. In the
large $N_c$ limit, there is an extra contracted spin-flavor symmetry which
further constrains the LECs.  The HB$\chi$PT
Lagrangian constructed to incorporate these constraints is given in
Ref.~\cite{Jenkins:1995gc}.
In the large $N_c$ limit, the spin-1/2 and spin-3/2 states become degenerate, leaving the decuplet as stable asymptotic states.
Their explicit inclusion is also necessary to observe the predicted suppression of certain radiative corrections as well~\cite{FloresMendieta:2000mz}.

The combined $SU(3)$-flavor and large-$N_c$ expansions lead to a predicted scaling of linear combinations of octet and decuplet masses.  These were compared to LQCD results over a range of pion masses from 300--800 MeV, with the strange-quark mass held fixed near its physical value~\cite{WalkerLoud:2008bp} with good qualitative agreement with these predictions~\cite{Jenkins:2009wv}.  This was studied in more detail in Ref.~\cite{Fernando:2014dna}.
Another interesting feature to emerge was the observation of evidence for non-analytic light-quark mass dependence in the spectrum.
A restriction to a subset of the mass relations led to a determination of the $SU(3)$ axial couplings consistent with phenomenological determinations, $D=0.70(5)$ and $F=0.47(3)$, and a convergence pattern that is not immediately problematic~\cite{WalkerLoud:2011ab}.
Such studies led to a renewed interest in performing
LQCD calculations with varying numbers of
colors~\cite{DeGrand:2012hd,DeGrand:2013nna,DeGrand:2016pur}, which is required
for a more extensive study of such mass relations, the convergence of
HB$\chi$PT, and a quantitative understanding of the patterns of $SU(3)$ flavor
and large $N_c$ symmetry breaking in QCD.

A practical approach to understand properties of hyperons has been to formulate
$SU(2)$ EFTs for these states with external strange
quarks~\cite{Roessl:1999iu,Beane:2003yx,Tiburzi:2008bk,Jiang:2009sf,Jiang:2009fa}.
The convergence pattern for the baryons should improve as the strangeness is
increased since the value of the axial couplings to these hyperons decreases,
thus reducing the strength of the pion-cloud corrections to observables. Such an
$SU(2)$ flavor formulation for hyperons was not favorable during the advent of
HB$\chi$PT since there was insufficient experimental data to reliably constrain
the extra LECs, which result for each strangeness channel. In the era of
precision results from LQCD with near-physical pion masses, we can now reliably
constrain these EFTs and use them to make a number of predictions.

Shortly after HB$\chi$PT was formulated, the chiral Lagrangian for nucleons and pions based upon the approximate $SU(2)$ chiral symmetry was constructed~\cite{Bernard:1992qa}. The delta-resonances were not explicitly included and later, it was argued that below the resonance threshold, the dominant effects of these states could be incorporated in certain LECs through resonance saturation~\cite{Bernard:1995dp}. We denote this theory $SU(2)$
HB$\chi$PT($\nd$).
Subsequently, $SU(2)$ HB$\chi$PT including explicit delta-resonances was formulated, which has come to be known as the small-scale expansion (SSE)~\cite{Hemmert:1996xg,Hemmert:1997ye}, in which the delta-nucleon mass splitting, $\D = M_\D - M_N \approx 2 \, m_\pi$, is treated as a small scale with the same counting,
\begin{equation}
	Q = \frac{q}{\L_\chi}
	\sim \frac{m_\pi}{\L_\chi} \sim \frac{\D}{\L_\chi}\, .
\end{equation}

There is a long list of literature discussing the impact of including or not
including explicitly delta-resonances in the EFT without a clear consensus; especially, one
interesting idea of counting $\d = \frac{m_\pi}{\D} \sim
\frac{\D}{\L_\chi}$ in the power counting~\cite{Pascalutsa:2002pi}.  Certainly,
when the energy scale of the external probe, such as in Compton scattering,
approaches the resonance region, a static treatment of the cross sections with
the effects of the delta only contained within LECs is bound to fail. On the
other hand, if one does not utilize large $N_c$, the delta-resonances are not
asymptotic states and so care must be taken to incorporate them in the EFT~\cite{Beneke:2003xh}.
See Ref.~\cite{Griesshammer:2012we} for a review of nucleon Compton scattering in
EFT which touches on these issues.

Including the delta-resonances in HB$\chi$PT is not the only
challenge the theory faces. One issue is the determination of the scalar matrix
element at non-zero momentum transfer, which is often referred to as the sigma
term. In this process, a new kinematic small scale enters the problem, $1 -
\frac{t}{4m_\pi^2}$ with the Mandelstam parameter $t = q^2$. Neglecting this
small parameter leads to an unphysical kinematic singularity.  To remedy the
problem, one is required to re-sum the NLO kinetic corrections to the nucleon
such that the leading recoil term is included and the propagator is given by
Eq.~\eqref{eq:nlo_hb_prop}.

Alternatively, an IR regularization was introduced for
baryon $\chi$PT. The full relativistic nucleon field and an infrared regulator
are used to separate the long-range pion effects with non-analytic dependence
upon the quark mass from the short-range contributions, which exhibit only a
polynomial quark-mass dependence~\cite{Becher:1999he}. Issues with carrying out this IR regularization beyond one-loop led to a generalization that works for multi-loop integrals and is called the extended-on-mass-shell
regularization~\cite{Fuchs:2003qc,Schindler:2003xv,Schindler:2003je}. It has
been observed, but not understood formally, that these new regularization
schemes tend to lead to improved convergence patterns as compared to HB$\chi$PT
for pion masses heavier than the physical values, and also for $SU(3)$
HB$\chi$PT~\cite{MartinCamalich:2010fp,Ren:2012aj}. It has also been observed
that the use of a finite-range regulator, in the form of a dipole
regulator~\cite{Donoghue:1998bs}, also improves the convergence pattern of
HB$\chi$PT~\cite{Leinweber:1999ig,Leinweber:2003dg}.

Lattice QCD is the perfect tool to help quantify which of these many EFTs and/or
regularization schemes is preferred by QCD near the physical value of the pion
mass. Over the last few years, this process has begun and most recently, we now
have constraints from LQCD with precise calculations at the physical pion mass.
In one of the earliest comparisons of baryon $\chi$PT with LQCD results at
relatively light pion masses, an interesting phenomenon was observed: to a very
good approximation, independent of the discretization method, the nucleon mass
scales linearly with the pion mass%
%FOOTNOTE
\footnote{Pseudo-numerologists will be amused to note that all the octet baryon
masses exhibited linear pion-mass dependence, at a fixed strange quark mass,
with slopes consistent at 1-sigma with $2/3$, $1/2$ and $1/3$ for the $\L$, $\S$ and
$\Xi$ baryons respectively~\cite{WalkerLoud:2008bp}.} %%%%%%%%%%
\begin{equation}
	M_N \simeq 800 \MeV + m_\pi \, ,
\end{equation}
correctly predicting the nucleon mass at the physical point, $M_N = 938\pm9$~MeV~\cite{WalkerLoud:2008bp,WalkerLoud:2008pj}.
New results showed the same tendency with pion masses as light as $m_\pi\sim175$~MeV~\cite{Walker-Loud:2013yua,Walker-Loud:2014iea}.

We know that this linear pion mass dependence cannot persist to arbitrarily
small pion masses.  The leading prediction for the pion mass dependence of the
nucleon mass is $m_\pi^2$, since $m_\pi^2 \propto m_l$, in all known, well
founded theoretical descriptions of the nucleon. This linear pion mass
dependence has at least two significant implications, however. If this
dependence holds between the physical pion mass and $m_\pi\sim200$~MeV, the
predicted value of the pion-nucleon sigma term is $\s_{\pi N}\simeq
67\pm5$~MeV~\cite{Walker-Loud:2013yua}, which is in sharp contrast with other
LQCD determinations~\cite{Bali:2012qs,Durr:2015dna,Yang:2015uis,Bali:2016lvx,Alexandrou:2019brg} and consistent with the phenomenological
determination that uses low-energy pion-nucleon scattering
constraints~\cite{Hoferichter:2015dsa}.  There might be underestimated
systematic uncertainties in the LQCD determination of the sigma term, or it
might be demonstrated that this phenomenological \textit{ruler} fit is not
consistent with sufficiently precise computations of the nucleon mass.

The more problematic implication is that in order to
reproduce such a linear pion-mass dependence, the contributions to the nucleon
mass from different orders in the chiral expansion have to conspire in such a
way that there is almost no curvature.  This can only happen through a delicate
cancellation, which also arises from the cancellation of large
terms~\cite{WalkerLoud:2008bp,WalkerLoud:2008pj}. This does not bode well for
$SU(2)$ HB$\chi$PT having a well-converging expansion.

We can use a recent determination of $g_A$ over a range of pion masses $130 \MeV
\lesssim m_\pi \lesssim 400 \MeV$~\cite{Chang:2018uxx} to explore the convergence
in more detail. For example, downloading the Jupyter Notebook provided with this
work,%
% FOOTNOTE ---------------------------------------------------------------------
\footnote{Clone the CalLat github repository at \url{https://github.com/callat-qcd/project_gA}.}
%-------------------------------------------------------------------------------
one can run the \texttt{xpt\_3}
and \texttt{xpt-full\_4} analyses by selecting them from the
\texttt{switches["ansatz"]["type"]} list. These two analyses use the complete
\nxlo{2}~\cite{Kambor:1998pi} and \nxlo{3}~\cite{Bernard:2006te} $SU(2)$
HB$\chi$PT($\nd$) extrapolation formulae.  The \nxlo{2} expression is given by
the first line of the following \nxlo{3} expression,
\begin{align}
	g_A &= g_0 +c_2 \e_\pi^2 - \e_\pi^2 (g_0 + 2g_0^3)\ln (\e_\pi^2)
	+ g_0 c_3 \e_\pi^3
\nonumber\\
	&\phantom{=} +\e_\pi^4 \bigg[ c_4  + \tilde{\g}_4 \ln(\e_\pi^2)
%\nonumber\\&\phantom{=} \qquad
	+\left( \frac{2}{3}g_0 + \frac{37}{12}g_0^3 +
	4g_0^5\right) \ln^2 (\e_\pi^2) \bigg]\, .
\end{align}
%-------------------------------------------------------------------------------
% gA convergence table
\begin{wraptable}{R}{0.58\textwidth}
%\begin{table}
\caption{\label{tab:ga_expansion}
Order-by-order contribution to $g_A$ at the physical pion mass resulting from
\nxlo{2} and  \nxlo{3} $SU(2)$ HB$\chi$PT extrapolation  of the  results in
Ref.~\cite{Chang:2018uxx} without explicit delta
degrees of freedom.}
\centering
	\begin{tabular}{c|cccc}
		\hline\hline
		N${}^n$LO& LO& NLO& \nxlo{2}& \nxlo{3}\\
		\hline \nxlo{2}& 1.237(34)& -0.026(30)& 0.062(14)& -- \\
		\nxlo{3}& 1.296(76)& -0.19(12)\phantom{0}& 0.045(63)& 0.117(66)\\
		\hline\hline
	\end{tabular}
%\end{table}
\end{wraptable}
%-------------------------------------------------------------------------------
In Table~\ref{tab:ga_expansion}, we list the order-by-order contributions from
these two extrapolations at the physical pion mass.
As with the nucleon mass, the coefficient of the NLO $\ln$ term is also large $g_0 + 2g_0^3 \simeq 5$.
At face value, these results are indicative of a failing perturbative
expansion. In the \nxlo{2} analysis, while the NLO contribution is significantly
smaller than the LO contribution, the \nxlo{2} contribution has opposite sign
and is twice as large as the NLO contribution. Adding a higher order
contribution makes the convergence pattern worse.  The NLO contribution
becomes an order of magnitude larger, the \nxlo{2} contribution stays roughly
the same, but the \nxlo{3} term is opposite in sign from NLO and almost as
large.
The \nxlo{3} analysis is a zero-degree-of-freedom fit, however, so definite conclusions can not be drawn, but if these results hold, it signals a breakdown of
$SU(2)$ HB$\chi$PT($\nd$), even at the physical pion mass.

\begin{wrapfigure}{R}{0.5\textwidth}
\centering
%\begin{figure}
\includegraphics[width=0.5\textwidth]{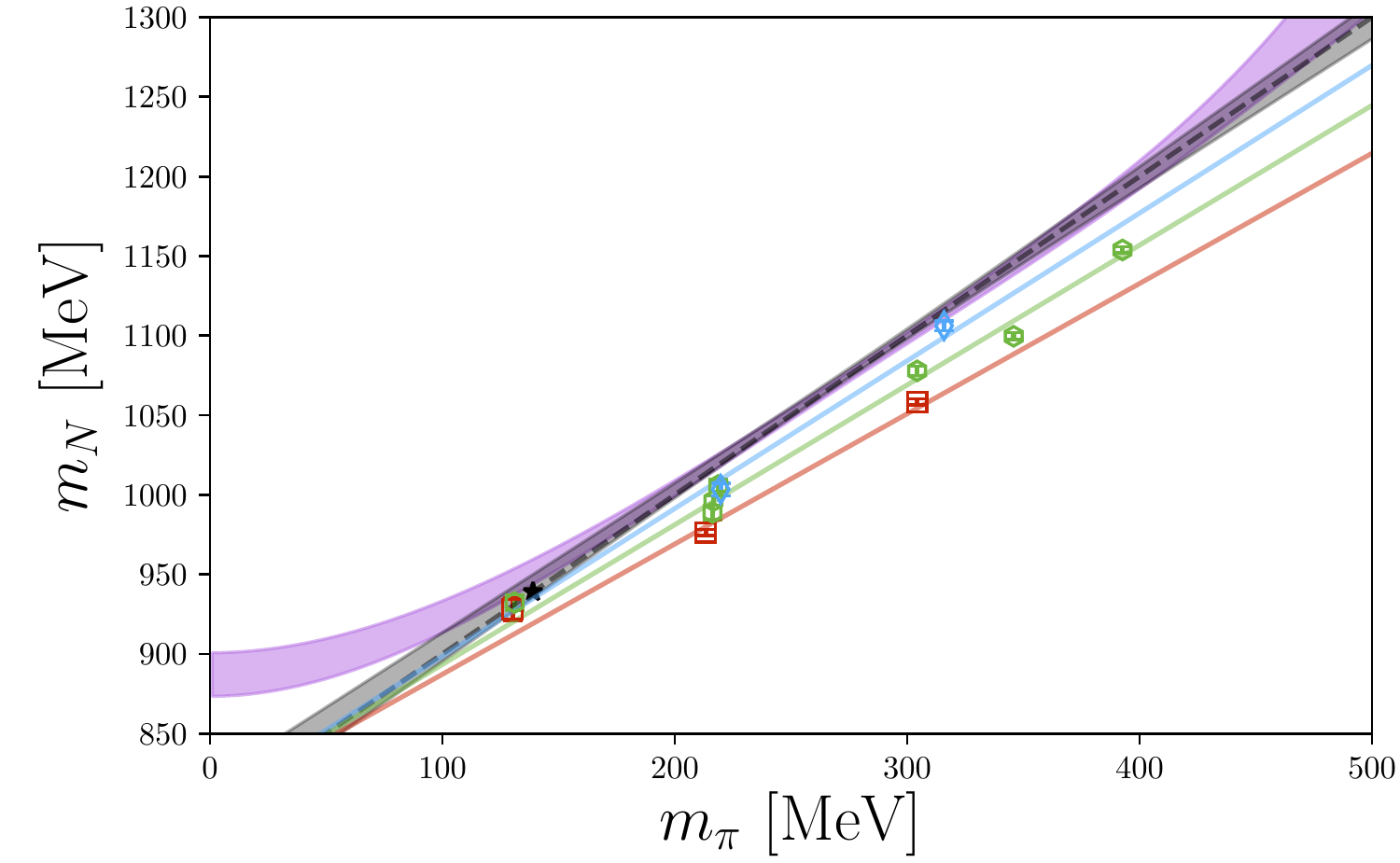}
\caption{\label{fig:mn_v_mpi}
Extrapolation analysis of $M_N$ versus the pion mass.
The nucleon mass results are those determined in the published analysis of $g_A$~\cite{Chang:2018uxx}, and the two-point correlation functions used are available with that publication.  However, there has not yet been a dedicated publication of the nucleon mass results themselves, but one is in preparation by CalLat with updated statistics and new ensembles.
The black dashed line is the fit from
Refs.~\cite{WalkerLoud:2008bp,WalkerLoud:2008pj} in 2008.  The curved magenta band is
the result of an \nxlo{2} $SU(2)$ HB$\chi$PT($\nd$).  The gray band is the
continuum-extrapolated result arising from a linear-in-$m_\pi$ analysis using
Eq.~\eqref{eq:mn_ruler_enhanced}.  The solid red, green, and blue lines are the
central values from this analysis plotted at the three lattice spacings used in
Ref.~\cite{Chang:2018uxx}. The black star is the physical value of the nucleon
mass, which is not used in any fit.
}
%\end{figure}
\end{wrapfigure}

The yet-to-be published results generated with those of
Ref.~\cite{Chang:2018uxx} can also be used to explore the convergence pattern of
the nucleon mass and explore the linear pion mass dependence, which we show in
Fig.~\ref{fig:mn_v_mpi}. We observe that these LQCD results are consistent with
both the \nxlo{2} $SU(2)$ HB$\chi$PT without delta extrapolation as well as the
linear (in $m_\pi$) extrapolation.  In this case, the slope is allowed to vary
as a function of the discretization scale
\begin{equation}\label{eq:mn_ruler_enhanced}
	M_N = M_0 + \a_1 m_\pi + \b_1 m_\pi a^2\, ,
\end{equation}
where $a$ is the lattice spacing. As
before~\cite{WalkerLoud:2008bp,WalkerLoud:2008pj}, it is observed that over a
large range of pion masses, the HB$\chi$PT fit conspires to produce this very
linear dependence observed in the LQCD results.
In order to place better constraints on the range of convergence of $SU(2)$
HB$\chi$PT($\nd$), a combined analysis of $M_N$ and $g_A$ is currently being
performed, as the LECs of the two observables become coupled at NLO and beyond.

%%%%%%%%%%%%%%%%%%%%%%%%%%%%%%%%%%%%%%%%
% MN vs Mpi figure
%%%%%%%%%%%%%%%%%%%%%%%%%%%%%%%%%%%%%%%%
%%%%%%%%%%%%%%%%%%%%%%%%%%%%%%%%%%%%%%%%

An exception to this poor convergence pattern is the QCD (vs QED) corrections to $M_n-M_p$.  In this quantity, the NLO contributions cancel, leaving~\cite{Tiburzi:2005na,WalkerLoud:2009nf}
\begin{align}
\d M_N &= \d \bigg\{ \a_N \left[
	1 - \e_\pi^2 \frac{6g_A^2+1}{2} \ln(\e_\pi^2)
	\right]
\nonumber\\&\phantom{=}\quad
	+\b_N \e_\pi^2
	\bigg\} ,
\end{align}
where $2\d \equiv m_d-m_u$.  The coefficient of the \nxlo{2} $\ln$ term is $\frac{1}{2}(6g_A^2+1) \simeq 5.3$.
A LQCD calculation found strong evidence in support of this predicted non-analaytic quark mass dependence as well as a converging expansion~\cite{Brantley:2016our}.

If $SU(2)$ HB$\chi$PT($\nd$) does not converge, it is still possible that $SU(2)$
HB$\chi$PT($\D$) will be a converging EFT at low orders. For $g_A$, there is a
partial cancellation between virtual corrections involving nucleon and delta
loops, a consequence of the large $N_c$ limit~\cite{Dashen:1993as,Dashen:1993ac}.
However, for the nucleon mass, the virtual nucleon and delta corrections add
with the same sign and are of a comparable magnitude.
It is essential to have a set of LQCD results which determine the nucleon and
delta spectrum as well as the matrix elements and transition matrix elements,
such that a quantitative comparison with both $\D$-full and $\D$-less HB$\chi$PT
can be performed. Such results will be possible with exascale computing.

%% Chiral EFT
\subsection{Chiral Effective Field Theory \label{sec:nn_eft}}
Going beyond the single-nucleon sector, EFTs are the key ingredients for
connecting LQCD results to \abinitio nuclear structure and reaction studies
across the nuclear chart. Rather than aiming for direct applications of LQCD to
these nuclei~\cite{Briceno:2014tqa,McIlroy:2017ssf}, the basic strategy is to
constrain \efts{} from LQCD calculations in the two- and few-body sector
and to use the derived potentials and external currents combined with \abinitio
many-body frameworks to solve the nuclear many-body problem. In this section, we
give a brief overview of chiral EFT in Weinberg power counting and of
RG-invariant chiral EFT, before discussing HOBET, an effective theory directly
constructed in the harmonic oscillator
basis~\cite{Haxton:1999vg,Haxton:2006gw,Haxton:2007hx}, in the next Section.

Chiral EFT~\cite{Epelbaum:2008ga,Machleidt:2011zz,Epelbaum:2012zz,Hammer:2012id,Machleidt:2016rvv,Hammer:2019poc} was pioneered by Steven Weinberg in the early
1990s~\cite{Weinberg:1990rz,Weinberg:1991um,Weinberg:1992yk}. Weinberg observed that, for
systems with more than one nucleon, the power counting rules of $\chi$PT cannot
be applied directly to scattering amplitudes, because of the infrared
enhancement of diagrams whose intermediate states consist purely of propagating,
almost on-shell, nucleons. These (``reducible'') diagrams are enhanced by small
energy denominators, scaling as $p^2/M_N$ rather than $p$, and need to be
resummed at all orders. The nonperturbative nature of nuclear scattering
amplitudes manifests itself in the existence of bound states, i.e., the nuclei,
which cannot be obtained in perturbation theory. On the other hand,
(``irreducible'') diagrams whose intermediate states contain interacting
nucleons and pions are not enhanced and follow the usual $\chi$PT power
counting. Weinberg realized that the $S$-matrix for scattering processes
involving $A$ nucleons is obtained by patching together irreducible diagrams
with intermediate states consisting of $A$ free-nucleon propagators; or,
equivalently, by solving the Schr\"odinger equation for $A$ nucleons with a
potential $V$ that is defined by the sum of irreducible diagrams.
Chiral EFT,
the EFT that extends $\chi$PT to the $A\geqslant2 $ sector, provides a perturbative
expansion of the nuclear potential and external currents in powers
\begin{equation} \label{INTRO:eq:Q_def}
%	Q(p) = \frac{ \max \left\{  p,\, m_\pi \right\} }{\Lambda_b} \, .
	Q = \frac{p}{\L_b} \sim \frac{m_\pi}{\L_b}\, .
\end{equation}

Just as with the single-nucleon, Eq.~\eqref{INTRO:eq:Q_n}, this is a double expansion in soft momentum and the pion mass, which are typically treated as the same order, as implied by this equation.
If $p$ is substantially different from $m_\pi$, in principle, one should work to different orders in $p/\L_b$ and $m_\pi/\L_b$ to have a fixed relative theoretical uncertainty.
In practice, this is not very practical but it is understood, for small momenta, the truncation uncertainty will be dominated by the $m_\pi$ expansion while at large momenta, the derivative truncation uncertainty will dominate~\cite{Epelbaum:2014efa}.
A more sophisticated treatment of the truncation error, with a smooth momentum interpolation, has been constructed in a Bayesian framework~\cite{Melendez:2017phj}.

Unfortunately, unlike the single-nucleon case, the breakdown scale $\L_b$, which is largely phenomenologically motivated from so-called
Lepage plots~\cite{Lepage:1997cs}, is found to be lower than $\L_\chi$ and even lower than $m_\rho \approx 770\MeV$;
typically, $\L_b\sim500\MeV$, so $Q$ is usually
larger than in the single-nucleon case,
\begin{equation} \label{eq:Q_approx}
	Q \sim \frac{m_\pi}{\Lambda_b}
	\approx \frac{140 \MeV}{500 \MeV} \approx \frac{1}{3}\, .
\end{equation}

When considering processes in which all momenta $p \ll m_\pi$, pionless EFT~\cite{Bedaque:1997qi,vanKolck:1998bw,Bedaque:1998mb,Chen:1999tn,Kong:1999sf,Bedaque:2002mn} is a powerful approach.
In pionless EFT, pions are treated as heavy degrees of freedom and
strong interactions are represented by contact two-, three- and
higher-body operators, supplemented by long-range electromagnetic interactions.
The double expansion in Eq. \eqref{INTRO:eq:Q_def} reduces
to a single expansion in $p/\Lambda_b$, with breakdown scale $\Lambda_b \sim m_\pi$,
which becomes increasingly accurate when lowering $p^2$.
The power counting of pionless EFT is built around the fine tunings observed in the $^1S_0$ and $^3S_1$ channels.
In the NN sector, the theory contains two $S$-wave LECs at LO that can be determined
in terms of the $S$-wave scattering lengths, or of the deuteron binding energy and the position of the $^1S_0$ virtual state in the complex plane.
At $\mathcal O(Q)$, there appear two two-derivative interactions, which are fit to the
$^1S_0$ and $^3S_1$ effective ranges.  All remaining interactions, including, e.g.,
contact interactions in the $P$-waves, arise at $\mathcal O(Q^2)$ or higher. In the three-body sector,
the integral equations for the $nd$ scattering amplitude in the doublet channel computed  with only short-range two-body forces do not have a
unique solution in the limit of infinite regulator
cutoffs~\cite{Bedaque:1998km,Bedaque:1998mb,Bedaque:1998kg}.
This demands the inclusion of a  single LO three-nucleon (3N)
force~\cite{Bedaque:1998km,Bedaque:1998mb,Bedaque:1998kg},
which  can be fit, e.g., to the triton binding energy or the $nd$ scattering length. A two-derivative
3N force appears at $\mathcal O(Q^2)$~\cite{Bedaque:2002yg}.
No four-nucleon (4N) force is needed at LO~\cite{Platter:2004zs}.
Pionless EFT has been shown to converge very well in the
NN~\cite{Chen:1999tn,Kong:1999sf} and 3N
sector~\cite{Vanasse:2013sda,Konig:2015aka}, while recently studies of nuclei as
heavy as $^{16}$O and $^{40}$Ca have been
performed~\cite{Contessi:2017rww,Bansal:2017pwn}. The determination of how far in $A$
and to which class of nuclei and observables the reach of pionless EFT extends is an open subject of investigation.
In addition to purely QCD processes, pionless EFT has been used to calculate electromagnetic
form factors~\cite{Vanasse:2015fph,Vanasse:2017kgh}, weak processes
such as  $n p \rightarrow d\gamma$, proton-proton fusion, tritium $\beta$ decays and di-neutron
two-neutrino double beta decay~\cite{Kaplan:1998xi,Ando:2008va,De-Leon:2016wyu,Tiburzi:2017iux},
and low-energy parity-violating NN scattering~\cite{Phillips:2008hn,Schindler:2009wd,Schindler:2013yua}. These calculations usually involve
new NN LECs, which can be fitted to data, or to LQCD simulations~\cite{Beane:2015yha,Chang:2015qxa,Savage:2016kon}.

%%%%%%%%%%%%%%%%%%%%%%%%%%%%%%%%%%%%%%%%
%%%% Chiral EFT forces
%%%%%%%%%%%%%%%%%%%%%%%%%%%%%%%%%%%%%%%%
\begin{figure*}[tb]
	\begin{center}
		\includegraphics[scale=0.75]{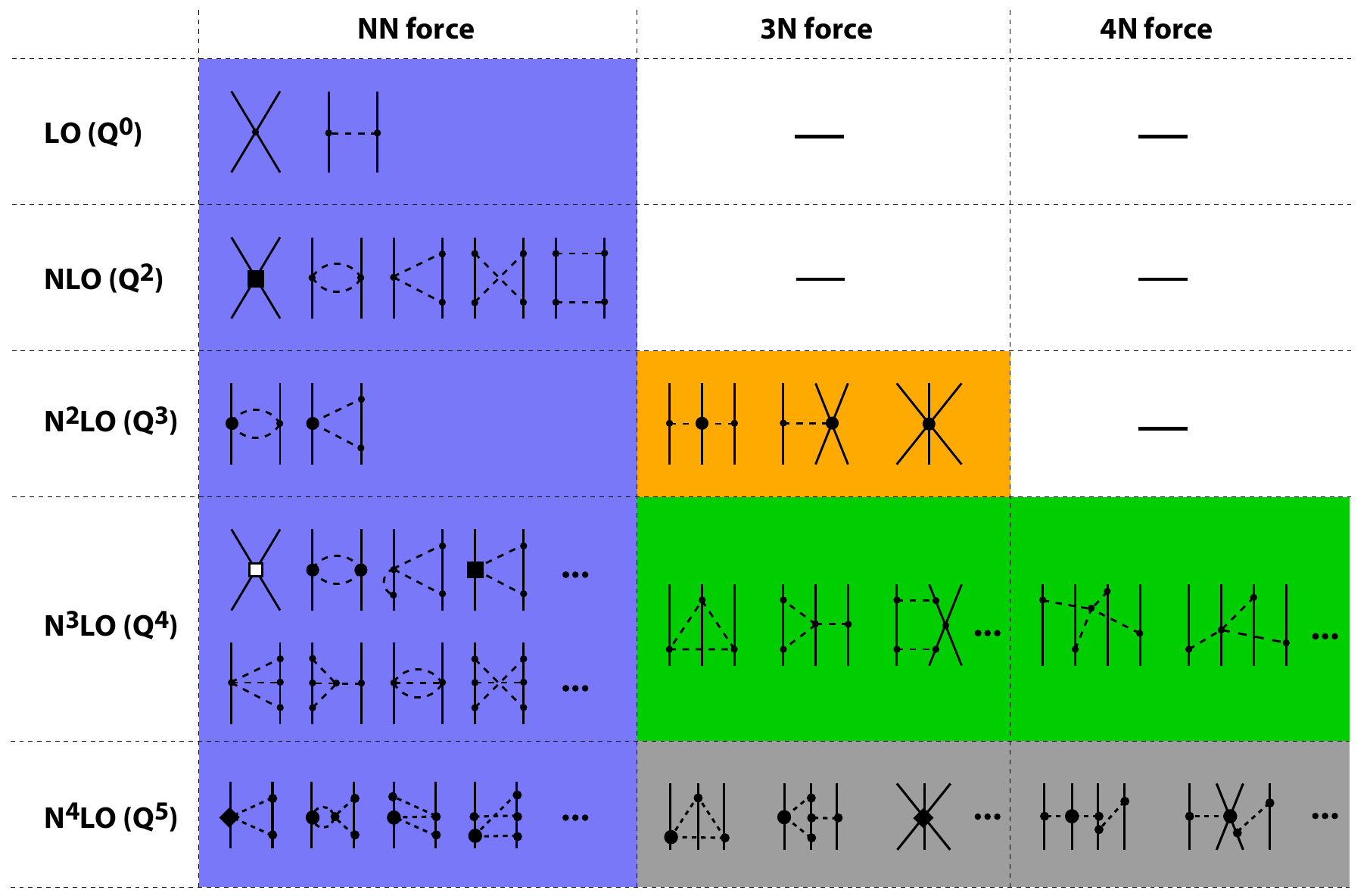}
	\end{center}
		\caption{\label{fig:Hierarchy_EFT} Hierarchy of chiral nuclear forces up to
			N$^4$LO in WPC. Nucleons (pions) are depicted by solid (dashes) lines. Blue-,
			orange-, and green-shaded diagrams are available for many-body calculations,
			whereas the gray-shaded interactions are under development (3N forces) or have
			not been worked out yet (4N forces). The figure has been modified from
			Ref.~\cite{Epelbaum:2015pfa}, see also Refs.~\cite{Machleidt:2011zz,Epelbaum:2012zz}.}
\end{figure*} %%%%%%%%%%%%%%%%%%%%%%%%%%%%%%%%%%%%%%%

Chiral interactions up to N$^4$LO are shown in Fig.~\ref{fig:Hierarchy_EFT}.
The potential receives long- and intermediate-range contributions from one- and
multiple-pion exchanges, respectively, and  short-range contributions from
physics at scales beyond $\Lambda_b$. While the importance of the long- and
intermediate range components follows from the power counting of $\chi$PT in the
one-nucleon sector, a crucial assumption of Refs.~\cite{Weinberg:1990rz,Weinberg:1991um,Weinberg:1992yk} is that also the size of
short-range interactions is determined by naive dimensional analysis (NDA)~\cite{Manohar:1983md,Weinberg:1990rz,Weinberg:1991um}. This assumption defines the
organization scheme of chiral EFT operators that is commonly named Weinberg
power counting (WPC). In WPC, the nuclear potential at LO consists of a
two-nucleon component $V_\text{NN}$, given by the long-range $1\pi$-exchange
potential and by two $S$-wave chiral-invariant contact interactions, $C_S$ and
$C_T$. Soon after Weinberg's seminal papers, the derivation of $V_{\text{NN}}$
was extended to N$^{2}$LO, where 7 contact operators (up to $P$-waves) as well
as the $2\pi$ exchange potential enter the expansion
\cite{Ordonez:1992xp,Ordonez:1993tn,Ordonez:1995rz,Epelbaum:1998ka}. This
N$^{2}$LO potential yields a good description of NN scattering data. With the
inclusion of  $\mathcal O(Q^4)$~\cite{Entem:2003ft,Epelbaum:2004fk,Machleidt:2011zz} and
$\mathcal O(Q^5)$ corrections~\cite{Epelbaum:2014efa,Entem:2017gor,Reinert:2017usi},  chiral
EFT potentials are
now able to fit scattering data up to laboratory energy  $E_\text{lab} \sim 300$
MeV with a $\chi^2/$datum $\sim
1$~\cite{Entem:2003ft,Machleidt:2011zz,Entem:2017gor,Reinert:2017usi}, comparable to
high-quality phenomenological potentials such as Argonne $v_{18}$~\cite{Machleidt:2011zz}
(see Refs.~\cite{Machleidt:2011zz,Epelbaum:2008ga,Hammer:2012id,Hammer:2019poc} for detailed
reviews). This impressive success, coupled with the theoretically appealing
connection to QCD, has generated great interest in the nuclear structure
community, and chiral potentials and currents are increasingly being used in
\abinitio calculations of nuclear properties, weak transitions, and BSM nuclear
matrix elements.

The unknown LECs in chiral and pionless EFT can, in principle, be determined
directly from QCD, but are  usually fit to experimental data due to the
challenging nonperturbative nature of low-energy QCD. There is, however, a subtle distinction as
pointed out in
Ref.~\cite{Contessi:2017rww}: when fitting to LQCD data, the predictions of the
resulting EFT are consequences of QCD itself, whereas when matching to
experiment, consequences of any underlying theory having the same
(low-energy) symmetries as QCD. The first LQCD calculations of light nuclei have been
performed a few years ago with pion masses of
$m_\pi\sim800$~MeV~\cite{Beane:2012vq,Beane:2013br},
510~MeV~\cite{Yamazaki:2012hi} and 300~MeV~\cite{Yamazaki:2015asa}. Pionless EFT
has been matched to the results at
$m_\pi\sim800$~MeV~\cite{Barnea:2013uqa,Kirscher:2015yda,Contessi:2017rww,Bansal:2017pwn}
and 510~MeV~\cite{Kirscher:2015yda} and used to compute lattice nuclei including
${}^{16}$O~\cite{Contessi:2017rww,Bansal:2017pwn} and even
${}^{40}$Ca~\cite{Bansal:2017pwn}.

This matching was performed by taking the binding energies computed from LQCD as
input to the pionless EFT.  For NN interactions, the \luscher{} method
can be used to convert the finite volume energy levels to infinite volume phase
shifts, which can be used to constrain parameters of the EFT directly, or through
an effective range expansion, see for example Ref.~\cite{Beane:2013br}.

Alternatively, for non-relativistic systems, one can construct the \eft{} in a
finite volume. If done properly, this will include all the unphysical partial-wave mixing that occurs
in a periodic volume with cubic instead of the full $\mathcal{O}(3)$ symmetry of the infinite volume.
These
effects
become significantly more important for boosted systems where the center-of-mass
momentum is not zero and the partial-wave mixing is enhanced. Instead of
processing the results through the \luscher{} method, or the equivalent for
three or more nucleons, one directly matches the energy levels in the finite
volume by tuning the volume of the \eft{} to match that of the LQCD calculation.
See Sec.~\ref{sec:hobetbox} for more details. For two-baryon systems, this
direct matching will likely involve an equivalent amount of effort as applying
the \luscher{} method, but when LQCD calculations of three or more baryons
become common, particularly those which include scattering states as well as the
bound nuclei, it will likely be simpler to match the \eft{} calculations to LQCD
results of the energy spectrum and directly determine the LECs, rather than
process them through the multi-body \luscher{} formalism that is being
developed~\cite{Hansen:2014eka,Hansen:2015zga,Briceno:2017tce,Doring:2018xxx}.

A first step in this direction explored the two-neutron system, using
auxiliary-field diffusion Monte
Carlo~\cite{Carlson:2014vla,Gandolfi:2015jma,Lynn:2019rdt}, in an $S$-wave at
zero total momentum in finite volumes of various sizes~\cite{Klos:2016fdb}.  The
volumes were sufficiently large that the partial-wave mixing was irrelevant at
the precision computed, but also, out of reach for present and projected near
future LQCD calculations, except for the smallest one (see
Ref.~\cite{Gandolfi:2016bth} for how this will work for three- and four-neutron systems).
The determination of LECs in the nuclear \efts{} will become
particularly exciting as LQCD computations progress and can be used to compute
properties of systems in which experimental data is controversial, limited or
unavailable, such as few-neutron systems, see, e.g.,
Ref.~\cite{Gardestig:2009ya}, and few-baryon systems with one or two hyperons.

One particular scheme that could be imagined is a direct matching between LQCD calculations of few-nucleon systems (the spectrum and reactions to external probes) and the lattice regularized nuclear EFT (NLEFT) we mentioned earlier~\cite{Lee:2016fhn,Lahde:2019npb}.
NLEFT~\cite{Muller:1999cp,Lee:2004si} is the formulation of chiral EFT with a lattice UV regulator.  As such, it is natural to simulate on a finite lattice and one could imagine constructing the lattices to match the volumes in size and boundary conditions directly to the LQCD calculations.  On its own, NLEFT has been successfully utilized to describe properties of light and medium nuclei, beginning with constraints in the two- and three- nucleon sectors to constrain the experimental phase shifts and binding energies.  One of the first significant successes was the description of the Hoyle state~\cite{Hoyle:1954zz}, one of the more famous fine-tunings in nuclear physics critical to stellar nuclear fusion~\cite{Burbidge:1957vc} which is related to the clustering of $\a$ particles~\cite{Epelbaum:2011md}.  These methods were also utilized to describe alpha clustering in $^{16}$O~\cite{Epelbaum:2013paa}, showing the predictive reach of NLEFT.  Then, BSM currents constrained with LQCD in two- and three- nucleon systems could be propagated into calculations of nuclei using this scheme.

To gain confidence in a theory of nuclear structure and reactions based on this
coupling between LQCD and \eft{}, it is important that the theoretical
uncertainties in the \eft{} expansion are well understood and under control. The
size of the expansion parameter in Eq.~\eqref{eq:Q_approx} is such that the
perturbative expansion will converge, albeit slowly, provided the LECs are natural and one can
work at low-enough orders. As is common in \xpt{} and baryon \xpt{}, the
theoretical truncation error can be estimated from the first order not included
in the analysis~\cite{Epelbaum:2014efa,Epelbaum:2014sza}. Such theoretical error
estimates were applied to few-body
calculations~\cite{Binder:2015mbz,Lynn:2015jua,Lynn:2017fxg,Lonardoni:2017hgs,Binder:2018pgl,Epelbaum:2018ogq}
and nuclear-matter calculations~\cite{Drischler:2016cpy,Drischler:2017wtt}. More
recently, Bayesian methods have been developed for
parameter~\cite{Wesolowski:2015fqa,Wesolowski:2018lzj} and truncation-error
estimation~\cite{Furnstahl:2015rha,Melendez:2017phj,Melendez:2019izc}.\footnote{The BUQYEYE
collaboration provides the well-documented Python
	package \texttt{gsum}, which allows analyzing the convergence pattern of
	EFT observables using Bayesian methods. See\\ \url{https://buqeye.github.io} for more details.}
	Bayesian inference of
breakdown scales, model checking and selection give important insights
into the efficacy of EFT expansions. A comprehensive recipe for
Bayesian uncertainty quantification in nuclear physics has been worked out by the BUQEYE
collaboration~\cite{Furnstahl:2014xsa}.
The connection with NLEFT also would provide a natural means to assess the convergence pattern and associated theory uncertainty since it is formulated in chiral EFT, and thus also prescribes a specific quark mass dependence, which could hopefully be used to match to LQCD computations at heavier than physical quark masses before those at the physical quark mass are available.

Chiral interactions, moreover, require an UV regularization scheme
(that cuts off high momenta) associated with a cutoff
scale $\Lambda_{\rm UV}$~\cite{Epelbaum:2014efa}.%; or several, if many-body forces are regularized differently.
Although the regulator choice should be arbitrary, due to the
renormalization issues discussed below, in practice the actual functional form
can have important consequences at a given order in the chiral expansion due to
induced regulator artifacts~\cite{Dyhdalo:2016ygz}. This freedom has recently led to
the development of various new families of chiral NN potentials with
conceptually different regulators schemes:
nonlocal~\cite{Ekstrom:2013kea,Ekstrom:2015rta,Carlsson:2015vda,Entem:2014msa,Entem:2017gor},
local~\cite{Gezerlis:2013ipa,Gezerlis:2014zia}, and semilocal regularization in, both,
coordinate~\cite{Epelbaum:2014efa,Epelbaum:2014sza} and momentum
space~\cite{Reinert:2017usi}. An overview of these NN potentials including a
comprehensive Weinberg eigenvalue analysis can be found in
Ref.~\cite{Hoppe:2017lok}. Recently, the first study of selected properties of
light nuclei up to $A = 16$ based on semilocal coordinate-space regularized NN
and 3N forces at N$^2$LO was presented by the LENPIC
collaboration~\cite{Epelbaum:2018ogq}.\footnote{See also \url{http://www.lenpic.org}.}  There is evidence for the convergence of chiral EFT for
NN scattering observables at the physical pion mass, for the moderate values of the ultraviolet
cutoff ($\L_{\rm UV}$) typically used in nuclear calculations ($\L_{\rm
UV}\sim400-600$~MeV)~\cite{Epelbaum:2014sza,Melendez:2017phj}

An important point to take into account when discussing theoretical
uncertainties is the consistency of the EFT itself, and in particular whether
the LECs included at each order are sufficient to guarantee that observables are
insensitive  to unphysical regulators introduced in intermediate steps of the
calculation. Weinberg's original papers did not address the subtle issue of
renormalizability of scattering amplitudes. The singular nature of the LO
NN potential, arising either from the contact interactions, which
parameterize the ``hard-core'' of the nuclear force, or from the $1/r^3$ behavior
of the tensor component of the 1$\pi$ exchange potential, requires the aforementioned UV
regulator in the solution of
the Schr\"odinger equation. If the theory is correctly renormalized, observables
should depend on inverse powers of $\Lambda_{\rm UV}$, so that by taking
$\Lambda_{\rm UV} \gtrsim \Lambda_b$ one can ensure that the theoretical error
is dominated by missing orders in the chiral EFT expansions, rather than by
regulator artifacts. Kaplan, Savage, and Wise first pointed out that the
requirement of nonperturbative renormalizability, essential for a consistent
EFT, leads to conflicts with the NDA expectations of the scaling of contact
interactions, which are based on perturbative arguments~\cite{Manohar:1983md}.
In Ref.~\cite{Kaplan:1996xu}, they realized that the interference between
contact interactions and pion-exchange leads to logarithmic divergences
proportional to $m_\pi^2 \log \Lambda_{\rm UV}$ in the LO scattering amplitude
for two nucleons in the $^1S_0$ channel. This divergence can be absorbed by a
counterterm proportional to the quark mass, which, however, only appears at NLO in WPC.
Renormalization requires the promotion of this counterterm to LO; or,
alternatively, the demotion of the pion-exchange to subleading order
\cite{Kaplan:1998tg,Kaplan:1998we}. As we will discuss, this problem appears in essentially the
same form for all long-range potentials with $\sim 1/\mathbf q^2$ behavior (at
large $|\mathbf q|$), acting in the $^1S_0$ channel, e.g., photon exchange,
relevant for charge-independence breaking in NN scattering, or
light-Majorana-neutrino exchange, important for $0\nu\beta\beta$. In all these
cases, the renormalization of scattering amplitudes requires the
long-range potential to be complemented with a short-range contact interaction,
enhanced by two powers of $1/Q$ with respect to NDA
\cite{Kong:1998sx,Kong:1999sf,Cirigliano:2017tvr,Cirigliano:2018hja}.

In the case of the strong interaction, the issue pointed out in Ref.
\cite{Kaplan:1996xu} led  Kaplan, Savage, and Wise~\cite{Kaplan:1998tg,Kaplan:1998we} to
propose a different power counting scheme, in which pions are perturbative
(KSW). Such a scheme works well in the $^1S_0$ channel, where the N$^2$LO
calculation of Ref.~\cite{Fleming:1999ee} satisfactorily reproduces phase
shifts. However, the KSW power counting dramatically fails to converge in the
$^3S_1-{}^3D_1$ and $^3P_0$ channels~\cite{Fleming:1999ee,Kaplan:2019znu}, where
pion-exchange plays a more important role. For the $S$-waves, an expansion around the chiral limit was proposed~\cite{Beane:2001bc} in which the
difference between the 1$\pi$ exchange contribution at non-zero quark masses and in the
chiral limit occurs at the same order as the $\mathcal O(m^2_\pi)$ counterterms.
This expansion coincides with KSW in the $^1S_0$ channel, and with WPC in the
$^3S_1-{}^3D_1$ channel, capturing the most desirable features of both schemes.

The iteration of the $1\pi$ exchange potential, however, leads to a second
problem, highlighted by Nogga, Timmermans, and van Kolck in Ref.~\cite{Nogga:2005hy} (see also Ref.~\cite{PavonValderrama:2005uj}). The authors
found that in each attractive triplet wave where the 1$\pi$ exchange is iterated,
the $r^{-3}$ singularity leads to divergences that require a contact
interaction for renormalization. With the exception of the $^3S_1-{}^3D_1$
channel, these contact interactions only appear at higher orders in WPC. Their
promotion would, of course, destroy the predictive power of the theory, since an
infinite number of LECs would need to be fitted already at LO. Fortunately, in
most higher partial waves the pion exchange is a small effect that can be treated in
perturbation theory~\cite{Nogga:2005hy}. There are strong indications that
$1\pi$ exchange is perturbative in the $^3P_2-{}^3F_2$ coupled channel, in the
$D$-waves, and in all partial waves with $l >
3$~\cite{Fleming:1999ee,Birse:2005um,Wu:2018lai,Kaplan:2019znu}. This would imply
that a local counterterm needs to be promoted to LO only in the $^3P_0$ wave.
The need for a LO contact interaction in the $^3P_0$ channel also emerges in
Lorentz-invariant formulations of chiral EFT~\cite{Epelbaum:2012ua,Ren:2016jna}.

This work forms the basis of a renormalizable version of chiral EFT (i.e., RG-invariant chiral EFT) aimed at curing the shortcomings of WPC and KSW discussed above,  which is an active field of research.
The issue of how to incorporate
subleading corrections has been discussed in  Refs.~\cite{Valderrama:2009ei,Long:2011xw,Long:2012ve}, which included $\mathcal{O}
(Q^2)$ corrections, while studies of the 3N sector have been carried out in
Refs.~\cite{Nogga:2005hy,Song:2016ale}, and processes involving external weak
currents have been studied in Ref.~\cite{Valderrama:2014vra}. We stress that the
consistency checks of the power counting of chiral operators  that are provided
by renormalization studies are particularly important for observables sensitive
to BSM physics. In these cases, the ambiguities in the power counting
cannot be resolved by including high-enough orders and fitting to data, but, to
retain predictive power, it is necessary to identify all the needed operator
structures at a given order.
See Sec.~\ref{sec:0nubb_light} for an explicit example.

While providing a theoretically robust framework, RG-invariant chiral EFT has
not achieved the same level of success as WPC. The main challenge for extending
RG-invariant chiral EFT to higher orders and to systems with more nucleons is
that subleading potentials become increasingly more singular than $r^{-3}$. While this is not an
issue when the higher-order potentials are treated in perturbation theory
\cite{Valderrama:2009ei,Long:2011xw,Long:2012ve}, it might destroy
renormalizability when the entire potential is iterated at all orders
\cite{Valderrama:2009ei}, as customarily done in WPC and in standard nuclear-structure
calculations. On the other hand, a perturbative approach complicates
the interface with \abinitio many-body methods, to which the chiral
potential in WPC can be applied straightforwardly. The success of WPC
\cite{Entem:2003ft,Machleidt:2011zz,Entem:2017gor,Reinert:2017usi}, raises the
possibility that WPC  with cutoff variation in a small window $ Q <  \Lambda_{\rm UV}
\lesssim \Lambda_b$, with the inclusion of sufficiently high orders, and with the
appropriate choice of regulators might be equivalent to RG-invariant chiral EFT,
an issue that is not fully settled~\cite{Hammer:2019poc}.
We now return  to the description of
chiral EFT in WPC, its successes, and the role that LQCD can play in improving
its predictive power.

Coming back to Fig.~\ref{fig:Hierarchy_EFT}, we see that, in addition to
organizing interactions in the NN sector, one success of chiral EFT,
both in WPC and in its RG-invariant version, is that many-body forces appear
naturally but are suppressed, i.e., $V_\text{NN} \gg V_\text{3N} \gg V_\text{4N}
\ldots$, so the total Hamiltonian up to N$^4$LO reads
\begin{equation}
	H(\Lambda; \, \lambda)= T + V_\text{NN}(\Lambda; \, \lambda) +
	V_\text{3N}(\Lambda; \, \lambda) + V_\text{4N} (\Lambda; \, \lambda) \,,
\end{equation}
where $\Lambda$ ($\lambda$) is a momentum cutoff scale (resolution scale). The
first nonvanishing 3N forces appear at N$^2$LO (orange-shaded). The long-range
$2\pi$ exchange is governed by the LECs $c_{1,\,3,\,4}$, consistent with the NN
forces at this order. Different determinations of these LECs from $\pi$N or NN
data show significant deviations but agree within the uncertainties (see Table~I
in Ref.~\cite{Hammer:2012id}). Only a few years ago, Hoferichter~\etal have
significantly improved this through a low-energy analysis of $\pi$N scattering
in the framework of Roy-Steiner equations~\cite{Hoferichter:2015tha,Hoferichter:2015hva},
with remarkably precise constraints on the $\pi$N LECs. The unnaturally large
values for $c_{3,\,4}$ are due to resonance
saturation~\cite{Bernard:1996gq,Epelbaum:2001fm,Krebs:2007rh} since $\Delta$-resonances,
are not explicitly included in the EFT.
Including $\Delta$-resonances, in general, is found to improve the rate of convergence of chiral
EFT, so progress is made in deriving delta-full EFT to high
orders~\cite{Krebs:2007rh,Epelbaum:2007sq,Krebs:2018jkc}.

The 3N LECs, $c_D$ and $c_E$, emerge from the $1\pi$-exchange-contact and the pure 3N
contact interaction, respectively, which are usually fit to the \isotope[3]{H}
binding energy combined with, e.g., the charge radius of
\isotope[4]{He}~\cite{Hebeler:2010xb}, the \isotope[3]{H} $\beta$-decay
half-life~\cite{Klos:2016omi}, or the $nd$ scattering cross
section~\cite{Epelbaum:2019zqc,Epelbaum:2018ogq}. Also constraints from the
empirical saturation point of nuclear matter have been studied
recently~\cite{Drischler:2017wtt}.
There are no additional contact interactions
from N$^2$LO NN forces. At N$^3$LO, subleading 3N interactions as well as first
4N forces (green-shaded) contribute without introducing unknown parameters. They
depend, however, on the LO LECs $C_{S,\,T}$. Furthermore, 12 contact
interactions\footnote{Three out of 15 LECs could be eliminated in recent chiral
NN potentials using unitary transformations~\cite{Reinert:2017usi}.} (up to
$D$-waves), contribute to the $2\pi$ exchange, and the leading $3\pi$
exchange are present. The derivation of the N$^4$LO 3N interactions is work in
progress~\cite{Krebs:2012yv,Girlanda:2011fh,Krebs:2013kha}, whereas the N$^4$LO
4N forces have not been worked out yet (gray-shaded), making N$^3$LO the highest
order for consistent calculations including all many-body contributions up to
this order.

In addition, significant progress in developing new optimization methods for
nuclear interactions~\cite{Ekstrom:2013kea,Ekstrom:2015rta,Carlsson:2015vda,Ekstrom:2014dxa},
investigating $\D$-full chiral
potentials~\cite{Piarulli:2014bda,Piarulli:2016vel,Ekstrom:2017koy}, studying the
chiral convergence order-by-order up to
N$^4$LO~\cite{Entem:2014msa,Epelbaum:2014sza,Epelbaum:2014efa,Entem:2017gor}, even
partly up to N$^5$LO~\cite{Entem:2015xwa}, and applying Similarity
Renormalization Group (SRG)~\cite{Bogner:2006pc} techniques to systematically soften nuclear potentials for
improving the convergence of many-body
calculations~\cite{Bogner:2009bt,Hebeler:2010xb,Hebeler:2012pr,Furnstahl:2013oba} along with
advances in \abinitio frameworks~\cite{Hagen:2015yea} has led nuclear physics to
an era of precision~\cite{Epelbaum:2015pfa}.

The quantitative study of the weak sector of the SM and of nuclear observables
sensitive to BSM physics requires the construction of external currents and
symmetry-violating potentials. The strength of the chiral EFT approach is that
currents and BSM potentials can be constructed consistently with the nuclear
interactions, allowing for systematic and controlled calculations of weak and
BSM processes.  The electromagnetic currents and the SM  vector and axial
currents have been derived at N$^3$LO
\cite{Pastore:2009is,Baroni:2015uza,Krebs:2016rqz,Krebs:2019aka}, resulting in
precise calculations of magnetic moments and weak transitions in light nuclei
\cite{Pastore:2012rp,Pastore:2017uwc}. In particular, Ref.
\cite{Pastore:2017uwc} showed that the ``quenching'' of $g_A$, required for
shell-model calculations to reproduce the observed Gamow-Teller strengths in
light nuclei, can be explained by the inclusion of enough correlations in the
few-body wavefunctions and of two-body weak currents. Ref.~\cite{Gysbers:2019uyb} found a similar solution to the $g_A$ quenching problem
in medium-mass nuclei, by combining chiral EFT potentials and currents with
many-body methods such as coupled-clusters and in-medium SRG (IM-SRG). Chiral EFT currents for
DM-nucleus scattering have been considered at NLO ~\cite{Hoferichter:2015ipa},
and have been used for calculations in light
\cite{Korber:2017ery,Andreoli:2018etf} as well as heavy systems
\cite{Hoferichter:2016nvd,Hoferichter:2018acd}. At higher orders, the currents
depend on new LECs, which cannot be extracted from data. LQCD calculations of,
for example, scalar and tensor currents in light nuclei~\cite{Chang:2017eiq}
will allow the determination of such LECs, an important task for precision
calculations.

The NN chiral PV potential~\cite{Zhu:2004vw,deVries:2013fxa,Viviani:2014zha} and
TV potential
\cite{Maekawa:2011vs,deVries:2012ab,Bsaisou:2014zwa,Bsaisou:2014oka,Gnech:2019dod}
have also been calculated at NLO, extending and completing phenomenological
derivations~\cite{Desplanques:1979hn,Barton:1961eg,Towner:1994qe}.  Already at LO
these potentials contain unknown LECs, whose first principle determination is
crucial for testing the SM and for connecting nuclear EDMs with microscopic
sources of CP violation.
We conclude this section by summarizing the impact that LQCD calculations in the
NN and 3N sector can have:
\begin{itemize}
\item LQCD will enable a study of the dependence of nuclear observables on fundamental QCD parameters, in particular the light quark masses. This study will directly address the issue raised in Ref.~\cite{Kaplan:1996xu}.
If the LECs accompanying the short-distance NN operators
are found to have significant pion mass dependence, this is indicative that the short distance quark mass dependent operator should be promoted to LO to have a consistently renormalized theory.
If the quark mass dependence is mild, this is indicative that the perturbative pion treatment is more appropriate~\cite{Kaplan:1998tg,Kaplan:1998we}.
Resolving this question
will enable an understanding of fine tunings observed in nuclear physics, such as the large scattering length and large effective range in the $^1S_0$ channel or the small binding energy of the deuteron, and guide us towards the establishment of a power counting which incorporates these fine tunings;

\item Calculations of pion-nucleon and pion-nucleus scattering, of the $^3$H GT matrix element, and of three-nucleon scattering will allow a first  principle determination of low-energy constants such as $c_{1,\,3,\,4}$, $c_D$ and $c_E$ that enter chiral 3N forces at N$^2$LO, validating existing  extractions or shedding light on how the theory may need to be improved.
Related, $c_1$ is the LEC that gives the dominant contribution to the pion-nucleon sigma term ($\s_{\pi N}$) relevant to direct dark matter detection in which the phenomenological determination of $\s_{\pi N}$~\cite{Hoferichter:2015dsa} is in significant tension with the LQCD determination~\cite{Bali:2012qs,Durr:2015dna,Yang:2015uis,Bali:2016lvx,Alexandrou:2019brg}.  A direct determination from LQCD may clear up this discrepancy or shed light on further issues.
Finally, LQCD can provide a window on three-neutron interactions, which are not well determined experimentally;

\item LQCD will provide a first principle determination of the YN and YNN interactions that are challenging to determine due to the limited experimental data. The limited constraints on our understanding of these hyper-nuclear interactions means we do not know if there are similar  fine tunings or other issues present in these interactions that must be understood to form a  converging EFT.  For example, there is an indication from LQCD that the $n\S^{-}$ interaction in the ${}^3S_1$ channel is strongly repulsive~\cite{Beane:2012ey}, although the conclusions rely on an expansion about the $SU(3)$ chiral limit.  There also exist several calculations of YN interactions using the HAL QCD potential method including results at the physical pion mass~\cite{Iritani:2018sra}.  Significant progress in understanding YN and YNN interactions will come from the comparison of the $SU(2)$ chiral EFT~\cite{Beane:2003yx} with forthcoming LQCD results extrapolated to the physical pion mass point;

\item For applications to searches of BSM physics, LQCD can determine LECs in the two-nucleon sector, which, in the case of BSM currents, cannot be extracted from data. Examples that we already mentioned are the short-range components of the two-body scalar and tensor charges, relevant for DM searches or searches of non-standard contributions to $\beta$ decays, the short-range component of the neutrino potential for $0\nu\beta\beta$, and the parity-violating as well as parity- and time-reversal violating NN interactions.
These calculations are necessary for the interpretation of low-energy probes of BSM physics. In addition, as we will further discuss in Sec.~\ref{sec:0nubb}, they will allow to check the power counting of the EFT, and to quantitatively assess the impact of the renormalization issues discussed earlier.
\end{itemize}

%% HOBET
\subsection{Nuclear Effective Theory: HOBET \label{sec:hobet}}
Harmonic Oscillator Based Effective Theory (HOBET)~\cite{Haxton:1999vg, Haxton:2006gw, Haxton:2007hx, Haxton:2000bi, Haxton:2002kb, Luu:2004xc, McElvain:2019ltw}  is an ET designed for traditional nuclear physics and its well-developed machinery -- specifically, large-scale diagonalizations of Hamiltonians in an explicitly anti-symmetric Slater determinant basis over single particle harmonic oscillator (HO) states.    HOBET was built to
allow one to proceed from QCD to the nuclear scale in a single step, so that the effective interaction one constructs is precisely that needed for many-body calculations,
with no cutoffs entering other than those that define the nuclear space.  HOBET avoids the construction of a high-momentum NN interaction entirely.

%As an ET, HOBET begins with the exact Bloch-Horowitz (BH) equation for an effective Hamiltonian defined in a restricted subspace by the projection operator $P$, and the excluded subspace projector $Q=1-P$.  To implement $P$, it is defined as a function of the HO parameter $b$, and a quanta cut off of the energy $\Lambda_{\xSM}$.
%The resulting BH effective Hamiltonian, obtained by projecting the full Hamiltonian, reproduces the spectrum of the full Hamiltonian with the projected eigenstates that overlap with $P$.  This well known property~\addcite{some ref?} is independent of $b$ and $\L_{\xSM}$.
%The rate of convergence of the BH effective Hamiltonian will depend upon the choices of $b$ and $\L_{\xSM}$, as discussed in more detail below.  While poor choices will delay convergence, they will not prevent it.  HOBET is the systematic organization of the tower of operators in the BH effective Hamiltonian that faithfully reproduce the eigenstates and eigenenergies of the full Hamiltonian, and is thus an ET of the non-relativistic description of nuclear physics.}

HOBET starts with the Bloch-Horowitz equation, which is a formally exact
expression for the effective interaction in a potential theory.  The ET expansion is constructed from the well-known Talmi integrals~\cite{DeShalit1963}, generalized for a nonlocal interaction,  systematically removing $r^n$-weighted Gaussian moments of the interaction (with the known long range pion contribution removed).  More formally, this expansion is systematic in the lowering operators of the harmonic oscillator, and is complete in any finite HO space. HOBET truncates this short-range expansion at a designated order. The pionful version represents all higher-order Gaussian moments -- Talmi integrals lacking an LEC --
by the pion contribution.  The ET expansion has the attractive feature that the pion plays no role at short range, removing the difficulties that arise in EFTs that treat the tensor interaction as a $1/r^3$ potential between point nucleons, requiring care near $r=0$.

HOBET preserves all symmetries of the original non-relativistic theory,
including Galilean invariance. HOBET results are independent of the length scale defined by the oscillator parameter $b$ and the quanta cutoff $\Lambda$ defining the maximum energy of Slater determinants in the included space $P$. The rate of convergence does depend on the choice of these parameters, and is typically optimized by selecting $b$ near $\sim 1/m_\pi \sim R$, where $R$ is the nuclear
radius. The theory is highly convergent for values of $b$ in this natural
range. This convergence has been explored in potential models and
shown numerically to correspond to powers of $b {\kern 1pt} m_{\rho}$ where $m_\rho$ represents the inverse range of the omitted short-range physics -- as would be expected from the Talmi integrals themselves.

The HOBET expansion can be shown to correspond quite closely to those of conventional chiral EFTs, though with differences that arise because the included space is defined in terms of the total quanta in a multi-nucleon Slater determinant, while the chiral EFT cutoff is typically given by the momentum of individual nucleons.  Chiral EFTs typically yield soft interactions appropriate for a momentum scale $\sim$ 500 MeV, so that additional transformations are needed to derive an interaction appropriate for the still softer Hilbert spaces used in HO-based nuclear calculations.
HOBET is a non-relativistic ET, as described in \secref{sec:net}, that takes one directly from QCD to the Hilbert space most commonly used in non-relativistic many-body calculations, using phase shifts or a finite volume spectrum derived from NN LQCD calculations to fix the LECs in the ET expansion.

Unlike the EFT approaches previously discussed, which involve an operator expansion around $\mathbf{q}=0$, HOBET's expansion is around an intermediate scale
defined by the HO oscillator parameter $b$, mimicking the balance nature strikes between kinetic energy minimization via delocalization, and potential energy minimization by
grouping nucleons together, near the peak of the midscale nuclear attraction.  The coordinate space/momentum space conjugacy of the HO captures this
physics well.  The HO has another property that is crucial, the ability to exactly preserve (Galilean) translational invariance.  But with these properties comes a wrinkle that makes the ET quite interesting: a discrete HO basis excludes
both UV and IR contributions.  This requires
one to organize the effective interaction to isolate short-range terms from longer range ones associated with the kinetic energy operator.  This separation
allows one to incorporate chiral symmetry into HOBET quite simply, through pionic interactions that act exclusively at relatively long range.  This comes about because
HOBET's effective interaction can be expressed in terms of a single, complete operator expansion:  the low-order operators associated with UV physics are governed
by LECs that can be determined from scattering observables, while high-order IR operators are governed by the pion.
Low-momentum observables can then be predicted with quantifiable errors, determined from the rate of convergence of the operator expansions.

HO basis functions are the unique discrete basis that preserves translational invariance, a crucial requirement in ET:
a properly chosen included space ($P$-space) can be exactly factorized into relative and center-of-mass wave functions.
The $P$-space plays another essential
role: as we discuss below, finite HO bases require correction in the UV (omitted short-range interactions)
and in the IR (as the harmonic oscillator is too confining).  By using larger $P$-spaces, one can further separate
the omitted UV and IR physics scales, improving the convergence.   When HOBET is executed, it generates results
that are simply related to the full solutions, namely exact eigenvalues and exact restrictions of the true wave
functions to $P$.   The theory is analytically continuous in energy, and specifically produces the continuum
solutions for $E>0$.  This has the important property that the theory's parameters -- LECs
that determine the strength of short-range operators -- can be precisely related to phase shifts and other
scattering parameters, which depend sensitively on $E$ near threshold.

HOBET provides Galilean-invariant solutions of the Schr{\"o}dinger equation and thus is an ET,
in contrast to an EFT, which generates relativistically covariant solutions in quantum field theory.  (This
distinction between \efts{} is not universal, with others describing both as EFTs.)   But whether the
formulation is Galilean and quantum mechanical, or covariant and field theoretic, both follow the same rules.
Both are formulated in a
subset of a complete basis: chiral EFT interactions employ a momentum regulator with a scale
$\Lambda_{\text{UV}}$,
while HOBET employs a discrete Slater determinant basis spanning $P$ with a total energy cutoff
$\Lambda_{\xSM}$, as well as a length
scale $b$, the harmonic oscillator size parameter.   Both employ controlled expansions in effective operators that respect the underlying symmetries of the full theory
(such as translational invariance/covariance, parity, time reversal, Hermiticity): the operator expansion produces an effective interaction that
systematically corrects for the omission of degrees of freedom excluded by the chosen cutoffs.   Properly converged solutions should be independent
of the cutoffs employed -- though cutoffs appropriately chosen for a given application can speed the rate of convergence,
and thus the efficiency of the ET/EFT.    Convergence is achieved below some momentum or energy scale, related to the cutoffs and to the order to
which the expansion is carried out.

HOBET respects all of the rules of modern, systematic ET/EFTs, while also making
use of the technology and intuition that has been accumulated through years of model-based approaches to
nuclear structure and effective interactions.

Early attempts to produce an effective interaction recognized the critical role of repeated short-range scattering
through the nucleon-nucleon hard core, which led to summation of the two-nucleon ladders to produce the bare two-nucleon
reaction matrix $G_0$~\cite{Bethe:1956zz}.   Attempts were made to expand the effective interaction perturbatively in $G_0$
through generation of intermediate particle-hole excitations~\cite{Kuo:1966zz}.   In the early 1970s, Barrett and Kirson~\cite{Barrett:1970ner}, working in a shell model (SM\footnote{Throughout this section we use the abbreviation SM for shell model as opposed to Standard Model in the other sections.}) basis, evaluated the effective interaction for $^{18}$F, finding large third-order contributions to $G$ that largely canceled against second-order contributions.   The failure to converge was consistent with arguments made by Shucan and Weidenm\"uller~\cite{SCHUCAN1972108, SCHUCAN1973483} that identified intruder states -- states primarily residing outside of $P$, but with eigenvalues lying in the spectrum of $PHP$ -- as a generic source of non-perturbative behavior.   The strong mixing of such states with nearby states in $P$ is clearly problematic.

Such early perturbative efforts to derive a softened effective interaction appropriate to a low-momentum
$P$-space, starting from a full-space NN hard core interaction,
have been replaced by more modern non-perturbative techniques.
One of these is the use of unitary transforms to
reduce the strength of the coupling between $P$-  and the excluded- or $Q$-space.
The SRG approach employs a continuous sequence of unitary transforms, indexed by a continuous flow parameter $s$ ~\cite{Glazek:1994qc,Glazek:1993rc,Bogner:2006pc,Jurgenson:2007td},
see also Ref.~\cite{Glazek:2002jg}.
When the approach is used with the common choice of $T$, as the generator in a momentum basis, it suppresses off diagonal matrix elements, softening the interaction, and  reducing that source of non-perturbative behavior, but if the transformation is pushed too far, the induced three- and higher-body contributions become  dominant.   Importantly, it does not address the role of $T$ in the HO basis where it strongly connects $P$ and $Q$.
The IM-SRG extension of this idea operates directly on the  $A$-body Hamiltonian expressed relative to a Slater determinant reference state to drive towards a block-diagonal form~\cite{Hergert:2015awm}.

Another method in frequent use is the no-core shell model (NCSM).   In this approach an improved interaction is constructed with multiple cutoffs, observables are computed for each cutoff, and extrapolated to no cutoff.    If the improved interaction were an ET according to our definition, then extrapolation would not be needed,
as results must be independent of the choice of cutoff.      Various methods are used to produce the improved interaction including embedding a momentum-based EFT, pionless or chiral for example, in the HO basis, with the LECs fitted to phase shifts~\cite{Stetcu:2006ey,Yang:2016brl,Bansal:2017pwn}.   Such an embedding is a great improvement over simply taking matrix elements of the interaction in the truncated basis and exhibits much more rapid convergence as the cutoff is raised.   However, the computational cost rises extremely rapidly with the cutoff, and the results obtained are not fully independent of the choice of $P$, e.g.,
varying with the choice of oscillator parameter $b$.

From an ET perspective, the treatment of the NN interaction in the approaches above is somewhat unnatural.   The underlying UV theory is QCD, and the resulting ET we want to
determine is formulated in $P$.  The most straightforward approach would be a direct step from QCD to $P$, with available experimental (or, in the near
future, LQCD) input used in $P$ to fix the LECs of the operator expansion.   Instead, the available experimental information -- including detailed high-momentum
information that is not really needed in $P$ -- is encoded in a rather ill-behaved hard-core NN potential, only to be decoded later in
a difficult renormalization step.

HOBET, in contrast, is an ET formulation designed to allow such a direct, one-step transition from QCD to $P$
\begin{enumerate}
\item It employs the same cutoffs -- $\Lambda_{\xSM}$ and $b$ -- that SM practitioners use, thereby
avoiding the issue that arises in chiral interactions, which employ a momentum basis that is not orthogonal to the SM basis (which  limits the scale $\Lambda$ to which the potential
can be softened in momentum space, through techniques like $V_{\text{low}\,k}$
\cite{Nogga:2004ab,Bogner:2005sn,Bogner:2005fn,Bogner:2006ai,Bogner:2006tw}).
\item HOBET is explicitly continuous in energy, and thus treats bound states and continuum states on an equal footing.  This
is important for multiple reasons.  First, it allows one to use scattering data directly, seamlessly connecting phase shifts to bound-state properties: much of the
relevant information in phase shifts and mixing angles is connected with
their rapid evolution with $E$ near threshold: in the $^1S_0$ channel, the NN system is unbound by less than $100 \keV$.   HOBET allows one to take a measured phase shift $\delta(E)$ in a region
of rapid variation and relate it precisely to HOBET's LECs.
Second, HOBET's energy dependence allows one to employ a finite HO basis -- necessary for practical diagonalizations in $P$ -- while at the same time
generating an infinite number of solutions, including continuum solutions that vary continuously with $E$.  Third, it yields solutions that are simply related to the exact solutions, namely
the exact eigenvalues and the exact projections of eigenfunctions onto the chosen HO basis.  As projection does not preserve norms, this requires nonorthogonal solutions,
which are a natural consequence of HOBET's energy dependence.
\item HOBET manifests these attractive properties of energy-dependent ETs, at a very modest cost in additional complexity.  Specifically, due to a clever reorganization of the
underlying Bloch-Horowitz equation, HOBET's LECs are energy independent.
\item HOBET builds on wisdom gained in the field over decades from model-based approaches like the SM.
It does not implant a contact-gradient expansion appropriate to a plane-wave chiral EFTs onto a SM-like HO basis, as others have
done~\cite{Stetcu:2006ey,Stetcu:2009ic,Rotureau:2011vf}, but instead employs an
operator expansion built from HO raising and lowering operators.  This connects HOBET's LECs to familiar SM quantities, the Talmi
integrals and their nonlocal generalization.   The operator basis is complete.
\item  The LECs for low-order operators, which encode the effects of missing UV physics on $P$, are fitted to experiment (e.g., phase shifts and mixing angles).
One knows a priori that this UV expansion will be rapidly convergent, because Talmi integral expansions in
nuclear physics are quite efficient.
\item  The remaining LECS are determined by chiral symmetry, implemented as an operator expansion that begins in the IR and progresses to smaller $\langle r \rangle$, where it
joins up with the UV expansion.  HOBET's implementation of chiral symmetry is simpler than that of
\figref{fig:Hierarchy_EFT}.
Chiral potentials separate out the entire contribution of the pion, idealizing it as an exchange between point nucleons.
This introduces a singular $r^{-3}$ tensor interaction that requires regularization, with attendant difficulties in renormalization as described in Sec. \ref{sec:hbchipt}.
In contrast, in HOBET
all short-range physics is treated as unknown, pionic or otherwise, and absorbed into the UV LECs.   Chiral symmetry is used to fix the values of LECs beyond those determined
in the short-range expansion.  These IR LECs are all computable functions of experimentally known parameters,
$f_\pi$ and $m_\pi$.  The dimensionless parameters controlling the convergence of the UV and IR expansions depend on $b$, which then can be adjusted to optimize
the joining of the two expansions.
\end{enumerate}
While HOBET developments have been reported in a series of technical papers ~\cite{Haxton:1999vg, Haxton:2006gw, Haxton:2007hx, Haxton:2000bi, Haxton:2002kb, Luu:2004xc, McElvain:2019ltw} and in Ref.~\cite{McElvain2017thesis}, here we present an overview focused on the broad issues summarized above.

%%%%%%%%%%%%%%%%%%%%%%%%%%%%%%%%%%%%%%%%%%%%%%%%%%%%%%%%%
\subsubsection{The Bloch-Horowitz Equation}
In HOBET one constructs a Galilean-invariant effective interaction in a HO basis directly from observables.
The theory's cutoffs, $\Lambda_{\xSM}$ and $b$, are defined at the nuclear scale, where multi-nucleon calculations can be done.
The observables can be traditional experimental ones such as binding energies and phase shifts,
but can also be the energy spectrum of two or three nucleons in a finite volume, computed in LQCD.  Results, if fully converged,
will be independent of the cutoff choices made.

$P\left(\Lambda_{\xSM}, b\right)$ typically consists of all Slater determinants up to an energy $\Lambda_{\xSM} \hbar \omega$, measured relative to the lowest energy configuration.
This exploits the exact separability of such spaces, guaranteeing that the effective interaction is translationally invariant.  The parameters describing $P$, namely $\Lambda_{\xSM}$
and the oscillator parameter length scale $b$, are the theory's cutoffs.  Here $b^2=\hbar/\left(M \omega\right)$ with $M$ the nucleon mass and $\omega$ the oscillator angular frequency.
Dimensionless energies are given in units of $\hbar \omega$, while the dimensionless two-nucleon Jacobi coordinate is $\mathbf{r}=(\mathbf{r}_1-\mathbf{r}_2)/(\sqrt{2} b)$.  We employ nodal
quantum numbers $n=1, \, 2,\dots$.  Various details can be found in Appendix A
of Ref.~\cite{McElvain2017thesis}.

HOBET's starting Hamiltonian is $H=T+V$, where $T$ is the relative kinetic energy and $V$ the
potential, with Schr{\"o}dinger
equation full-space solutions
\begin{equation} \label{eqn:FullHSpectrum}
    H \ket{\psi_i} = E_i \ket{\psi_i} \, .
\end{equation}
Defining $Q=1-P$, with
$(P+Q)=1$, the solution of the effective interactions problem is given by the Bloch-Horowitz (BH) equation
\begin{eqnarray}
\label{eqn:BH}
P H^\mathrm{eff}(E_i) P |\Psi_i \rangle &=& E_i P |\Psi_i \rangle \, , \nonumber \\
H^{\mathrm{eff}}(E_i) &=&  H + H~G_{QH}(E_i)~ QH  \, ,\nonumber \\
G_{QH}(E_i) &\equiv& \frac{1}{E_i-QH}\, .
\end{eqnarray}
The equation must be solved self-consistently, as $H^\mathrm{eff}(E_i)$ depends on the eigenvalue
to be found: in practice solutions quickly converge on simple fixed-point iteration, typically in $5-6$
cycles.
As noted above, all non-zero $P\ket{\psi_i}$ are eigenstates of $H^{\mathrm{eff}}(E)$.
In particular, projections of continuum states can be generated continuously with $E$, apart
from isolated eigenvalues where $P\ket{\psi_i}$ vanishes.

Solutions corresponding to projections imply an attractive evolution of eigenfunctions with increases in the
energy cutoff $\Lambda_{\xSM}$: the eigenfunctions change through the
addition of new components corresponding to the added shells, with previous components staying fixed.
Correspondingly, one can define a growing norm
for the wave function that reflects the new components added with increasing
$\Lambda_{\xSM}$,
reaching unity as  $\Lambda_{\xSM} \rightarrow \infty$.  Similarly, orthonormality of the eigensolutions
is restored in this limit.

\Eqnref{eqn:BH} involves a potential that spans $P$ and $Q$.  The rewriting of this
equation in a form more suggestive of an ET came about from a study of the full
Green's function in perturbation theory,
\begin{equation}
    \frac{1}{E{-}QH} = \frac{1}{E{-}H_0} + \frac{1}{E{-}H_0}Q \left(V-V_0\right) \frac{1}{E{-}H_0}+\cdots \, ,
\end{equation}
where $H_0=T+V_0$ is the HO Hamiltonian.  Note that $T$ and $V_0$ in the HO are hopping operators, connecting
nearest neighbor HO shells of the same parity, but not distant shells.  When this expansion was employed in $H^\mathrm{eff}$,
certain effective interaction matrix elements in $P$ continued to oscillate even after hundreds of orders of perturbation theory.
These matrix elements involved HO ``edge states," configurations in the last included shell in $P$, that
are coupled to $Q$ by $T$~\cite{Haxton:2000bi}, indicative of convergence difficulties in the IR.
This led to the Haxton-Luu reorganization of the BH equation in which IR and UV components are separated through a summation
of $QT$ to all orders,
\begin{eqnarray} \label{eqn:HLBH}
&&H^{\mathrm{eff}}{=}P E G_{TQ} \!\left[T{-} T\frac{Q}{E}T+ V {+} VG_{QH} QV\right] E G_{QT} P \,, \nonumber \\
&&~~~~~~~~~G_{TQ}{\equiv}\frac{1}{E-TQ}\,
 ,~~~~~~G_{QT}{\equiv}\frac{1}{E-QT}\, .
\end{eqnarray}
This reorganization isolates almost all of energy dependence of the BH equation in the kinetic energy Green's functions,
which can be re-expressed in terms of the known free Green's function $G_T$,
\begin{equation}
\label{eq:BHreorg}
    E G_{QT}P = G_T \left\{P G_T P\right\}^{-1} \, ,~~~~~~~G_{T}{\equiv}\frac{1}{E-T}\, ,
\end{equation}
at the modest numerical cost of a matrix inversion in $P$.

\begin{figure*}[ht]
\centering
\includegraphics[scale=0.6]{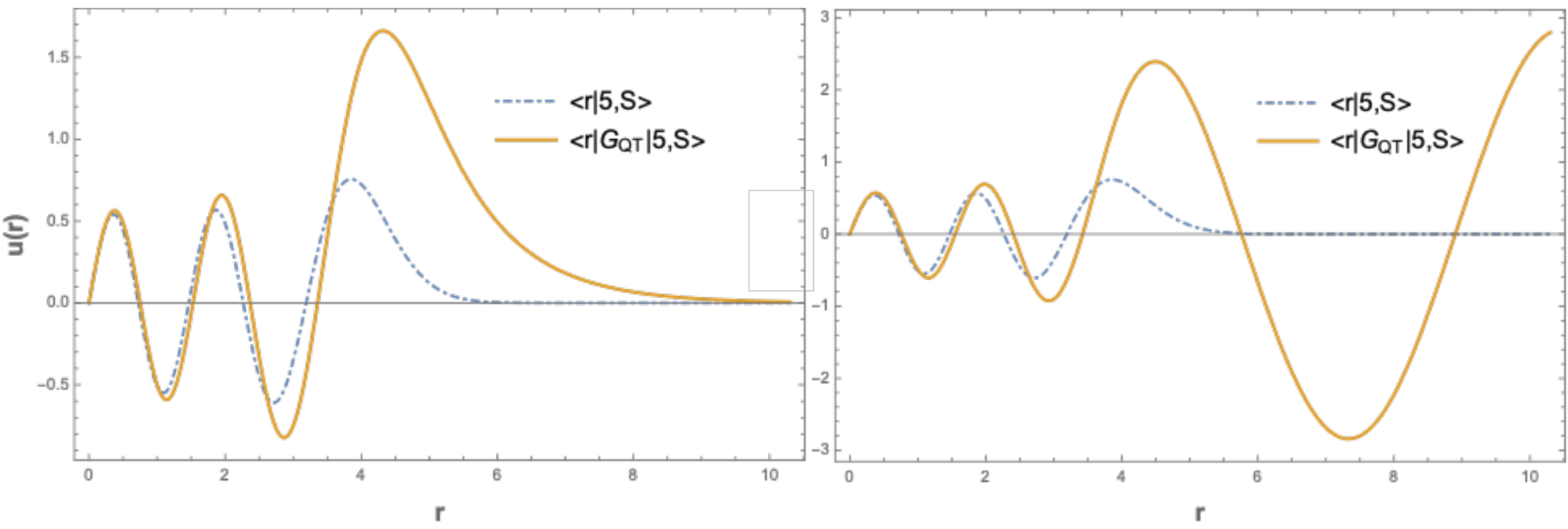}
\caption{\label{fig:GQT_Edge}
Transform of the fifth and highest $S$-channel state in $P$ with $E=-1/2 \, (1/2)$ on the left (right) (in $\hbar\omega$ units).   The transformed state has been scaled to match the amplitude of the original basis function near
the origin (dot-dashed lines).  The exponential decay of a bound state and oscillations of a scattering state are reflected in the edge-state behavior.}
\end{figure*}

As $T$ is a hopping operator, the action of the Green's function on the spatial components of basis states comprising $P$ -- for the NN case this would be
$EG_{QT} | n \ell m_\ell \rangle$ --  has no effect on components that do not reside in the last included shell.
However, its action on an edge state modifies the wave function as in \figref{fig:GQT_Edge}: at small $r$ the shape remains that of the HO, while at larger $r$, on the left, with $E<0$ the wave function takes on the shape of a bound state with an exponentially decaying wave function.
At larger $r$, on the right, with $E>0$ the wave function takes on the shape of a scattering wave.
For $E<0$ the extended tail becomes increasingly pronounced as $\left| E\right| \rightarrow 0$.
The net effect is to alter matrix elements of short-range operators by a normalization, while building into
$H^\mathrm{eff}$ long-range strong-interaction corrections (dominated by the pion) that would otherwise be missing, as we will describe in more detail later.  One should not misconstrue $EG_{QT}$ as creating  a ``tail"  on the wave function.   This operator is part of an exact
rewriting of the BH $H^{\mathrm{eff}}$ that operates within a compact, translationally invariant HO space defined by $P$.

The first two terms within the bracket in \eqnref{eqn:HLBH} correct the kinetic energy to all orders.  The remaining
two terms provide the starting point for formulating the effective theory.   They can be divided as follows
\begin{eqnarray}
\label{eq:split}
&&V+ V G_{QH}QV \equiv V^{IR} + V^{UV}  + V G_{QH}QV \,, \nonumber \\
&&V^{IR} \rightarrow V^{IR}_\pi \,, ~~~~~~~~~   V^{UV}  + V G_{QH}QV \rightarrow V_\delta \, .
\end{eqnarray}
Here $V_\delta$ represents the systematic operator expansion describe in the next section, which accounts
for all of the short-range physics residing in $Q$, that in a potential theory would be generated by the repeated
scattering of $QV$.  It also accounts for any short-range physics associated with
$V$ operating in $P$.  This leaves the long range contributions of $V$ in $P$, denoted $V^{IR}$, defined as physics beyond the range
of our operator expansion.  Thus $V^{IR}$ depends on the order of the UV expansion.   Under the (testable) assumption that the pion dominates the long-range
physics in $P$, we equate $V^{IR}$ to $V^{IR}_\pi$.

\begin{wrapfigure}{R}{0.5\textwidth}
%\begin{figure}[h]
\centering
\includegraphics[width=0.5\textwidth]{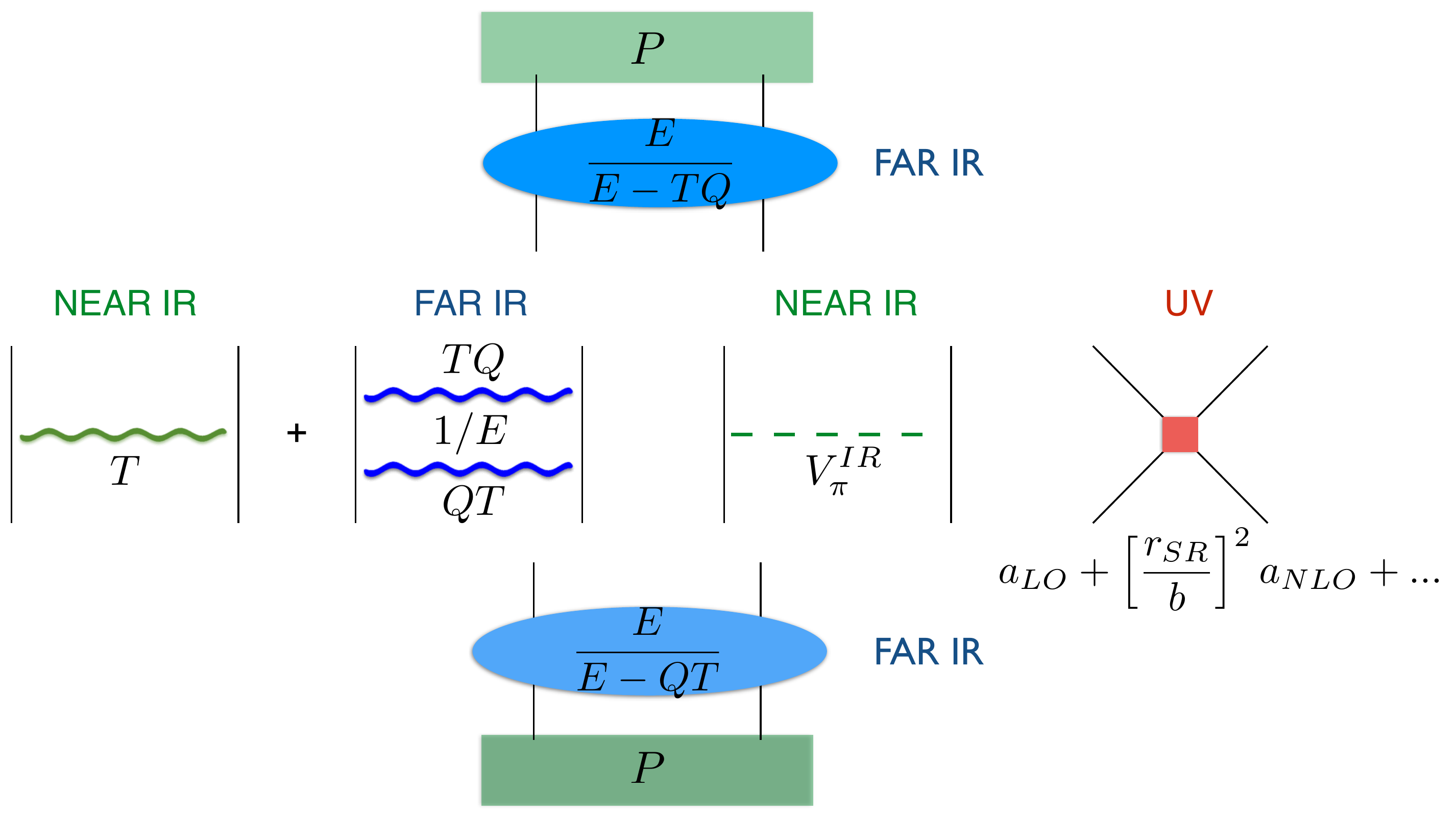}
\caption{\label{Fig_Heff}
HOBET's pionful effective interaction, appropriate to a HO where translational invariance
requires $P$ to be defined in terms of total quanta (in contrast to chiral interactions employing a momentum regulator). (Color: blue, green, red
indicate far-IR, near-IR, and UV corrections.)}
%\end{figure}
\end{wrapfigure}

With these replacements, pionful HOBET becomes a true effective theory: no reference to phenomenological potentials
remains.  The resulting $H^\mathrm{eff}$ is depicted in \figref{Fig_Heff}.   Chiral symmetry and the kinetic energy operator govern the IR physics.  The LECs for
HOBET's operator expansion are obtained by matching observables (UV) or computed from a chiral operator expansion (IR), as described below.
Consequently HOBET is an effective theory in which QCD is reduced to the nuclear scale in one step.

%-------------------------------------------------------------------------------
% HOBET

\subsubsection{HOBET's Operator Expansion}
\label{sec:operators}
While in principle one has freedom in defining an ET's short-range expansion, in practice it is
important to pay attention to the underlying physics to find the most suitable and efficient expansion.  In chiral EFT
the natural operator choice is a contact-gradient expansion, employing $\delta(\boldsymbol{r})$ and scalars built from $\boldsymbol{\nabla}$:
the expansion is around $\mathbf{q}=0$ and operators built from $\boldsymbol{\nabla}$ select out powers of $\boldsymbol{q}$.  HOBET is built on
the HO, where excitations are created by raising operators acting on the zero mode.  Coordinate and momentum space representations
of the HO are conjugate, with the coordinate structure governed by the dimensionless parameter $r_{12}/b$, and the momentum-space
structure by $qb$.  The momentum expansion is about an intermediate scale
determined by the inverse nuclear size.  The natural operator basis is that built from the HO raising and lowering
operators which act on the nodal quantum number, usually designated with $n$ and beginning with $n{=}1$, as well as the angular momentum, usually designated with $\ell$.  This choice produces a correspondence between the progression of low-energy modes and the tower of short-range effective operators.
There is a linear mapping between the LECs and the effective interaction matrix elements between the different HO states, e.g. the $n^\prime=1 \leftrightarrow n=1$ matrix element is uniquely determined by the LO LEC (see the lower-left matrix element depicted in Fig.~\ref{fig:Swave}b).
It also leads to a seamless division between
the UV operators associated with $V_\delta$ and the IR operators associated with $V^{IR}_\pi$, greatly simplifying HOBET's
UV and IR expansions, as described below.

This form of HOBET operators was first used in Ref.~\cite{Haxton:2007hx}.  Here we follow Ref.~\cite{McElvain:2019ltw}, where
a more conventional normalization was adopted.

We introduce the HO creation operators
 $(a^\dagger_x,a^\dagger_y,a^\dagger_z) \equiv a_i^\dagger$  and their conjugates
 \begin{equation} \nonumber
 a_i^\dagger \equiv \frac{1}{\sqrt{2}} \left( -\frac{\partial}{\partial r_i} + r_i \right), \;\quad a_i\equiv \frac{1}{\sqrt{2}} \left( \frac{\partial}{\partial r_i} + r_i \right), \vspace{1pt}
 \end{equation}
which satisfy the usual commutation relations.
Here $\boldsymbol{r}=(\boldsymbol{r}_1-\boldsymbol{r}_2)/\sqrt{2} b$ is the dimensionless Jacobi coordinate.
Defining projections with good angular momentum,
 $a^\dagger_M = \hat{\boldsymbol{e}}_M \cdot \boldsymbol{a}^\dagger$ and $\tilde{a}_M=(-1)^{M} a_{-M}$, where
 $\hat{\boldsymbol{e}}_M$ is the spherical unit vector, we can form the scalar HO nodal
 raising/lowering operators $\hat{A}^\dagger \equiv {\bf a}^\dagger \odot {\bf a}^\dagger$,  $\hat{A} \equiv {\bf \tilde{a}} \odot {\bf \tilde{a}}$
 \begin{equation}
  \hat{A}~ \ket{n \ell m } = -2\; \sqrt{\left(n-1\right)\left(n+\ell-{1/ 2}\right) }  \; \ket{n-1 \, \ell m }, \nonumber
 \end{equation}
where $\ket{n \ell m}$ is a normalized HO state.  Using
\begin{align}
 \delta(\boldsymbol{r})  &= \sum_{n^\prime n} d_{n^\prime n} \ketbra{ n^\prime 0 0 }{ n 00 } \, %,~~~~ \hfill
 \nonumber \\
 d_{n^\prime n}  &\equiv \frac{2}{\pi^2} \left[ \frac{\Gamma(n^\prime +{\frac{1}{2} }) \Gamma(n+\frac{1}{2}) }{ (n^\prime-1)! \, (n-1)!} \right]^{1/2} ,% \hfill
 \end{align}
 HOBET's short-range expansion can be carried out~\cite{Haxton:2007hx}.
 We obtain for the $^1S_0$ channel at \nxlo{3}
\begin{align}
\label{eqn:HobetExpansion}
V_\delta^{\mathrm{S}} &=\sum_{n^\prime n} d_{n^\prime n} \, \Big[ a^{\mathrm{S}}_{LO} \, \ketbra{ n^\prime \, 0 }{ n \, 0}
\nonumber \\&\phantom{=}
%&~~~
    + a_{NLO}^{\mathrm{S}} \, \left( \hat{A}^\dagger  \ketbra{ n^\prime \, 0 }{ n \, 0 }  +  \ketbra{ n^\prime \, 0 }{ n \, 0} \hat{A} \right)
\nonumber \\&\phantom{=}
%&~~~
    + a_{NNLO}^{\mathrm{S} , 22} \, \hat{A}^\dagger  \ketbra{ n^\prime \, 0 }{ n \, 0 } \hat{A}
\nonumber \\&\phantom{=}
%&~~~
    + a_{NNLO}^{\mathrm{S} ,40} \, \left( \hat{A}^{\dagger \, 2} \ketbra{n^\prime \, 0 }{ n \, 0} +  \ketbra{n^\prime \, 0 }{n \, 0} \hat{A}^2 \right)
\nonumber \\&\phantom{=}
%&~~~
    +  a_{N^3LO}^{\mathrm{S} ,42} \, \left( \hat{A}^{\dagger \, 2} \ketbra{n^\prime \, 0}{n \, 0} \hat{A} + \hat{A}^\dagger \ketbra{n^\prime \, 0 }{ n \, 0} \hat{A}^2 \right)
\nonumber \\&\phantom{=}
%&~~~
    +   a_{N^3LO}^{\mathrm{S} ,60} \, \left( \hat{A}^{\dagger \, 3} \ketbra{n^\prime \, 0}{n \, 0} + \ketbra{n^\prime \, 0 }{ n \, 0} \hat{A}^3 \right) \Big] .
\end{align}
where the LECs $a_{LO}, a_{NLO}, ...$ carry units of energy.  The HO matrix elements are
\begin{multline} \label{eq:deltaVMe}
\langle n^\prime (\ell^\prime=0\,S)JM | V_\delta^\mathrm{S} \ket{ n(\ell=0\,S)JM } =
d_{n^\prime n}~
    \biggl[
    a_{LO}^{\mathrm{S}}
    -2 c_{n^\prime n}^S\, a_{NLO}^{\mathrm{S}}
\\
    +4 c_{n^\prime n}^{S,22}\, a_{NNLO}^{\mathrm{S},22}
    +4 c_{n^\prime n}^{S,40}\, a_{NNLO}^{\mathrm{S},40}
%\\
    -8 c_{n^\prime n}^{S,42}\, a_{N3LO}^{\mathrm{S},42}
    -8 c_{n^\prime n}^{S,60}\, a_{N3LO}^{\mathrm{S},60} \biggr] \,,
\end{multline}
with the coefficients
\begin{align}
    c_{n^\prime n}^S &=
        (n^\prime{-}1)+(n{-}1)\, ,
\nonumber\\
    c_{n^\prime n}^{S,22} &=
        (n^\prime{-}1)(n{-}1)\, ,
\nonumber\\
    c_{n^\prime n}^{S,40} &=
        (n^\prime{-}1)(n^\prime{-}2)+(n{-}1)(n{-}2)\, ,
\nonumber\\
    c_{n^\prime n}^{S,42} &=
        (n^\prime{-}1)(n^\prime{-}2)(n{-}1)+ (n^\prime{-}1)(n{-}1)(n{-}2)\, ,
\nonumber\\
    c_{n^\prime n}^{S,60} &=
        (n^\prime{-}1)(n^\prime{-}2)(n^\prime{-}3)+(n{-}1)(n{-}2)(n{-}3)\, .
\end{align}

In tensor channels the angular momentum raising and lowering operators are needed, formed
from the aligned coupling of the spherical creation and annihilation operators
%\begin{equation}
%\begin{gathered}
\begin{align}
\langle n \ell || \left[ {\bf {a}}^\dagger  \otimes \cdots \otimes{\bf {a}}^\dagger \right]_\ell  || n 0 \rangle
    &=(-1)^\ell \langle n 0 || \left[ {\bf \tilde{a}} \otimes  \cdots \otimes{\bf \tilde{a}}\right]_\ell  || n \ell \rangle%  \hfill
\nonumber\\&=
    2^{\ell/2}  \sqrt{ \frac{l! }{ (2 \ell-1)!!} \frac{\Gamma[n+\ell+ \frac{1 }{ 2}] }{ \Gamma[n+ \frac{1 }{ 2}]}} , %\hfill
\end{align}
%\end{gathered}
%\end{equation}
where $||\ldots||$ denotes a reduced matrix element.  The needed operator expansion can then be constructed, e.g.,
\begin{align}
V_\delta^{\mathrm{SD}} = &\sum_{n^\prime n} d_{n^\prime n}
    \Big[ a_{NLO}^{\mathrm{SD}} \big\{
        [{\bf a}^\dagger \otimes {\bf a}^\dagger ]_2
        | n^\prime 0 \rangle \langle n 0 |
%\nonumber\\&+
    \ +\ | n^\prime 0 \rangle \langle n 0 |
        [{\bf \tilde{a}} \otimes {\bf \tilde{a}} ]_2
    \big\}
    \odot
    [ \boldsymbol{\sigma}_1 \otimes \boldsymbol{\sigma}_2 ]_2
    +\cdots
    \Big]\, .
\end{align}
Full results through N$^3$LO for all contributing channels can be found in Ref.~\cite{Haxton:2007hx}.

Equation (\ref{eq:deltaVMe}) shows that HOBET's  ladder operator expansion generates a characteristic dependence on nodal quantum
numbers $n,n^\prime$: $a^S_{LO}$ is the only LEC contributing to the HO
$1s$-$1s$ ($n=n^\prime=1$) matrix element, $a^S_{NLO}$ is the
only additional LEC contributing to the $1s$-$2s$ matrix element, etc.   Consequently if one starts with an NN potential --
the two-step process described previously for either a hard potential like Argonne $v_{18}$ or a softer one like
$V_{\mathrm{low}~k}$ -- the LECs can be fixed in a scheme-independent way, once one computes
individual matrix elements of the effective interaction.  If $a^S_{LO}$ is determined from the $1s$-$1s$ matrix element in a LO calculation, that
value will not change at NLO, and so on.
The $s$-wave LECs are determined by expanding the HO wave functions as a Gaussians times
a finite polynomial in $r^2$, then equating terms
\begin{eqnarray} \label{eq:TalmiInt}
&& \int r^{\prime \, 2} dr^\prime r^2 dr  ~ r^{\prime \, 2 p^\prime} e^{-{r^\prime}^2/2} \,
 V_{\textrm{sr}}(r^\prime, r) \, r^{2 p} e^{-{r}^2/2}  \nonumber  \\
&& \quad \equiv\int r^{\prime \, 2} dr^\prime r^2 dr  ~ r^{\prime \, 2 p^\prime} e^{-{r^\prime}^2/2} \,
 V_\delta^S \, r^{2 p} e^{-{r}^2/2} \,,
 \end{eqnarray}
 where $p^\prime=n^\prime-1$, $p=n-1$, and $V_{\textrm{sr}} \equiv V^{UV}+ V G_{QH}QV$: the LECs are determined
 by this nonlocal generalization of Talmi integrals~\cite{DeShalit1963}.

\subsubsection{Pionful HOBET's Power Counting for the NN System \label{sec:hobet_pc}}

\begin{figure}
\centering
\includegraphics[width=0.97\textwidth]{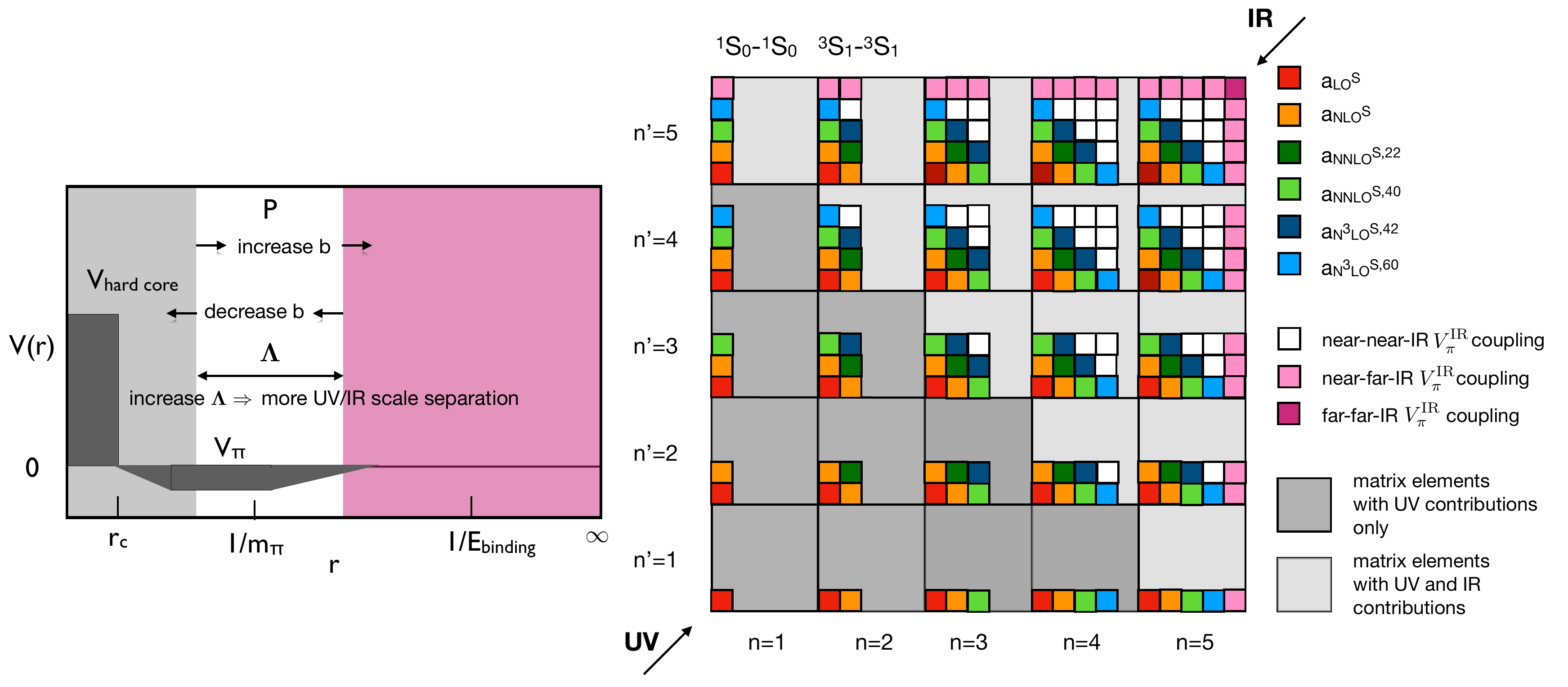}
\caption{\label{fig:Swave}
a) A schematic of HOBET's Hilbert space decomposition into UV (shaded), $P$ (white), and IR (pink)
components.  $\Lambda_{\xSM}$ controls the separation of the UV-IR spaces, while $b$ can be used to reposition
$P$ relative to the underlying NN potential.   b) HOBET's N$^3$LO $^1S_0{-}^1S_0$/$^3S_1{-}^3S_1$ operator structure
for $\Lambda_{\xSM}=8$.  The large squares label HO matrix elements;
the small ones indicate contributing operators/LECs.
LECs determined
in the UV expansion correct the short-range contributions to matrix elements.  The IR expansion begins with the edge state-edge state LEC
indicated in darker pink, moving toward the UV (lower orders) -- which may require elaboration of the chiral expansion,
to capture the needed physics. The chiral expansion generates all other LECs. Large squares with darker gray background indicate
matrix elements entirely determined by the UV expansion at N$^3$LO; those in light gray require in addition IR corrections.}
\end{figure}

HOBET can be summarized as follows:
\begin{enumerate}
\item HOBET's effective interaction within $P$ is completely defined by the set of LECs accompanying the
 ladder operators of Sec.~\ref{sec:operators}.  This includes a specific linear combination of operators/LECs associated with IR
 corrections in $Q$ that are introduced through the edge states, and thus depend sensitively on $E$.
 \item HOBET's short-range or UV expansion fixes the lowest order LECs.  This
expansion quickly accounts for essentially all of the UV physics missing from $P$, encoding this information
 in a small set of LECs for the low-order operators, which are then determined by matching experiment (or LQCD).
 \item The IR has a complementary expansion
 that exploits our knowledge of the long-distance
 nuclear problem.  HOBET's far-IR behavior is governed by the kinetic energy operator and by the binding energy $E$,
 which HOBET makes explicit.  HOBET's near-IR behavior is governed by chiral symmetry.  All non-UV
 LECs are computable functions of $E$, $f_\pi$, and $m_\pi$.
 \item HOBET's IR treatment requires a chiral operator expansion analogous to
 that of \figref{fig:Hierarchy_EFT}.  HOBET's use of a single set of operators
 guarantees seamless matching of the UV and IR expansions.  The IR chiral expansion needs to be accurate only beyond the range
 of the UV expansion.
 \item Both expansions are controlled by dimensionless parameters that depend on $b$.  Thus $b$ can be tuned to optimize the
 intermediate point at which the two expansions meet.
 \end{enumerate}

Panel a) of \figref{fig:Swave} illustrates the connection between HOBET's UV/IR operator structure  to
 its cutoffs $\Lambda_{\xSM}$ and $b$.  For typical choices of $b$ very strong ``hard core"  interactions -- often modeled in terms
 vector meson exchange, Pauli repulsion, etc. -- as well as short-range contributions of pion  exchange (including the
 sigma meson) are not resolved in $P$.  HOBET treats all such short-range
 physics equivalently, determining its effects on the effective interaction in $P$ in a few LECs for the lowest order
 operators in the
 expansion, fitting these
 to experimental observables.  This contrasts with chiral EFT,
 where an attractive and singular
 potential associated with pion exchange is treated explicitly and must be regulated, with its short-range contributions separated out from other
 short-range physics (even though the point-pion potential has little relevance to the physical potential at short range).   We would argue that HOBET's
 agnostic treatment of short-range physics is more appealing in an ET.
 The diagram also shows that for a typical choice of $b$ near the nuclear radius, $m_\pi b \sim 1$, so that over pion ranges,
 the bulk of interactions are absorbed into $P$.  What remains is a distant tail of the pion that extends into the IR, which is
 treated in HOBET through edge-state matrix elements of $V_\pi^{IR}$.

 The choice of $\Lambda_{\xSM}$ and $b$ influence how much and what type of physics is  missing from $P$.
 HOBET's answers are technically
 independent of $b$ and $\Lambda_{\xSM}$ -- though a poor choice of these parameters can  slow convergence,
 making it impractical to obtain a precise answer.  $\Lambda_{\xSM}$ determines the separation  of the scales of UV and IR physics:
 Increasing $\Lambda_\xSM$ decreases, for example, the amount of UV physics omitted from $P$, thus decreasing the complexity
 of the effective operator expansion.   By adjusting $b$ to
 smaller values one re-centers the peak resolution of $P$ to smaller values of $\langle r \rangle$, pulling more short-range
 into $P$, where it will be iterated to all orders in solving the Schr{\"o}dinger equation, and consequently putting more stress on the
 near-IR chiral expansion we describe below.  An adjustment to larger values does
 the opposite.  Thus in HOBET $b$ can be tuned to optimize the meeting point between the short-range expansion
 and the chiral expansion for long-distance physics.
 The fact that ``optimal" values for $b$ are $\sim 1/m_\pi \sim R_N$ where $R_N$ is the
 nuclear radius is not surprising because Nature does a similar optimization: nuclear radii reflect
 a balance between kinetic energy (minimized by complete delocalization) and potential energy (the midrange attraction  influenced by the pion).

%%%%%%%%%%%%%%%%%%%%%%%%%%%%%%%%%%%%%%%%%%%%%%%%
% TALMI FIGURE - moved here from above
%\begin{figure}
\begin{wrapfigure}{R}{0.5\textwidth}
%\centering
\resizebox{0.45\textwidth}{!}{%
\includegraphics[width=0.5\textwidth]{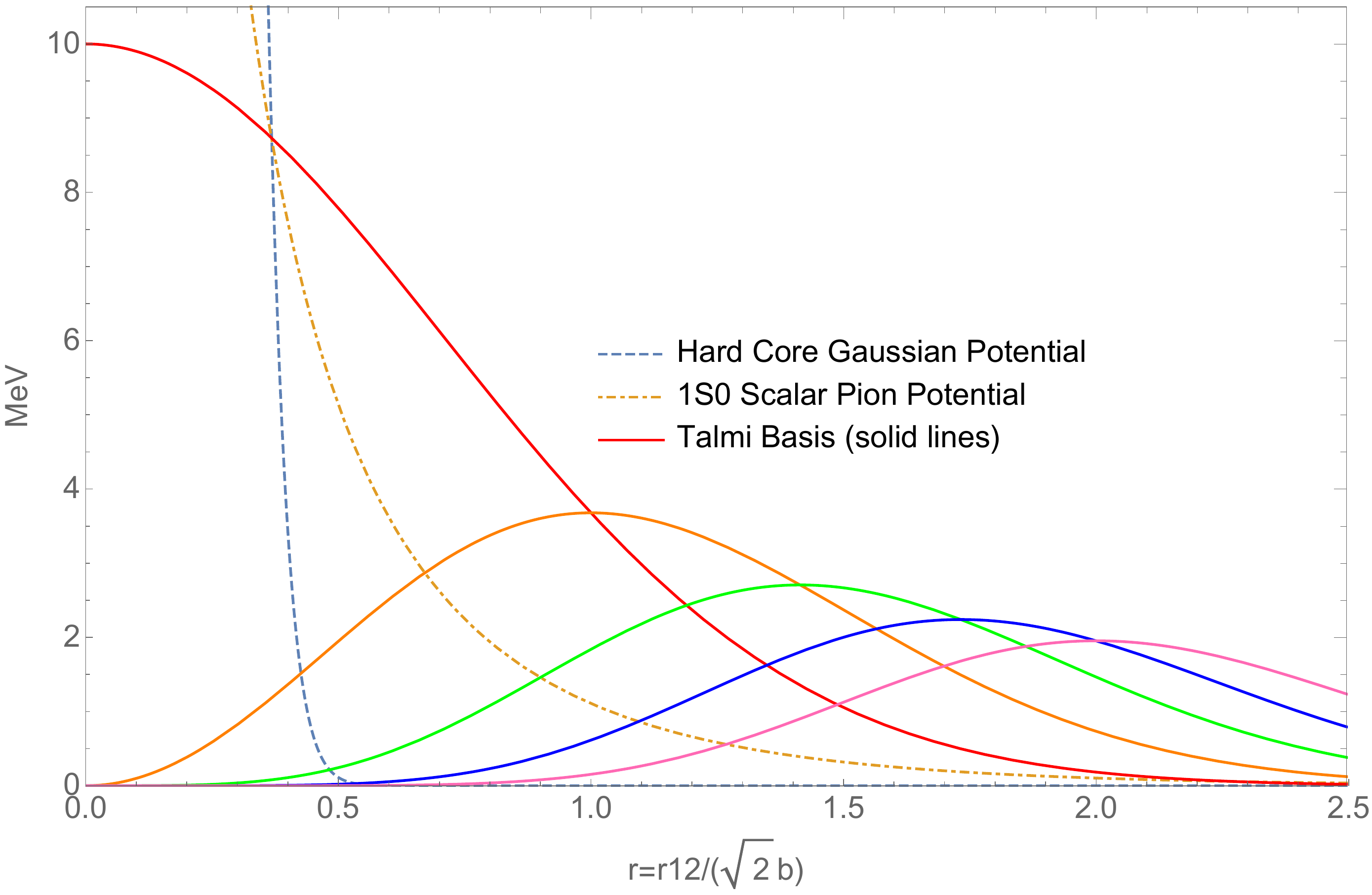}
}
\caption{\label{fig:TalmiBasis}
Talmi basis functions $e^{-r^2} r^{2(p^\prime+p)}$ with $p^\prime+p=(n^\prime-1)+(n-1)$ ranging from 0 to 4 (corresponding to (LO, ..., \nxlo{4}) and scaled with $10/p!$ for viewing.  The peak of each curve is located at $r=\sqrt{p}$.   Here $r=r_{12}/(\sqrt{2}b)$ is the dimension Jacobi coordinate.  Superimposed on these are a Gaussian short-range $^1S_0$ NN potential characteristic the of UV physics that lies outside of the $P$-space, as described in the text,
as well as the one-pion exchange potential contribution to the $^1S_0$ channel.}
\end{wrapfigure}
%\end{figure}
%%%%%%%%%%%%%%%%%%%%%%%%%%%%%%%%%%%%%%%%%%%%%%%%

\Figref{fig:TalmiBasis} includes a local Gaussian potential representative of the UV physics not already captured in $P$, that must be taken into account through the effective operator expansion.  It has the form
 \begin{align*}
 V_{UV}(r) &= V_0 e^{-r_{12}^2/r_{\textrm{sr}}^2}\,,
 	V_0 \sim -1.5 \mathrm{~GeV}\,,
\nonumber\\&
	r_{\textrm{sr}} \sim 0.39 \mathrm{~fm} \,.
\end{align*}
This potential reproduces reasonably well the pattern of $^1 S_0$ LECs derived in a  detailed HOBET calculation that  used the Argonne $v_{18}$ potential~\cite{Haxton:2007hx}.  (The parameters used in that work, $b=1.7$ fm and $\Lambda_{\xSM}=8$, are also employed in examples we show here.)   This comparison of this potential to the Talmi integrands demonstrates the essentials of the UV convergence.  The Talmi integrands peak, in our dimensionless radial units, at $\sqrt{p}$, $p=0,\,1,\,2\ldots$ -- so the peaks move apart more rapidly at small $p$.  The UV  potential (its absolute value is plotted) overlaps significantly with the Talmi integrals for $p=0,\,1$ (LO, NLO), a bit with $p=3$ (N$^2$LO), and negligibly  with $p=4$ (N$^3$LO).
With the representative Gaussian UV potential a dimensionless parameter $r_{\textrm{sr}}^2/(r_{\textrm{sr}}^2+2 b^2)$ governs the convergence: the LECs drop off by this factor, from order to order.  One
 has made an appropriate choice of $b$:  the UV physics has been well captured, and there are enough contributing Talmi
 integrals to provide adequate resolution of its detailed structure.

 If one moves $b$ to small values, or if the short-range expansion were terminated at LO or NLO, there would be UV physics
 not adequately corrected in $P$ by the short-range expansion.  By HOBET's rules, all physics outside the range of the UV
 expansion would be treated in the chiral expansion.  This would stress that expansion -- at short ranges multi-pion corrections
 dominate -- and not be adequate, as non-pionic contributions would be missed.   If one moves $b$ to large values, resolution
 is lost unnecessarily, and the short-range expansion would now penetrate to larger radii where we know the interaction,
 due to chiral symmetry, wasting effort.

\begin{figure*}
\centering
\includegraphics[width=\textwidth]{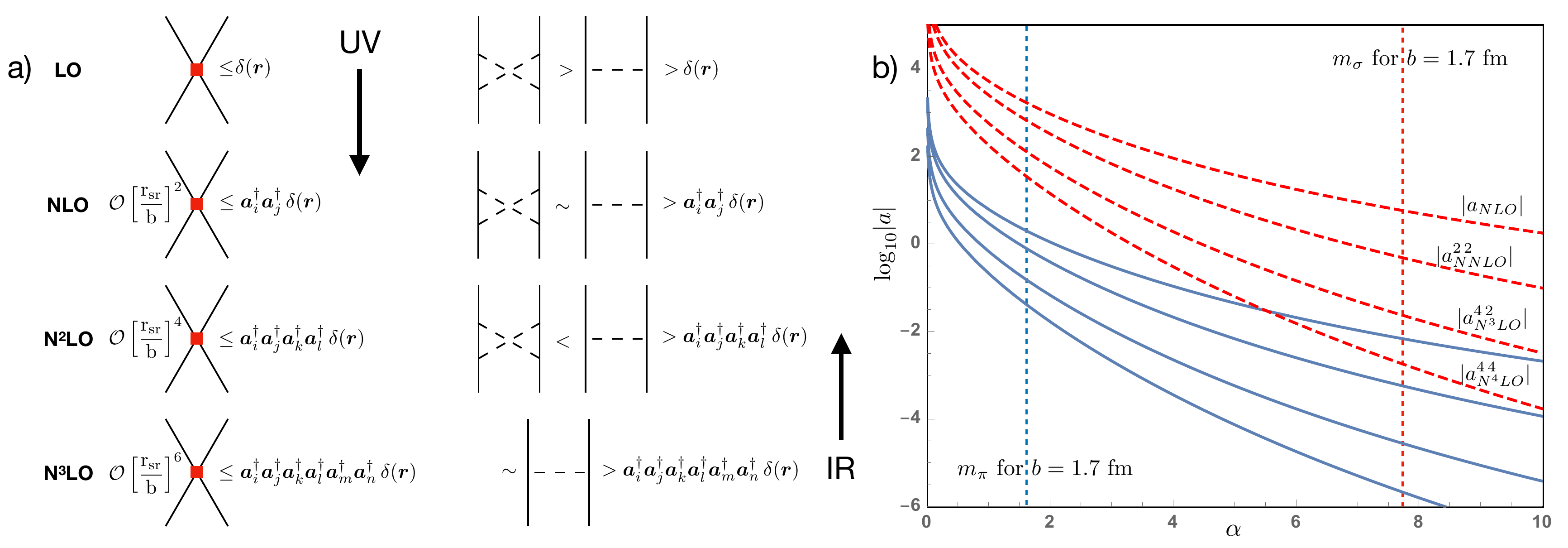}
\caption{\label{fig:Interaction}
a) HOBET's ${}^3S_1-{}^3S_1$ effective interaction.  The UV expansion (first column) proceeds from the top down, from the UV toward larger $r$,
with convergence controlled by the dimensionless parameter $r_{\textrm{sr}}^2/b^2$, discussed in the text.  Various scalar operators are formed by combining the
indicated ladder operators. The IR expansion (second column) proceeds from the bottom up,
from the IR (where OPEP is sufficient) to smaller $r$, where more complicated operators might be needed in the chiral expansion.    Each row thus indicates the theory for a
given UV order.  b) If one approximates the isoscalar contribution of  $V_{2 \pi}^{IR}$ as $\sigma$ exchange, the IR LECs for both $V_\pi^{IR}$ and $V_{2 \pi}^{IR}$ can be evaluated
analytically: they depend on the dimensionless parameter $\alpha = \sqrt{2} \, m b$, where $m$ is the exchanged mass. UV calculations at LO, ..., N$^3$LO require the IR
expansion to provide LECs of orders $\geqslant$ NLO, ..., N$^4$LO.  The leading LECs are plotted for $V_\pi$ (blue solid) and $V_{2 \pi}$ (red dashed).  The vertical dashed lines indicate the values
for $\alpha$ and $\alpha_\sigma$ when $b$= 1.7.  While relativity suppresses the $V_\pi$ contribution,
the Yukawa falloff suppresses $V_{2 \pi}$, with effects larger at higher order.
One finds that $V_{2 \pi}^{IR}$ is only a 5\% correction in a N$^3$LO calculation, for the lowest order IR operator.  In contrast, it is dominant at LO.
}
\end{figure*}

 \Figref{fig:TalmiBasis} also shows the absolute value of the Yukawa-like $^1S_0$ potential from  tree-level one-pion exchange (orange dashed line).
 It is apparent that a great deal of this interaction is absorbed into the first few Talmi integrals (more correctly, into HOBET's first few
 effective operators through the values assigned to their LECs).   The pion is a contributor to $V_{UV}$ in \eqnref{eq:split} that,
 together with $V G_{QT} QV$, is represented by the short-range expansion.
 Yet tree-level pion contributions to higher Talmi integrals
 decline slowly, leaving Talmi integrals for which, in our N$^3$LO example,
 we will have no fitted LECs: the dimensionless parameter governing this convergence is $\alpha \equiv \sqrt{2} b m_\pi \sim 1.68$
 (see \figref{fig:Interaction}b).
 There are in fact an infinite number of contributing Talmi integrals, because in the edge states, $V_\pi^{IR}$
 is sandwiched between the Green's functions $E G_{QT}$.  One thus sees that the IR corrections that must be added to $P$,
 to correct from strong interactions occurring at long distances, are controlled by $E$: this physics is a major advantage of an
 energy-dependent theory, as one has access to the parameter that determines the specific linear combination of states in $Q$
 that must be summed to give the needed IR correction.  For bound states, as $|E| \rightarrow 0$, the IR strong interaction
 contribution are enhanced by the $E G_{QT}$ propagator.

 HOBET's IR or chiral expansion is not an LEC expansion, because the LECs for a technically infinite set of operators beyond
 N$^3$LO are completely determined by $E$, $f_\pi$, and $m_\pi$.   It is instead a pion operator expansion, similar to
 (but we will argue simpler than) the corresponding operator expansion in chiral EFT of  Fig.~\ref{fig:Hierarchy_EFT}.

 Just as the UV expansion starts at the UV corner of $P$ -- the $n^\prime=1 \leftrightarrow n=1$ (so $p^\prime=p=0$ in Eq.~(\ref{eq:TalmiInt})) matrix element is determined by
 $a_{LO}$ -- and then continues to larger $\langle r \rangle$, moving upward along the diagonal, it is helpful to view the IR chiral expansion
 in a complementary way, as a series that starts in the opposite IR corner of $P$ -- the $n^\prime = 5 \rightarrow n=5$ matrix element,
 determining the remaining LECs in a similar, step-wise fashion.   \Figref{fig:Swave} guides our
 discussion: it graphically
 depicts an N$^3$LO $^1S_0$ calculation for $\Lambda_{\xSM}=8$.  The large boxes represent
 the needed $P$-space HO matrix elements.
 Within each box the small red, orange, green, and blue boxes represent the LO, NLO, NNLO, and N$^3$LO LECs determined in
 the short-range operator expansion.  The pink (edge states -- involving both interactions in $P$ and corrections for long-range
 strong interactions in $Q$) and white boxes (interactions in $P$) represent the HO ladder operators beyond N$^3$LO,
 that we have argued (with $b$ properly chosen) are determined by $V^{IR}_\pi$.  The purpose of this discussion is to determine
 the operator form of $V^{IR}_\pi$, specifically what steps we must take to evolve that operator as we move from the most IR
 contribution to $P$ -- the upper right corner of \figref{fig:Swave} -- downward along the diagonal,
 and eventually joining our UV
 ladder operator expansion.   The starting point is the edge-edge state contribution indicated in dark pink in the figure,
 which contributes to the $P$-space $n^\prime=5 \leftrightarrow n=5$ matrix element.  Because
 this is an edge state, it involves an infinite series of Talmi integrals for which $p^\prime+p \geqslant 8$, starting with
 $p^\prime=p=4$.  This would be an N$^8$LO contribution in our UV operator terminology, very long ranged.
 For $p=8$ the Talmi integrand peaks at $r = \sqrt{8}$
 or $r_{12} = 4b \sim 6.8$ fm $\sim 4.8/m_\pi$.  It is abundantly apparent that $V^{IR}_\pi$ at N$^3$LO is very well approximated
 by one-pion exchange.

Moving inward, the next step with be the $p^\prime=4 \leftrightarrow p=3$, $p^\prime=3 \leftrightarrow p=4$ contributions -- or N$^7$LO.  This
operator contributes to transitions involving one edge state (light pink in \figref{fig:Swave}), being the
shortest range component.
This is the leading order chiral (longest range) contribution to the $n^\prime=5 \leftrightarrow n=4$, $n^\prime=4 \leftrightarrow n=5$ $P$-space
matrix element, and an additional contribution to the $n^\prime=5 \leftrightarrow n=5$ HO matrix element.
For $p=7$ the Talmi integrand peaks at $r_{12} = \sqrt{14} b \sim 6.4$ fm:  large $\sqrt{p^\prime+p}$ values correspond
to contributions peaked at long distance.

One would continue this process through N$^4$LO.  At each step one corrects $P$-space HO oscillator matrix elements further
down the diagonal of \figref{fig:Swave} for the first time, and adds new corrections to matrix element above the diagonal.
At N$^4$LO one joins onto the UV operator expansion, correcting for the first time the set of matrix elements immediately above
the gray background area of \figref{fig:Swave} (matrix elements entirely determined by the N$^3$LO UV expansion).

The key question to ask is whether or not one-pion exchange (iterated to all orders in the Schr{\"o}dinger equation, in
combination with HOBET's short-range and kinetic energy operators) is an adequate description of HOBET's $V_\pi^{IR}$, or whether
instead we need a more complicated treatment resembling the NN expansion of \figref{fig:Hierarchy_EFT}.   A full and careful discussion
of HOBET's chiral power counting, including the treatment of tensor contribution in triplet NN channels, will appear elsewhere~\cite{KenWick2}.
HOBET has an intrinsically simple power counting, which we will elaborate on.
First, the theory requires self-consistent solutions,
which guarantees there is a fitted LEC in every channel.  Thus, for example, the s-wave one-pion-exchange contact interaction that
contributes to the $^1S_0$ channel is trivially absorbed into the fitted $a_{LO}$.   More generally, while one could insist on treating pion
exchange at all length scales as a point-nucleon exchange, it makes no sense to do so, because all short-range LECs are fitted to observables.
All one would accomplish is to alter the numerical values of LECs, while leaving the physics unchanged.  Finally, pion operator expansions are
simpler in HOBET because $P$ and $Q$ are defined in terms of the total energy of a Slater determinant, a many-nucleon
condition.  This contrasts with the momentum cut in chiral EFTs, which allow one nucleon to scatter into $Q$ (represented by a
contact operator) while interacting with a second in $P$.  For NN interactions in HOBET, either both nucleons are in $P$ or
both are in $Q$.

A simple channel to elucidate the counting is the ${}^3S_1-{}^3S_1$, which is isoscalar and has no tensor contribution.
In isoscalar channels one
can reasonably approximate two-pion-exchange contributions by an isoscalar sigma exchange.   We do so here, treating the
sigma as elementary, because this approximation will allow us to isolate the dimensionless parameters governing the
competition between one- and two-pion exchange.  In the $T=0$ channel the two-pion exchange is relativistically enhanced over
one-pion exchange, allowing us to evaluate this enhancement against the suppression that accompanies HOBET's treatment
of the pion as exclusively a long-range contribution,
$V_\pi^{IR}$. The potentials are
\begin{eqnarray}
~V_\pi(\mathbf{r})& =& \left( \frac{g_A}{\sqrt{2}f_\pi}\right)^2
    \frac{m_\pi^3}{12 \pi} \Big[
        \frac{e^{-\alpha r}}{\alpha r} \mathbf{\sigma}_1 \cdot \mathbf{\sigma}_2
        +\mathrm{tensor}+ \mathrm{contact}
    \Big] \mathbf{\tau}_1 \cdot \mathbf{\tau}_2
\nonumber \\
~~V_{2 \pi} (\mathbf{r}) &=&
    -\frac{g_s^2}{4\pi} m_\sigma \frac{e^{-\alpha_\sigma r}}{\alpha_\sigma r}
    \left( \mathbf{\tau}_1 \cdot \mathbf{\tau}_2 \right)^2\, ,
\end{eqnarray}
where the tensor contribution cannot contribute to the $^3S_1$ channel while the contact term cannot contribute to $V_\pi^{IR}$
as the LO operator must be part of the UV expansion.  Here $\alpha = \sqrt{2} b \, m_\pi $ and $\alpha_\sigma=\sqrt{2} b \, m_\sigma$,
and the various constants are assigned values $g_A=1.276$, $f_\pi=130.4$~MeV, $m_\pi=138.03$~MeV, $g_s^2=1.45$, and $m_\sigma=650$~MeV.

The general formula for the LECs is
\begin{align}
a_{\ell+p^\prime+p}^{\ell,2p^\prime+ \ell, 2p+\ell} &=
    \frac{(2\ell+1)!!}{2^\ell \ell!}
    \langle p^\prime (\ell S)J; T | V(\mathbf{r}) | p (\ell S) J; T \rangle
\nonumber \\
R_{p \,  \ell} &\equiv
    \frac{\pi}{p! \Gamma[\ell+p+\frac{3}{2}]}
    r^\ell e^{-r^2/2} \left( \frac{r^2}{2} \right)^p
\end{align}
The superscripts of $a$ give the angular momentum channel, and the number of derivatives applied to the left and right by the operator, and the subscript gives the operator order,
yielding for example the correspondence $a^{0,4,2}_{3} = a^{S,42}_{N^3LO}$, as used in Eq.~\eqref{eqn:HobetExpansion} for example.
The relevant values of the operator order here are those beyond the order of the UV expansion.
Using the monomial radial wave functions the matrix elements can be evaluated analytically for $V_\pi$ and $V_{2 \pi}$, yielding expressions involving error functions,
governed by the dimensionless parameters $\alpha$ and $\alpha_\sigma$.  This formula determines
the LECs corresponding to the the white and pink boxes in \figref{fig:Swave}, panel b).

 With one underlying set of operators and two expansions that approach each other from the UV and IR sides.
 one has the opportunity to assess errors by studying the matching.   This provides an empirical test of the adequacy
 of the chiral operator structure employed.  For example, the standard HOBET calculations
 of Ref.~\cite{McElvain:2019ltw}
 were done in N$^3$LO, a choice that yields about a keV error in the binding energy of the deuteron.   At the
 outset the simplest IR operator structure -- just $V_\pi^{IR}$ -- was assumed, then this assumption was tested.
 An ideal testing ground is the $^1F_3$ partial wave, which has one contributing N$^3$LO operator.  The LEC
 was determined from the fit (see below) to phase shift data.   That value was then matched to the value predicted by
 $V_\pi^{IR}$, with $f_\pi$ treated as an adjustable variable: the standard value of $f_\pi$ was obtained to within a few percent.
 This gave great confidence that the IR expansion -- which describes N$^4$LO LECs and higher -- would be
 reliable at tree-level.  This result is consistent with the conclusions drawn from \figref{fig:Interaction}.

%\begin{figure}
\begin{wrapfigure}{R}{0.45\textwidth}
\centering
\includegraphics[width=0.4\textwidth]{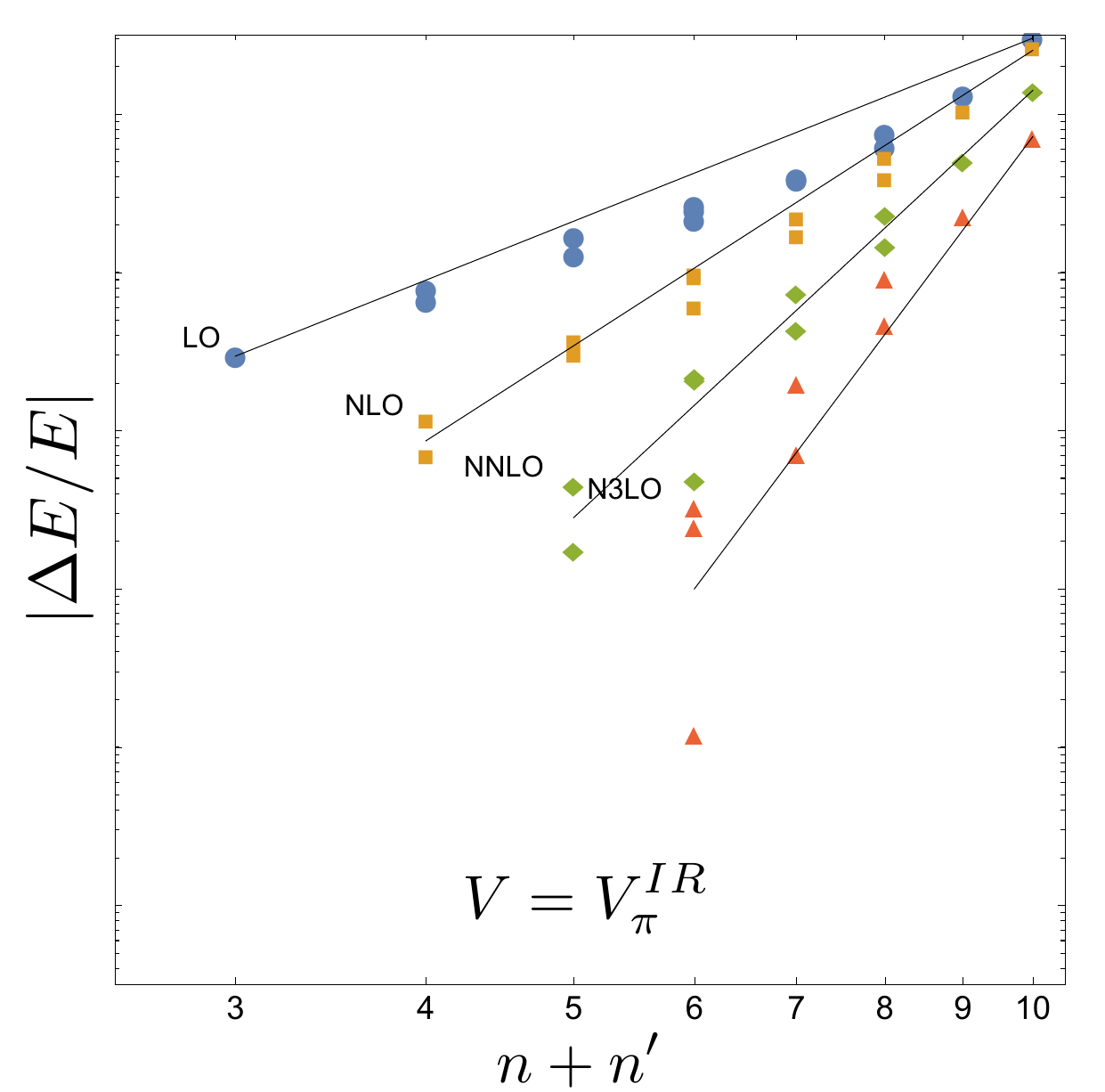}
\caption{\label{fig:HobetLepage}
The Lepage plot of $S$-wave matrix elements of $H^{\mathrm{eff}}$ computed in HOBET
in the deuteron ($^3S_1-^3D_1$) channel at the deuteron binding energy,
starting from Argonne $v_{18}$
phase shifts.  The fractional error in $H^{\mathrm{eff}}$ -- calculated by
comparing the results to exact Argonne $v_{18}$  $H^{\mathrm{eff}}$ matrix elements -- is plotted vs.
the sum of the nodal quantum numbers.  All points plotted are predictions unconstrained by the fit.}
\end{wrapfigure}
%\end{figure}

The Lepage plot of \figref{fig:HobetLepage} shows that HOBET is a rapidly convergent, predictive ET.  HOBET was executed order by order using phase shifts taken from the Argonne $v_{18}$ potential, then
all of the matrix elements not used in the fit were predicted and compared to results computed directly from the potential.  The scheme-independent procedure is described in
Ref.~\cite{Haxton:2007hx}.  The figure shows the fractional errors as a function of nodal quantum numbers, $n^\prime+n$, of the matrix element.  The lines show the trends.
The rapid steepening of the lines with order demonstrate the kind of systemic improvement characteristic of a well-behaved ET.

%%%%%%%%%%%%%%%%%%% Fitting LECs %%%%%%%%%%%%%%%%%%%%%%%%%%%%%%
\subsubsection{Fitting HOBET Short-Range LECs}
The Green's function in $V G_{QH} Q V$ results in a small residual energy dependence of the term being expanded.
However, the expansion~\eqnref{eqn:HobetExpansion} contains lowering operators that are closely related to gradient operators and are therefore sensitive to the energy of the wave function.    It was demonstrated in Ref.~\cite{McElvain:2019ltw} that the small residual energy dependence can be absorbed by the highest order operators in a fit, leaving the LECs constant, making the fitting and evaluation of $H^{\mathrm{eff}}\left(E\right)$ much simpler.

The energy independent LECs of the expansion in \eqnref{eqn:HobetExpansion} can be fit to observables such as phase shifts in a straightforward way.
Given a set of phase shift samples $\left\{ E_i, \delta_i \right\}$ and LECs we can construct an approximate effective Hamiltonian
\begin{equation}
H^{\mathrm{eff}}\left(E_i, \, \delta_i, \, \mathrm{LECs} \right) P \ket{\psi^\prime_i} = \epsilon_i P \ket{\psi^\prime_i}\;,
\end{equation}
where the eigenvectors and eigenvalues differ from $E_i$ and $P\ket{\psi_i}$ due to the LEC values and order cutoff on the expansion.
We minimize a cost function of the difference in eigenvalues while taking into account the impact of omitted LEC orders.
\begin{equation} \label{eqn:HobetFit}
\chi^2_{\mathrm{order}~N} = \sum_{i \in \{ \mathrm{sample} \} } \frac{(\epsilon^N_i - E_i)^2 }{ \sigma_{N+1}(i)^2 } ,
\end{equation}
where $\{ \mathrm{sample} \}$ represents the set of energy points used.

The variance $\sigma_i^2$ is an estimate of the contributions of omitted higher-order LECs not included
in the fit.
\begin{equation}
\sigma^2_{N+1}(i) \sim  \kappa^2_{N+1} \sum_{ \{ a_j^{N+1} \} } \left(  \frac{ \partial \epsilon_i^{N+1} }{ \partial a_j^{N+1} } \biggr\rvert_{a_j^{N+1} =0} \right)^2 .
\label{eq:var}
\end{equation}
Here $\epsilon_i^{N+1}$ is the eigenvalue at one order beyond that being employed in the fit, $\kappa$ the scale of the LECs of the next order, and
$\{ a_j^{N+1} \}$ is the set of LECs that contribute in that order.     At low energies, high order contributions are suppressed.
%The variance automatically selects the relevant set of samples in a soft way for the order of the fit.
The theory variance automatically deemphasizes the samples where the as yet unknown values of higher-order LECs make large
shifts in the $H^{\mathrm{eff}}$ eigenvalue.  Using the variance in the fit function replaces the need to trim the sample set to the appropriate energy range for the fit's order.

\begin{wrapfigure}{R}{0.5\textwidth}
%\resizebox{0.48\textwidth}{R}{%
\includegraphics[width=0.5\textwidth]{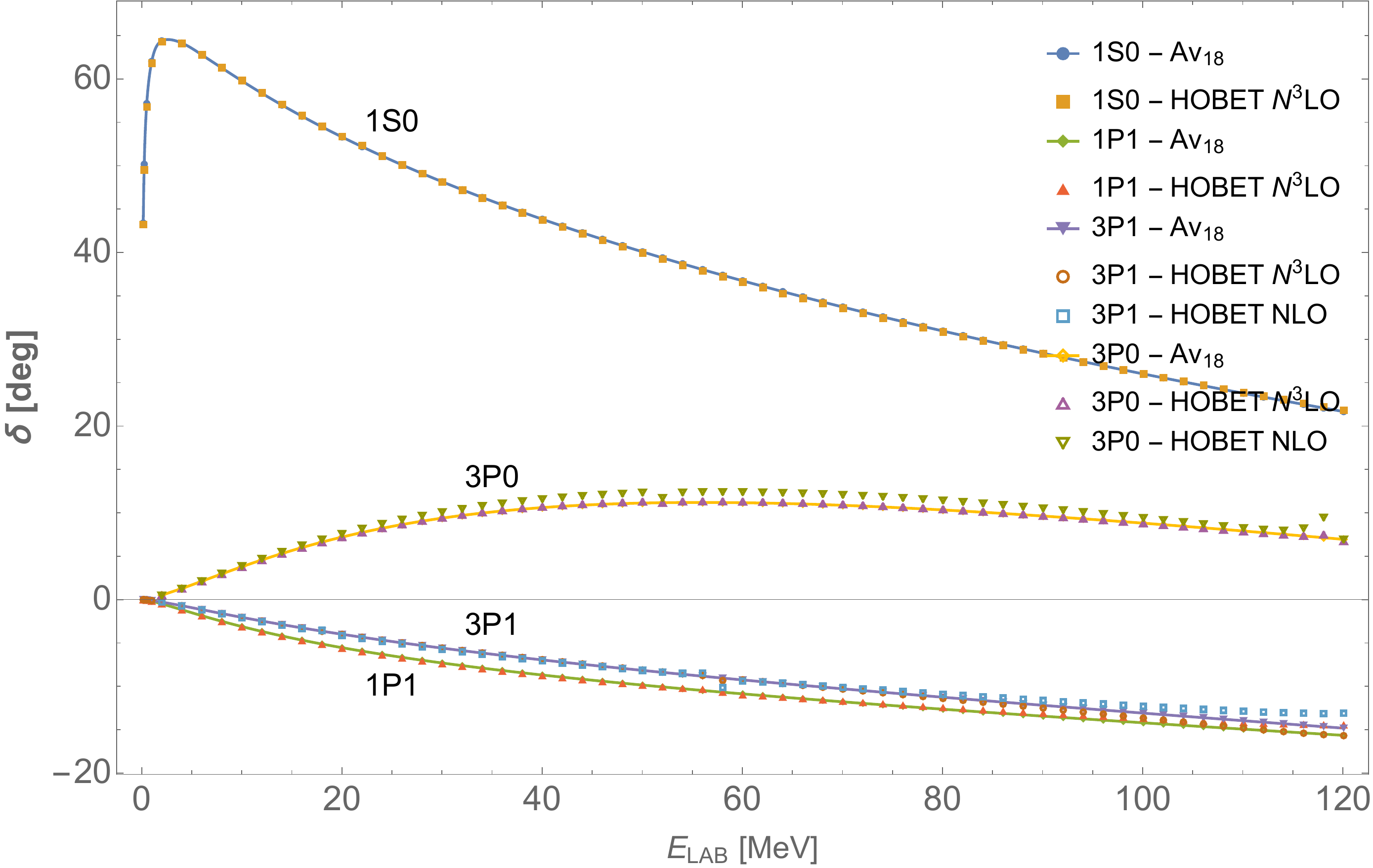}
%}
\caption{\label{fig:HobetPhaseShifts}
Phase shifts regenerated from LECs fit to data from 1 to $80 \MeV$ and compared to the original phase shifts from Argonne $v_{18}$.  In the $^1S_0$ channel the low energy behavior down to 50 keV associated with a resonance at $\sim 74 \keV$ is reproduced from data above $1 \MeV$.   In the $^3P_0$ and $^3P_1$ channels even NLO results based on a single LEC reproduce phase shifts quite well.  }
\end{wrapfigure}

Using cutoffs of $\Lambda_{\xSM}=8$ and $b=1.7$ fm,  a one-pion-exchange potential (OPEP) for $V_{IR}$, and 40 phase shifts samples derived from the Argonne $v_{18}$ potential~\cite{Wiringa:1994wb} evenly spaced in $k$ from 1  to $80 \MeV$ in various channels, LECs up to N$^3$LO were fit.
From a small number of LECs parameterizing the small basis $H^{\mathrm{eff}}$ in each channel phase shifts are regenerated and compared to the original phase shifts in \figref{fig:HobetPhaseShifts}.

This convergence answers a persistent effective interactions problem first clearly analyzed in the early
1970s~\cite{Barrett:1970ner,Schucan1972,Schucan1973}.
The exemplar used was of two nucleons outside an $^{16}$O core.
Various perturbative expansions of the effective interaction in a $P$-space were tried without success.
The convergence problem was traced to a spectrum overlap between $PHP$ and $QHQ$, preventing convergence of perturbative expansions.
In HOBET all states of $H$ overlapping $P$ are included, avoiding the spectrum overlap problem.
Energy dependence enables HOBET to carry much more information about the spectrum of $H$ in a small $P$-space.

%\begin{figure}

Early LQCD calculations of NN phase shifts with non-physical pion masses of $800$~ \cite{Beane:2013br,Berkowitz:2015eaa} and $450 \MeV$~\cite{Orginos:2015aya} demonstrated the feasibility of computing phase shifts directly from QCD.    Recent advances in supercomputers and techniques will soon give us access to physical pion mass results.  Thus in the relatively near future, the HOBET program can
be executed with such input.   Section~\ref{sec:hobetbox} describes how to adapt HOBET to use LQCD input in a more elegant way.

\subsubsection{A-body HOBET \label{sec:hobetAbody}}
While the HOBET discussions here have focused on the two-body system -- the operators, power counting, and LEC fitting --
the goal of HOBET is to build a systematic effective theory of nuclei in which
two and higher-body input data from experiment or
LQCD can be propagated into more complicated $A$-body systems, to produce results with quantifiable uncertainties.   The
soundness of this approach is predicated on many years of shell model phenomenology,
where a similar approach based on fitted potentials has been successful.  While a full discussion of this topic is beyond the
scope of this paper, we will describe how to embed the two-body effective interaction in an $A$-body system.  This exercise
establishes some interesting connections between HOBET and old concepts, such as the healing distance.

The discussion parallels that for the two-body interaction: we motivated results by using potential theory, but in the end sever
all connections to a potential, creating a true ET.
We assume an $A$-body Hamiltonian that is a sum of two-body terms
\begin{equation}
\label{eq:Ham2}
\hat{H}= \sum_{i<j}^A \left( \frac{2}{A} \hat{T}_{ij} + \hat{V}_{ij} \right) \equiv  \sum_{i<j}^A \hat{H}_{ij} .
\end{equation}
Here we employ ``hats" on operators to stress that they are operators acting over the full Hilbert space.
The factor of $2 / A$ is needed to correctly produce the A-body relative kinetic energy from the two-body kinetic energy.
We will in fact take $A$=3, to simplify the notation, as this is sufficient to illustrate the procedure for general $A$.
The two- and  three-body $P$-spaces $P_{12}$ and $P_{123}$ are defined by the conditions
\begin{align}
P_{12}:&\ \dot{\Lambda}_{1} \leqslant \Lambda_\mathrm{SM} ,  \nonumber \\
P_{123}:&\ \dot{\Lambda}_{1}+\dot{\Lambda}_2  \leqslant \Lambda_\mathrm{SM} ,
\end{align}
where $\dot{\Lambda}_1$ and $\dot{\Lambda}_2$ correspond to the dimensionless Jacobi coordinates $(\boldsymbol{r}_1-\boldsymbol{r}_2)/\sqrt{2}b$ and
$(2 \boldsymbol{r}_3-\boldsymbol{r}_1-\boldsymbol{r}_2)/\sqrt{6} b$, respectively.
In the following development we will use $\Lambda_i$ without the dot to denote an operator measuring the spectator quanta to the pair not involving particle $i$.

The solution to the three-body effective interaction problem is given by the BH equation
\begin{equation}
\label{eq:BH3Body}
P^{\Lambda_{SM}}_{123} \hat{H}^{\mathrm{eff} \, \Lambda_{SM}}_{123}P^{\Lambda_{SM}}_{123} =
    P^{\Lambda_{SM}}_{123} \hat{H}
    \frac{E}{E -Q^{\Lambda_{SM}}_{123} \hat{H}}
    P^{\Lambda_{SM}}_{123} .
\end{equation}
which for a two-body $\hat{H}$ becomes
\begin{equation}
\hat{H}^{\mathrm{eff} \, \Lambda_{SM}}_{123} =
    \left[ \hat{H}_{12}+\hat{H}_{13}+\hat{H}_{23}\right]
    \frac{E}{E- Q^{\Lambda_{SM}}_{123} \hat{H}}.
\end{equation}
One can expand this as a series in two-body scattering.
Starting with representative entry $\hat{H_{12}}$ from the brackets above
\begin{multline}
P^{\Lambda_{SM}}_{123} \hat{H}_{12}
    \frac{E}{E -Q^{\Lambda_{SM}}_{123} \hat{H}}
    P^{\Lambda_{SM}}_{123}
=
    P^{\Lambda_{SM}}_{123} \Bigg[
    \hat{H}_{12}
    \frac{E}{E-Q_{12}^{\Lambda_{SM}-\Lambda_3} \hat{H}_{12}}
%\nonumber \\&
\\
    + \hat{H}_{12}
    \frac{E}{E-Q_{12}^{\Lambda_{SM}-\Lambda_3} \hat{H}_{12}}
    Q^{\Lambda_{SM}}_{123}
    \frac{\hat{H}_{23}}{E}
    \frac{E}{E-Q_{23}^{\Lambda_{SM}-\Lambda_1} \hat{H}_{23}}
%\nonumber \\&
\\
    + \hat{H}_{12}
    \frac{E}{E-Q_{12}^{\Lambda_{SM}-\Lambda_3} \hat{H}_{12}} Q^{\Lambda_{SM}}_{123}
    \frac{\hat{H}_{13}}{E}
    \frac{E}{E-Q_{13}^{\Lambda_{SM}-\Lambda_2} \hat{H}_{13}} \Big)
%\nonumber \\&
    +  \cdots \Big] P^{\Lambda_{SM}}_{123} .
\label{eq:bigspace}
\end{multline}
We emphasize that multiple re-scattering terms always involve at least 3 particles.
We will insert unity, e.g., $P_{12}+Q_{12} \equiv 1$, in the series above.  For example, projector $P^{\Lambda_{SM}}_{123}$
can be written as
\begin{equation}
 P^{\Lambda_{SM}}_{123} = P^{\Lambda_{SM}}_{123} (Q_{12}^{\Lambda_{SM}} + P_{12}^{\Lambda_{SM}}) = P_{12}^{\Lambda_{SM}-\Lambda_3} ,
 \end{equation}
If $\Lambda_3$ exceeds $\Lambda_{SM}$ then the superscript becomes negative, yielding an empty $P_{12}$, and limiting the three-body total quanta to $\Lambda_{SM}$ .
For the projector appearing in the multiple re-scattering terms
\begin{eqnarray}
 Q^{\Lambda_{SM}}_{123} &=& (Q_{12}^{\Lambda_{SM}} + P_{12}^{\Lambda_{SM}})Q^{\Lambda_{SM}}_{123} (Q_{23}^{\Lambda_{SM}} +  P_{23}^{\Lambda_{SM}}) \nonumber \\
 &{\underset{\mathrm{retain}}\longrightarrow}& P_{12}^{\Lambda_{SM}} Q^{\Lambda_{SM}}_{123}  P_{23}^{\Lambda_{SM}} \nonumber \\
 &\equiv& Q_{123}^{\Lambda_{SM}} [\Lambda_3,\Lambda_1] .
 \label{eq:approx}
 \end{eqnarray}
Dropping the two-body $Q$ projectors will discard some three-body physics that we will later argue is both small and can be absorbed in the fitting of a three-body interaction.
Here $Q_{123}^{\Lambda_{SM}} [\Lambda_3,\Lambda_1]$
a matrix whose basis can be practically limited to quanta of $\sim (A-1) \Lambda_{SM}$, and
whose nonzero elements in $Q_{123}$ are
\begin{equation}
\ketbra{\alpha^\prime_{12},\beta^\prime_{3} } { \alpha_{23}, \beta_{1} },
\end{equation}
where $\alpha$ represents the quantum numbers of the first Jacobi coordinate, and $\beta$ the second Jacobi coordinate.
The $\beta_i$ subscript corresponds to the subscript  of $\Lambda_i$ which measures the spectator quanta of the state.
While the Jacobi coordinates used in the ket and bra states differ in the ordering of the underlying single particle coordinates, they can be easily transformed into each other.    The quanta associated with the two Jacobi coordinates are constrained by.
\[ \Lambda_{SM} \ge \Lambda_{12} > \Lambda_{SM}-\Lambda_3,~~~~~\Lambda_{SM} \ge \Lambda_{23} > \Lambda_{SM}-\Lambda_1 . \]
After such replacements, the retained linked two-body amplitudes that involve intermediate summations over an infinite $Q_{123}$ can be
replaced exactly by effective operators of  feasible dimension, the needed matrix elements of which can be determined from our existing
ET treatment of the two-body problem.
The retained part of Eq. (\ref{eq:bigspace}) reduces to
``soft" operators
\begin{multline}
\hat{H}_{12}  \frac{E}{E -Q^{\Lambda_{SM}}_{123} \hat{H}} {\underset{\mathrm{retain}}\longrightarrow} %\nonumber \\
    \Big[ \hat{H}^{\mathrm{eff} \, \Lambda_{SM}-\Lambda_3}_{12}
    + \Big(\hat{H}^{\mathrm{eff} \, \Lambda_{SM}-\Lambda_3} _{12}
    \frac{Q^{\Lambda_{SM}}_{123} [\Lambda_3,\Lambda_1]}{E}
    \hat{H}^{\mathrm{eff} \, \Lambda_{SM}-\Lambda_1}_{23}
\\
    + \hat{H}^{\mathrm{eff} \, \Lambda_{SM}-\Lambda_3} _{12}
    \frac{Q^{\Lambda_{SM}}_{123} [\Lambda_3,\Lambda_2]}{E}
    \hat{H}^{\mathrm{eff} \, \Lambda_{SM}-\Lambda_2}_{13}  \Big)  +  \cdots \Big]
\label{eq:smallspace}
\end{multline}
Note that the needed matrices can all be obtained from the two-body interaction $\hat{H}^{\mathrm{eff} \, \Lambda_{SM}} _{12}$
via
\begin{equation}
\hat{H}^{\mathrm{eff} \, \Lambda_{SM}-\Lambda_3}_{12} = \frac{E}{E - \hat{H}^{\mathrm{eff} \, \Lambda_{SM}} Q^\Lambda_{{SM}_{123}} [\Lambda_3,\Lambda_3]} \hat{H}^{\mathrm{eff} \, \Lambda_{SM}}_{12}
\end{equation}

It should be apparent that the above steps identify every term in the three-body BH equation that can
be determined from the two-body effective interaction:  every term discarded includes at least one two-body scattering
in $Q_{123}$ that cannot be obtained from $\hat{H}^\mathrm{eff}_{12}$.  Consequently if we are executing HOBET as an ET,
e.g., obtaining its two-body LECs for a cutoff $\Lambda_{SM}$ from phase shifts, we do not have access to that
information.  Conversely, discarded terms involve consecutive scatterings of distinct nucleon pairs, separated
by high-energy states.  These terms are then three-body for our choice of $\Lambda_{SM}$ and expected to be dominantly short-range.
HOBET distinguishes these UV unconstrained three-body terms from other three-body terms that can be computed from the
two-body LECs and our IR operators, $T$ and $V_\pi^{IR}$.

To build the next level of HOBET one would replace the UV three-body terms by an expansion as was done in the two-body case.
\begin{eqnarray}
H_{123}^{eff} =P^{\Lambda_{SM}}_{123} \frac{E}{{E - {T_{123}}Q_{123}^{SM}}}\Big[ T_{123} - T_{123} \frac{Q_{123}^{SM}}{E} T_{123}  %\nonumber \\
    + V_{123}^{IR} + {V_{\delta ,123}} \Big]\frac{E}{{E - Q_{123}^{SM}{T_{123}}}}P^{\Lambda_{SM}}_{123}
\end{eqnarray}
\Eqnref{eq:smallspace} encodes our knowledge of induced three-body physics and  determines $V_{123}^{IR}$.  The residual plus the fundamental three-body interaction would then be fit by the expansion $V_{\delta,123}$, whose LECs would be determined from observables such as the $A=3$ binding energy.  Once this is done, no connection to a potential would remain: HOBET becomes a true ET at the  three-body level, with experimental input encoded entirely in the LECs.

By faithfully following HOBET's principles, a connection has been made to an important concept
in nuclear physics, the healing distance \cite{Gomes:1957zz,COESTER1960477}.  Practitioners in the 1950s were greatly puzzled
by the success of mean field descriptions of the nucleus, given the violence of the short-range scattering.  This was
resolved by recognizing that a nucleon pair within a nucleus undergoing repeated
GeV hard-core scattering
is limited to a natural distance/time scale given by the uncertainty principle of $c \Delta t \sim $ fm.  Thus pair wave functions
quickly ``heal" to their mean-field forms.  One can envision integrating out the short-range scattering, to define the mean-field
$A$-body result.

This is effectively the picture that describes the terms retained above - the terms HOBET was designed to cleanly
isolate.  Each two-body term embeds hard-core pair scattering
between IR operators $E/E-QT$ which connect the short range scattering to the long-range physics of $P_{12}$: by replacing
the hard-core scattering by effective operators, we remove the UV two-nucleon physics, while retaining the IR physics (the BH restriction
of the true wave function to $P$).  In pionful HOBET the IR contributions of $V_\pi^{IR}$ are also included. The first term in Eq. (\ref{eq:smallspace})
has a spectator dependence that has been previously noted as necessary in any calculation employing more than
one oscillator shell \cite{Zheng:1995td}.  When this term is used in the Schr{\"o}edinger equation, the scattering within
$P$ by $H^{eff}_{ij}$ is iterated to all orders.  The remain terms involving scattering on multiple nucleon pairs in $Q_{123}$ are three-body, but
not short-range, as the associated two-nucleon short-range physics has been integrated \new{out}, and the scattering begins and ends
in $P_{12}$.  That is, our careful separation of IR and UV effects have allowed us to divide the three-body physics into
two components: the first corresponds to isolated two-nucleon short-range scattering that ``heals" back to $P_{12}$ and
thus is IR from a three-body perspective, while the second has three nucleons interaction at short range, and is UV.

In actual calculations the interactions have been summed to all orders, which is more conveniently done
using an antisymmetrized basis. (We avoided this choice above in favor of one that brings out the physics associated with
spectator nucleons more clearly.)
In principle, repeated applications of $H_{ij}^\mathrm{eff}$ can generate high quanta states.
For example, one can see how the second Jacobi oscillator could reach $\Lambda_{SM}$ quanta and then an application of $H_{12}^\mathrm{eff}$ could raise the first oscillator to $\Lambda_{SM}$ quanta.   An anti-symmetric state with $2\Lambda_{SM}$ quanta will have unbalanced components that will allow higher quanta states to be reached, but the amplitudes for those unbalanced components will be progressively smaller.   In practice, we have seen saturation of results at a few quanta over $2\Lambda_{SM}$, reflecting the replacement the hard core with a much softer effective operator.

The calculation uses the fact that in an anti-symmetric basis with basis members $x$ and $y$,
\begin{equation} \nonumber
\braopket{x}{\sum\limits_{i<j=2}^A{ \hat{O}_{ij} }}{y} = c \braopket{x}{\hat{O}_{12}}{y},~c=A(A-1)/2 .
\end{equation}
Using this fact, the complete result can be obtained with a small rewrite of \eqnref{eq:BH3Body} at the cost of matrix
inversion in the enlarged space.    Following \eqnref{eq:approx} we substitute $H_{12}^{eff}$ for $\hat{H}_{12}$
and evaluate in an anti-symmetrized basis carrying more than $(A -1) \Lambda_{SM}$ quanta for intermediate results.
\begin{eqnarray}
P^{\Lambda_{SM}}_{123} \hat{H}_{123}^{\mathrm{eff}} P^{\Lambda_{SM}}_{123}
\quad{\underset{\mathrm{retain}}\longrightarrow}\quad
%\\
P^{\Lambda_{SM}}_{123} c P_{12} \hat{H}_{12}^{\mathrm{eff}} P_{12}  \frac{E}{E -Q^{\Lambda_{SM}}_{123} c P_{12} \hat{H}_{12}^{\mathrm{eff}} P_{12} } P^{\Lambda_{SM}}_{123} .% \nonumber
\end{eqnarray}

One can then evaluate the residual -- the terms omitted by the replacement in Eq. (\ref{eq:approx}) -- by subtracting
the retained terms from a numerically evaluated exact solution.  As described above, we associate the residual with
the three-body UV physics that would be absorbed by three-body contact operators and their LECs.
We have done this evaluation for a variety of choices of $b$ and
$\Lambda_{SM}$ in the 8 to 12 range.   The three-body residual UV contribution to the binding energy is quite small, about 200 keV  and decreasing with increasing $\Lambda_{SM}$ \cite{WickKen} , and significantly smaller than the known three-body shift of the triton.
This is the first step in verifying the basic premise of HOBET: that if one
carefully separates UV and IR physics, removing the effects of $T$ and $V_\pi^{IR}$, one will find rapid convergence
in the number of nucleons interacting at one time through the strong interaction at short distance.  That is, nuclei are dilute Fermi
systems.  The large three-body terms are the ones easily calculated -- because they are soft from a three-body perspective
involving propagation in $P_{12}$ between successive scatterings.

The procedures described here are now being applied to a series of light nuclei \cite{WickKen}.  If these results substantiate what has
been found for $A=3$ -- that almost all of the relevant three-body physics arises from two-body hard interactions sandwiched between IR
physics governed by $T$ and $V_\pi^{IR}$ -- then accurate predictions for light nuclei can be based on high-quality
NN lattice results.

\subsubsection{Effective Operators in HOBET \label{sec:hobeteffop}}

The BH equation also gives a prescription for operator evaluation.   Given a short-range operator $O$,
\begin{equation} \nonumber
 O_{ji}^{\mathrm{eff},\Lambda }\left( E \right) = P\frac{{{E_j}}}{{{E_j} - HQ}} O\frac{{{E_i}}}{{{E_i} - QH}}P\,.
\end{equation}
In the above, $i$ and $j$ label eigenstates of the Hamiltonian, or of the effective Hamiltonian.
As discussed before, the Green's functions used above with appropriate boundary conditions reconstruct the full wave function from the projections of eigenstates.
For bound states this boundary condition is simply that the wave function exponentially decays outside the range of the interaction.

While the above statement is formally correct, when using HOBET with LQCD input, the full Hamiltonian is not available.
Instead, we would like to express the operator in a parallel way to the HOBET effective interaction, containing a lowering operator expansion with each operator paired with an LEC:
\begin{align*}
O_{ji}^{\mathrm{eff}} &= P\frac{{{E_j}}}{{{E_j} - HQ}} O \frac{{{E_i}}}{{{E_i} - QH}}P \,,\\
	&= P\frac{{{E_j}}}{{{E_j} {-} TQ}}\Big[ {O + VQ\frac{{{E_j}}}{{{E_j} - HQ}} O + O \frac{{{E_i}}}{{{E_i} - QH}}QV}
\\ &\quad 
    + {VQ \frac{{{E_j}}}{{{E_j} - HQ}} O\frac{{{E_i}}}{{{E_i} - QH}}QV} \Big]\frac{{{E_i}}}{{{E_i} - QT}}P\,,\\
&\to P\frac{{{E_j}}}{{{E_j} - TQ}}\left[ { O + {{ O}_\delta }} \right]\frac{{{E_i}}}{{{E_i} - QT}}P\,.
\end{align*}
The last step depends on the operator $O$ being short range so that the last three terms are capped on each end by a short-range operator, either $O$ or $V$.
The sum of the three terms is then short range and can be replaced by the same form of effective theory expansion as before.
The expansion can then be fit to observables in the same way as $H^{\mathrm{eff}}$.
With $H$ and $O$ in hand, the matrix elements for $O^{\mathrm{eff}}$ can be calculated directly for different eigenstates $i$ and $j$.
Then the LECs can be fit in a scheme-independent way, showing a small energy dependence.
The energy dependence is small because most of the energy dependence has been captured in the Green's function for $G_{QT}$.
As was also the case with $H^{\mathrm{eff}}$ this residual energy dependence can be absorbed into the highest order LECs when fitting to observables across a range of energies, yielding an effective operator based on energy independent LECs.

Since the operator may not respect the same symmetries, channels omitted in $H^{\mathrm{eff}}$ such as ${O}_\delta^{\mathrm{SP}}$ which violate parity may have to be included in the expansion.
When the two-body effective operator is applied in an $A$-body context, the two-body $P$ space is reduced by the quanta of the spectators in much the same way as the effective interaction becomes spectator dependent.

The next section, Sec.~\ref{sec:hobetbox},  will show how the effective interaction LECs can be fit directly to the two-nucleon spectrum in a finite volume, as can be produced from LQCD.
The effective operator now has the same expansion form and we expect that much the same process can be used to fit the operator LECs to two nucleon operators evaluated in a finite volume.
An important example where this can be applied is the $\Delta I=2$ hadronic parity violating matrix element in proton-proton scattering.

\subsubsection{HOBET in a Box \label{sec:hobetbox}}
A way to make a deeper and more direct connection between LQCD and the HOBET effective interaction is to fit the LECs directly
to finite volume observables such as the spectrum of two nucleons.
A few key properties of HOBET make this possible.
First, the only place that boundary conditions appear in HOBET is  the Green's functions for $G_{QT}$.
As long as the cut off HO basis and $V_\delta$  are shorter range than the volume, then the boundary conditions have been segregated into the Green's function, with the result that
$V_\delta$ and the LECs controlling the expansion are independent of the volume, finite or infinite.
Instead of using the \luscher{} method to convert the finite volume LQCD spectrum to infinite volume phase shifts at the energies specified by the quantization condition, one can tune one set of HOBET LECs across one or more volumes of the same physical sizes as are used in the LQCD computations, until the FV spectrum is reproduced.
These LECs can then be used to predict the infinite volume physics, be it bound state energies or the scattering phase shifts.
Note, this also alleviates the need to perform an effective range expansion fit, or some other parameterization of the phase shift results, which can be problematic if there are deeply bound states.

Second, a 3D harmonic oscillator can be described equivalently in either spherical or Cartesian form.
Spherical-Cartesian brackets give a unitary transformation between sets of states with the same number of quanta.
In a finite volume or box, the effective theory is much easier to write in Cartesian form, where we
can write the expansion for $V_\delta$ in terms of 1D HO lowering operators $a_x$, $a_y$, $a_z$, acting in the $\hat{x}$, $\hat{y}$, and $\hat{z}$ directions with their own LECs.    Example operators with Cartesian LEC names can be found in \tabref{table:CartesianLECs}.
%The numeric fields in the names express the number of lowering operators and directions to the left and right.
%Operators that are transformed into each other via cubic rotations or reflection share the same LEC.
The LEC names begin with c to indicate that they are Cartesian LECs.   The letter d simply separates the digit triples.
To construct the name for an operator the left and right hand digit triples first encode the number of lowering operators to each side in the $\hat{x}$, $\hat{y}$, and $\hat{z}$ directions.
The triples are swapped and the digits for $\hat{x}$, $\hat{y}$, and $\hat{z}$ ordered (in the same way in both triples) so as to make the resulting 6 digit number as large as possible.
This gives all operators that are equivalent under cubic rotations or reflection the same LEC.
%\begin{table}[ht]
\begin{wraptable}{R}{0.55\textwidth}
\caption{\label{table:CartesianLECs}
LECs and Cartesian operators}
\centering
\begin{tabular}{l @{\;\;\;} c}
\hline\hline
LEC & operators \\
\hline
$\textrm{c000d000}$ & $\delta(r)$ \\
$\textrm{c100d100}$ & $\left(a_x^\dagger \delta(r) a_x + a_y^\dagger  \delta(r) a_y + a_z^\dagger  \delta(r) a_z\right)$ \\
$\textrm{c100d010}$ & $\left(a_x^\dagger  \delta(r) a_y + a_x^\dagger  \delta(r) a_z + a_y^\dagger  \delta(r) a_z\right) + \text{h.c.}$ \\
$\textrm{c200d000}$ & $\left(a_x^{\dagger 2} + a_y^{\dagger 2} + a_z^{\dagger 2}\right) \delta(r) + \text{h.c.}$ \\
$\textrm{c110d000}$ & $\left(a_x^\dagger a_y^\dagger + a_x^\dagger a_z^\dagger + a_y^\dagger a_z^\dagger \right) \delta(r) + \text{h.c.}$ \\
$\textrm{c200d200}$ & $\left(a_x^{\dagger 2} \delta(r) a_x^{2} + a_y^{\dagger 2} \delta(r) a_y^{2} +a_z^{\dagger 2} \delta(r) a_z^{2} \right)$ \\
\hline
\end{tabular}
\end{wraptable}
%\end{table}

It is harder to express rotational invariance in the Cartesian form so we do
not bother, generating a larger set of operators and associated LECs.    While
this expansion is written in terms of its own set of LECs it must describe the
same $V_\delta$.     Using  spherical-Cartesian brackets the matrix
elements on non-edge states of $P$ of the Cartesian form can be
transformed to the spherical basis where the two expansions must be equal
at the same order.   $V_\delta^{\mathrm{cart}} =
V_\delta^{\mathrm{sph}}$.    This equality establishes a linear relation
between the spherical LECs and the Cartesian LECs, which can now be
replaced by linear combinations of the spherical LECs:
\begin{equation}
\begin{aligned} \nonumber
\textrm{c000d000} &=    a_{\textrm{LO}}^{\textrm{1S0}}  \,,\\
\textrm{c200d000} &=     a_{\textrm{NLO}}^{\textrm{1S0}} \,,  \\
\textrm{c200d200} &=     a_{\textrm{NNLO22}}^{\textrm{1S0}}   + (2/3) a_{\textrm{NNLO}}^{\textrm{1D2}} \,, \\
\textrm{c200d020} &=      a_{\textrm{NNLO22}}^{\textrm{1S0}}  - (1/3) a_{\textrm{NNLO}}^{\textrm{1D2}}  \,,\\
\textrm{c110d110} &=       2 a_{\textrm{NNLO}}^{\textrm{1D2}}  \,.\\
\end{aligned}
\end{equation}
A nice side effect of the substitution is the automatic imposition of rotational invariance on the Cartesian $V_\delta$.
In the above equation you can see that the last three Cartesian LECs are functions of just two of the spherical LECs.

The periodic boundary conditions are enforced by working in a basis that respects them.   The periodic box has side lengths $L_i$ (coordinates from $-L_i/2$ to $+L_i/2$, and $i = 1\ldots 3$ for direction).   A complete basis of states in the periodic box is a set of sine and cosine waves in each dimension such that $\phi(-L/2)=\phi(L/2)$.   For calculation purposes we impose a cutoff $N$ on the size of the basis, which is enlarged until results converge.    We use normalized odd and even real wave functions.
\begin{align}
\phi_{i,s,m}(x) &= \sqrt {2/L_i}  \sin(\alpha_{i,m} x),
	& n &= 1, \ldots ,N/2
\nonumber\\
\phi_{i,c,0}(x) &= \sqrt {2/L_i} (1/\sqrt{2}),
	& n&=0 \,,
\nonumber\\
\phi_{i,c,m}(x) &= \sqrt {2/L_i}  \cos(\alpha_{i,m} x),
	& n&= 1, \ldots ,N/2 \,,
\end{align}
with $\alpha_{i,m_i} = 2\pi \left| m \right| / L_i$.
Letting $m$ range from $-N/2$ to $N/2$, the negative indices indicate the sine members and positive indices to indicate the cosine members of the basis.     We can then write the  basis of 3D solutions as
\begin{equation}
	{\phi _{\mathbf m}}\left( {x,y,z} \right) = \phi_{m_x}\left( x \right) \phi _{m_y}\left( y \right)\phi _{m_z}\left( z 	\right) \,.
\end{equation}
Overlaps between these momentum states and HO states, specified with label $n$ can be analytically calculated (see Chapter~9 in Ref.~\cite{McElvain2017thesis}):
\begin{equation}
	\chi_{\mathbf{n},\mathbf{m}} = \braket{\mathbf{n}}{\mathbf{m}}\,.
\end{equation}
The kinetic energy operator depends on the possibly different side lengths $L_i$.
For example, if the LQCD calculation involves a boosted system along an axis, with non-zero total momentum, this can be handled in part by reducing the box size in HOBET by the Lorentz contraction factor of the boost.
\begin{equation}
	\hat T{\phi _{\mathbf m}}\left( {x,y,z} \right) = 2{\pi ^2}
	\left( {{\text{ }}\sum\limits_i {\frac{{m_i^2}}{{L_i^2}}} } \right){\phi _{\mathbf m}} = {\lambda _{\mathbf m}}{\phi_{\mathbf m}}\left( {x,y,z} \right)\,.
\end{equation}
The Green's functions in \eqnref{eqn:HLBH} can be computed by expansion over the periodic momentum basis.
\begin{equation}
	E G_{QT} P  =E G_T {\left\{ {P E G_T P} \right\}^{ - 1}}P\,.
\end{equation}
Using a bilinear eigenfunction expansion we can write the Green's function $G_T$ as
\begin{equation} \nonumber
	E {G_T}\left( E;\; \mathbf{r}, \mathbf{r}^{\,\prime} \right) = \sum\limits_{\mathbf m} {\frac{E}{{E - {\lambda _{\mathbf m}} + i\varepsilon }}} {\phi _{\mathbf m}}\left( { \mathbf{r}^{\,\prime}} \right){\phi_{\mathbf m}}\left( {\mathbf{r}} \right) \;.
\end{equation}
Applying this expansion to an HO state involves integrating over $\mathbf{r}^\prime$ and yields a sum over momentum states.
\begin{equation}
E G_T \ket{\mathbf{n}} =  \sum\limits_{\mathbf{m}} { \frac{E}{E-\lambda_{\mathbf{m}}+ i\varepsilon} \chi_{\mathbf{m},\mathbf{n} } \ket{\mathbf{m}} } \,.
\end{equation}
Inserting this into matrix elements for $\left\{P E G_T P\right\}$ yields
\begin{equation}
	\braopket{\mathbf{n}^\prime}{E G_T}{\mathbf{n}} = \sum\limits_{\mathbf{m}}{
	\frac{E}{E-\lambda_{\mathbf{m}}+ i\varepsilon}
	 \chi_{\mathbf{n}^\prime,\mathbf{m}} \chi_{\mathbf{n},\mathbf{m}}}\;,
\end{equation}
which is a key component in the evaluation of the effective kinetic energy.  See Section~5.2 in Ref.~\cite{McElvain2017thesis}  for a derivation of the final expression below.
\begin{align*}
T^{\mathrm{eff}}_{\mathbf{n}^\prime,\mathbf{n}}=& \braopket{\mathbf{n}^\prime}{E G_{TQ} \left[T + T\frac{Q}{E}T\right] EG_{QT} }{\mathbf{n}} \,, \\
=&E \left(\delta_{\mathbf{n}^\prime,\mathbf{n}} -\braopket{\mathbf{n}^\prime}{E G_T}{\mathbf{n}}^{-1}\right)\, .
\end{align*}
Other pieces of \eqnref{eqn:HLBH}  can also be evaluated via the same process.    The most expensive calculation is obtaining the matrix elements of $V^{LR}$, which involves a double sum over the momentum basis.

%%%%%%%%%%%%%%%%%%%%%%%%%%%%%%
% long range potential figure
\begin{wrapfigure}{R}{0.48\textwidth}
\includegraphics[width=0.48\textwidth]{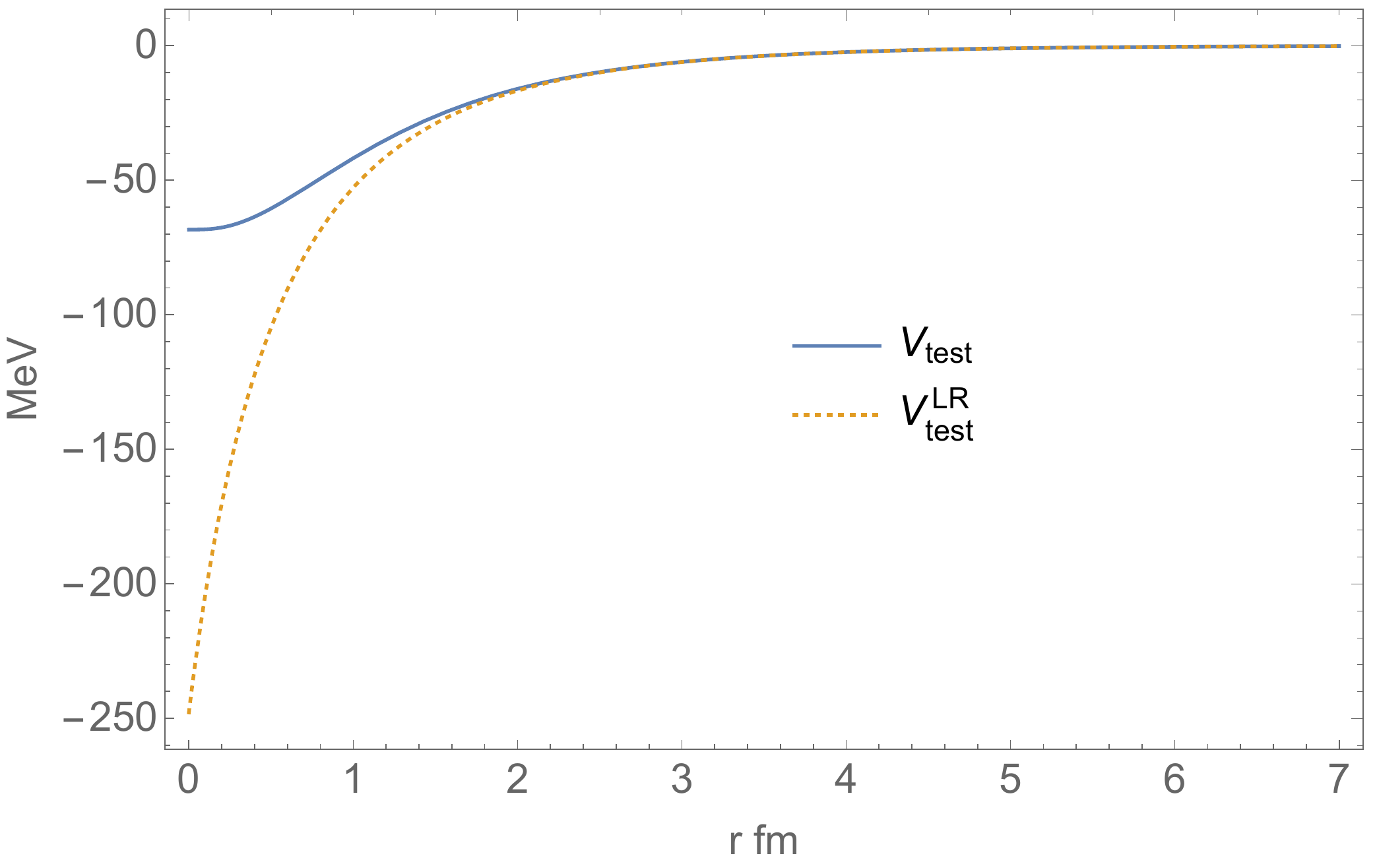}
\caption{The full $V_{\textrm{test}}$ (solid, blue) which reaches HOBET purely through the FV spectrum in \tabref{table:HobetBoxTestStates}, and the long range part $V^{\textrm{LR}}_{\textrm{test}}$ (dashed, gold) takes the role of the OPEP in earlier sections. }
 \label{fig:HobetBoxTestPot}
\end{wrapfigure}
%%%%%%%%%%%%%%%%%%%%%%%%%%%%%%
As before, LECs are fit using \eqnref{eqn:HobetFit}.    Some care must be taken to ensure that when fitting a channel such as $^1S_0$ that the finite volume states used are from cubic representations overlapping the angular momentum of the channel.

Since a sufficient data set from LQCD is not yet available we generate a set of eigenstates of a known potential $V_{\mathrm{test}}$ in a finite volume.    Then the method can be tested by first fitting the LECs of the finite volume $H^{\mathrm{eff}}$ to reproduce the eigenvalues.   Then the LECs are used to construct infinite volume $H^{\mathrm{eff}}\left(E_j\right)$ for a set of energies $\left\{E_j\right\}$, solving for phase shifts that result in energy self consistency.      Given that the finite volume spectrum was generated from a known potential, the phase shifts can then be compared to phase shifts derived directly from $V_{\mathrm{test}}$.   A sample $V_\mathrm{test}$ is given below and shown in \figref{fig:HobetBoxTestPot}.   This potential has a long range OPEP behavior:
\begin{align}
	V_{\mathrm{test}}^{\mathrm{LR}}\left(r\right) &= -130 \, \frac{\exp\left(-m_\pi r \right) }{m_\pi \left(r + 0.75\;\mathrm{fm}\right)}\;\MeV \,, \\
	V_{\mathrm{test}}\left(r\right) &= V_{\mathrm{test}}^{\mathrm{LR}}\left(r\right) + 179.4 \exp\left( -4 	m_\pi r \right)\MeV .
\end{align}

A $\Lambda=500$, $b=1.7$ fm calculation in infinite volume yields a bound state at $-4.0518 \MeV$, which we expect the HOBET interaction to predict.   The spectrum in an $m_\pi L = 10$ box with $L=14.3$ fm and $m_\pi=138.039 \MeV$ was determined on lattices with sizes $350^3$, $400^3$, and $450^3$, enabling a continuum extrapolation of the energies of states shown in \tabref{table:HobetBoxTestStates}.
\begin{wraptable}{R}{0.48\textwidth}
\caption{\label{table:HobetBoxTestStates}
The positive parity spectrum of $H=T+V_\mathrm{test}$ in a periodic volume with $L=14.3\;\mathrm{fm}$ including the overlap with angular momentum states.}
\centering
\begin{tabular}{l r @{\;\;\;} r r r r}
\hline\hline
Rep & MeV & L=0 & L=2 & L=4 & L=6 \\
\hline
$A_1^{+}$ & $-4.4428$ &   $0.5$  & $0$ & $0.866$ & $0$ \\
$A_1^{+}$ & $2.0314$ &  $0.155$ & $0$ &  $0.988$ &$0$ \\
$E^{+}$   & $7.5995$ & $0$ & $0.424$ & $0.361$ & $0.830$ \\
$E^{+}$   & $15.2980$ & $0$ & $0.474$ & $0.393$ & $0.788$ \\
$A_1^{+}$ & $21.6167$ &   $0.326$ & $0$ & $0.265$ & $0.908$ \\
 $E^{+}$   & $23.2423$ & $0$ & $0.468$ & $0.597$ & $0.651$ \\
$A_1^{+}$ & $29.4041$ & $0.521$ & $0$ & $0.853$ & $0.023$ \\
$E^{+}$   & $30.9457$ & $0$ & $0.567$ & $0.428$ & $0.704$ \\
$A_1^{+}$ & $35.2449$ &  $0.655$ & $0$ & $0.189$ & $0.732$ \\
$E^{+}$   & $38.4043$ & $0$ & $0.882$ & $0.176$ & $0.437$ \\
\vspace{1pt} $A_1^{+}$ & $45.1402$ & $0.526$ & $0$ & $0.576$ & $0.625$ \\
\hline
\end{tabular}
\end{wraptable}
For the fit at N$^3$LO in the $S$-channel we choose the six states overlapping $L=0$, yielding the LECs in \tabref{table:HobetBoxN3LOLECs}.
An error estimate from the truncation of the expansion can be made for HOBET results by exploring the natural range of the next order or in this case N$^4$LO LECs using \eqnref{eq:var}.
%%%%%%%%%%%%%%%%%%%%%%%%%%%%%%%%%%%%%%
% TABLE HOBET LECs
\begin{table}[b]
\caption{\label{table:HobetBoxN3LOLECs}
$S$-channel LECs from fit to six states in $L=14.3$ fm box.   The LECs (given in MeV) show the expected order by order magnitude reduction.
}
\centering
\begin{tabular}{c|cccccc}
\hline\hline
LEC& $a_{\textrm{LO}}^S$& $a_{\textrm{NLO}}^S$& $a_{\textrm{NNLO}}^{40,S}$& $a_{\textrm{NNLO}}^{22,S}$& $a_{\textrm{N3LO}}^{60,S}$& $a_{\textrm{N3LO}}^{42,S}$\\
\hline
Value& $70.0274$& $10.2596$& $1.94882$& $3.12887$& $0.13716$& $0.10177$ \\
\hline\hline
\end{tabular}
\end{table}
Based on these LECs, a self consistent infinite volume bound state energy of $-4.066 \MeV$ was found by fixed point iteration,
which should be compared to the $-4.052\MeV$ value found in a large basis, yielding a difference similar in size to the omitted next order LECs.

Phase shifts have been generated directly from the potential, from the HOBET ET, and from a first order and second order L\"uscher's method calculation based on Equation~8 of Ref.~\cite{Luu:2011ep}.  The L\"uscher's formula results are used to fit an effective range expansion, including terms up to $k^6$, to the phase shifts for five positive energies, which are in turn used to generate the phase shifts in \tabref{table:HobetBoxPhaseShifts}.    Column $V$ should be considered the reference value.   The second order L\"uscher results depend on the accurate calculation of $\delta_4$ directly from $V_\mathrm{test}$.   In normal practice the second order  L\"uscher formula only establishes a relationship between $\delta_0(E)$ and $\delta_4(E)$, so one would combine it with a multi-channel generalization of the effective range expansion, fit to all the data, to tease out the two values at each energy.   In addition, the next order formula represents a cutoff on the angular momenta considered.
In the HOBET formulation, no such cutoff is imposed.   A last source of error for the L\"uscher formula is that the tail of the potential does overlap the edge of the box, violating slightly the requirement for a region of free propagation where free propagating waves in Cartesian and spherical bases can be matched.   HOBET does not have this requirement and images of $V^{LR}$ from neighboring periodic iterations are included.

In actual practice, there are other sources of error such as finite volume effects on particle masses that are exponentially suppressed in $L$ and which are common to both the HOBET and \luscher{} approaches, but the relative impact of these error sources and the previous ones have not been evaluated
(the leading exponential corrections to the \luscher{} relation have been derived for the isospin-2 $\pi\pi$ scattering length~\cite{Bedaque:2006yi} and for the NN phase shifts~\cite{Sato:2007ms}, but they have not been compared in detail to LQCD calculations.)   A study of these effects would be of interest.
\begin{wraptable}{R}{0.58\textwidth}
\caption{\label{table:HobetBoxPhaseShifts}
Phase shifts in degrees from the potential $V_{\textrm{test}}$, HOBET, and L\"uscher's method from a $L=14.3\;\text{fm}$ periodic volume.}
\centering
\begin{tabular}{c c c c c}
\hline\hline
             &  From   &   From   & Leading   & Next Order \\
E [MeV] & $V_{\textrm{test}}$ &HOBET & L\"uscher & L\"uscher \\
\hline
$1$ & $142.023$ & $141.931$ & $142.498$ &  $142.269$ \\
$2$ & $128.972$ & $128.860$ & $129.515$ & $129.166$ \\
$4$ & $113.602$ & $113.464$ & $114.159$ & $113.552$ \\
$8$ & $96.919$ & $96.752$ & $97.552$ & $96.330$ \\
$10$ & $91.473$ & $91.296$ & $92.212$ & $90.651$ \\
$15$ & $81.672$ & $81.480$ & $82.849$ & $80.398$ \\
$20$ & $74.876$ & $74.691$ & $76.670$ & $73.303$ \\
\hline
\end{tabular}
\end{wraptable}
Fitting the HOBET LECs directly to the states found in an LQCD volume yields an accurate effective interaction which can be used for more than simply generating phase shifts.   The interaction is also suitable for $A$-body calculations leading to projected wave functions that can be used for operator evaluation.

HOBET effective operators have a similar expansion that can also be related to a Cartesian finite volume form.   This means that the operator LECs can be fit to finite volume LQCD results and the resulting infinite volume operator with the same LECs can then be used in a shell model context.

One exciting prospect is the fitting of a three-body effective interaction following the model of \eqnref{eqn:HobetExpansion}  that can be formulated in terms of lowering operators for two Jacobi coordinate oscillators.      Fitting such an interaction in a finite volume would give a straightforward path to infinite volume three- and higher-body observables, a problem yet to be solved in the L\"uscher formalism.

\subsubsection{HOBET Outlook}

HOBET's effective interaction corresponds to a single, complete set of operators whose LECs can be fit on the UV side to experiment,
and determined on the IR side by chiral symmetry.  These two expansions meet seamlessly at intermediate scales.  HOBET's
philosophy that all short-range physics should be treated as unknown and absorbed into the LECs of low-order operators, significantly simplifies HOBET's
treatment of the pion, which is used only for long-distance corrections associated with high-order operators.  The pionic physics simplifies as
the order of the UV expansion increases, with the pion treatable at tree-level at and beyond N$^3$LO.

HOBET forms a direct bridge from QCD to nuclear structure and observables.    The effective interaction can be directly fit to the finite volume spectrum of nucleons calculated in LQCD.
The LECs are independent of the volume, finite or otherwise, so the interaction matched to LQCD in several finite volumes can be immediately used in infinite volume.

The fact that the effective wave functions are simply projections of the full wave functions gives a straightforward implementation of effective operators.
These effective operators also have a volume independent short-range expansion like the effective interaction and can also be fit to LQCD finite volume observables.

The resulting two-body effective interaction can be expanded into an $A$-body effective Hamiltonian where we expect rapid convergence in the number of interacting nucleons.
Work is underway constructing a HOBET-enabled shell model implementation.    NN scattering spectra from LQCD calculations are eagerly awaited to enable light nuclei calculations firmly anchored in QCD.

%%% 0nuBB
\section{Neutrinoless Double Beta Decay \label{sec:0nubb}}
The search for lepton-number-violating (LNV) neutrinoless double-beta decay ($0\nu\beta\beta$), in which two neutrons are converted into protons with the emission of two electrons and no neutrinos, was originally suggested by Racah~\cite{Racah:1937qq} and Furry~\cite{Furry:1939qr} as a way to test Majorana's proposal that the neutrino, thanks to its lack of electric charge, may be its own antiparticle.
However, this process
has not been observed to this day. In addition to being a second-order weak process, its rate is
proportional to an effective absolute mass scale for the neutrinos. Current observational and
experimental bounds indicate that the masses of active neutrinos have to be smaller than $0.1 - 1.0$~eV~\cite{Tanabashi:2018oca,Aker:2019uuj}, leading to another large suppression of this decay mode
compared to typical nuclear decay energies of a few MeV and momenta of the order of $\sim$ 100 MeV.

Despite these overwhelming challenges, an enormous world-wide effort to detect $0\nu\beta\beta$ is underway \cite{Albert:2017hjq,Martin-Albo:2015rhw,KamLAND-Zen:2016pfg,Shirai:2017jyz,Aalseth:2017btx,Abgrall:2017syy,
Agostini:2013mzu,Albert:2014awa,Alfonso:2015wka,Andringa:2015tza,Alduino:2017ehq,Agostini:2017iyd,Aalseth:2017btx,Albert:2017owj,Azzolini:2018dyb,Agostini:2018tnm},
with the goal of answering fundamental questions on the nature of massive neutrinos that cannot be addressed by oscillation experiments alone.
This experimental program has led to the precise measurement of the lepton-number-conserving two-neutrino double beta ($2\nu\beta\beta$) half-lives of various isotopes \cite{Elliott:1987kp,Argyriades:2009ph,Ackerman:2011gz,Agostini:2012nm,Alduino:2016vtd,Arnold:2016ezh,KamLAND-Zen:2019imh,Arnold:2018tmo,NEMO-3:2019gwo}.
Ranging between $10^{19}$ and $10^{21}$ years, these are the rarest nuclear processes ever observed. In addition, the current generation of experiments has placed
limits on the $0\nu\beta\beta$ half-lives at the level of $10^{25}-10^{26}$ years~\cite{Agostini:2013mzu,Albert:2014awa,KamLAND-Zen:2016pfg,Alduino:2017ehq,Agostini:2017iyd,Aalseth:2017btx,Albert:2017owj,Azzolini:2018dyb,Agostini:2018tnm}.
The observation of $0\nu\beta\beta$ will determine that neutrinos are Majorana particles~\cite{Schechter:1980gr}, and give insight on the absolute scale of neutrino masses and on the high-energy dynamics responsible for their generation.
In addition,  $0\nu\beta\beta$ will shed light on one of the most important open problems in contemporary physics, the origin of the matter-antimatter asymmetry.
The observation of lepton number violation in $0\nu\beta\beta$ would provide support for ``leptogenesis'' scenarios~\cite{Fukugita:1986hr}, in which the lepton asymmetry
induced by LNV and CP-violating decays of very heavy right-handed neutrinos
is converted into a baryon asymmetry by nonperturbative SM processes. A recent discussion of the implications of $0\nu\beta\beta$ experiments on leptogenesis is given in Ref.~\cite{Deppisch:2017ecm}.

Experimentalists utilize special nuclei for which double-beta decay is the only energetically allowed decay mode, e.g., $^{76}$Ge, $^{130}$Te, or $^{136}$Xe. In the U.S., a next-generation ton-scale experiment is currently planned as one of the flagship enterprises outlined in the 2015 Long Range Plan for Nuclear Science~\cite{Geesaman:2015fha}. Several different technologies, materials, and designs are being tested at WIPP and Sanford Laboratory, as well as at other laboratories around the world with large U.S. involvement.  The European $0\nu\beta\beta$ effort is similarly denoted in the Astroparticle Physics European Consortium report~\cite{Giuliani:2019uno}.
Theoretical input is key not only for the planning stages behind these next-generation experiments, but also for understanding different potential mechanisms behind the decay and the interplay with neutrino oscillation experiments,
and for constraining models of BSM physics.

\begin{wrapfigure}{R}{0.5\textwidth}
\includegraphics[width=0.5\textwidth]{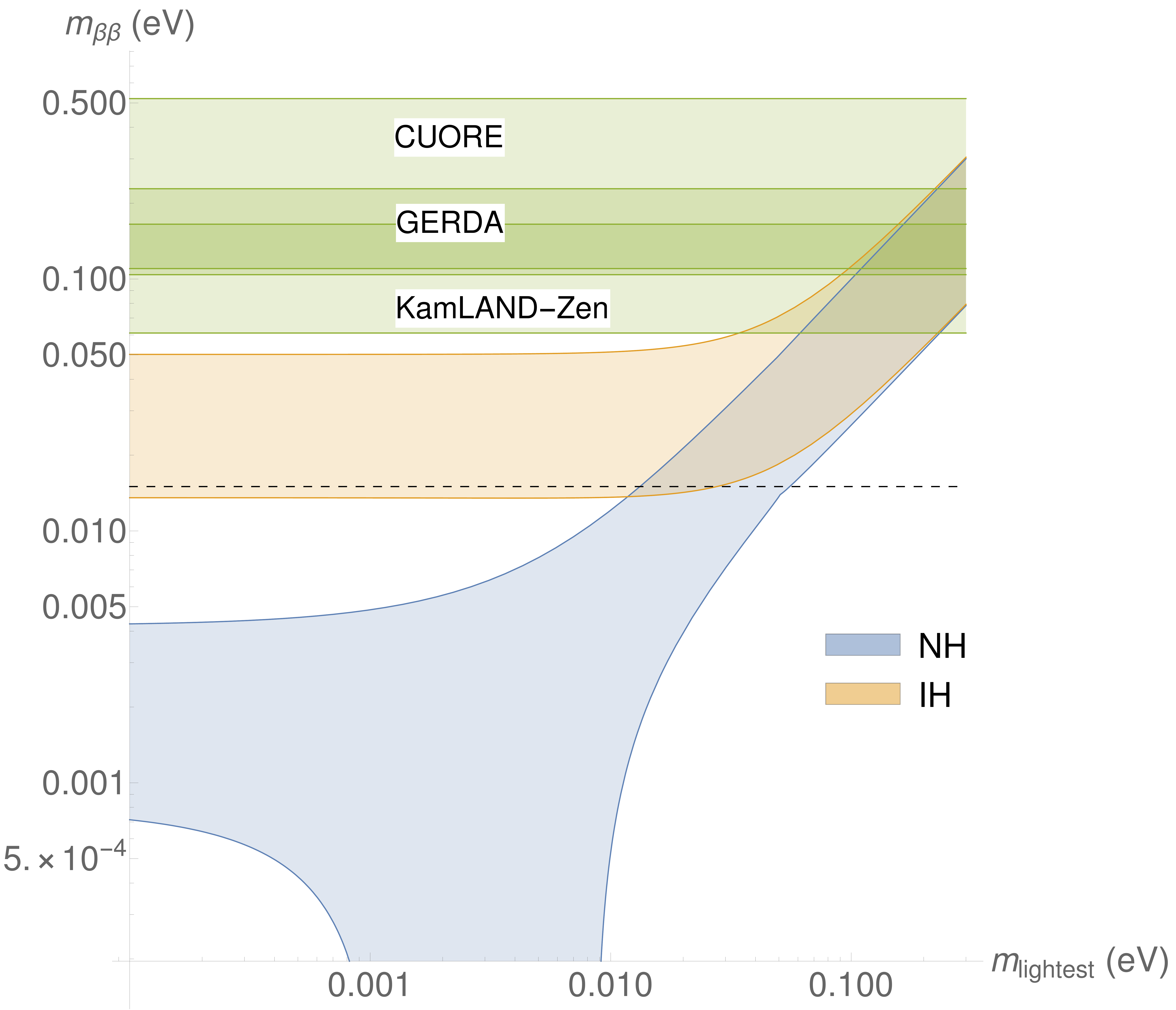}
\caption{Effective neutrino Majorana  mass $m_{\beta\beta}$ as a function of the lightest neutrino mass. The blue and orange regions are the predictions
based on the neutrino oscillation parameters reported in Ref. \cite{Tanabashi:2018oca} in the normal hierarchy (NH) and inverted hierarchy (IH). The bands
include 3$\sigma$ uncertainties on the oscillation parameters, and are obtained by marginalizing over the Majorana phases. The horizontal green bands show the
current $90\%$ confidence level upper limits on $m_{\beta\beta}$ from the KamLAND-Zen $^{136}$Xe experiment \cite{KamLAND-Zen:2016pfg},
the GERDA $^{76}$Ge experiment \cite{Agostini:2018tnm} and the CUORE $^{130}$Te experiment \cite{Alduino:2017ehq}.
$m_{\beta\beta}$ was extracted using the nuclear matrix elements in Refs.~\cite{Menendez:2008jp,Rodriguez:2010mn,Meroni:2012qf,Simkovic:2013qiy,Mustonen:2013zu,Vaquero:2014dna,Yao:2014uta,Neacsu:2014bia,Barea:2015kwa,Hyvarinen:2015bda,Horoi:2015tkc}.
The dashed black line is a projection of the sensitivity of the next generation of $0\nu\beta\beta$ experiments \cite{Geesaman:2015fha}.
}\label{mbbvsml}
\end{wrapfigure}

Figure~\ref{mbbvsml} shows the effective $0\nu\beta\beta$ mass parameter $m_{\beta\beta}$ as a
function of the lightest neutrino for the two possible neutrino mass hierarchies, inverted and normal
(where the pair of neutrinos with the smallest mass splitting are the heaviest and lightest, respectively).
The green bands denote the bounds from existing $0\nu\beta\beta$ experiments.
The next generation of experiment aims at probing $|m_{\beta\beta}| \gtrsim 0.015$~eV, thus covering the full parameter space of the inverted mass hierarchy.
However, Fig.~\ref{mbbvsml} was obtained assuming that $0\nu\beta\beta$ is dominated by the
exchange of light Majorana neutrinos, a minimal scenario realized  in only a fraction of BSM models.
In addition, the calculations of the nuclear matrix elements, even though performed with state-of-the-art many-body methods, are still affected by
unquantified but necessarily large theoretical uncertainties,
which affect our ability to extract microscopic LNV parameters, and thus falsify models of  neutrino mass generation.
In the next two Subsections we discuss the role LQCD and nuclear EFTs can play towards the reduction of these theoretical uncertainties, in both  standard and non-standard $0\nu\beta\beta$ scenarios.

\subsection{Light-Majorana neutrino exchange \label{sec:0nubb_light}}
The most studied scenario for $0\nu\beta\beta$ is the so called ``standard mechanism'', in which LNV is mediated by the Majorana masses of three left-handed neutrinos.
In the SM, a renormalizable Majorana mass term is forbidden by the  $SU(2)_L \times U(1)_Y$ gauge symmetry of the theory. It is however possible to
introduce a dimension-five gauge-invariant operator~\cite{Weinberg:1979sa}, suppressed by one power of the high-energy scale $\Lambda$, which, after electroweak symmetry breaking, gives rise to neutrino masses and mixings.
To explain neutrino masses in the $0.1-1$ eV range, the scale associated with the operator has to be very high, $\Lambda \sim 10^{14}$ GeV. In the next section we will discuss additional scenarios, in which the effective LNV scale is lower, and higher-dimensional operators become competitive.
In the standard mechanism there is a direct relation between the parameters constrained in
oscillation experiments and those responsible for $0\nu\beta\beta$. Indeed, $0\nu\beta\beta$ rates are proportional to
$m_{\beta\beta} = \sum U_{ei}^2 m_i$, where $m_i$ are the masses of the neutrino eigenstates, and $U_{ei}$ elements of the Pontecorvo-Maki-Nakagawa-Sakata (PMNS) mixing matrix \cite{Pontecorvo:1957qd,Maki:1962mu}.
This relation is illustrated by the blue and orange bands in Fig.~\ref{mbbvsml}.
The extraction of $m_{\beta\beta}$, and thus of an absolute neutrino mass scale, from experiment requires precise calculations of the $0\nu\beta\beta$ half-lives, which in turn
are proportional to the matrix elements of the nuclear $0\nu\beta\beta$ transition operator.  The nuclei of interest are medium and heavy open-shell nuclei with complex nuclear structure,
challenging to describe with nuclear theory. As a consequence,  the nuclear matrix elements (NMEs)
are affected by large theoretical uncertainties, giving rise to variations of up to a factor of $\sim$ 2 depending on the many-body method employed (see Ref.~\cite{Engel:2016xgb} for a recent review and Ref.~\cite{Dolinski:2019nrj} that also provides a synopsis of current and future experiments).

The calculation of $0\nu\beta\beta$ NMEs with controlled theoretical uncertainties is one of the main goals of the nuclear physics community and is supported by a DOE Topical Collaboration\footnote{See the website of the DBD Collaboration:\\ \url{https://a51.lbl.gov/~0nubb/webhome/}.}.
A large thrust in this effort has been towards improving existing methods,
by, for example, extending the configuration space and constructing consistent operators for shell-model calculations, or adding correlations to quasiparticle random phase approximation, interacting boson model, and energy density functional calculations. These efforts are reviewed in detail in Ref.~\cite{Engel:2016xgb}.
In addition, the goal of obtaining a direct connection between $0\nu\beta\beta$ rates and microscopic LNV parameters has spurred
a new focus on \abinitio calculations, in which the NMEs are computed taking chiral Hamiltonians fit to two- and few-nucleon data as a starting point for many-body methods
such as coupled-clusters (CC) and IM-SRG~\cite{Yao:2019rck}.
As in the case of  $\beta$ decays discussed earlier, in the \abinitio framework it is important to have a derivation of the transition operator $V_\nu$
that is consistent with the nuclear interaction. At LO, $V_\nu$ is a two-body operator, with a well known long-range component, represented by the exchange of a light Majorana neutrino between nucleons (see for example Ref.~\cite{Haxton:1985am}).
The neutrino interacts with the nucleons via the vector and axial form factors, and modern derivations of $V_\nu$ include the contributions of the induced pseudoscalar
and weak magnetic form factors~\cite{Simkovic:1999re}.  In the $^1S_0$ channel, the most relevant for $0\nu\beta\beta$, $V_\nu$ has a long-range Coulombic component given, at LO, by
\begin{equation}
V^{^1S_0}_{\nu\, L} = \frac{\tau^{(a) +} \tau^{(b) +}}{\vec q^{\,2}} \left(1 + 2 g_A^2 + \frac{g_A^2 m_\pi^2}{(\vec q^{\, 2} + m_\pi^2)^2}\right),
\end{equation}
where $G_F^2$ and $m_{\beta\beta}$ have been factored out~\cite{Cirigliano:2019vdj}.

At higher orders, in addition to the form factors, one has to consider two-body weak currents, which induce
three-body corrections to $V_\nu$~\cite{Menendez:2011qq,Wang:2018htk}.  The three-body $0\nu\beta\beta$ operator involves the LECs $c_{1,3,4,6}$
and $c_D$, which also enter the chiral EFT three-body force, and whose determination was discussed in
Sec.~\ref{sec:nn_eft}.
Ref.~\cite{Cirigliano:2017tvr} furthermore pointed out that in chiral EFT there arise non-factorizable
corrections, that is genuine two-hadron contributions that are not captured by multiplying single-hadron
weak form factors.
These operators are induced by high-energy neutrinos, with $|\mathbf q\,| \gtrsim \Lambda_b$, which
are not resolved in the EFT. They come with three unknown LECs, $g_\nu^{\pi\pi}$, $g_\nu^{\pi \rm N}$,
and $g_{\nu}^{\rm NN}$, which can be determined by computing LNV scattering amplitudes, such as
$\pi^- \rightarrow \pi^+ e^- e^-$
or $n n \rightarrow p p e^- e^-$ with LQCD. Results for the amplitude
 $\pi^- \rightarrow \pi^+ e^- e^-$ and thus for  $g_\nu^{\pi\pi}$
have been recently obtained by two lattice groups~\cite{Feng:2018pdq,Detmold:2018zan,Tuo:2019bue}.
%-------------------------------------------------------------------------------
% FIG: Anu
\begin{wrapfigure}{R}{0.5\textwidth}
\includegraphics[width=0.5\textwidth]{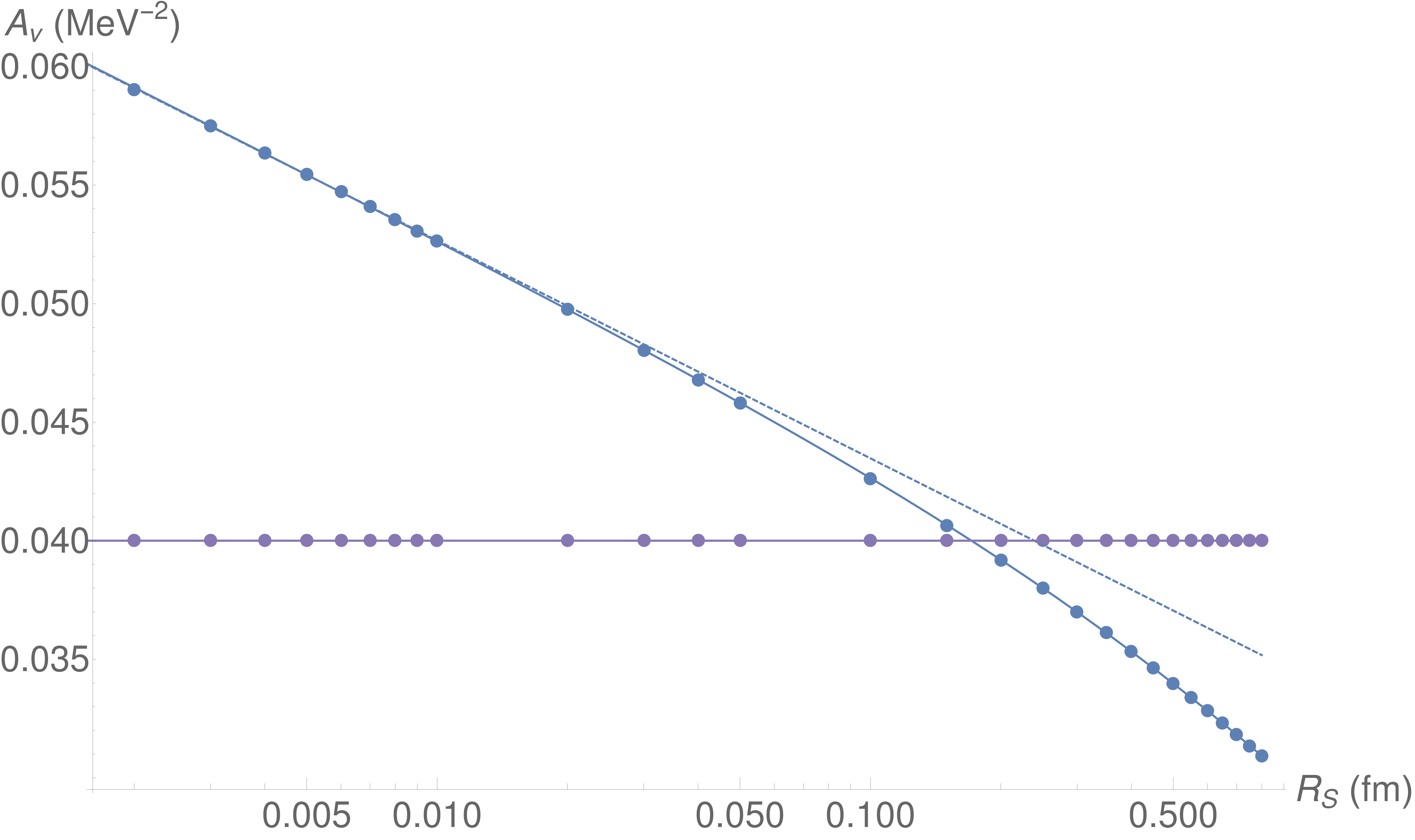}
\caption{\label{Fig:Anu}
Absolute value of the $nn \rightarrow p p e^- e^-$ scattering amplitude $\mathcal A_\nu$,
defined in Eq.~\eqref{Anu}, as a function of a Gaussian cut-off $R_S$. The amplitude is evaluated at the
kinematic point $|\mathbf p |=1 $ MeV, $|\mathbf p^\prime| = 38$ MeV, where $\mathbf p$ and $\mathbf
p^\prime$ are the relative momenta of the $nn$ and $pp$ pair, and the electrons are emitted at zero
momentum.
The blue points denote the matrix elements of the long-range operator $V_{\nu\, L}$, which, for small $R_S$, are well described by $\mathcal A_\nu = a + b \log R_S$ (denoted by the dashed blue line).
The purple points include the short-range counterterm $g_{\nu}^{\rm NN}$, which makes $\mathcal A_\nu$ regulator independent.
}
\end{wrapfigure}
%-------------------------------------------------------------------------------

In WPC $g_\nu^{\pi\pi}$, $g_\nu^{\pi \rm N}$ and $g_{\nu}^{\rm NN}$ contribute at NLO, so their values are important for precision calculations but they would not significantly shift the NMEs.
However, as previously discussed, WPC fails for Coulomb-like potentials in the $^1S_0$ channel, leading to the enhancement of short-range effects.
The problem can be identified by studying the $n n \rightarrow p p e^- e^-$ scattering amplitude \cite{Cirigliano:2018hja,Cirigliano:2019vdj}. At first order in the very weak perturbation
$V^{^1S_0}_{\nu\, L}$, the scattering amplitude is given by
\begin{equation}\label{Anu}
\mathcal A_\nu = -\int d^3 \mathbf r \, \psi^{- \,*}_{\mathbf p^\prime}(\mathbf r) V^{^1S_0}_{\nu\,
L}(\mathbf
r) \psi^{+}_{\mathbf p}(\mathbf r),
\end{equation}
where $\psi^{\pm}_{\mathbf p}(\mathbf r)$ are scattering wavefunctions for two neutrons or two protons
in the $^1S_0$ channel, obtained with
the chiral EFT potential, which, at LO, consists of a contact interaction $C$ that must be regulated (Sec.~\ref{sec:nn_eft}) and a Yukawa one-pion-exchange potential.
In Fig.~\ref{Fig:Anu} we show results obtained with a local Gaussian
regulator, with width $R_S$, but similar results are obtained in dimensional regularization or with a
non-local momentum cutoff \cite{Cirigliano:2018hja,Cirigliano:2019vdj}.
For each value of $R_S$, $C(R_S)$ is fit to the $^1S_0$ $np$ scattering length. Once $C$ and the
wavefunctions are determined, it is straightforward to evaluate Eq.~\eqref{Anu}.
Since $|\mathcal A_\nu|^2$ is an observable (though in practice unmeasurable), the matrix element
should not sensitively depend on the cut-off. However, as shown in Fig.~\ref{Fig:Anu},
$\mathcal A_\nu$ diverges logarithmically as $R_S$ is removed. To absorb the logarithmic divergence in the LO scattering amplitude
it is sufficient to promote the counterterm $g_{\nu}^{\rm NN}$ from NLO to LO, as shown by the purple
dots in Fig.~\ref{Fig:Anu}.
A well defined observable requires $V_\nu$ to have a short-range component appearing at the same order as the usual Coulombic long-range piece~\cite{Cirigliano:2018hja,Cirigliano:2019vdj},
\begin{equation}\label{Vnu}
V^{^1S_0}_\nu %&=& V^{^1S_0}_{\nu\, L } + V^{^1S_0}_{\nu\, S} \\
	=\tau^{(a) +} \tau^{(b) +}  \bigg\{
		\frac{1}{\mathbf q^{\,2}} \left(
			1
			+ 2 g_A^2 + \frac{g_A^2 m_\pi^2}{(\mathbf q^{\,2} + m_\pi^2)^2}
			\right)
%\\
		- 2 g_\nu^{\rm NN}  \bigg\}\, .
\end{equation}
While the analysis was performed on scattering states, the same problem affects
transitions between bounds states, as long as the nuclei are described in chiral EFT.
The situation for the $0\nu\beta\beta$ operator mirrors what is observed in the study of charge-independence breaking (CIB) in nucleon-nucleon scattering.
The charge-dependence of the nuclear force is manifest in the difference between the scattering lengths for $pp$, $np$ and $nn$ in the $^1S_0$ channel.
In WPC, the CIB potential has a long-range component, mediated by photon exchange, and a pion-range component, induced by the electromagnetic pion mass splitting,
which, as $V_{\nu\, L}$, behave as $1/{\mathbf q^{\, 2}}$ at large $|\mathbf q\,|$.
Both in pionless EFT \cite{Kong:1998sx,Kong:1999sf} and in chiral EFT \cite{Cirigliano:2019vdj}, the Coulombic nature of the CIB potential induces logarithmic divergences and,
in order to obtain regulator-independent phase-shifts and to reproduce the observed scattering lengths, it is necessary to introduce a CIB short-range interaction at LO in the EM coupling $e^2$, enhanced by two orders with respect to WPC.
We stress that the importance of short-range CIB effects is not peculiar to chiral EFT,
but appears also in high-quality phenomenological potentials such as Argonne $v_{18}$
and CD-Bonn
\cite{Wiringa:1994wb,Machleidt:2000ge}.

While the analysis of the renormalization of LNV scattering amplitudes
and the analogy between CIB and LNV, which can be made formal by using the approximate chiral and isospin invariance of the QCD Lagrangian \cite{Cirigliano:2018hja,Cirigliano:2019vdj},
provide strong indication of the need of a LO short-range component of $V_\nu$,
they are not sufficient to determine the finite piece of $g_\nu^{\rm NN}$ and thus its numerical impact on $0\nu\beta\beta$ NMEs.
This is a perfect opportunity to highlight the symbiotic relationship between LQCD and EFT.
This $0\nu\beta\beta$ $nn\rightarrow pp$ amplitude can be computed with LQCD in the exascale era and analyzed with EFT which will both determine the strength of the short range contribution, and thus the overall strength of the NMEs, as well as establish which power counting scheme describes the physics.
It will also be interesting to match the LQCD calculations to HOBET, with an extension of the finite volume formulation (Sec.~\ref{sec:hobetbox}) that includes matrix elements of external currents, and see if such a short range operator is relevant in this ET.

Such calculations are computationally and theoretically challenging due to the need for the insertion of
two vector and axial current operators, leading to a so-called four-point function calculation.
Computationally, this requires the calculation of propagators from all points on the lattice and ending on
all points (all-to-all propagators), to ensure proper momentum projection at all times. These volume by
volume computational operations become daunting when coupled with the nucleon signal-to-noise
problem discussed in Sec~\ref{sec:lattice}. In Refs.~\cite{Tiburzi:2017iux,Shanahan:2017bgi} an LQCD
calculation was performed for $2\nu\beta\beta$, in which the effects of background axial fields were
folded into modified propagators. This method avoids the calculation of expensive all-to-all propagators,
but currently does not have a clear extension to $0\nu\beta\beta$. Interestingly, this calculation found
that the effects of two-nucleon contact operators were enhanced in these processes relative to
expectations from power-counting. While this calculation was performed with $m_\pi\sim800$~MeV, it
highlights the need for full, non-perturbative LQCD input for reliable double beta decay predictions.
The formalism required to relate these four-point calculations to the infinite volume amplitudes of interest is more complex and has not been completely developed yet.  Work along the lines of Refs.~\cite{Briceno:2015tza,Briceno:2019opb} will provide this connection.
Two promising LQCD calculations \cite{Feng:2018pdq,Tuo:2019bue,Detmold:2018zan} have addressed some of these computational and theoretical issues in the much simpler $\pi^- \rightarrow \pi^+ e^- e^-$ process for $0\nu\beta\beta$, paving the way for future calculations of the full $n n \rightarrow p p e^- e^-$ process.

\subsection{Short-range NMEs}

The standard mechanism, while minimal, is by no means the only possible scenario for $0\nu\beta\beta$.
In well-motivated models of BSM physics, the dimension-five Weinberg operator
\begin{equation}
\mathcal L^{(5)} = \frac{\mathcal C^{(5)}}{\Lambda}\left(\tilde\varphi^{\dagger} L\right)^T \, C  \left(\tilde\varphi^{\dagger} L\right),
\end{equation}
which generates a neutrino Majorana mass when the Higgs field $\varphi$ gets its vacuum expectation value,
is suppressed by additional small couplings,
such that the exchange of new heavy particles are important.
For example, left-right symmetric extensions of the SM lead naturally to the well-known seesaw mechanism, a favored explanation for the relative lightness of the neutrinos compared to all other SM particles~\cite{Mohapatra:1974hk,Senjanovic:1975rk}. In these models, heavy right-handed  neutrinos can also drive $0\nu\beta\beta$, and may even be dominant depending on the values of the masses and mixing parameters. While na\"ively one might expect heavy particle exchange to be suppressed by the large mass in the heavy propagator, light neutrino exchange is proportional to the tiny light neutrino mass, and is therefore also heavily suppressed. In a simple seesaw scenario, both the heavy neutrino propagator and the light neutrino mass scale roughly as $1/M_h$, where $M_h$ is the mass of the heavy partner neutrino; which mechanism dominates depends upon the details of a particular model.

In addition to left-right symmetric models, there is an array of proposed new physics models containing heavy beyond the SM particles that may also drive these short-range processes. As an example, the decay may involve the exchange of heavy charged leptonic superpartners in R-parity violating supersymmetric extensions of the SM~\cite{Mohapatra:1986su,Vergados:1986td,Hirsch:1995ek}. R-parity is imperative for the stability of the lightest superpartner, which is a dark matter candidate. Therefore, if theorists can relate experimental bounds to these supersymmetric models, we may be able to constrain potential R-parity violating parameters \cite{Prezeau:2003xn}, even if $0\nu\beta\beta$ is never observed experimentally.

Mechanisms in which LNV arises at scales $\Lambda \gg v$ can be captured by higher dimensional operators in the SM-EFT. LNV operators have odd dimension~\cite{Kobach:2016ami},
and they have been studied in several papers (see for example Refs. \cite{Babu:2001ex,Prezeau:2003xn,deGouvea:2007qla,Lehman:2014jma,Graesser:2016bpz}). In particular, Ref.~\cite{Lehman:2014jma}
identified all the dimension-seven SM-EFT operators. The full list of dimension-nine $SU(2)_L \times U(1)_Y$-invariant operators is still unknown, but Ref.
\cite{Graesser:2016bpz} constructed the complete set of operators with four-quark and two-leptons, clarifying several issues in the literature.
To calculate $0\nu\beta\beta$ rates, the LNV operators at the EW scale are matched onto $SU(3) \times U(1)_{\rm em}$-invariant operators at the QCD scale \cite{Cirigliano:2017djv,Cirigliano:2018yza},
and then onto LNV interactions between pions, nucleons, electrons and neutrinos \cite{Pas:1999fc,Pas:2000vn,Prezeau:2003xn,Cirigliano:2017djv,Cirigliano:2018yza,Graf:2018ozy}.
Earlier derivations of the $0\nu\beta\beta$ transition operators~\cite{Pas:1999fc,Pas:2000vn}
relied on unwarranted assumptions, such as factorization of four-nucleon matrix elements, which lead to large, uncontrolled errors in the $0\nu\beta\beta$ half-lives.
More modern constructions \cite{Prezeau:2003xn,Cirigliano:2018yza} take into account the constraints from the symmetries of QCD,
but still they require non-perturbative input from LQCD to determine the couplings in the $0\nu\beta\beta$ transition operators, and thus to model-independently connect
various models and experimental detection rates. Lattice QCD can thus make an immediate and quantitative impact on the study on non-standard $0\nu\beta\beta$ scenarios.

The lowest-order chiral Lagrangian induced by SM-EFT LNV operators up to dimension-9 is given in
Ref.~\cite{Cirigliano:2018yza},
which built upon the seminal paper~\cite{Prezeau:2003xn}. After integrating out heavy SM degrees of
freedom, one finds two kinds of corrections. First, there are LNV
charged-current interactions between quark bilinears, electrons and neutrinos. These induce long-range contributions to the transition operator, parameterized
by the nucleon axial, vector, scalar, pseudoscalar and tensor form factors. As discussed in
Sec.~\ref{sec:lattice}, these form factors can and have been calculated in LQCD, with good accuracy.
The one exception is the recoil-order tensor form factor $g_{T}^\prime$, which, playing a small role for non-standard $\beta$ decay searches, has not yet been computed.
As for the standard mechanism, a consistent power counting requires the transition operators induced by most non-standard charged-currents to have a short-range component as well \cite{Cirigliano:2018yza}, to be determined via a dedicated LQCD calculation. Only in the case of pseudoscalar charged-current the LO $0\nu\beta\beta$ transition operator is purely long-range.

\begin{figure}
\begin{center}
\includegraphics[width=0.7\textwidth]{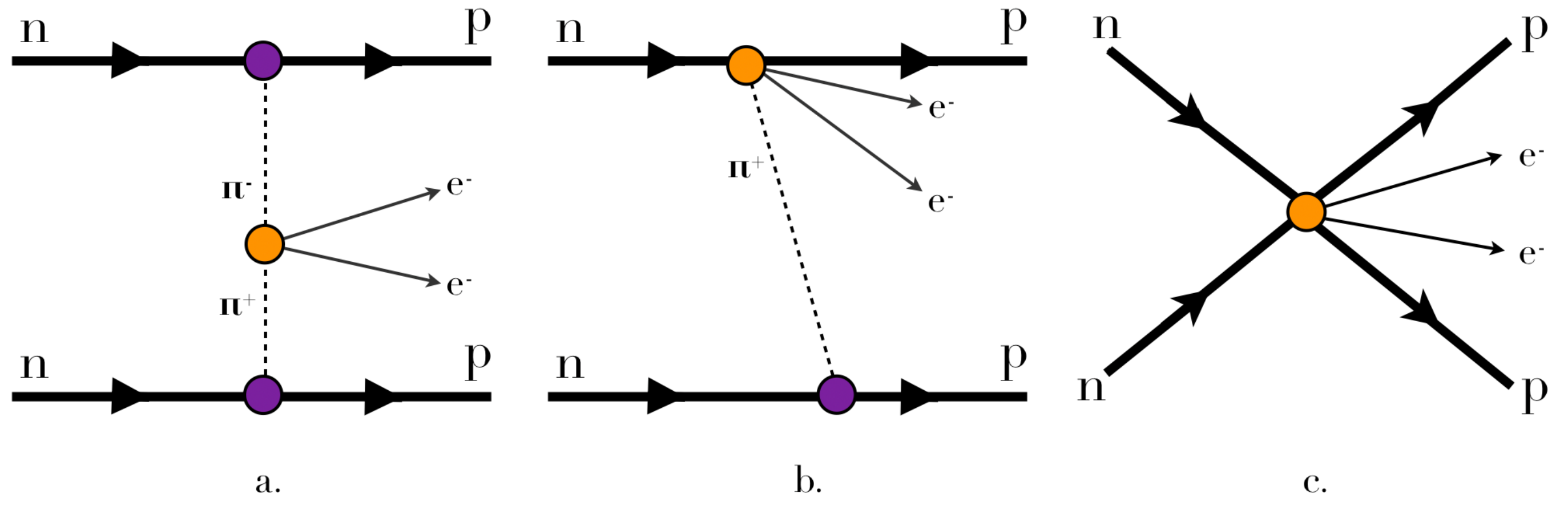}
\caption{\label{fig:0nubbEFT}Diagrams contributing to two-nucleon $0\nu\beta\beta$ at (a) LO, (b) NLO,
and (c) NNLO (from left to right), according to chiral EFT \cite{Prezeau:2003xn}. Yellow dots represent an
insertion of any of the 4-quark operators from Eq.~\eqref{eq:Ops}. Purple dots indicate pion-nucleon
couplings. Note that the NLO diagram will not contribute at this order to the $0^{+} \to 0^{+}$ nuclear
transitions studied by experiment; they will, however, contribute at NNLO with a single derivative at the
vertex.}
\end{center}
\end{figure}

The second kind of correction is represented by local dimension-nine LNV four-quark two-electron operators.
The operators have been tabulated in Refs.~\cite{Prezeau:2003xn,Graesser:2016bpz},
and include ``scalar'' and ``vector'' operators, depending on the Lorentz transformation properties of the electron bilinear.
Focusing on the scalar operators, one has
\beq
\label{eq:Ops}
\mathcal{O}_{1+}^{++} &=&
	\left(\bar{q}_L \tau^+ \gamma^{\mu}q_L\right)
	\left[\bar{q}_R \tau^+\gamma_{\mu} q_R \right]
	\ , \cr
\mathcal{O}_{1+}^{'++} &=&
	\left(\bar{q}_L \tau^+ \gamma^{\mu}q_L\right]
	\left[\bar{q}_R \tau^+\gamma_{\mu} q_R \right)
	\ , \cr
\mathcal{O}_{2+}^{++} &=&
	\left(\bar{q}_R \tau^+ q_L\right)\left[\bar{q}_R \tau^+ q_L \right]
	+ \left(\bar{q}_L \tau^+ q_R\right)\left[\bar{q}_L \tau^+ q_R \right]
	\ , \cr
\mathcal{O}_{2+}^{'++} &=&
	\left(\bar{q}_R \tau^+ q_L\right]\left[\bar{q}_R \tau^+  q_L \right)
	+ \left(\bar{q}_L \tau^+q_R\right]\left[\bar{q}_L \tau^+ q_R \right)
	\ , \cr
\mathcal{O}_{3+}^{++} &=&
	\left(\bar{q}_L \tau^+ \gamma^{\mu}q_L\right)
	\left[\bar{q}_L \tau^+ \gamma_{\mu} q_L \right]
	+ q_L\rightarrow q_R
	\ ,
\eeq
where $(\ldots)$ and $[\ldots]$ denote color contractions and the $\mathcal{O}_{i+}^{++}$ and
$\mathcal{O}_{i+}^{'++}$ mix under renormalization.
The contributions of these operators to two nucleon $0\nu\beta\beta$ have been categorized according
to chiral EFT (see Fig.~\ref{fig:0nubbEFT}), where the LO contributions occur via light pion exchange, and
the N$^2$LO contributions include both a pion-nucleon contact operator,
and a two-nucleon contact operator.

A complete calculation of the LO (in Weinberg power-counting) contribution from these operators, at the physical point with all systematics controlled has been presented in Ref.~\cite{Nicholson:2018mwc}. The lattice QCD calculation determined the matrix elements for the operators within a single pion state undergoing a $\pi^{-} \to \pi^{+}$ transition, which are then inserted into a modified pion potential for the two-nucleon system.
The results from Ref. \cite{Nicholson:2018mwc}  confirm that the pion matrix elements of the operators
$\mathcal O_{1+}^{++}$, $\mathcal O_{1+}^{\prime\, ++}$, $\mathcal O_{2+}^{++}$,
$\mathcal O_{2+}^{\prime\, ++}$ are of  $\mathcal O(F_\pi^2 \Lambda_\chi^2)$, while $\mathcal O_{3+}^{++}$
induces much smaller contributions, $\sim \mathcal O(F_\pi^2 m_\pi^2)$, in agreement with expectations
from NDA and vacuum saturation.
While the relative sizes of the matrix elements are also similar to the vacuum saturation predictions, the LQCD matrix elements
are typically smaller, roughly by a factor of $0.3$ or $0.4$, indicating the importance of QCD dynamics.
These results~\cite{Nicholson:2018mwc} enable a demonstration that the ``factorization'' assumption used in some of the literature, where matrix elements of the operators Eq.~\eqref{eq:Ops} between two neutrons and two protons are reduced to products of nucleon currents,
does not accurately describe the hadronization and can lead to severe underestimates of the transition operator~\cite{Cirigliano:2018yza}.

This pion exchange contribution Ref.~\cite{Nicholson:2018mwc} is LO in WPC.
As noted in the previous subsection, such light-particle exchange effects can become enhanced and also require a short range operator to properly renormalize the amplitude (Fig.~\ref{fig:0nubbEFT}(c)), an effect that may also be important with pion-exchange~\cite{Cirigliano:2018yza}.  Furthermore, in some models, such as left-right symmetric models with no mixing between left- and right-handed W bosons, the single pion LO contribution vanishes exactly and only higher-order contributions are present.
Therefore, it is also important to perform the full LQCD calculation of the $nn\rightarrow ppe^-e^-$ amplitude in order quantify the importance of the various contributions appearing in Fig.~\ref{fig:0nubbEFT} and understand which power counting is more correct.  Such calculations will be possible in the exascale era.

%%% Nuclear EoS
\section{Nuclear-Matter Equation of State \label{sec:neos}}
In this section, we review recent advances in nuclear-matter calculations based
on chiral interactions and highlight potential connections to LQCD. We focus the discussion on the
status of 3N forces as well as how such important contributions can be included in state-of-the-art
calculations up to N$^3$LO (and beyond). These developments have gained importance because
of the dawning of multi-messenger astronomy. This new era began with the first
direct detection of gravitational waves by the LIGO-Virgo collaboration in 2016~\cite{Abbott:2016blz}.
One year later followed the first multi-messenger observation of the binary neutron-star
merger GW170817~\cite{GBM:2017lvd,TheLIGOScientific:2017qsa,Abbott:2018exr}. Multi-messenger
astronomy sheds
light on long-standing problems, such as the synthesis of heavy elements in the
universe through the r-process~\cite{Kasen:2017sxr}.

Neutron-rich matter plays a central role in this setting, covering physics over
an extreme range of densities~\cite{Hebeler:2015hla,Gandolfi:2015jma,Gandolfi:2019zpj}. Constraining
the EOS simultaneously from terrestrial experiments (e.g., at RIBF, FRIB, and FAIR at
GSI), observations, and theory is therefore an active field of research.
Advances on the theory side are of particular interest given novel
observational constraints, such as precise radius measurements from the
Neutron Star Interior Composition Explorer
(NICER)~\cite{Arzo14NICER,Gend16NICER}, or the planned Large Observatory for
X-ray Timing (LOFT)~\cite{Feroci:2011jc,Zane:2014vya} and Enhanced X-ray Timing and
Polarimetry (eXTP) missions~\cite{Watts:2018iom}. LQCD calculations of two- and
few-nucleon systems can bridge chiral EFT and QCD to provide truly QCD-based predictions
for the EOS in the foreseeable future. In particular, matching chiral LECs to
lattice data for which experimental data is either limited or unavailable (see
Sec.~\ref{sec:intro_fs}) will help guide extrapolations of the EOS towards the high densities relevant
for neutron stars. Reducing theoretical uncertainties in the low-density
regime has important consequences for astrophysical applications. For instance, the
neutron-star radii are most sensitive to the EOS at about twice saturation density~\cite{Lattimer:2012xj}.

Homogeneous nuclear matter is an ideal environment for testing nuclear forces
derived in EFTs in medium. Additionally, it provides tight constraints of the EOS
at nuclear densities and the structure of neutron
stars~\cite{Lattimer:2012nd,Watts:2016uzu,Gandolfi:2019zpj}. Neutron and \mbox{(isospin-)}symmetric
matter have been studied with chiral NN and 3N interactions in numerous
many-body frameworks: coupled-cluster
theory~\cite{Ekstrom:2013kea,Baardsen:2013vwa,Hagen:2013yba,Ekstrom:2015rta}, quantum Monte Carlo
methods~\cite{Gandolfi:2011xu,Gezerlis:2013ipa,Gezerlis:2014zia,Lynn:2015jua,Tews:2015ufa},
self-consistent Green's function
method~\cite{Rios:2008fz,Carbone:2013eqa,Carbone:2013rca,Drischler:2016djf,Xu:2019ouo,Carbone:2019pkr,Carbone:2018kji},
 and many-body
perturbation theory (MBPT)~\cite{Hebeler:2010xb,Tews:2012fj,Kruger:2013kua,Holt:2013fwa,Coraggio:2014nva,Wellenhofer:2014hya,Drischler:2015eba,Drischler:2016djf,Holt:2016pjb,
	Wellenhofer:2016lnl,Drischler:2017wtt}  (see, e.g., Ref.~\cite{Hebeler:2015hla} for a
review). On the other hand, only a few direct (and more involved) calculations of isospin-asymmetric matter are
available~\cite{Drischler:2013iza,Drischler:2015eba,Wellenhofer:2016lnl}. This section focuses on recent
advances in MBPT. Neutron matter is well-suited for deriving
tight constraints on the neutron-rich EOS as well as testing different power counting schemes since all
many-body forces up to N$^3$LO are completely determined by the NN forces. The N$^2$LO 3N forces
associated with the 3N LECs $c_{D,\,E}$ vanish due to the coupling of pions to spin and the Pauli
principle,
respectively~\cite{Hebeler:2009iv}. In addition, the term proportional to $c_4$ in the 3N two-pion
exchange does not contribute.

The energy per nucleon $E/A(n,\, \beta)$ at zero temperature is a
function of the total density $n = n_n + n_p$ and isospin-asymmetry $\beta = (n_n - n_p)/n$. Here, $n_n$
($n_p$) denotes the neutron (proton) density. Asymmetric matter can reasonably well be
interpolated between symmetric ($\beta=0$)
and neutron matter ($\beta=1$) using the quadratic expansion in
$\beta$~\cite{Drischler:2013iza,Drischler:2015eba,Wellenhofer:2016lnl},
\begin{equation} \label{eq:quadr_exp}
	 \frac{E}{A} \left(\beta, \, n\right) \approx \frac{E}{A}
	 \left(\beta=0,\, n \right) + \beta^2 \, E_\text{sym}(n) \, ,
\end{equation}
or similar parametrizations, such as the one inspired by energy-density
functionals in Ref.~\cite{Hebeler:2013nza}.  A typical proton fraction in ($\beta$-equilibrated) neutron-star matter is $x = n_p/n \sim 5\%$, corresponding to $\beta = 1 - 2x \sim 0.9$. Omitted higher-order corrections~\cite{Gonzalez-Boquera:2017uep}, such as the quartic term
$\propto \beta^4$, give rise to
non-analyticities~\cite{Kaiser:2015qia,Wellenhofer:2016lnl}.

The nuclear symmetry energy in the expansion~\eqref{eq:quadr_exp},
\begin{equation}
	 E_\text{sym} (n) \approx \frac{E}{A}
	 \left(\beta=1,\, n \right) - \frac{E}{A} \left(\beta=0,\, n \right) \,,
\end{equation}
corresponds to the energy required to convert symmetric matter into neutron matter. It is a key
quantity
for the EOS and the structure
of neutron stars~\cite{Heiselberg:2000dn,Lattimer:2012nd}. For instance, its slope at saturation density,
\begin{equation}
	L = 3 \, n_0 \, \frac{\text{d}E_\text{sym}}{\text{d}n} \bigg|_{n=n_0} \, ,
\end{equation}
is correlated with the neutron-star radius~\cite{Lattimer:2012xj} and with
properties of heavy nuclei measurable in laboratory experiments, such as the
electric-dipole polarizability or the neutron-skin thickness of
\isotope[208]{Pb}~\cite{Roca-Maza:2013mla,Lattimer:2014sga,Roca-Maza:2018ujj}. The Lead (Pb) Radius EXperiment (PREX) measured a remarkably
large neutron-skin thickness of \isotope[208]{Pb}, $R_\text{skin} = 0.33_{-0.18}^{+0.16}\fm$~\cite{Abrahamyan:2012gp}, but the uncertainties are too
large to draw a definite conclusion. If the more accurate PREX-II confirms PREX's central value, the EOS
at nuclear densities is expected to be stiff, whereas the relatively small radius constraints from GW170817
suggest a soft EOS at about twice saturation density. This transition from a stiff
to a soft EOS could indicate a phase transition (e.g., to exotic matter) inside neutron stars~\cite{Fattoyev:2017jql}. As pointed out in
Sec.~\ref{sec:intro_fs}, future LQCD data can provide the constraints necessary to improve EFT
interactions for strange matter in order to address (and eventually solve), e.g., the the long-standing
\emph{hyperon puzzle}.

Predictions of $E_\text{sym}(n_0)-L$ based on the neutron-matter calculations by
Hebeler~\etal~\cite{Hebeler:2010jx,Hebeler:2013nza,Hebeler:2009iv} using MBPT with chiral NN
and 3N interactions, and by Gandolfi~\etal~\cite{Gandolfi:2011xu} using QMC methods
with NN and 3N Hamiltonians adjusted to a range of symmetry energies are in remarkable
agreement with empirical constraints (see Ref.~\cite{Hebeler:2010jx,Hebeler:2013nza,Lattimer:2012xj} for details).
In fact, theory provides the most precise determinations; especially, for
the relatively uncertain $L$ parameter.

Constraining the symmetry energy further from both, theory and experiment is an
important task~\cite{Tsang:2012se,Horowitz:2014bja}. The doubly magic
\isotope[48]{Ca} is, to this end, of particular interest since it can be studied
in \abinitio calculations and
experiments~\cite{Hagen:2015yea,Birkhan:2016qkr,Simonis:2019spj} such as the upcoming Calcium Radius Experiment (CREX)~\cite{Horowitz:2013wha}. On the theory side, this requires, in particular, an improved treatment of 3N forces in many-body calculations~\cite{Drischler:2015eba} and reliable fit values for the two 3N LECs up to N$^3$LO~\cite{Drischler:2017wtt,Hoppe:2019uyw}.

Many-body frameworks for computing the EOS are generally formulated for NN interactions. Implementing 3N (and higher-body) forces directly in these
frameworks is therefore not straightforward. Normal ordering with respect to a given reference state has become the standard approach for including dominant 3N contributions as density-dependent effective two-body potentials in many-body calculations of finite nuclei and nuclear matter (see Refs.~\cite{Bogner:2009bt,Hebeler:2015hla} for details).
In infinite matter, the Hartree-Fock reference state is a natural choice.

Applying Wick's theorem decomposes a general 3N Hamiltonian \emph{exactly} into
its normal-ordered zero-, one-, and two-body contributions and an irreducible
(residual) three-body term~\cite{Bogner:2009bt}. Contributions from the latter can be assumed
relatively small and are thus typically neglected (see also
Refs.~\cite{Hagen:2007ew,Roth:2011vt,Carbone:2014mja}); besides, it would require
implementing 3N forces explicitly in many-body frameworks. At second order, this approximation has been assessed later by direct calculations for a set of chiral 3N
interactions~\cite{Drischler:2017wtt} up to N$^3$LO.

Normal-ordering for 3N forces sums one particle over the (occupied) states in
the Fermi sphere~\cite{Holt:2009ty,Hebeler:2009iv},
\begin{equation} \label{eq:normord_singpart}
	 \overline{V}_{\rm{3N}} = \text{Tr}_{\sigma_3}
	 \text{Tr}_{\tau_3} \int \frac{\text{d} \mathbf{k}_3}{(2 \pi)^3} \,
	 n_{\mathbf{k}_3}^{\tau_3} \, \mathcal{A}_{123} V_{\rm{3N}} \, ,
\end{equation}
which involves the initial antisymmetrized 3N interaction $\mathcal{A}_{123}
V_{\rm{3N}}$, sums over spin and isospin quantum numbers $\sigma_3$ and
$\tau_3$, respectively, and an integration over all momentum states weighted by
the neutron or proton distribution function $n^{\tau_3}_{\bf k}$; e.g., the
Fermi-Dirac distribution function at zero temperature, $n^{\tau_3}_{\bf k} =
\theta(\kf^{\tau_3} - |{\bf k}|)$, with the Fermi momentum $\kf^{\tau_3}$ being related
to the proton and neutron density of the system, respectively.

The effective two-body potential~\eqref{eq:normord_singpart} depends on the total momentum $\mathbf{P}$ of
the two remaining particles, in contrast to Galilean-invariant NN potentials. This additional momentum-dependence makes it conceptually and computationally difficult to combine effective potentials with
genuine NN potentials in many-body calculations.

To elude these difficulties (and to make the derivation actually tractable),
Refs.~\cite{Holt:2009ty,Hebeler:2009iv}  evaluated the effective two-body
potential~\eqref{eq:normord_singpart} semi-analytically by fixing $\mathbf{P} = 0$, an approximation
that was assessed by comparing 3N Hartree-Fock energies. This approximation makes combining the effective potential $\overline{V}_{\rm{3N}}$ with a genuine NN potential straightforward, $V_\text{NN+3N} = V_\text{NN} +
\xi \, \overline{V}_{\rm{3N}}$. The combinatorial normal-ordering factor $\xi$
depends on the specific calculation of interest, and can be derived using Wick's theorem as discussed, e.g., in Refs.~\cite{Hebeler:2009iv,Drischler:2015eba,Drischler:2016cpy}.
The derivations of the effective two-body potential in
Refs.~\cite{Holt:2009ty,Hebeler:2009iv} are based on the operator
definition of the N$^2$LO 3N in neutron and symmetric matter. This approach ties
the derivation to a specific isospin asymmetry and 3N interaction term. Changing
one of them requires a rederivation of the effective potential, which is very
tedious, especially, in view of the operator-rich N$^3$LO 3N forces (see also
Refs.~\cite{Kaiser:2018ige,Kaiser:2019yrc,Kaiser:2019jvc}).

A generalized normal-ordering framework for
asymmetric matter and arbitrary partial-wave decomposed 3N forces has been
introduced along with an improved $\mathbf{P}$ angle-averaging approximation in
Ref.~\cite{Drischler:2015eba}. A comparison of
3N Hartree-Fock energies demonstrated that the new approximation is generally
more stable beyond $n = 0.13 \fmiq$; however, reasonable overall agreement was found up
to about saturation density. Due to the development of an efficient method for the
partial-wave decomposition of chiral 3N forces~\cite{Hebeler:2015wxa},
normal-ordered 3N interactions can now be included in all partial-wave based infinite-matter calculations
up N$^3$LO without any additional efforts. In neutron matter, first
applications
to the EOS~\cite{Drischler:2016djf} and BCS pairing gaps~\cite{Drischler:2016cpy}  followed soon after. This
paves the way for infinite-matter
studies at consecutive orders in the chiral expansion as well as
consistently-evolved many-body forces~\cite{Hebeler:2012pr,Hebeler:2013ri} with controlled approximations. Such order-by-order
calculations are mandatory for truncation-error analyses of the chiral expansion.

MBPT applied to soft(ened) potentials is a computational efficient framework, capable of estimating the
many-body uncertainties through
order-by-order comparisons. Despite these advantages, MBPT for infinite matter at zero
temperature had only been applied up to third order in the many-body expansion, including the involved
particle-hole
diagram~\cite{Coraggio:2014nva,Holt:2016pjb,Kaiser:2017xie}; at finite temperatures,
yet only at second order~\cite{Holt:2013fwa,Wellenhofer:2014hya,Wellenhofer:2015qba}.  Improving the treatment of 3N
forces beyond the Hartree-Fock
approximation~\cite{Kruger:2013kua,Drischler:2016djf} and the rapidly increasing number of
diagrams~\cite{Stev03autgen,Arthuis:2018yoo} towards higher orders, especially the ones with
particle-hole excitations, were among the serious challenges.

The novel Monte Carlo framework for MBPT developed in
Ref.~\cite{Drischler:2017wtt} has overcome these challenges efficiently using
automatic code generation based on the analytic expressions of chiral
interactions and many-body diagrams directly in a single-particle basis.
Lepage's adaptive Monte Carlo algorithm VEGAS~\cite{Lepage:1977sw} allows for the accurate computation of
multi-dimensional momentum integrals. In a first
application~\cite{Drischler:2017wtt}, NN (normal-ordered 3N) interactions up to
fourth (third) order in MBPT were considered in addition to the residual 3N term
at second order, which had only been evaluated for contact
interactions~\cite{Kaiser:2012mm}. These residual 3N contributions turned out
to be small relative to overall uncertainties from the chiral EFT expansion, as
expected. Whether or not that turns out to be the case at third order will be interesting to see.

In a second application, the Monte Carlo framework was used to compute the
complete fourth-order term in the Fermi-momentum (or $\kf a_s$) expansion of the
ground-state energy of a dilute Fermi gas for the first time~\cite{Well18kfas}.
At this order, logarithmic divergences need to be renormalized, which gives rise
to a non-analytic term. Comparisons against quantum Monte-Carlo results showed
that the Fermi-momentum expansion converges well for spin one-half fermions in
the region $|\kf a_s| \lesssim 0.5$. In addition, the range for the Bertsch parameter obtained from two Pad{\'e}
approximants, $\xi_\text{Pad{\'e}} = 0.33 - 0.54$, was found to be consistent with
the value extracted from experiments with cold atomic gases, $\xi \approx 0.376$, as well as the extrapolated value for the normal (i.e., non-superfluid) Bertsch parameter, $\xi_n \approx 0.45$~\cite{Ku563}. Resummation techniques (such as Pad{\'e}
approximants) greatly benefit from
these high-order calculations. Future applications to the nuclear-matter EOS are
therefore promising in order to extract ``converged'' results in calculations of
slow many-body convergence.

Thanks to the automated generator of many-body diagrams recently developed by
Arthuis~\etal \cite{Arthuis:2018yoo}, the Monte Carlo framework can be applied to
push MBPT for infinite matter to (arbitrary) high orders. Preliminary proof-of-principle calculations at fifth (and even sixth)
order based on chiral NN potentials have already demonstrated the efficacy of
the approach. The Monte Carlo framework has achieved a high level of parallelization and performance optimization by taking advantage of GPU acceleration. These advances set the stage for
future applications to the nuclear-matter EOS at finite temperatures and
arbitrary proton fractions. They moreover enable order-by-order studies of the
chiral as well as the many-body convergence up to high orders for Bayesian
uncertainty quantification~\cite{Melendez:2017phj,Melendez:2019izc}. Work along
these lines~\cite{Drischler:2020hwi,Drischler:2020yad} is currently in progress (see also Ref.~\cite{Hu:2019zwa}).

The advances in many-body methods have revealed major challenges
with chiral interactions: none of the presently available interactions is capable of describing both,
experimental ground-state energies and charge radii of nuclei, over a wide range of the nuclear chart~\cite{Hoppe:2019uyw,Ekstrom:2015rta}.
Most chiral NN and 3N interactions that were constrained in the two- and
few-body sector are able to predict light nuclei in agreement with experiment;
however, theoretical uncertainties increase with mass number (see, e.g.,
Ref.~\cite{Carlsson:2015vda}), and significant discrepancies have been observed in
heavy nuclei~\cite{Binder:2013xaa,Carlsson:2015vda,Tichai:2016joa}. A simultaneous
reproduction of experimental ground-state energies and charge radii has yet to be achieved. \Abinitio
calculations of medium-mass
and heavy nuclei empirically indicate that realistic saturation properties of symmetric matter are important~\cite{Hebeler:2010xb,Ekstrom:2015rta,Simonis:2015vja,Hagen:2015yea,Simonis:2017dny,Ekstrom:2017koy}. Optimizing the chiral potential ``NNLOsat'' with
respect to heavier nuclei~\cite{Ekstrom:2015rta} improves,
as expected, the agreement with medium-mass nuclei as well as the empirical
saturation point~\cite{Hagen:2015yea,Ruiz:2016gne,Hagen:2016uwj,Morris:2017vxi}.
Scattering phase shifts, on the other hand, can only be reproduced up to relatively
low energies.

In a first attempt to fit nuclear interactions directly to the empirical saturation point,
Ref.~\cite{Drischler:2017wtt} studied saturation properties of chiral NN and 3N
interactions up to N$^3$LO as a function of $c_D$, which was related to $c_E$ by
matching the \isotope[3]{H} binding energy. Sets of Hamiltonians with reasonable
saturation properties, close to the empirical point but typically slightly
underbound, could be identified for all considered NN interactions;
specifically, the potentials by Entem~\etal~\cite{Entem:2017gor} with momentum
cutoffs $\Lambda = 420,\,450$, and $500 \MeV$.

Subsequent IM-SRG~\cite{Hergert:2016etg,Hergert:2015awm} calculations of selected closed-shell nuclei up to \isotope[68]{Ni} with these potentials found that ground-state energies and charge radii have a significantly weaker dependence on $c_D$ as compared to nuclear-matter calculations~\cite{Hoppe:2019uyw}. Furthermore, charge radii did not follow
the trend expected from the corresponding saturation densities. The $\Delta$-full N$^2$LO potential of Ref.~\cite{Ekstrom:2017koy}, on the other hand, leads to better agreement with these observables, which might indicate similar issues with $SU(2)$HB$\chi$PT($\nd$) as discussed in Sec.~\ref{sec:hbchipt}.
This clearly shows that a
quantitative understanding of the empirical correlation between nuclear
saturation and properties of medium-mass to heavy nuclei needs further
investigation.

The remarkable (phenomenological) success of the NN and 3N interactions fit by
Hebeler~\etal, foremost the softest potential labeled ``1.8/2.0'', in reproducing
ground-state energies of closed-shell nuclei from \isotope[4]{He} to
\isotope[100]{Sn} may provide useful hints in this direction, even though their
charge radii also tend to be too small~\cite{Simonis:2017dny,Morris:2017vxi}. These
SRG-evolved N$^3$LO NN potentials combined with bare N$^2$LO 3N forces exhibit
reasonable saturation properties (see the Coester band in
Ref.~\cite{Drischler:2017wtt}), but were only fit to two- and few-body
observables. Quantifying theoretical uncertainties in calculations with these
potentials is, however, very difficult since NN and 3N forces are treated at
different orders (and not consistently evolved). The use of such inconsistent
potentials has been greatly reduced, though, in favor of full N$^2$LO
calculations, while N$^3$LO 3N potentials are being actively developed
(e.g., by the LENPIC collaboration).
\begin{wrapfigure}{R}{0.45\textwidth}
\begin{center}
    \includegraphics[width=0.42\textwidth]{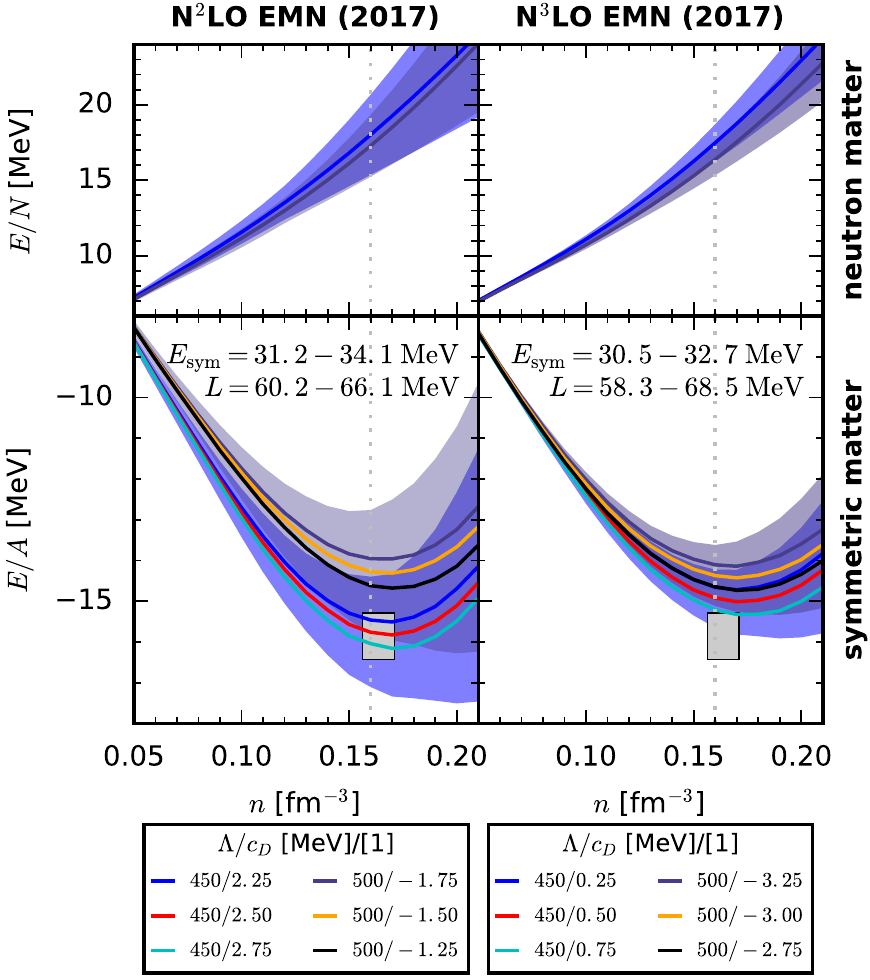}
\end{center}
\caption{\label{fig:eos_new}
Energy per particle in neutron matter (top row) and symmetric nuclear matter (bottom row) based on consistent chiral NN and 3N at N$^2$LO (first column) and N$^3$LO (second column) that were fit to the empirical saturation region (taken from Ref.~\cite{Drischler:2017wtt}). The fitted Hamiltonians are labeled by $\Lambda/c_D$ in the legend. The blue ($\Lambda=500~\MeV$) and gray ($\Lambda=450~\MeV$) bands estimate the theoretical uncertainty following Ref.~\cite{Epelbaum:2014efa}. The gray box denotes the empirical saturation point as obtained from a set of energy-density functionals.}
\end{wrapfigure}

Figure~\ref{fig:eos_new} shows the state-of-the-art MBPT calculations of
Ref.~\cite{Drischler:2017wtt} in neutron matter (top row) and symmetric nuclear
matter (bottom row). The 3N forces were fit to the triton as well as the
empirical saturation point (gray box). For each momentum cutoff, $\Lambda = 450$
and $500\MeV$, three Hamiltonians with different $c_D/c_E$ values were
considered (see legend). These six Hamiltonians collapse to two lines
(corresponding to the two momentum cutoffs) in neutron matter because the
shorter-range leading 3N interactions do not contribute, as already discussed.
The 4N Hartree-Fock energies are small compared to the overall uncertainties,
only $\approx -(150-200) \keV$ for the two cutoffs and isospin asymmetries, and thus
were not included in Fig.~\ref{fig:eos_new}.
The annotated values represent the predictions for the symmetry energy and the
$L$ parameter at saturation density. Both are predicted with remarkable
accuracy.

In contrast to N$^3$LO, the results for the energy per particle of symmetric matter are clearly
separated for the two momentum cutoffs at N$^2$LO, indicating a reduced cutoff
dependence at N$^3$LO. The uncertainty bands estimate the chiral convergence
following the EKM approach introduced in Ref.~\cite{Epelbaum:2014efa}, with $Q =
p/\Lambda_b$, breakdown scale $\Lambda_b = 500\MeV$, and average Fermi momentum
$p = \sqrt{3/5} \, \kf$. Although the bands overlap and get smaller from N$^2$LO
to N$^3$LO, as expected, the uncertainties remain relatively large with several MeV at saturation
density. The dominant uncertainty in these calculations comes from the chiral convergence. A few years
back, the uncertainties in the neutron-matter EOS were dominated by $c_i$ variation in the 3N forces (see
Fig.~10
in Ref.~\cite{Hebeler:2015hla}). This has considerably changed with the precise Roy-Steiner equation
analysis performed by Hoferichter~\etal Comparing these values to LQCD
extractions will be possible in the exascale computing era.

Determining LECs robustly from two- and few-body data is important for model checking
and selection of chiral EFT derived within different power countings.
Nuclear matter allows for this validation to be performed in medium, with important consequences for nuclear-structure applications. Future LQCD
results will provide here guidance as well as insights into the
weakly-constrained isospin $\mathcal{T} = 3/2$ components of the 3N forces~\cite{Wynen:2018zbt},
which are crucial for the neutron-rich matter EOS (as discussed in
Sec.~\ref{sec:intro_fs}).

\begin{wrapfigure}{R}{0.5\textwidth}
	\begin{center}
		\includegraphics[page=1,scale=0.58, clip]{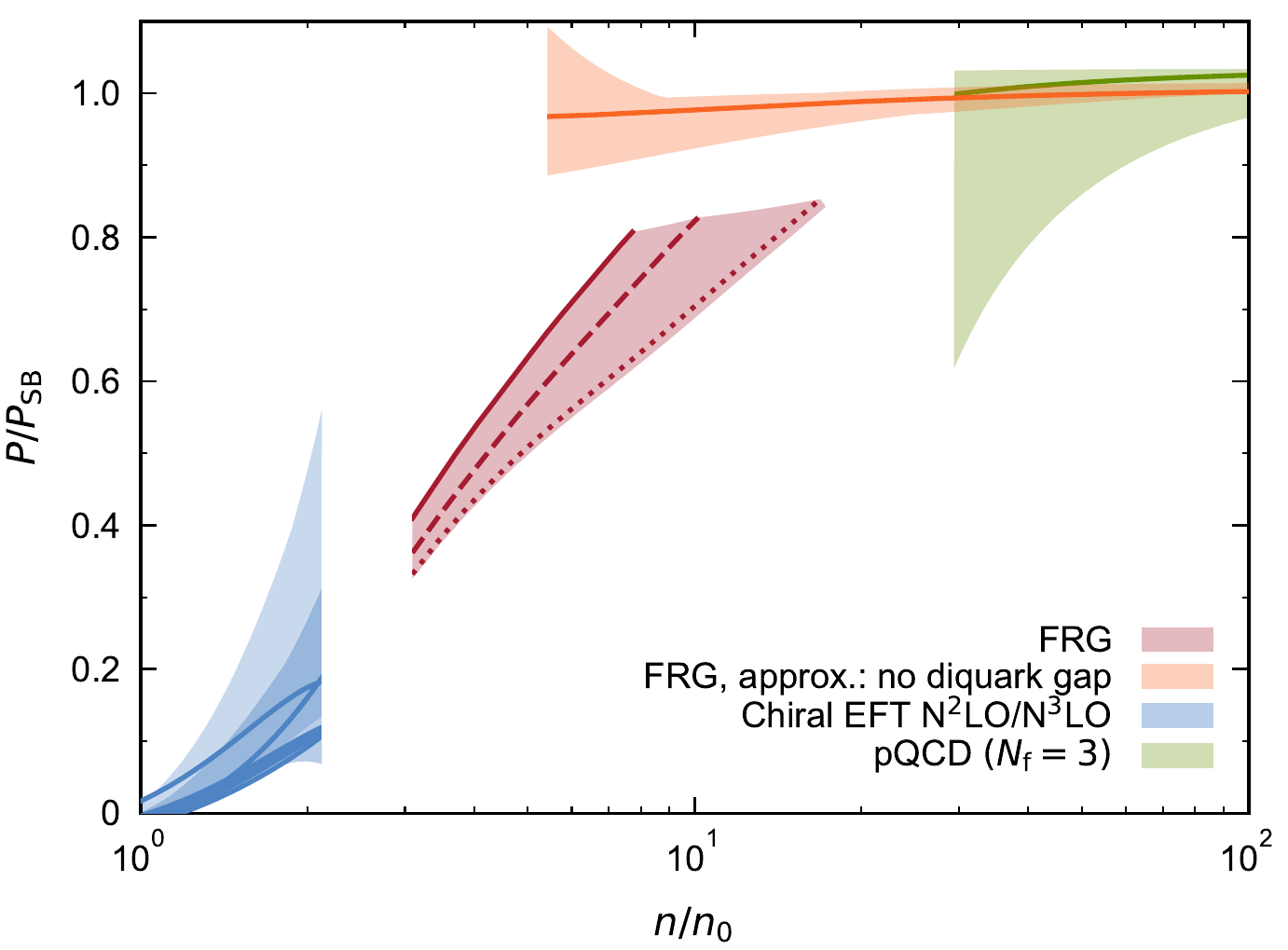}
	\end{center}
	\caption{\label{fig:eos_frg_match}
    Pressure $P$ of symmetric nuclear matter divided by the pressure of the free quark
	gas $P_\text{SB}$ as a function of the normalized density $n/n_0$ as
	obtained from chiral EFT (blue bands), FRG (red band), including results from an approximation
	without
	accounting for a diquark gap (orange band), and pQCD (green band). The blue bands correspond to
	the results shown in Fig.~\ref{fig:eos_new}. This figure was taken from
	Ref.~\cite{Leonhardt:2019fua}.}
\end{wrapfigure}

Astrophysical applications of the nuclear-matter EOS necessitate a wide range of
densities that is usually beyond the reach of nuclear EFTs; for
instance, typical central densities of neutron stars exceed nuclear saturation
density many times. The nuclear-matter EOS thus might undergo phase transitions (e.g., to
exotic matter). Up to which densities chiral EFT converges is \apriori unknown and should be manifested
in the uncertainty bands for the observables. Bayesian methods are powerful in inferring EFT truncation
errors and breakdown scales from data.

Piece-wise polytropic expansions~\cite{Read:2008iy} and speed-of-sound parametrizations of the
EOS~\cite{Tews:2018kmu}, guided by causality and observational constraints, have been studied extensively
to construct the high-density EOS without making assumptions on the composition
of matter. The speed of sound is not necessarily continuous
when using polytropic expansions. The precise mass measurements of two neutron
stars with $\sim 2 \Msol$~\cite{Demorest:2010bx,Antoniadis:2013pzd,Fonseca:2016tux} provide strong constraints
for the extrapolations
since all realistic EOS have to be compatible with these lower maximum-mass limits (see Fig.~11 in
Ref.~\cite{Hebeler:2013nza}).
Recently, Cromartie~\etal raised the lower maximum-mass limit by measuring a
neutron star with $2.17_{-0.10}^{+0.11} \Msol$~\cite{Cromartie:2019kug}. The
uncertainties, however, are still relatively large compared to the other
measurements. On the other hand, precisely measuring neutron-star radii is much
more challenging, e.g., due to the small number of suitable neutron stars~\cite{Watts:2016uzu}. Other
observational constraints come from, e.g., gravitational-wave
detections (tidal deformabilities)~\cite{GBM:2017lvd,TheLIGOScientific:2017qsa,Abbott:2018exr} or
neutron-star radius
measurements. GW170817 has been used to constrain
the radius of a $1.4 \Msol$ neutron star to be $\lesssim 14
\km$~\cite{Tews:2018chv,Most:2018hfd,Lim:2018bkq,Abbott:2018exr}(see also Refs.~\cite{Carson:2018xri,Carson:2019xxz}). The most stringent constraints on neutron-star radii were recently determined, $R_{1.4\Msol} = 11.0_{-0.6}^{+0.9}$~\cite{Capano:2019eae}. A comprehensive Bayesian
analysis of the EOS exploring the impact of future direct radius measurements can be found in Ref.~\cite{Greif:2018njt}.

The Tolman-Oppenheimer-Volkoff (TOV) equations~\cite{Tolman:1939jz,Oppenheimer:1939ne}, two coupled
differential
equations, connect the mass-radius ($M$--$R$) relation of nonrotating neutron stars
to the EOS bidirectionally. That means, the $M$--$R$ relation can be calculated
given an EOS, and vice versa, the EOS mapped given simultaneous $M$--$R$
measurements of neutron stars.
Such $M$--$R$ constraints have been inferred (e.g., in Ref.~\cite{Steiner:2012xt}), while NASA's NICER mission has recently provided its first joint $M$--$R$ measurement of the millisecond pulsar PSR J0030+0451~\cite{Bogd19NICER2, Riley19NICER, Mill19NICER, Raai19NICER}. More data is to be expected in the near future. This will tighten the uncertainties of the high-density
extrapolations considerably. Observations, however, provide only indirect
constraints on the EOS in contrast to microscopic calculations.

Leonhardt~\etal~\cite{Leonhardt:2019fua} used two complimentary microscopic
approaches to study the zero-temperature EOS of symmetric nuclear matter over a
wide range of densities as shown in Fig.~\ref{fig:eos_frg_match}: at $n
\leqslant 2 \, n_0$, based on the Monte Carlo framework applied to a
set of modern chiral NN and 3N
interactions (blue bands), while at $n \geqslant 3 \, n_0$, directly based on
QCD using the FRG (red band), which matches the high-density limit of pQCD
(green band).

Specifically, Fig.~\ref{fig:eos_frg_match} shows the pressure $P$ divided by the
pressure of the free quark gas $P_\text{SB}$ as a function of density in
units of nuclear saturation density, $n/n_0$. The blue uncertainty band estimates
the EFT truncation error using order-by-order calculations with consistent NN
and 3N up to N$^3$LO. Although the two approaches break down individually at $n
\sim (2-3) \, n_0$, the results in the different regimes indicate a
promising consistency that seems to allow smooth interpolations for the construction of the EOS
over a wide range of densities.

A similar comparison in neutron-rich matter is essential for astrophysical
applications. That requires an extension of the FRG framework in
Ref.~\cite{Leonhardt:2019fua} to matter with arbitrary
proton fractions. On the other hand, the temperature dependence of the EOS can
be studied once the Monte Carlo framework has been extended to finite
temperatures. Work along these lines is in progress.

This is an exciting era for nuclear physics and nuclear astrophysics. Once chiral EFT has been
connected
to LQCD, the outlined approach combining EFT and FRG calculations will enable truly QCD-based (i.e., \abinitio) predictions of the
nuclear-matter EOS with reduced theoretical uncertainties, from the low to the high densities inside neutron
stars.
This will enable the construction of the EOS as a function of density, isospin-asymmetry and temperature that can be used in large-scale astrophysical simulations, aiming at understanding the process of nucleosynthesis in the universe.

%%% Outlook
\section{Outlook \label{sec:outlook}}
The exascale computing era will be particularly exciting for the field of
nuclear physics and nuclear astrophysics.  The disruptive growth in
computational power (see Ref.~\cite{Berkowitz:2018gqe} as a lattice QCD example) will allow us to construct a
predictive theory of nuclear structure and reactions, rooted in the Standard
Model of particle physics and organized in a \emph{tower of effective (field)
theories}, built on distinct degrees of freedom, and matched in the regions of overlap.
In turn, this will enable us to make robust predictions of nuclear
processes with fully quantifiable theoretical uncertainties, which can be
confronted with upcoming experimental measurements aimed at searching for BSM
physics with unprecedented sensitivity, as well as understanding the behavior of
nuclear matter under extreme conditions within the era of multi-messenger
astronomy.

High-performance computing has already enabled us to reach milestones, some that were
hard to foresee just a few years ago, such as high-precision lattice QCD calculations of single nucleon observables at the physical point, with progress made toward developing systematic control over multi-nucleon scattering and reactions, and the first \abinitio calculations of medium-mass nuclei based on chiral NN and 3N
interactions.
For example, as shown in Figure~1 of Refs. \cite{Hagen:2015yea,Simonis:2019spj}, the recent
rapid growth of the reach in mass number $A$ of \abinitio nuclear-structure
calculations has been fueled by algorithmic advances and high-performance
computing. This makes it necessary to revisit and improve uncertainties in the
input Hamiltonians derived from (chiral) EFT. As lattice QCD progresses to
deliver two- and three-nucleon observables, we will see an earnest connection
between QCD and EFT/ETs such that a truly \abinitio understanding of nuclear
physics will become a reality in the foreseeable future.

\section*{Acknowledgements}

We are grateful to our collaborators for many stimulating discussions over the
years. We especially thank R.~Brice{\~n}o, V.~Cirigliano, R.J.~Furnstahl, B.~H{\"o}rz,
R.~Machleidt, J.~Men{\'e}ndez, and W.~Weise for their very insightful comments on this article.
C.D. acknowledges support by the Alexander von Humboldt Foundation through a
Feodor-Lynen Fellowship. E.M. acknowledges support from the LDRD program of Los Alamos National Laboratory.
This work is supported in part by the US Department of Energy,
the Office of Science, the Office of Nuclear Physics, and SciDAC under awards
DE\_SC00046548,  DE-AC52-06NA25396 and DE\_AC0205CH11231, under DOE LLNL Contract No. DE-AC52-07NA27344 and under National Science Foundation Award 1630782.

% BIBLIOGRAPHY
% no white space allowed on overleaf between file names
%\clearpage
\biboptions{numbers,sort&compress}
\bibliographystyle{elsarticle-num-names}
\bibliography{master}

\end{document}